\documentclass[final,5p,times,twocolumn,sort&compress]{elsarticle}

\usepackage{graphicx}
\usepackage{subfigure}
\usepackage[usenames]{color}
\usepackage{simplemargins}
\usepackage{amssymb}
\usepackage{amsmath}
\usepackage{amsfonts}
\usepackage{paralist}

\usepackage{multicol}
\usepackage{placeins}
\usepackage{url}
\usepackage{amsthm}

\newcommand{\porb}{\mathbf{\textnormal{Pr}}}
\newcommand{\kmax}{k_{\textnormal{max}}}

\journal{PLOS ONE}

\begin{document}
\begin{frontmatter}

\title{Estimation of global network statistics from incomplete data}
 
 \author{Catherine A. Bliss}
 \ead{catherine.bliss@uvm.edu}
     \author{Christopher M. Danforth}
     \ead{chris.danforth@uvm.edu}
    \author{and Peter Sheridan Dodds}
 \ead{peter.podds@uvm.edu}
 \address{Department of Mathematics and Statistics, Vermont Complex Systems Center,\\ The Computational Story Lab, and the Vermont Advanced Computing Core, 
 University of Vermont, Burlington, VT, 05401} 
\begin{abstract}
\noindent 
Complex networks underlie an enormous variety of social, biological, physical, and virtual systems. A profound complication for the science of complex networks is that in most cases, observing all nodes and all network interactions is impossible. Previous work addressing the impacts of partial network data is surprisingly limited, focuses primarily on missing nodes, and suggests that network statistics derived from subsampled data are not suitable estimators for the same network statistics describing the overall network topology. We generate scaling methods to predict true network statistics, including the degree distribution, from only partial knowledge of nodes, links, or weights. Our methods are transparent and do not assume a known generating process for the network, thus enabling prediction of network statistics for a wide variety of applications. We validate analytical results on four simulated network classes and empirical data sets of various sizes. We perform subsampling experiments by varying proportions of sampled data and demonstrate that our scaling methods can provide very good estimates of true network statistics while acknowledging limits. Lastly, we apply our techniques to a set of rich and evolving large-scale social networks, Twitter reply networks. Based on 100 million tweets, we use our scaling techniques to propose a statistical characterization of the Twitter Interactome from September 2008 to November 2008. Our treatment allows us to find support for Dunbar's hypothesis in detecting an upper threshold for the number of active social contacts that individuals maintain over the course of one week. \end{abstract}
\end{frontmatter}
\section{Introduction}
Data collected for complex networks is often incomplete due to covert interactions, measurement error, or constraints in sampling. Particular individuals may wish to remain hidden, such as members of organized crime, and individuals who are otherwise overt may have some interactions that they wish to remain hidden because those interactions are of a sensitive nature (e.g., romantic ties). In other instances, links may be erroneously inferred from spurious or noisy interactions. Furthermore, extremely large networks necessitate an understanding of how network statistics scale under various sampling regimes~\cite{Leskovec2006sampling, morstatter2013sample}. Explorations of empirically studied networks have largely ignored these biases and consequently, characterizations of the observable (sub)networks have been reported as if they represent the ``true'' network of interest. 

When members of a population are drawn at random, each with equal selection probability, the sample statistic being studied is often a good estimate of the population statistic. Problematically, subsampling networks often induces bias: some individuals or interactions may be more likely to be selected~\cite{Kolaczyk2009}. Consider, for example, a network for which a random selection of links is observed. The collection of observed nodes in such a subnetwork is biased because large degree nodes are more likely to be included in the sample than nodes of small degree.  

The development of techniques to correct sample estimates of population statistics is needed to enable more accurate portrayals of empirically studied large -scale networks and aid in efforts to model dynamics such as cascading failures and complex contagion~\cite{weng2013, hines2009cascading, pahwa2014abruptness, cotilla2012predicting}. 

A central confounding issue is that the errors introduced by biases in sampling may be exacerbated both by particular sampling strategies and by various underlying network topologies of the true network from which the subsamples are chosen~\cite{costenbader2003, Han2005, Stumpf2005, kossinets2006effects, Wiuf2006, Stumpf2008, frantz2009robustness, Martin2006870}. Researchers have explored the effects of sampling by nodes~\cite{Stumpf2008, Han2005, desilve2006, Lakhina2003, Leskovec2006sampling, Lee2006}; sampling by edges or messages~\cite{morstatter2013sample, Leskovec2006sampling, Lee2006}; and graph exploration methods based on random walks, snowball sampling, and respondent driven sampling~\cite{Leskovec2006sampling, frank1994estimating, biernacki1981snowball}.

We organize our paper as follows. First, we outline some of the most common global network statistics. In Section 2, we describe our data and sampling strategies. In Section 3, we describe scaling methods for global network statistics and apply our methods to four classes of simulated networks and six empirical datasets. We provide a summary of all our estimates in Table~\ref{table:compiled_scaling}. In Section 4, we apply our methods to Twitter reply networks as both a case of scientific interest and demonstration of our methods. In Section 5, we discuss the implications of our findings and suggest further areas of research.

\subsection*{Global network statistics}
Real complex networks have come to be characterized by a range of functional network statistics. In this paper, we explore how descriptive measures such as the
\begin{compactitem}
\item the number of nodes, $N$, 
\item the number of edges, $M$, 
\item degree distribution, $P_k$, 
\item the average degree, $k_{\rm avg}$,
\item the max degree, $\kmax$, 
\item clustering coefficient, $C$,~\cite{newman2001clustering}, and 
\item the proportion of nodes in the giant component, $S$, 

\end{compactitem}
scale with respect to missing network data. Based on our observations, we suggest predictor methods for inferring these network statistics from subsampled network data. 

The most important structural feature of a network is the degree distribution, $P_k$, and this has been the focus of much previous work on subsampled networks. The classical Erd{\"o}s-R\'{e}nyi random graph model famously exhibits a Poisson degree distribution, $P_k=\frac{\lambda^k e^{-\lambda}}{k!}$~\cite{erdHos1960}. In contrast to Erd{\"o}s-R\'{e}nyi random networks, preferential attachment growth models describe a random process whereby new nodes attach with greater likelihood to nodes of large degree giving rise to a Power-law or Scale-free degree distribution, $Pr(k) \propto k^{-\gamma}$~\cite{de_Solla_Price30071965, Barabási15101999, simon1955class, yule1925mathematical}. Other distributions, such as lognormals and power-laws with exponential cutoffs may equally characterize the degree distributions of some empirical networks~\cite{Clauset2009}.

Previous work has explored how the degree distribution is distorted when the subnetwork is the induced subgraph on sampled nodes~\cite{Han2005, Lee2006, Stumpf2005, Stumpf2008, stumpf-2005, Lakhina2003, Frank1980_estimation, girvan2013robustness}. Han et al.~\cite{Han2005} investigated the effect of sampling on four types of simulated networks: random graphs with (1) Poisson, (2) Exponential, (3) Power-law, and (4) Truncated normal distributions.  They observed that degree distributions of sampled Erd{\"o}s-R\'{e}nyi random graphs appear to be linear on a log-log plot. Others have also suggested that subnetworks of Erd{\"o}s-R\'{e}nyi random graphs appear ``power-law-like'' and could be mistaken for a scale-free network~\cite{Lakhina2003, Han2005}. Typically, scale-free networks have degree distributions which span several orders of magnitude and thus, subnetworks of Erd{\"o}s-R\'{e}nyi random graphs would not be classified as scale-free networks by most researchers. As warned in~\cite{Clauset2009}, further errors may be incurred when attempting to use linear regression to fit a power-law. 

Stumpf and Wiuf~\cite{stumpf-2005} examined how degree distributions of Erd{\"o}s-R\'{e}nyi random graphs scale when subnetworks are obtained through uniform random sampling on nodes and ``preferential sampling of nodes,'' whereby large degree nodes have a greater probability of being selected. They showed that Erd{\"o}s-R\'{e}nyi random graphs exhibit closure\footnote{An Erd{\"o}s-R\'{e}nyi random graph sampled by nodes is an Erd{\"o}s-R\'{e}nyi random graph.} under subsampling by nodes, but not under preferential sampling of nodes.

Stumpf et al.~\cite{Stumpf2008} suggested that the degree distribution of the subnetwork induced on randomly selecting nodes is independent of the proportion of nodes sampled and that the true degree distribution can only be determined by knowledge of the generating mechanism for the network. Unfortunately, this is often not known or fully understood. 

Several researchers have explored techniques for estimating the true degree distribution from subnetwork data. We first examine the subnetwork degree distribution before examining attempts to solve for the true degree distribution in terms of the subnetwork degree distribution. We consider three cases. First, when links are sampled with probability $q$ and the subnetwork is taken to be the network generated on sampled links, the probability that a node of degree
$i$ in the true network will become a node of degree $k$ in the subnetwork
($k \leq i$) is given by $\porb(k \,|\,i) = \binom{i}{k} q^k (1-q)^{i-k}$. The subnetwork degree distribution can be determined by weighting these probabilities by $P_i$, the probability of node $i$ appearing in the true network~\cite{cohen2000resilience}. The subnetwork degree distribution is then given by
 \begin{align}
\label{eq:true_degree_subsampled}
\tilde{P}_k&= \begin{cases}
  \sum^{\kmax}_{i=k} \binom{i}{k} q^k (1-q)^{i-k} P_i, & \text{ if } k >0\\
  0, & \text{ if } k=0.
\end{cases}
\end{align}
\noindent Next, we consider subnetworks obtained by link failure. In these cases, all nodes are observed, only a proportion ($q$) of links are observed. This cases is nearly identical to Equation~\ref{eq:true_degree_subsampled}, except for the presence of nodes of degree zero.
\begin{align}
  \tilde{P}_k=\sum^{\kmax}_{i=k} \binom{i}{k} q^k (1-q)^{i-k} P_i, \text{ for } k \geq 0.
\end{align}
\noindent Lastly, we consider subnetworks obtained from the induced network on sampled nodes. In this case, the probability of observing a node is $q$. As such, 
\begin{align*}
\mathbf{Pr}\left(v \text{ is observed and } \deg(v) \text{ is } k \right) = q \sum^{\kmax}_{i=k}   \binom{i}{k} q^k (1-q)^{i-k} P_i.
\end{align*}
We note that this is not the observed subnetwork degree distribution because when a subnetwork obtained from the induced network on sampled nodes is observed, the frequencies of nodes of degree $k$ are computed relative to the number of observed nodes. This becomes 
\begin{align*}
\mathbf{Pr}\left( \deg(v) \text{ is } k \, \mid \, v \text{ is observed}\right)&=\frac{\mathbf{Pr}\left(v \text{ observed \& } \deg(v) \text{ is } k\right)}{q}\\
&=\sum^{\kmax}_{i=k}   \binom{i}{k} q^k (1-q)^{i-k} P_i,
\end{align*}
which is normalized. For added clarity, consider a network of $N$ nodes and $M=0$ edges. We observe that $\mathbf{Pr}(v \text{ is observed and } \deg(v)=0)=q\binom{0}{0}q^0(1-q)^0P_0=q$ whereas $\mathbf{Pr}\left( \deg(v)=k \, \mid \, v \text{ is observed}\right)=\frac{q\binom{0}{0}q^0(1-q)^0P_0=q}{q}=1$. The latter agrees with our observation, namely the (observed) network induced on sampled nodes will have all nodes of degree 0 and an observed probability distribution which is simply $P_0=1$. 

Viewing Equation (\ref{eq:true_degree_subsampled}) as a system of $k$ equations, we may derive an expression for the true degree distribution in terms of the observed subnetwork degree distribution. We refer the interested reader to the Appendix for the derivation of this result:

Given a network with degree distribution $P_j$, with sampling fraction $q$, and the subnetwork degree distribution $\tilde{P}_i=\sum^{\kmax}_{j=i} \binom{j}{i} q^i \left(1-q\right)^{j-i} P_j$, we may solve for $P_j$ in terms of the subnetwork degree distribution $\tilde{P}_i$. This yields
\begin{equation}
\hat{P}_{k} = \sum^{\kmax}_{i=k} \frac{(-1)^{i-k} \binom{i}{k} \left(1-q\right)^{i-k}}{q^i} \tilde{P}_i,
\end{equation}
where $\hat{P}_{k}$ represents the predicted degree distribution and nodes of degree 0 are handled appropriately.
\noindent Verification of this result is also presented in the Appendix.\\

Our derivation differs from Frank~\cite{Frank1980_estimation} by a factor of $\frac{1}{q}$,
\begin{equation}\label{eq:franks_prediction}
\hat{P}_k= \sum^{\kmax}_{i=k} \frac{(-1)^{i-k} \binom{i}{k} (1-q)^{i-k}}{q^{i+1}} \tilde{P}_i.
\end{equation}
Equation~\ref{eq:franks_prediction} solves ${P_k}'= q \sum^{\kmax}_{i=k}   \binom{i}{k} q^k (1-q)^{i-k} P_i$, for $P_i$ in terms of ${P_k}'$, however ${P_k}'$ is not the observed degree distribution.  Neither of these derivations, however, are guaranteed to be non-negative~\cite{Kolaczyk2009} and their practicality of use is limited.

Model selection methods provide a different approach by employing maximum likelihood estimates to identify which type of degree distribution characterizes a true network, given only a subnetwork degree distribution~\cite{stumpf2005statistical}. Although these methods are able to discern that some network degree distributions may be better characterized by lognormal or exponential cutoff models instead of power-laws, only models selected \textit{a priori} for testing form the candidate pool of possible distributions. 

In contrast to the model selection technique proposed by Stumpf et al.~\cite{stumpf2005statistical}, we explore a probabilistic approach which utilizes knowledge of the proportion of sampled network data ($q$) and the subnetwork degree distribution. In doing so, we desire an estimation that captures the qualitative nature of the degree distribution without making any assumptions about candidate models. We show that reasonably good estimates of $P_k$ can be achieved with no knowledge of the generating mechanism. With a reasonable estimate of the degree distribution available, we are able to overcome a previously noted obstacle identified by Kolaczyk~\cite{Kolaczyk2009} who notes that predictors for network statistics (sampled by links) have proven more elusive because of the need for knowledge of the true degree distribution~\cite{Kolaczyk2009}. Our method can be used in conjunction with Hortiz-Thompson estimators to reasonably predict network statistics for cases where node selection is not uniform (i.e., subnetworks generated by sampled links or weights).

In the subsequent sections, we summarize this work and show how our method surmounts this obstacle. To our knowledge, scaling techniques for networks generated by sampled interactions (e.g., weighted networks) have not been addressed in the literature and given the interest in large, social networks derived from weighted, directed interactions, we find this analysis timely and relevant. 

\section{Methods}

In this paper, we focus on four sampling regimes: (1) subnetworks induced on randomly selected nodes, (2) subnetworks obtained by random failure of links, (3) subneworks generated by randomly selected links, and (4) weighted subnetworks generated by randomly selecting interactions. Motivated by our work with Twitter reply networks~\cite{Bliss2012} for which we have a very good approximation of the percent of messages which are obtained, we base our work on the assumption that the proportion of missing data is known. This is a critical assumption and one that we acknowledge may not always be satisfied in practice. Efforts to estimate the proportion of missing nodes or links are intriguing, but are beyond the scope of this paper. 

\subsection{Unweighted, undirected networks}
Our data consist of simulated and empirical networks. We generate unweighted, undirected networks with $N=2 \times 10^5$ nodes and average degree $k_{\rm avg}=10$ according to four known topologies: Erd{\"o}s-R\'{e}nyi random graphs with a Poisson degree distribution~\cite{erdHos1960}, Scale-Free random graphs with a power-law degree distribution~\cite{price1976, Barabási15101999}, Small world networks~\cite{watts1998collective}, and Range dependent networks~\cite{grindrod2002}.\footnote{Erd{\"o}s-R\'{e}nyi, Scale-free, Small world, and Range dependent models were constructed with the CONTEST Toolbox for Matlab~\cite{Taylor_Contest}. We note that the small world networks were set to have random rewiring probability $p=0.1$ and preferential attachment networks were set to have $d=5$ new links when they enter the network. Range dependent networks were set to establish a link between nodes $v_i$ and $v_j$ with probability $\alpha \lambda^{\left| j-i \right|-1}$ where we set $\lambda=0.9$ and $\alpha=1$. As noted by~\cite{Taylor_Contest}, this choice of $\alpha$ ensures that nodes $v_i$ and $v_{i+1}$ are adjacent and $\lambda^{\left| j-i \right|-1}$ ensures that short range connections are more probable than long range connections.
} We also examine six well known empirical network datasets: \textit{C. elegans}~\cite{White1986, watts1998collective}, Airlines~\cite{woolley2013}, Karate Club~\cite{zachary1977information}, Dolphins~\cite{Lusseau2003}, Condensed matter~\cite{newman2001clustering}, and Powergrid~\cite{watts1998collective}. 

We sample each of these simulated and empirical networks and examine the subnetwork induced on sampled nodes (Fig.~\ref{fig:missing_nodes_pic}), the subnetwork obtained by failing links (Fig.~\ref{fig:failed_links_pic}), and the subnetwork generated by sampled links (Fig.~\ref{fig:induced_links_pic}). For a given network, 100 simulated subnetworks are obtained for a given sampling strategy and subsampling percentage $q$, as $q$ varies from 5\% to 100\% in increments of 5\%. 

\subsection{Weighted, undirected networks}
We examine the effects of uniformly increasing edge weight (Experiment 1, Cases I-V) as well as the distribution of edge weights (Experiment 2, Cases VI and VII) on the scaling of network statistics (Table~\ref{table:weighted_network_experiments}). 

\begin{table*} [!ht]
\caption[Summary of weighted network experiments]{Summary of weighted network experiments. Note: $w(e_j)$ refers to the weight of edge $e_j$, $s(v_j)$ refers to the strength of node $v_i)$ and $randi\left\{1..9\right\}$ refers to a randomly selected integers between 1 and 9 (inclusive). }
\centering
\begin{tabular}{|l|c|c|c|}
\hline
Case & $k_{\rm avg}$   & $w_{\rm avg}$   &   Distribution of weights\\ \hline
I  & 6 &   1.0 & $w(e_j)=w_{\rm avg}$ (uniform)\\ 
II & 6  &  2.0 & $w(e_j)=w_{\rm avg}$ (uniform)\\ 
III & 6  &  3.0 & $w(e_j)=w_{\rm avg}$ (uniform)\\ 
IV & 6  &  4.0 & $w(e_j)=w_{\rm avg}$ (uniform)\\ 
V & 6  &  5.0 & $w(e_j)=w_{\rm avg}$ (uniform)\\ 
VI & 6  &  5.0 & $s(v_i)=\left\lceil \frac{30}{k}\right\rceil$ (equal effort)\\ 
VII & 6  &  5.0 & $w(e_j)=randi\left\{1..9\right\}$ (randomized)\\ 
\hline\hline                            
\end{tabular}
\label{table:weighted_network_experiments}
\end{table*}

\subsubsection{Experiment 1: Uniform distribution of edge weights}
In this set of experiments, we generate Erd{\"o}s-R\'{e}nyi networks with $N=2000$ nodes and $k_{\rm avg}=6$. We assign each edge to have equal weight, $w$, where $w=1,2,3,4,$ or 5 (corresponding to Cases I-V). We similarly generate Scale-free networks with $N=2000$ nodes and $k_{\rm avg}=6$. We then sample each of the weighted, undirected networks by randomly selecting $q \sum_{e_i \in E(G)} w(e_i)$ interactions and examine the subnetwork generated by links with $w(e_j)>0$ (Fig.~\ref{fig:subsamp_weighted_pic}). This procedure is repeated to generate one hundred simulated networks for each class and varying proportions of sampled interactions ($q$).

\subsubsection{Experiment 2: Non-uniform distribution of edge weights}
In this set of experiments, we explore how the distribution of weights on edges can impact scaling of global network statistics. As in the previous case, we first generate an Erd{\"o}s-R\'{e}nyi network with $N=2000$ and $k_{\rm avg}=6$.  We then add weights to edges in one of two ways. In Case VI, we assume ``equal effort'' in that all nodes will have an equal number of interactions distributed equally among their incident edges. This requirement ensures that all nodes have equal node strength and that effort is equally distributed to each neighbor. More specifically, for node $\deg(v_i)=k$, we set each of the $k$ edges to have weight $\left\lceil \frac{30}{k}\right\rceil$. In Case VII, for each edge we select an integer weight between 1 and 9 from a uniform probability distribution. Certainly, other variants of the weight distribution exist and their analysis may provide additional insight in future studies. 

\subsection{Weighted, directed networks--Twitter reply networks}
Twitter reply networks~\cite{Bliss2012} are weighted, directed networks constructed by establishing a directed edge between two individuals if we have a directed reply from a individual to another during the week under analysis. These networks are derived from over 100 million tweets obtained from the Twitter streaming API service during September 2008 to February 2009.\footnote{We refer the interested reader to~\cite{Bliss2012} for more information. The data for these networks is provided at \url{http://www.uvm.edu/~storylab/share/papers/bliss2014a/}} During this time, we obtained between 25\% to 55\% of all tweets (Table~\ref{table:datawehave}). Using the scaling methods developed in Sections 4.1-4.4, we predict global network statistics for the Twitter interactome during this period of time by viewing in- and out-network statistics separately (e.g., two distinct networks) to account for directionality.

\section{Estimating global network statistics}
\subsection{Sampling by nodes}
\begin{figure}[!htp]
\centering
\subfigure[Sampled nodes]{\includegraphics[width=.20\textwidth]{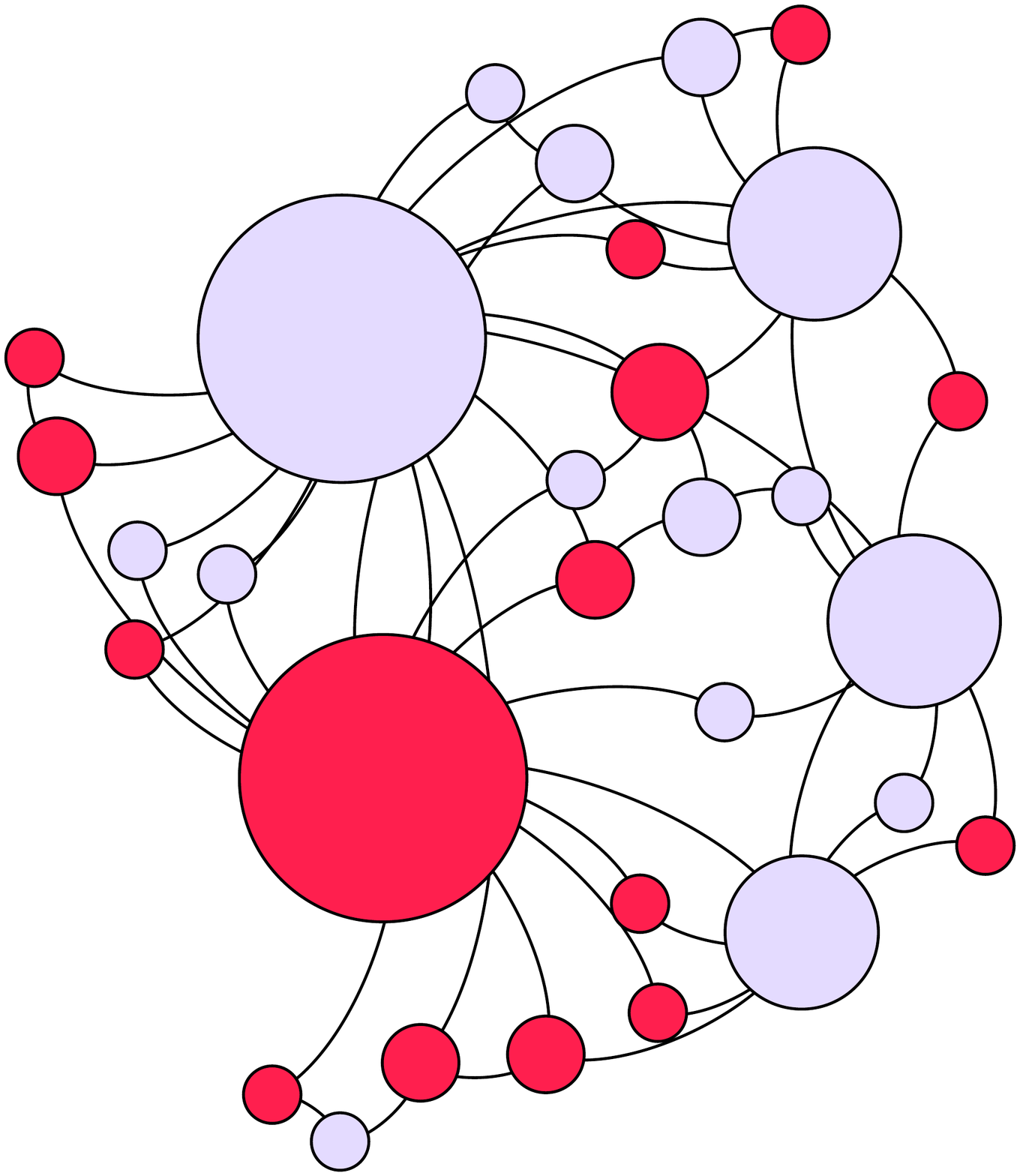}}
\subfigure[Nodes induced subnetwork]{\includegraphics[width=.20\textwidth]{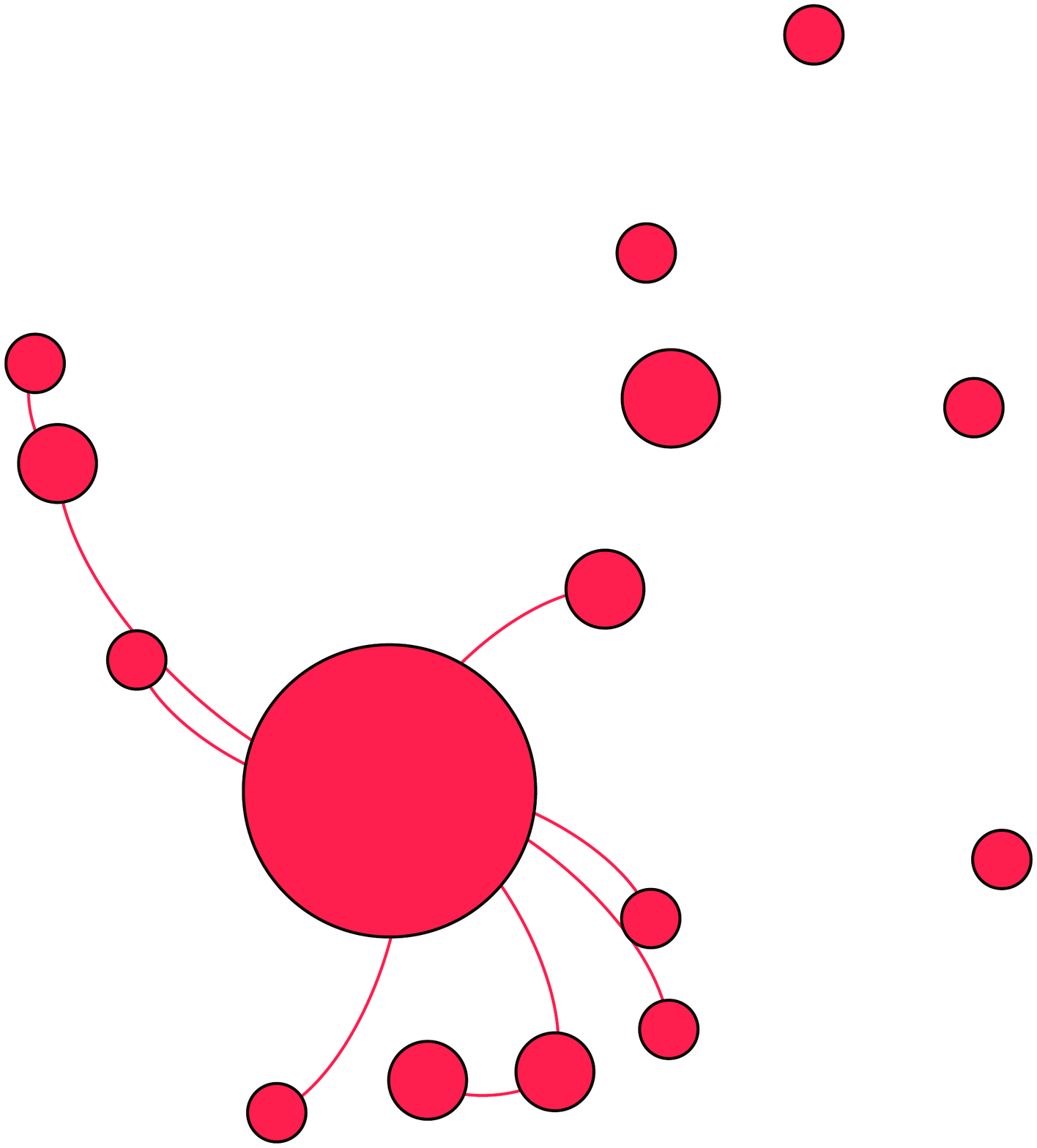}}
\caption[Node induced subnetwork on randomly sampled nodes]{Node induced subnetwork on randomly sampled nodes. (a) The true network is sampled by randomly selecting nodes (red). (b) The node induced subnetwork consists of sampled nodes and edges whose endpoints both lie in the collection of sampled nodes. }
\label{fig:missing_nodes_pic}
\end{figure}

Given a network, $G=(V,E)$, where $V$ is the collection of nodes (or vertices) and $E$ is the collection of links (or edges), we randomly select a portion of nodes $q$, where $0<q\leq 1$. The node induced subgraph on these randomly sampled nodes is given by $G^*=(V^*,E^*)$, where $V^*$ represents the randomly selected nodes and $E^*$ represents the edges in $E$ for whom both endpoints lie in $V^*$ (Fig.~\ref{fig:missing_nodes_pic}). This type of sampling occurs when a selected group, representative of the whole, is observed and all interactions between sampled individuals are known. This sampling strategy is well studied and we will only view key results here (see~\cite{Kolaczyk2009}). 

\subsubsection{Scaling of $N, M, k_{\rm avg}, C, \kmax,$ and $S$}
Given a subnetwork of size $n=qN$ known to be obtained by randomly selecting $qN$ nodes, the number of nodes in the subsample clearly scales linearly with $q$ (see Figs.~\ref{fig:sampling_by_nodes_scaling}a and~\ref{fig:sampling_by_nodes_empirical_scaling}a). The size of the true network is predicted by
\begin{equation} \label{eq:N_nodes}
\hat{N}=\frac{1}{q}n,
\end{equation}
which shows good agreement with the true network statistic (Table~\ref{table:error_N_sample_by_nodes}). Note that this result is independent of network type and is only dependent on $q$, the fraction of nodes subsampled, and $n$, the size of the subsample.

Given a network with $N$ nodes and a subnetwork of $n$ nodes, the probability of selecting edge $e_{ij}$ is given by $\frac{n(n-1)}{N(N-1)}$. This is simply the probability that the two nodes, $v_i$ and $v_j$, incident with the edge $e_{ij}$, are selected. 
The number of edges in the subnetwork is found by
\begin{equation} \label{eq:M_pred}
m=\frac{n(n-1)}{N(N-1)}   M,
\end{equation}
where $m$ represents the number of edges in the subnetwork and $M$ represents the number of edges in the true network. For large networks, $m \approx q^2 M.$ This agrees well with simulated results (Figs.~\ref{fig:sampling_by_nodes_scaling}b and~\ref{fig:sampling_by_nodes_empirical_scaling}b). The predicted number of edges is given by
\begin{equation} \label{eq:M_nodes}
\hat{M}=m   \frac{N(N-1)}{n(n-1)},
\end{equation}
which scales as $\hat{M} \approx \frac{1}{q^2} m$ for large networks. This predictor shows good agreement with actual values (Table~\ref{table:error_M_sample_by_nodes}).

The average degree, $k_{\rm avg}$, is found by
\begin{align*}
k_{\rm avg} &= \frac{2M}{N}.
\end{align*}
Given expressions for the expected number of edges \eqref{eq:M_nodes} and the expected number of nodes \eqref{eq:N_nodes}, the expected average degree of a true network, $\hat{k}_{\rm avg}$, based on an observed average degree of a subnetwork:

\begin{align}
\hat{k}_{\rm avg} &= \frac{2\hat{M}}{\hat{N}}\\
&=\frac{2 m   \frac{N(N-1)}{n(n-1)}}   {\frac{n}{q}}\\
&=\frac{2 m}{n}   \frac{N-1}{n-1}\\
&=k^{\text{obs}}_{\rm avg}   \frac{N-1}{n-1}\\
& \approx \frac{k^{\text{obs}}_{\rm avg}}{q},
\end{align}
where in line (10) we have assumed that $\hat{N} \approx N$, $N \gg 1$ and $n \gg 1$. Comparing this result to simulated subnetworks induced by subsampling nodes (Figs.~\ref{fig:sampling_by_nodes_scaling}c and~\ref{fig:sampling_by_nodes_empirical_scaling}c), we find very good agreement between the predicted average degree and true average degree (Table~\ref{table:error_avk_sample_by_nodes}), except for the small empirical networks (Karate club and Dolphins) sampled with low $q$. In these cases, we violate the assumption that $n \gg 1$ because subsamples of the Karate Club network degenerate to subnetworks of 3 edges or less when $q \leq0.20$. Similarly, subsamples of the Dolphin network degenerate to subnetworks of 3 edges or less when $q 
\leq 0.15$. When the observed number of edges in the subnetwork exceeds $3$, our predicted $\hat{M}$ has an error less than 5\% (Table~\ref{table:error_avk_sample_by_nodes}).

The scaling of the max degree is highly dependent on network type, or more precisely, the relative frequency of high degree nodes. For networks with relatively few large hubs and many small nodes of small degree, $\kmax$ scales linearly with $q$ and $\hat{k}_{\max} \approx \frac{\kmax}{q}$. For networks with many nodes of maximal degree\footnote{An example of this would be a regular lattice. All nodes have the same (and hence maximal) degree. This pathological example is not often seen in practice. Simulated Small world networks begin as a regular lattice with random rewiring probability, $p$. Since our Small world networks have $p=0.1$, our Small world networks exhibit this pathological behavior more so than several empirical Small world networks. We note that this is simply a matter of tuning $p$ and not indicative of all Small world networks.} $\kmax$ scales nonlinearly with $q$ (Figs.~\ref{fig:sampling_by_nodes_scaling}d and~\ref{fig:sampling_by_nodes_empirical_scaling}d). 

This distinction makes predicting the maximum degree more challenging since an accurate predictor ultimately relies on knowledge of the network type - knowledge one usually does not have in an empirical setting. Our proposed technique utilizes $\hat{k}_{\max} \approx \frac{k^{\textnormal{obs}}_{\max}}{q}$, unless our algorithm detects a large number of nodes with degree similar to $\kmax$ and are assured that the subnetwork that has not degenerated to a small network ($n<30$).\footnote{More specifically, if our algorithm detects $n_{\kmax-1}  \kmax-1 > n_{\kmax}   \kmax$, then we use the adjustment Equation~\ref{eq:kmax_predictor}, where $n_{\kmax-1}$ represents the number of nodes of degree $\kmax-1$.} In this case,  
\begin{equation}
\hat{k}_{\max} \approx \frac{k^{\textnormal{obs}}_{\max}}{1-\frac{q}{\theta}}, 
\label{eq:kmax_predictor}
\end{equation}
where $\theta=$the number of nodes with degree greater than 75\% of $\kmax$. 
The rationale for this rough approximation is that the nodes which have high degree ($>$ 75\% of the observed max. degree) may have been nearly equal contenders for losing a neighbor during subsampling. When all nodes have equal degree, the denominator of Equation~\ref{eq:kmax_predictor} tends to $\hat{k}_{\max} \approx k^{\textnormal{obs}}_{\max}$. Table~\ref{table:error_kmax_sample_by_nodes} presents the error for this predictor and demonstrates that our method performs reasonably well for most networks in our data set. To our knowledge, this is the first attempt to characterize how $\kmax$ scales with subsampling and we hope that future work improves upon our estimate.

We measure clustering using Newman's global clustering coefficient~\cite{newman2003clustering}
$C_G=\frac{3 \times \tau_{\Delta}(G)}{\tau^{+}_3(G)},$
where $\tau_{\Delta}(G)$ denote the number of triangles on a graph and  $\tau^{+}_3(G)=\tau_3(G)-3\tau_{\Delta}(G)$, which is the number of vertex triples connected by exactly two edges (as in the notation used by~\cite{Kolaczyk2009}). Since the probability of selecting a node is $q$, both the number of triangles and connected vertex triples scale as $q^3$. Thus, $\hat{\tau}_{\Delta}(G)= \frac{1}{q^3}\tau_{\Delta}(G^*)$ and $\hat{\tau}^{+}_3(G)=\frac{1}{q^3}\tau^{+}_3(G^*)$~\cite{Frank1978}. We then expect 
\begin{equation}
\hat{C}_G \approx C_G^*.
\end{equation} 
This is supported by simulations (Figs.~\ref{fig:sampling_by_nodes_scaling}e and~\ref{fig:sampling_by_nodes_empirical_scaling}e) and small errors in $\hat{C}_G$ (Table~\ref{table:error_cluster_sample_by_nodes}). We note that for small $q$, some subnetworks completely breakdown and no connected triples are present. In these situations, the clustering coefficient can not be computed nor can the true network's clustering coefficient be well predicted. 

We next explore how the size of the giant component scales with the proportion of nodes sampled (Fig.~\ref{fig:sampling_by_nodes_scaling}f and~\ref{fig:sampling_by_nodes_empirical_scaling}f). 
For the Erd{\"o}s-R\'{e}nyi and Scale-free random graphs, the giant component emerges when the subnetwork has $k^{\textnormal{sub}}_{\rm avg}>1$. This occurs when $q k_{\rm avg}>1$ and so for our simulated Erd{\"o}s-R\'{e}nyi and Scale-free networks, this occurs when $q=0.10$ because the true networks have $k_{\rm avg}=10$.  The thresholds for the emergence of the giant component in Small World and Range dependent networks are much higher. This may be due to the relatively large clustering coefficients of these networks. As suggested by Holme et al.~\cite{holme2002attack}, networks with a large clustering coefficient~\cite{watts1998collective} are more vulnerable to random removal of nodes. We observe the same trend with Newman's global clustering coefficient.

In the case of the empirical networks, we find that the giant component emerges for $q$ corresponding to $k^{\textnormal{obs}}_{\rm avg}>1$. \textit{C. elegans}, Airlines, and Condensed Matter networks are more resilient to random removal of nodes in that the giant component persists for small levels of $q$. This is most likely due to their relatively high average degrees, as compared to the other networks (heterogeneity of nodes' degrees in these networks). Heterogeneous networks demonstrate more resilience due to random removal of nodes at high levels of damage~\cite{barrat2008dynamical}. In general, it may be very difficult to predict the exact critical point at which the giant component emerges from subnetwork datasets. 

\subsubsection{Scaling of $P_k$}
The complementary cumulative degree distribution (CCDF) becomes more distorted as smaller proportions of nodes are sampled, as shown in Figure~\ref{fig:sampling_by_nodes_empirical_ccpdf} and given by Equation 1. Subnetworks obtained by the induced graph on sampled nodes will often have $\tilde{P}_0>0$. This occurs when $v_i$ is selected in sampling, but no neighbors of $v_i$ are selected in the sample. 

Our goal is to predict the degree distribution, given only knowledge of the proportion of nodes sampled ($q$) and the subnet degree distribution. We note that the probability that an observed node of degree $k$ came from a node of degree $j \geq k$ in the true network is given by
\begin{align*}
 \mathbf{Pr}(k \, \mid \,j)=\left\{ \begin{array}{l l} \binom{j}{k}q^k (1-q)^{j-k}, &\text{when }  j \geq k\\
0, &\text{when } j<k, \end{array}\right.
\end{align*}
where $q$ is the probability that a node's neighbor was included in the subsample and $1-q$ is the probability that a node's neighbor is not included in the subsample. 

After normalizing, we find $\psi(j) =  \frac{ \mathbf{Pr}(k \,\mid \,j)}{c} $ describes the normalized probability that an observed node of degree $k$ came from a node of degree $j$ in the true network, where $c= \sum^{\infty}_{j=k}  \mathbf{Pr}(k \,\mid  \,j)$. Note that when $|1-q|<1$ this series converges and we find $c= \sum^{\infty}_{j=k}  \mathbf{Pr}(k \, \mid  \,j)=\frac{1}{q}$. Thus,
\begin{align}
\label{eq:my_dist}
\psi(j)=\left\{ \begin{array}{l l} q \binom{j}{k}q^k (1-q)^{j-k}, &\text{when }  j \geq k\\
0, &\text{when } j<k. \end{array}\right.
\end{align}
 
Let $n_k$ represent the number of nodes of degree $k$.  We compute 
\begin{align}
\label{eq:dist_rollback}  
n_k   \psi(k) &= n_k   \left( \frac{\binom{j}{k}q^k (1-q)^{j-k}}{c} \right)\\
&=n_k   \left( q\binom{j}{k}q^k (1-q)^{j-k} \right),
\end{align}
where we use Stirling's approximation to estimate the binomial coefficients for large $j$. We have taken care to include observed nodes of degree zero in this process (e.g., $k=0$ in Equation~\ref{eq:dist_rollback}). 

For networks with nodes of large degree (e.g., hubs), one can further speed up the computation and reduce floating point arithmetic errors by mapping back observed nodes of degree $k$ to the expected value of the distribution obtained in Equation~\ref{eq:my_dist}:
\begin{align}
\label{eq:my_dist_shorter}   
E(j)&=\frac{1}{c}\sum^{\infty}_{j=k} j \binom{j}{k}q^k (1-q)^{j-k}\\
    &=q\frac{1-q+k}{q^2}\\
     &\approx \frac{k}{q}, \text{ for } k \gg 1,
\end{align}
where $c \approx \frac{1}{q}$. In making use of $E(j)\approx \frac{k}{q}$, we perform a separate calculation for nodes of degree zero\footnote{In all cases, we assume a finite network. We limit our calculations to $4  k^{\textnormal{obs}}_{\max}$ as a rough estimate on the upper bound needed for the sum in Equation~\ref{eq:my_dist}.}: $\left\{ n_0   \frac{(1-q)^j}{\sum_j (1-q)^j}\right\}^{4k^{\textnormal{obs}}_{\max}}_{j=1}$. 

Figure~\ref{fig:predicting_by_nodes_combined_ccpdf} reveals the predicted degree distribution for subnets induced on varying levels of randomly selected nodes. To test the goodness of fit for the estimated degree distribution and the true $P_k$, we apply the two sample Kolmogorov-Smirnov test. Figure~\ref{fig:one_over_q_nodes_D_stat_plot} shows the $D$ test statistics for the predicted degree distributions for both estimation methods (Equations~\ref{eq:dist_rollback} and~\ref{eq:my_dist_shorter}), as well as the $D_{\text{crit}}$ computed from $c(\alpha)\sqrt{\frac{n_1+n_2}{n_1 n_2}}$, where $c(0.05)=1.36, n_1=\kmax$ and $n_2=\hat{k}_{\max}$. For most networks, $D \leq D_{\text{crit}}$ for $q\geq 0.3$, suggesting that when at least 30\% of network nodes are sampled, our methods provide an estimated degree distribution which is statistically indistinguishable from the true degree distribution. Although we reject the null hypothesis for the preferential attachment case, for all $q \neq 1$, we wish to point out the potential for bias in the Kolmogorov-Smirnov test with large $n$~\cite{goldstein2004problems}. As shown, $D_{\text{crit}}$ values are quite low and the bias in this test is due to large $n_1$ and $n_2$. The statistical power in this test leads to the detection of statistically significant differences, even when the absolute difference is negligible. Thus, we caution the interpretation of this statistical test and place more interest in the value $D=\max\left| F_{i,\rm{true}}-F_{i,\rm{predicted}}\right|$, where $F_{\rm{true}}$ and $F_{\rm{prediction}}$ represent the true and predicted CDFs.

\subsection{Link failure} 
\begin{figure}[!htp]{
\centering
\subfigure[Failed links]{\includegraphics[width=.25\textwidth]{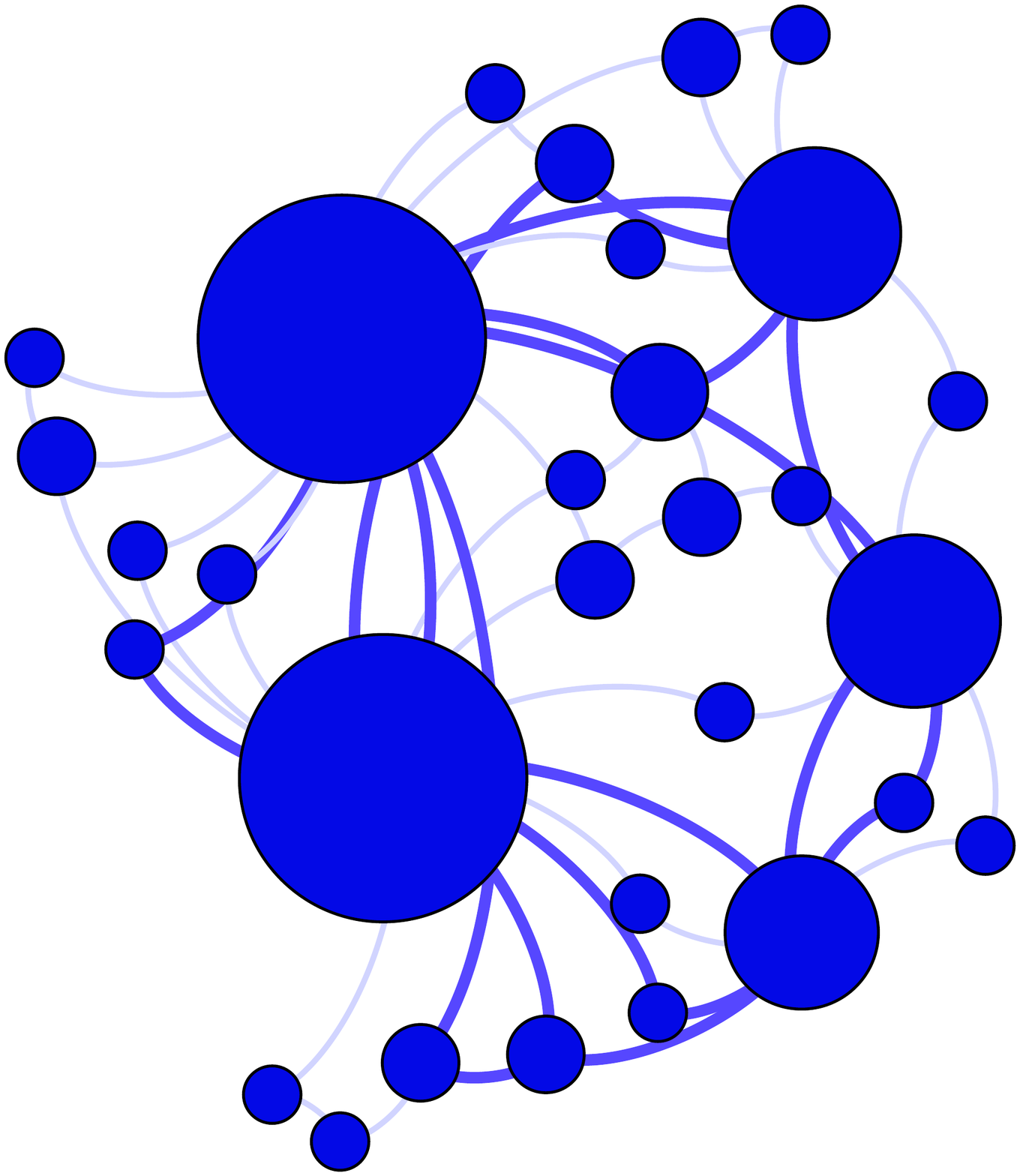}}
\caption[Failed link subnetwork]{Failed link subnetwork. Hidden or missing links are depicted in grey. All nodes remain in the subnetwork and only visible or sampled links remain.}
\label{fig:failed_links_pic}}
\end{figure}

We now turn our attention to link failure. As in the previous cases, we denote the true, unsampled network as $G=(V,E)$. Some proportion, $q$ of links remain ``on'' (or present in the sample) and $1-q$ are hidden or undetected by sampling. $E^* \subseteq E$ consists of precisely the links that remain ``on'' and $V^*=V$ (Fig.~\ref{fig:failed_links_pic}). Figures SI.5-SI.6 demonstrate how network statistics scale in this sampling regime.

In this case we may use the estimator to predict the number of nodes, $\hat{N}=n$ and we may predict the number of edges by $\hat{M}=\frac{m}{q}$. The average degree is found by
\begin{align}
\hat{k}_{\rm avg} &= \frac{2 \hat{M}}{\hat{N}}\\
&=\frac{2 m }      {q n}\\
&= \frac{ k^{\text{obs}}_{\rm avg}}{q}.
\end{align}
Using Newman's global clustering coefficient $C_G=\frac{3 \times \tau_{\Delta}(G)}{\tau^{+}_3(G)}$~\cite{newman2003clustering}, we note that  $q^3 \tau_{\Delta}(G)= \tau_{\Delta}(G^*)$ and $q^2 \tau^{+}_{3}(G)= \tau^{+}_{3}(G^*)$ because each edge is selected with probability $q$. Thus, 
\begin{align*}
C_G^* &=\frac{3 \times \tau_{\Delta}(G^*)}{\tau^{+}_3(G^*)}\\
&=\frac{3q^3 \times \tau_{\Delta}(G)}{q^2\tau^{+}_3(G)}\\
&=q C_{G}. 
\end{align*}
Thus, 
\begin{align}
\label{eq:cluster_failed_links}
\hat{C}_G=\frac{1}{q}C_G^*.
\end{align}

We compute the maximum degree with the same method as described in Section 4.4.1 because the number of neighbors of a node scales the same in both cases. Using these estimates, we find relatively low error in the predicted the network measures for $N, M, k_{\rm avg}, \kmax$, and $C_{G}$ (Tables~\ref{table:error_N_failed_links}--~\ref{table:error_kmax_failed_links}).

Several networks' giant components exhibit similar patterns of resilience when sampling by nodes or failing links. Comparing the resilience of the proportion of nodes in the giant component under sampling by nodes vs. failing links, we see that Erd{\"o}s-R\'{e}nyi random graphs, random graphs with preferential attachment, Airlines, Condensed matter, \textit{C. elegans}, and Powergrid networks all perform relatively similarly under the two sampling regimes. A noticeable difference is seen in Small world, Range dependent, Karate club, and Dolphin networks. In the case of Small world and Range dependent networks, the regularity of the underlying lattice in these networks means that each time a node is not observed, this also means that $k_{\rm avg}$ edges are also missing. Given that the majority of nodes have roughly the same degree for these networks, subsampling fractures the giant component quickly (i.e., for $q$ around 0.7 and 0.8 respectively). In the case of the small Karate club and Dolphins networks sampled by nodes, the proportion of nodes in the giant component increases with decreasing $q$. In these cases, the network consists of relatively few nodes, which are connected. In contrast, when examining the failing links case, we have all nodes present, but these nodes are missing almost all links and the network is highly disconnected. 

Figure~\ref{fig:failed_links_ccpdf} reveals the distortion of the CCDF when links fail in a network and all nodes remain known to the observer. Clearly, there are nodes of degree zero that are observed in this sampling regime. The predicted degree distribution is obtained by the methods described under sampling by nodes (including the treatment of observed nodes of degree zero) and presented in Fig. SI.8. The results of the two sample Kolmogorov-Smirnov test reveal that the estimated degree distribution and the true degree distribution are statistically indistinguishable for $q \geq 0.3$ for most networks (Fig.~\ref{fig:one_over_q_failed_links_D_stat_plot}). As previously noted, the large number of observations in degree distribution for the random graph grown with preferential attachment leads to high statistical power and a low $D_{\text{crit}}$.

\subsection{Sampling by links}
\begin{figure}[!htp]
\centering
\subfigure[Sampled links]{\includegraphics[width=.20\textwidth]{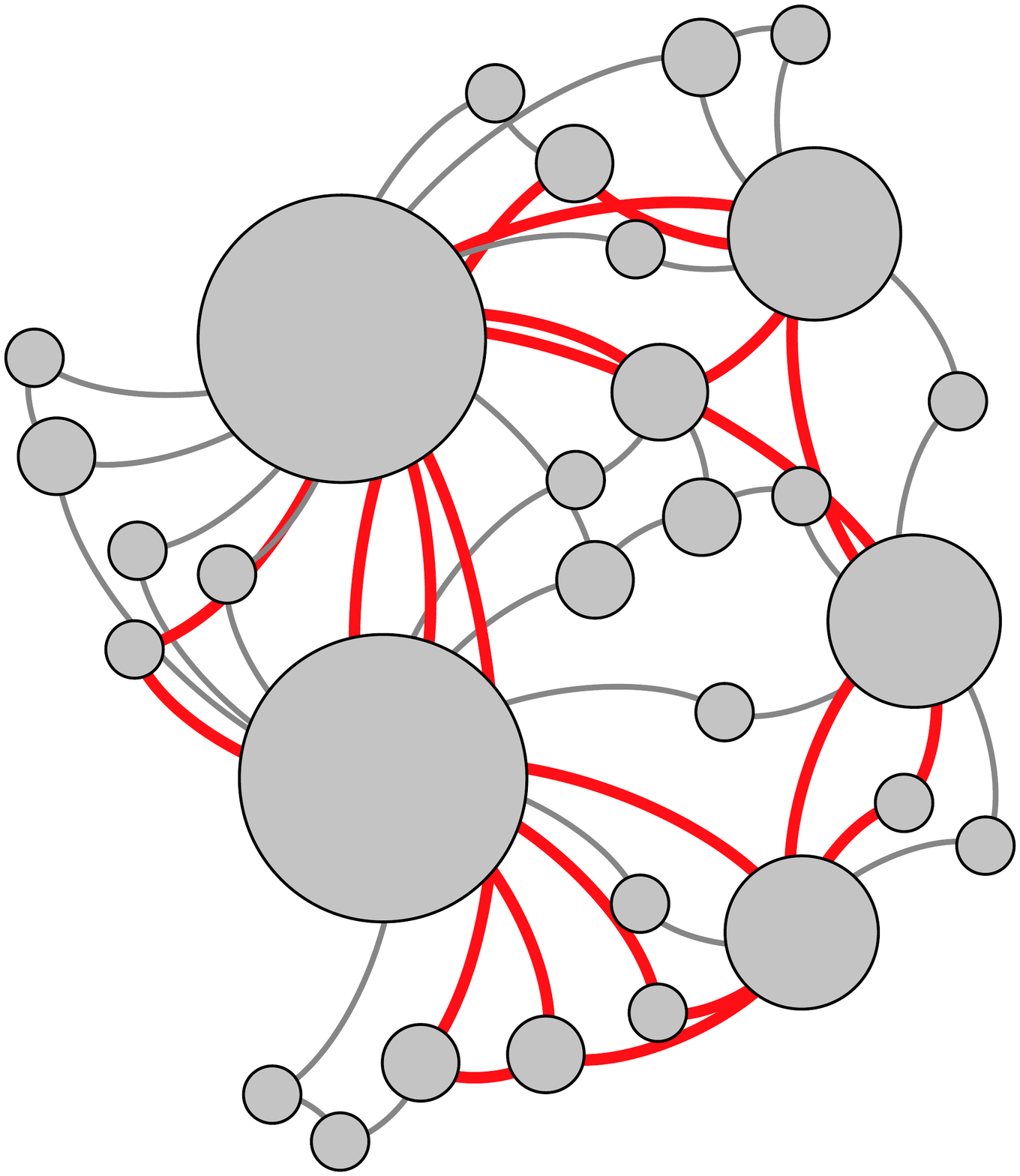}}
\subfigure[Link induced subnetwork]{\includegraphics[width=.20\textwidth]{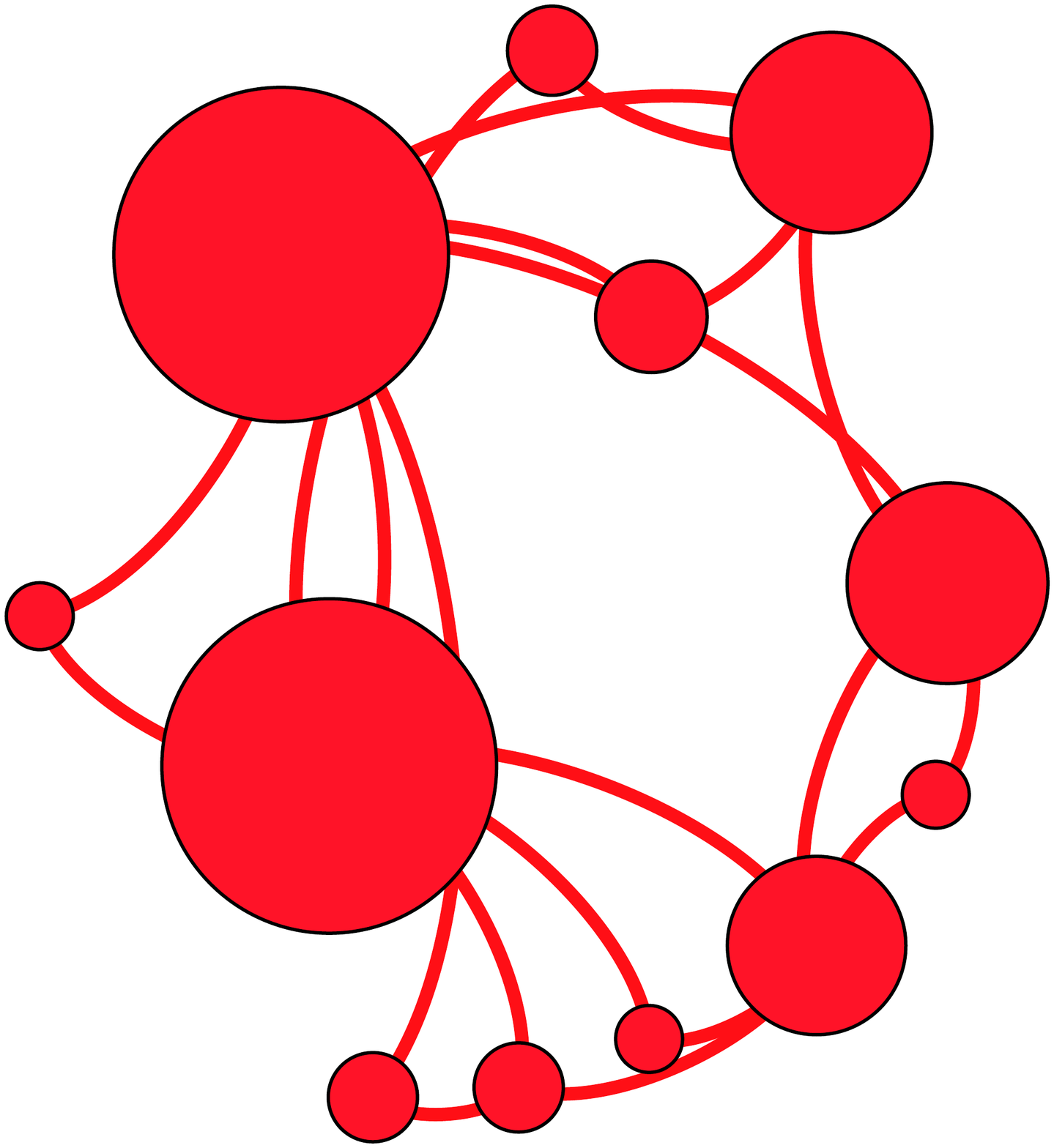}}\\
\caption[Subnetwork generated from sampled links]{Subnetwork generated from sampled links. (a) A network is sampled by randomly selecting links shown in red. (b) The subnetwork consists of all sampled links and only nodes which are incident with the sampled links. In this type of sampling, no nodes of degree zero are included in the network. Large degree nodes are more likely to be included in the subnetwork.}
\label{fig:induced_links_pic}
\end{figure}

The problem of missing links may also manifest itself in another manner. In contrast to the case when all nodes are known and some links are hidden, we now consider subnetworks generated by sampled links and the nodes incident to those links (Fig.~\ref{fig:induced_links_pic}). This type of sampling occurs in many social network settings, such as networks constructed from sampled email exchanges or message board posts. In this case, we have data pertaining to messages (links). Nodes (individuals) are only discovered when a link (email) which connects to them is detected. 

In this case, edges are sampled uniformly at random and we may use our previous estimator, $\hat{M}=\frac{m}{q}$. Node inclusion is biased, however, in that nodes of high degree will be detected with greater probability than nodes of low degree precisely because they are more likely to have an incident edge sampled. 

To motivate an appropriate predictor, we must first consider how the number of nodes in a subnetwork obtained by the subnetwork generated by sampled links scales with $q$ (Figs.~\ref{fig:sampling_by_links_incident_scaling}a and~\ref{fig:sampling_by_links_incident_empirical_scaling}a). To do this, let us consider the probability that a node is included in such a subsample. If the number of edges not sampled ($M-m$) is less than the degree $k(v_i)$ of node $v_i$, then we can be certain that our node of interest will be detected in sampling. On the other hand, if $M-m \geq k(v_i)$, then the probability of $v_i$ being in the subnetwork scales nonlinearly with $q$. Using the framework set forth by Kolaczyk~\cite{Kolaczyk2009}, observe that there are $\binom{M-k}{m}$ ways of choosing $m$ edges from the $M-k$ edges not incident with node $v_i$ and there are $\binom{M}{m}$ total ways of choosing $m$ edges from all $M$. Thus, we have 
\begin{align*}
\mathbf{Pr}(v_i \text{ is sampled})&=1-\mathbf{Pr}(\text{no edge incident to }v_i \text{ is sampled})\\
&=\begin{cases}
1 - \frac{\binom{M-k(v_i)}{m}}{\binom{M}{m}}, & \text{if $m \leq M-k(v_i)$}\\
1,  & \text{if $m > M-k(v_i)$}. 
\end{cases}
\end{align*}
The Horvitz-Thompson estimator given by
\begin{equation}
\hat{N}= \sum_{v_i \in V^*} \frac{1}{\pi_i},
\end{equation}
where $\pi_i =\mathbf{Pr}(v_i \text{ is sampled})$.

Kolaczyk~\cite{Kolaczyk2009} warns that this may not be a useful result, due to the fact that the true degree of a given node is likely to be unknown. We overcome this limitation by using our predicted degree distributions obtained by the techniques previously mentioned. Observe that when sampling by links, no nodes of degree zero will be observed. We also note that in the case when $k \ll M$ and $m$, we may make the following approximation which is less computationally burdensome:
\begin{align*}
\frac{\binom{M-k}{m}}{\binom{M}{m}} &= \frac{(M-k)!M-m)!}{M! (M-m-k)!}\\
&=\frac{(M-m)(M-m-1)(M-m-2) \hdots (M-m-(k-1))}{M(M-1)(M-2)\hdots (M-(k-1))}\\
&=\left(\frac{M-m}{M}\right)\left(\frac{M-1-m}{M-1}\right)\hdots \left(\frac{M-(k-1)-m}{M-(k-1)}\right)\\
&=\left(1-\frac{m}{M}\right)\left(1-\frac{m}{M-1}\right)\hdots \left(1-\frac{m}{M-(k-1)}\right)\\
& \approx \left(1-q \right)^{k(v_i)} \text{ for }k(v_i) \text{ relatively small compared to }m\text{ and }M.
\end{align*}
This is simply the probability that a node of degree $k(v_i)$ loses all edges during subsampling $q^0(1-q)^k$ and thus $\mathbf{Pr}(\text{not detecting }v_i) \approx (1-q)^{k(v_i)}.$ Thus,
\begin{align}
\hat{N}&= \sum_{v_i \in V^*} \frac{1}{\pi_i}\\
&=\sum_{v_i \in V^*} \frac{1}{1-(1-q)^{k(v_i)}}\\
\end{align}
We apply these methods to our simulated and empirical networks. 

Once $\hat{N}$ and $\hat{M}$ have been computed, the average degree is simply $\hat{k}_{\rm avg}=\frac{2 \hat{M}}{\hat{N}}$. The max degree scales roughly linearly for preferential attachment models and many of the empirical networks, however scales sublinearly in networks with a high proportion of nodes of similar degree (e.g. the regular lattice structure seen in Small world and Range dependent networks). Clustering scales approximately as $\hat{C}=\frac{c}{q}$ and the giant component shows a critical threshold which varies according to network type and average degree. The relative errors of our predictors are summarized in Tables~\ref{table:error_N_incident_links}--~\ref{table:error_kmax_incident_links}. The scaling of $P_k$ and the predicted degree distribution are presented in Figs. SI.11 and SI.12.

To test the goodness of fit for the estimated degree distribution and the true $P_k$, we again compute $D=\max \left|F_{i,\rm{true}}-F_{i,\rm{predicted}}\right|$, two sample Kolmogorov-Smirnov test statistic (Fig.~\ref{fig:one_over_q_incident_links_D_stat_plot}). This figure shows that reasonable results are achieved when $q>0.50$, a noticeable increase in the percent of network knowledge needed, as compared to other sampling strategies (sampling by nodes and failing links). 

\subsection{Sampling by interactions}
\begin{figure}[!htp]
\centering
\subfigure[Weighted network]{\includegraphics[width=.20\textwidth]{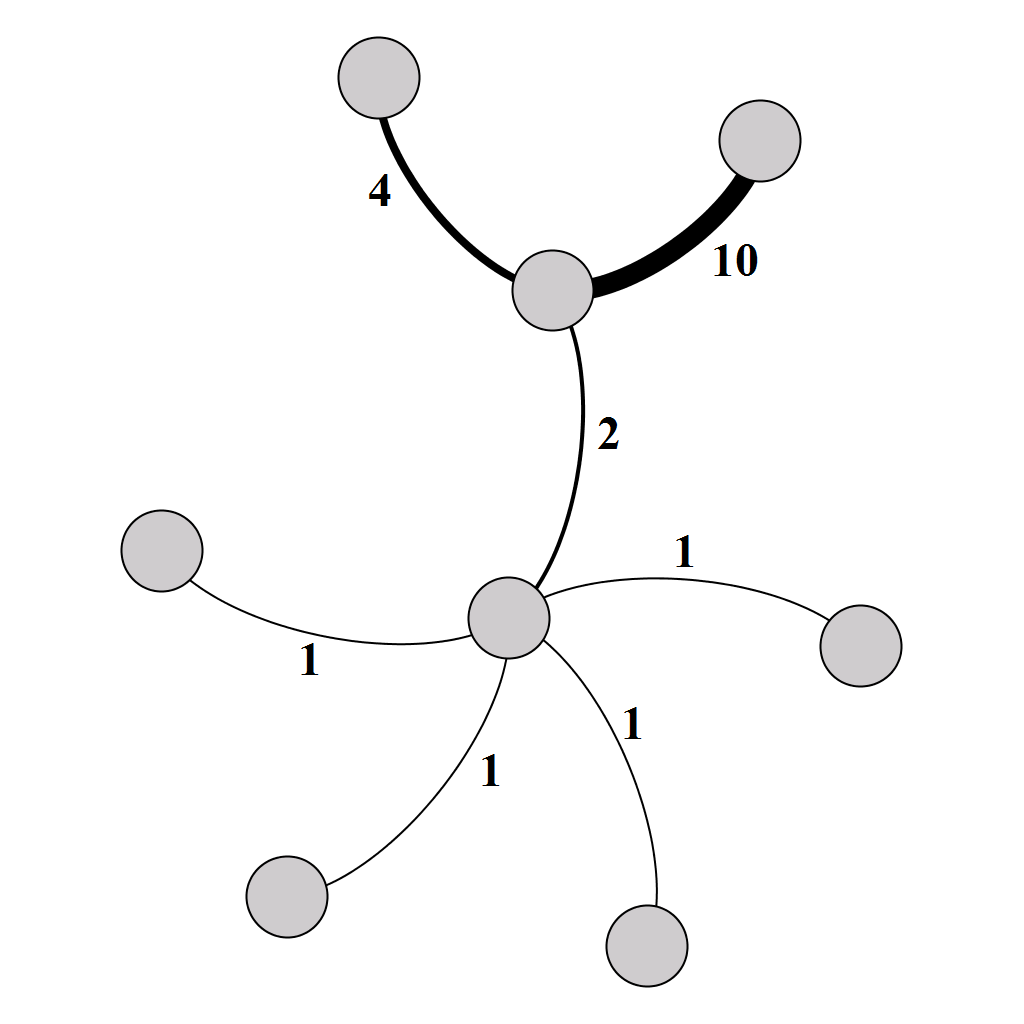}}  
\subfigure[Weighted subnetwork]{\includegraphics[width=.20\textwidth]{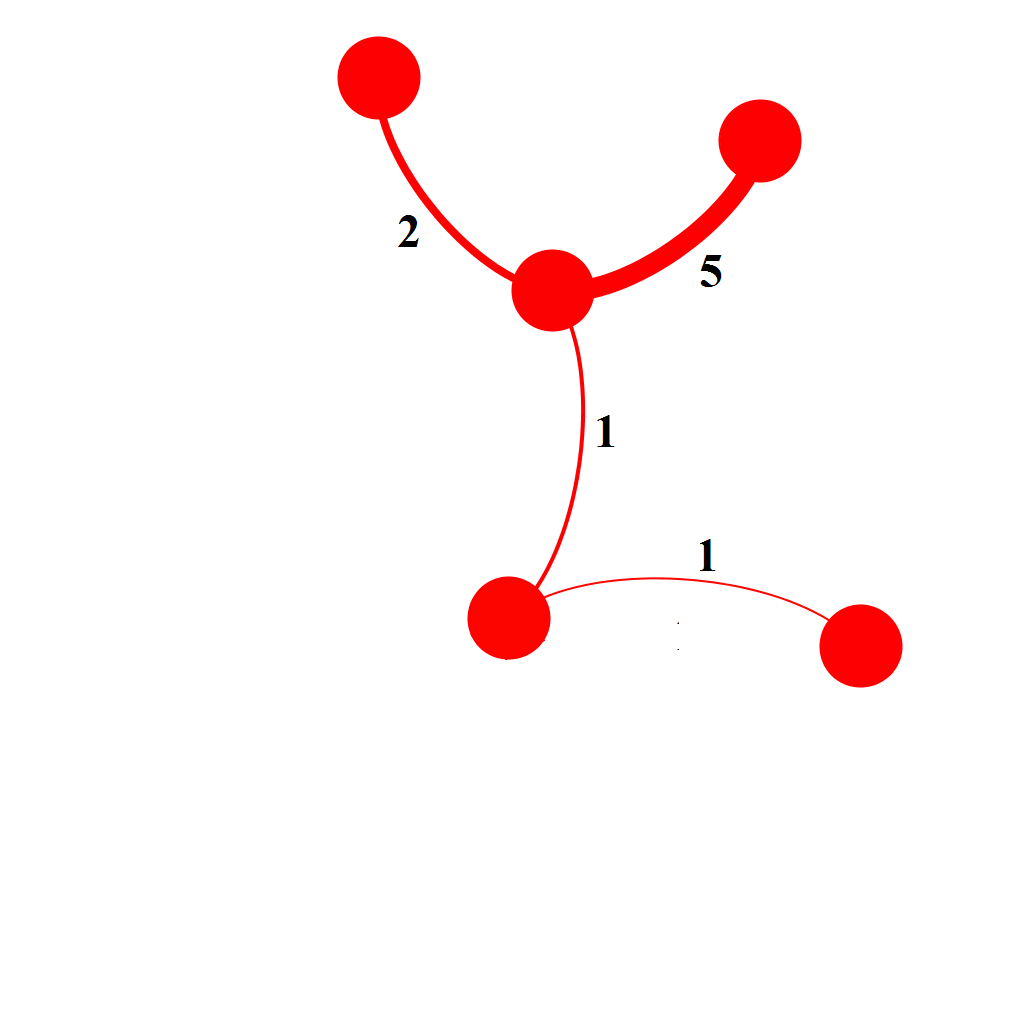}} 
\caption[Weighted subnetwork generated from sampled interactions]{Weighted subnetwork generated from sampled interactions. (a.) An unsampled weighted network consists of nodes, links and weights representing the number of interactions represented by the link. (b.) Sampling by interacting produces a subsample whereby links are included in the subsample only if at least one interaction has been sampled. The subnetwork is the induced subgraph on these links with $w_i \geq 1$.}
\label{fig:subsamp_weighted_pic}
\end{figure}

Lastly, we consider the case of sampling by interactions in the special case of a weighted network (Fig.~\ref{fig:subsamp_weighted_pic}). In this case, we begin with $G=(V,E)$, where $E$ is a set of edges\footnote{We will treat the directed case as a special case in the next section.}, $e_j$, with weight $w(e_j)$. The weight on an edge represents the number of interactions between two vertices. An alternative representation is simply a network with multiple edge between two such vertices, one for each interaction. A subnetwork generated by $q\sum_{e_j \in E} w(e_j)$ sampled interactions is simply a sampled collection of multi-edges and the nodes incident to these edges (e.g., the subnetwork generated by links with nonzero weight and nodes incident to those edges).

To consider how the number of nodes scales, we consider a similar formulation as discussed in the previous section for the probability that a given node is selected when sampling by links, however instead of the degree of a node, $k(v_i)$, we are now interested in the strength of a node. The strength of a node is given by $s(v_i)=\sum_{e_j \in \mathcal{N}(v_i)} w(e_j)$, where $\mathcal{N}(v_i)$ denotes the neighborhood of vertex $v_i$~\cite{barrat2004}. 
Let $L=\sum_{e_j \in E} w(e_j)$ represent network load and $\ell=qL$, the number of sampled interactions. If the number of interactions which are not sampled ($L-\ell$) is less than the strength of a node $s(v_i)$, then we can be certain that node $v_i$ will be detected in sampling. 

On the other hand, if $L-\ell \geq s(v_i)$, then there are at most $\binom{L-s(v_i)}{\ell}$ ways\footnote{As an upper bound, we assume that the $L-s(v_i)$ interactions are distributed over $L-s(v_i)$ edges (weight of 1 on each edge) which maximizes the number of ways these could be chosen.} of choosing $\ell$ interactions from the $L-s(v_i)$ interactions not involving node $v_i$ and there are at most $\binom{L}{\ell}$ total ways of choosing $\ell$ (distinct, labeled) interactions from all $L$. Letting $\mu(i)$ represent the probability that $v_i$ is sampled, we have 
\begin{align*}
\mu_i&=1-\mathbf{Pr}(\text{ no interaction incident to }v_i \text{ is sampled})\\
&=\begin{cases}
1 - \frac{\binom{L-s(v_i)}{\ell}}{\binom{L}{\ell}}, & \text{if $\ell \leq L-s(v_i)$}\\
1,  & \text{if $\ell > L-s(v_i)$}. 
\end{cases}
\end{align*}
Thus, our Horvitz-Thompson estimator is,
\begin{equation}
\label{eq:num_nodes_weighted_net}
\hat{N}= \sum_{v_i \in V^*} \frac{1}{\mu_i},
\end{equation}
where $\mu_i =\mathbf{Pr}(v_i \text{ is sampled})$.
This can be well approximated by 
\begin{equation}
\label{eq:pred_node_strength}  
\mu_i=1-(1-q)^{s(v_i)}.
\end{equation}
It should be noted that the strength of a node is merely predicted. Thus, effort must be made to predict the node strength distribution in the same spirit as was previously done for the degree distribution. To predict the node strength distribution, we modify Equation 17 and predict an observed node of strength $s$ to be of strength $\frac{s}{q}$ in the true network. Applying this corrector to our subsampled weighted networks, we find low relative error in the predicted number of nodes for most networks (Tables~\ref{table:error_N_weighted_ER} and~\ref{table:error_N_weighted_pref}). An exception to this is Case I (Erd{\"o}s-R\'{e}nyi) for $q<0.55$. We predict thenode strength to be $\frac{s}{q} \geq 2$ and yet in this case, the true network is unweighted (e.g., $w(e_j)=1, \forall e_j \in E$). If there is knowledge that the network is unweighted, this example shows that the techniques from sampling by edges subsection will yield much better results..

We now consider how the number of edges in the subnetwork scales with the proportion of sampled interactions. The probability of selecting an edge $e_{j} \in E$ is equal to 1-$\mathbf{Pr}(\text{ not selecting edge }e_j)$. Notice that when the $\ell>L-w(e_j)$, the edge $e_j$ is certain to be included in the subsample. When $\ell \leq L-w(e_j)$, the probability of not selecting edge $e_j$ is simply the number of ways of selecting the $L-w(e_j)$ interactions $\ell$ at a time, which are not on edge $e_j$ divided by the number of ways of selecting $\ell$ weights from $L$.
\begin{align*}
\mathbf{Pr}(e_j \text{ is sampled})&=1-\mathbf{Pr}(\text{ no interaction along }e_j \text{ is sampled})\\
&=\begin{cases}
1 - \frac{\binom{L-w(e_j)}{\ell}}{\binom{L}{\ell}}, & \text{if }\ell \leq L-w(e_j)\\
1,  & \text{if }\ell > L-w(e_j). 
\end{cases}
\end{align*}
Thus, our Horvitz-Thompson estimator is,
\begin{equation}
\label{eq:HT_estimator_sampled_interactions_edges}  
\hat{M}= \sum_{e_j \in E^*} \frac{1}{\lambda_j},
\end{equation}
where $\lambda_j =\mathbf{Pr}(e_j \text{ is observed})$, which is well approximated by 
\begin{equation}
\label{eq:approx_edges_sampled_interactions} 
\lambda_j=1-(1-q)^{w(e_j)}. 
\end{equation}
Again, we must have knowledge of the edge weights, or be able to predict them with reasonable accuracy. To do this, we predict an edge of weight $w(e_j)$ in the subnetwork to be of edge weight $\frac{w(e_j)}{q}$ in the true network.

As the weights on edges tends to 1 (the unweighted network case), we retrieve our result for how edges scale when links are sampled (synonymous with weights in the case where $w_i=1$): 
\begin{align*}
\lim _{w(e_j) \rightarrow 1} \mathbf{Pr}(e_j \text{ is observed})&=\lim _{w(e_j) \rightarrow 1} 1-\mathbf{Pr}(w(e_{j}))\\
&=\lim _{w(e_j) \rightarrow 1} 1 - \frac{\binom{L-w(e_i)}{\ell}}{\binom{L}{\ell}}\\
&= 1 - \frac{\binom{M-1}{m}}{\binom{M}{m}}\\
&= 1 - \frac{M-m}{M}\\
&=\frac{m}{M}\\
&=q,
\end{align*}
where $q$ is the proportion of sampled links. Thus, when the weights on edges tends to 1, our Horvitz-Thompson estimator is
\begin{align*}
\hat{M} &= \sum_{e_j \in E^*} \frac{1}{\lambda_j},\\
&= \frac{m}{q},
\end{align*}
which recovers our previous result for scaling of edges when sampling by links. The relative error incurred for the predicted number of edges is presented in 
Tables~\ref{table:error_M_weighted_ER} and~\ref{table:error_M_weighted_pref}.

Having found suitable predictors for $N$ and $M$, the average degree may be predicted by,
\begin{align*}
\hat{k}_{\rm avg} &= \frac{2 \hat{M}}{N}.
\end{align*}
Applying these scaling techniques, we obtain reasonably low error for both networks in both experiments 1 and 2 (Tables~\ref{table:error_avk_weighted_ER}--~\ref{table:error_avk_weighted_pref}).

To estimate $\kmax$, we recognize that the observed max degree will need to be scaled by roughly the proportion of missing edges. Using $\frac{\hat{M}}{m}$ as our scaling factor, we find relatively high error for both networks (Tables~\ref{table:error_kmax_weighted_ER}--~\ref{table:error_kmax_weighted_pref}). This is due to errors in $\hat{M}$ hindering accuracy in $\hat{k}_{\max}$.
 
\begin{table*} [!ht]
\caption[Summary of scaling techniques]{Summary of scaling techniques.}
\centering
\begin{tabular}{|l|c|c|c|c|}
\hline
 & Sampled  & Failed &  Sampled & Sampled\\
&  nodes            & links &  links & interactions\\[1ex]
\hline\hline                            
Predicted number of nodes ($\hat{N}$) & $\frac{n}{q}$ & $n$ & $\sum_{v_i \in V^*} \frac{1}{ 1 - (1-q)^{d(v_i)} }$ & $\sum_{v_i \in V^*} \frac{1}{1 - (1-q)^{s(v_i)}}$\\[.4cm]
Predicted number of edges  ($\hat{M}$) & $\frac{m}{q^2}$ & $\frac{m}{q}$ & $\frac{m}{q}$ & $\sum_{e_i \in E^*} \frac{1}{1 - (1-q)^{w(e_i)}}$ \\[.4cm]
Predicted average degree  $\left(\hat{k}^{\textnormal{obs}}_{\rm avg}\right)$ & $\frac{k^{\textnormal{obs}}_{\rm avg}}{q}$ & $\frac{k^{\textnormal{obs}}_{\rm avg}}{q}$ & $\frac{2\hat{M}}{\hat{N}}$ &  $\frac{2\hat{M}}{\hat{N}}$\\[.4cm] 
Predicted clustering   ($\hat{C}$) & $C$ & $ qC$ & $\frac{C}{q}$ & --\\[.4cm]
Predicted max. degree   $\left(\hat{k}_{\max}\right)$ & $\frac{k^{\textnormal{obs}}_{\max}}{q}$ & $\frac{k^{\textnormal{obs}}_{\max}}{q}$ & $\frac{k^{\textnormal{obs}}_{\max}}{q}$ & $\frac{\hat{M}}{m}   k^{\textnormal{obs}}_{\max}$ \\[.4cm]\hline
\end{tabular}
\label{table:compiled_scaling}
\end{table*}

\section{Estimating the size of the Twitter interactome}
We now consider the weighted, directed network of replies whereby a link from node $v_i$ to node $v_j$ represents the existence of at least one reply directed from $v_i$ to $v_j$ and the weight on this edge represents the number of messages sent in the time period under consideration. We apply our methods to reply networks constructed from tweets gathered during the ten week period from September 9, 2008 to November 17, 2008, a period for which we have a substantially higher percentage of all authored messages.

For each of these weeks, we receive between 20-55\% of all messages posted on Twitter and similarly believe that we receive approximately 20-55\% of all replies posted in this period (Table~\ref{table:datawehave}). We apply our previously developed methods to estimate the number of nodes, edges, strengths on these edges, average degree, max degree, and distribution of node strength. To help validate our predictions, we also predict the number of nodes, edges, average degree, and max degree by performing 100 sampling experiments in which a proportion $q$ of the observed messages used for subnetwork construction. These sampling experiments essentially ``hide'' some of the messages from our view and thus allow us to consider how further subsampling impacts the inferred networks statistics. Curve fitting over this region of $q$ allows us to extrapolate the network statistic to a predicted value over increased percentages of observed messages. We use this to validate with our estimated statistic using the methods from the previous section.

\subsection{Number of nodes}
\begin{figure*}[!htp]
\centering
\subfigure[$N_{\text{repliers}}$]{\includegraphics[width=.45\textwidth]{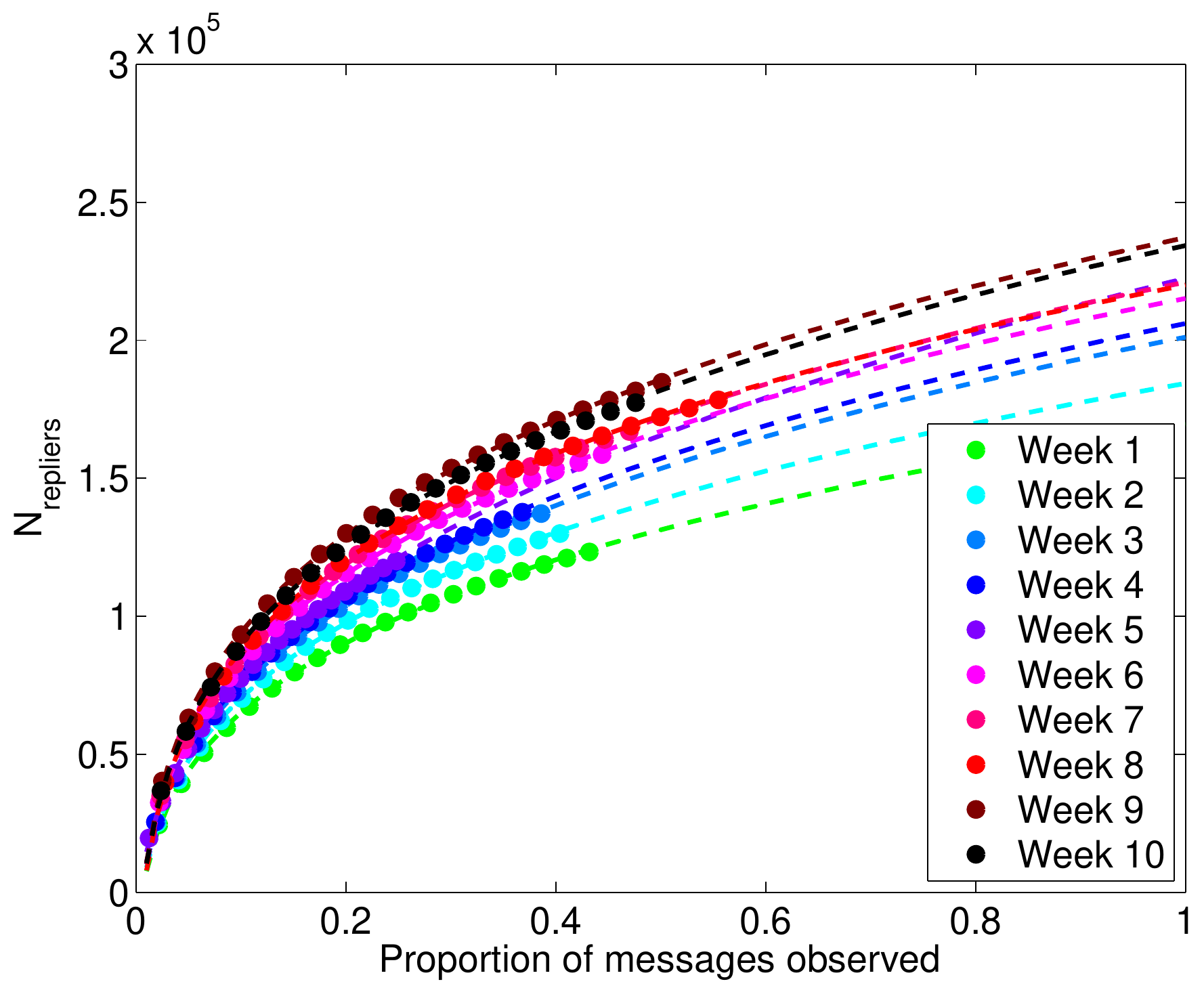}}
\subfigure[$N_{\text{receivers}}$]{\includegraphics[width=.45\textwidth]{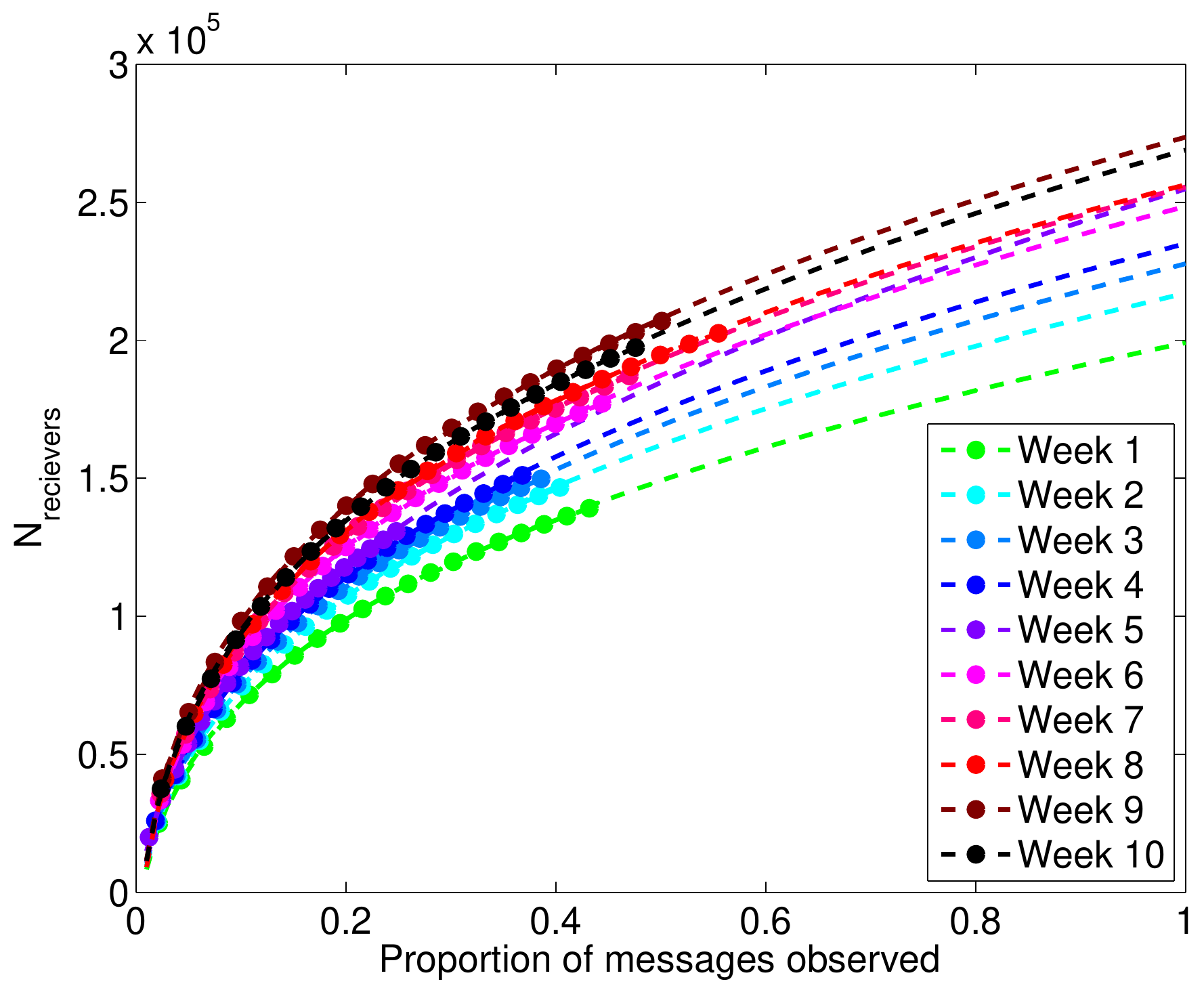}}
\caption[Number of nodes in Twitter reply subnetworks]{Number of nodes in Twitter reply subnetworks. (a.) The quantity $N_{\text{repliers}}$ is shown for Weeks 1 to 10, where each data point (dot) represents the average over 100 simulated subsampling experiments. The dashed line represents the best fitting model of the form $N_{\text{repliers}}=ax^b$ to the observed data. We extrapolate this model to predict $N_{\text{repliers}}$. (b.) The same as panel (a.), except for $N_{\text{receivers}}$. }
\label{fig:Twitter_RRN_subsample_stats}
\end{figure*}

\begin{figure*}[!htp]
\centering
\subfigure[$N_{\text{repliers}}$]{\includegraphics[width=.45\textwidth]{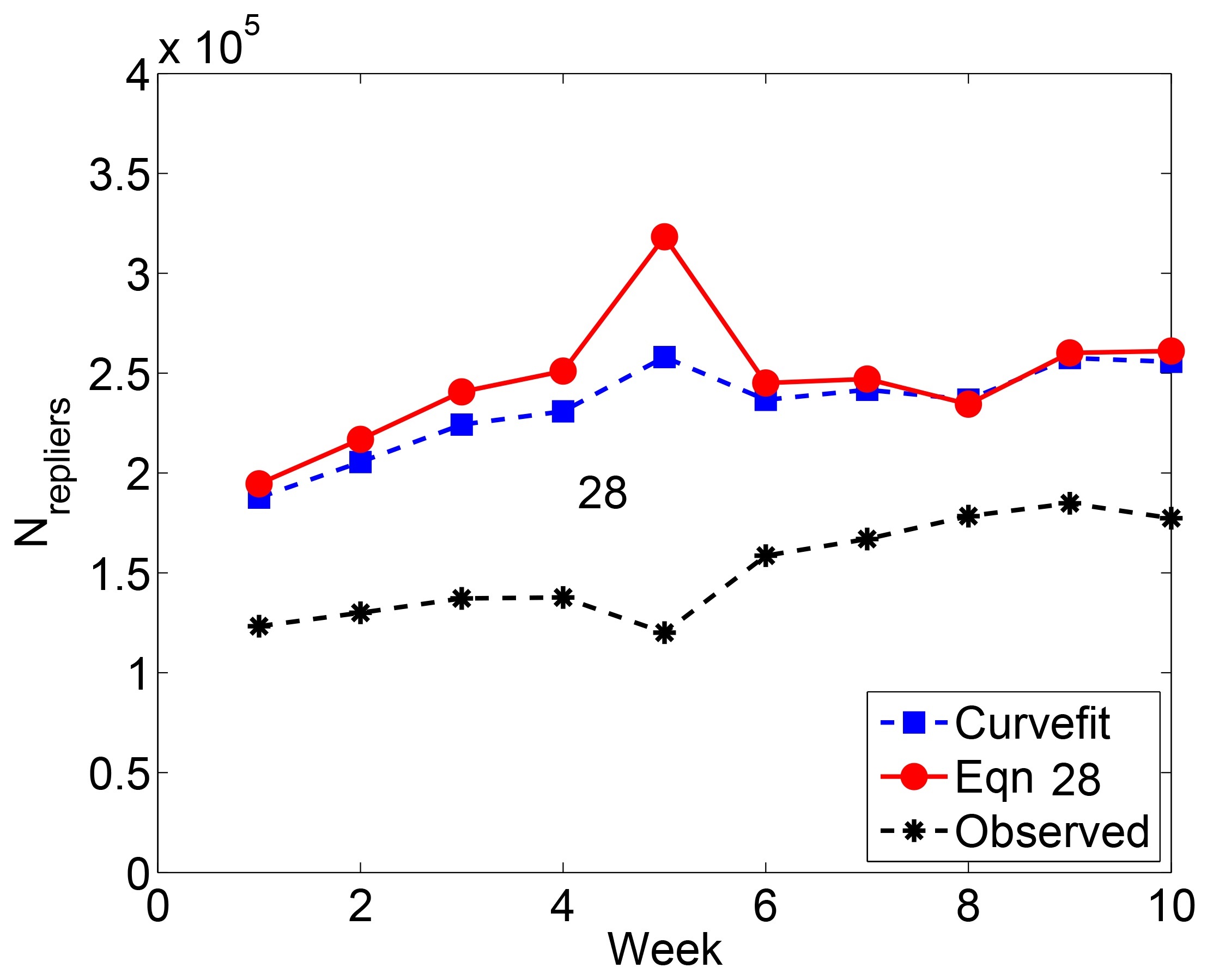}}
\subfigure[$N_{\text{receivers}}$]{\includegraphics[width=.45\textwidth]{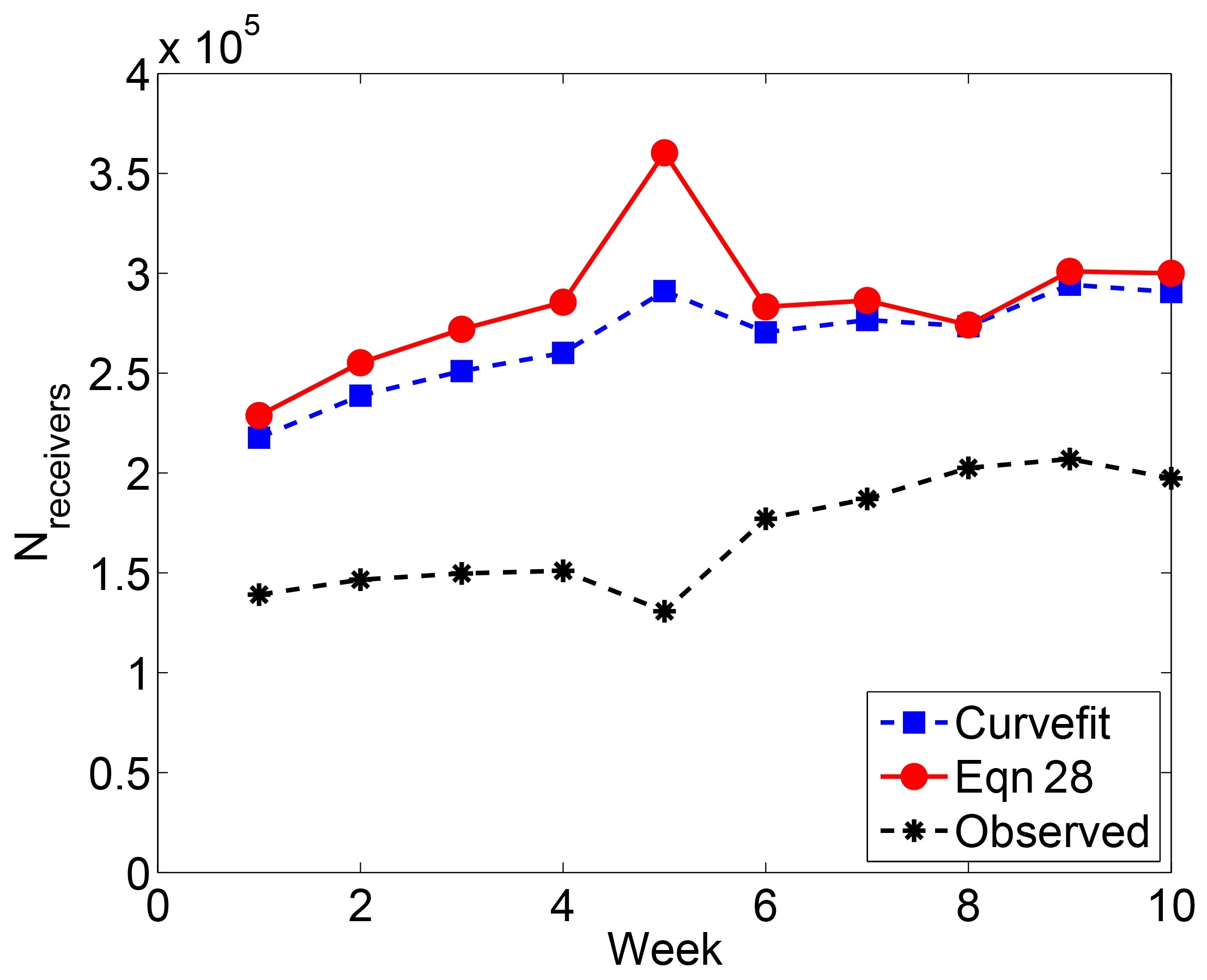}}
\caption[Predicted number of nodes in Twitter reply networks]{Predicted number of nodes in Twitter reply networks. The number of nodes observed for each week is depicted, along with the predicted number of nodes obtained from curve fitting (Fig. 5) and Equation 28. The predicted number of nodes is nearly double the number of observed nodes. The relatively low proportion of messages received for Week 5 ($<$ 25\%) may be creating greater inaccuracies in the predictors for that week. }
\label{fig:predicted_nodes_twitter}
\end{figure*}

Since our reply networks are directed, we consider both the number of nodes which make a reply ($N_{\text{replier}}$) and the number of nodes which receive a reply ($N_{\text{receiver}}$). As expected from our previous discussion, the number of nodes scales nonlinearly with the proportion of observed messages (Fig.~\ref{fig:Twitter_RRN_subsample_stats}).  We fit models of the form $N=ax^b$ to observed data and in doing so find an excellent fit ($R^2 \approx 0.99$) for all weeks over the subsampled region (Fig.~\ref{fig:Twitter_RRN_subsample_stats}). Extrapolating these fitted models to $q=1$, we find excellent agreement with our predicted number of nodes obtained from Equations~\ref{eq:num_nodes_weighted_net} and~\ref{eq:pred_node_strength}. The predicted number of nodes from both methods agree to within $\pm$ 5\%. Figure~\ref{fig:predicted_nodes_twitter} reveals that the predicted number of nodes is nearly double the number of observed nodes. 

\subsection{Strength of nodes}
\begin{figure*}[!htp]
\centering
\subfigure[In-coming]{\includegraphics[width=.45\textwidth]{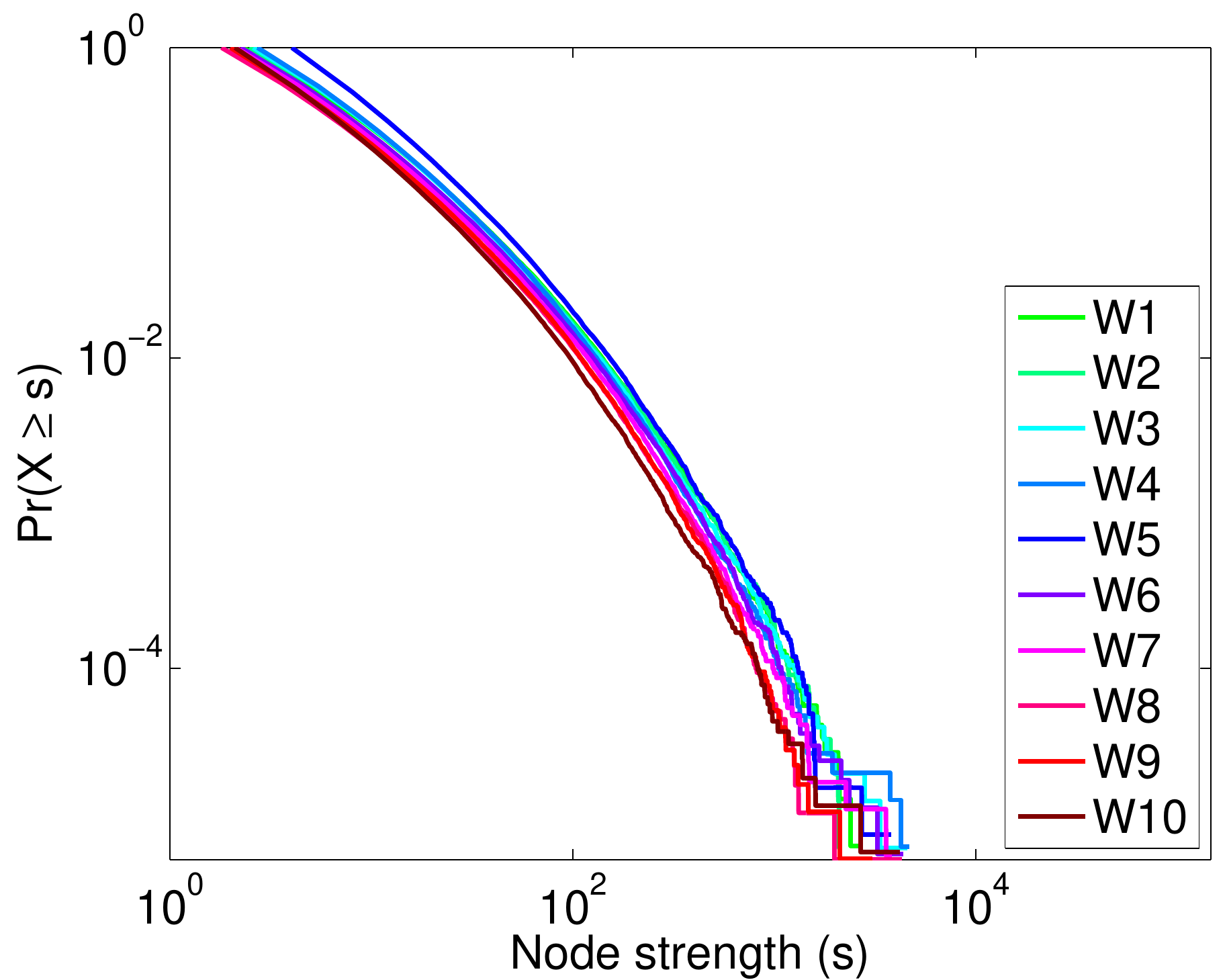}}
\subfigure[Out-going]{\includegraphics[width=.45\textwidth]{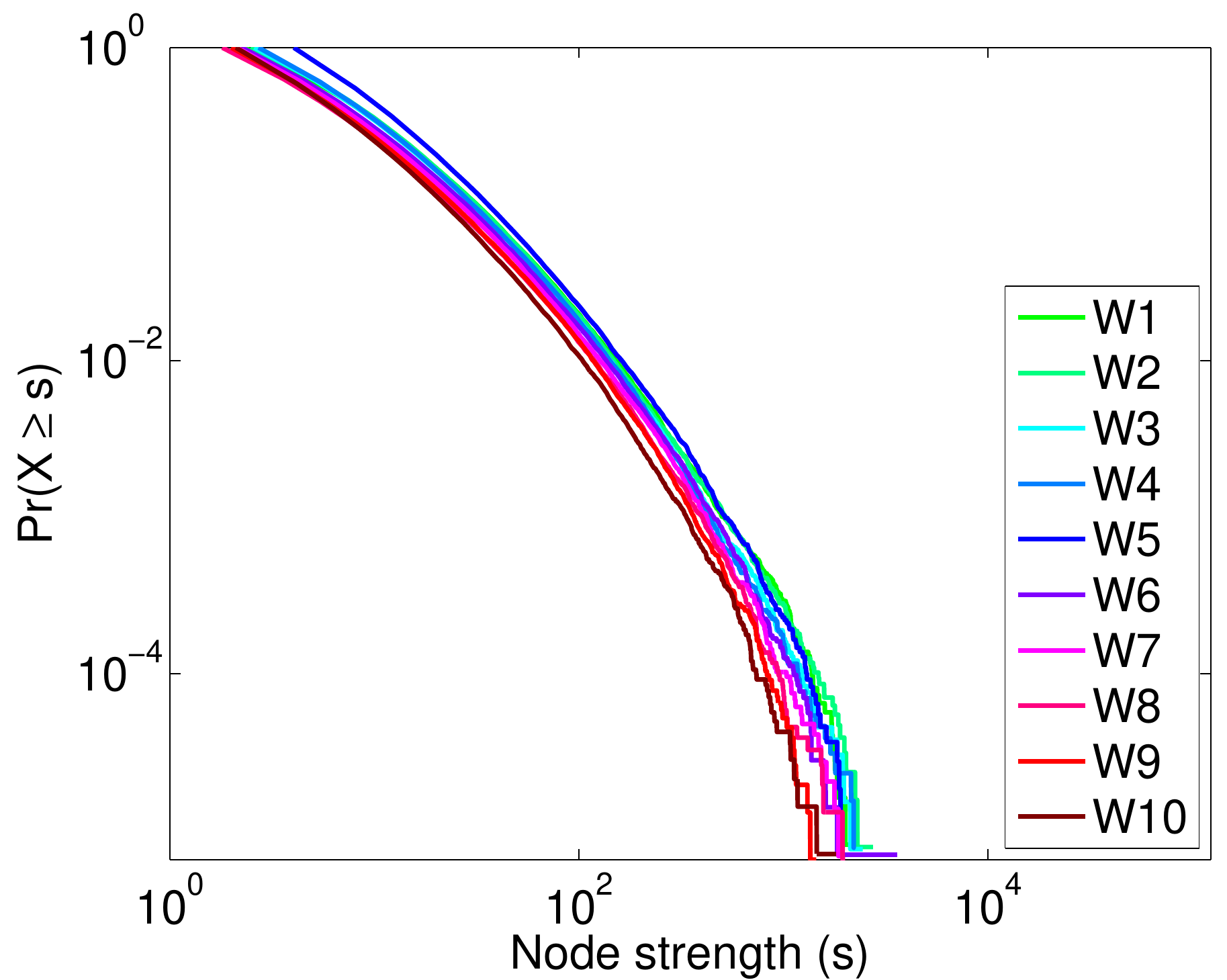}}
\caption[Predicted $P_s$ for Twitter reply networks]{Predicted $P_s$ for Twitter reply networks. (a.) The node strength distribution for in-coming interactions. (b.) The node strength distribution out-going interactions. In both cases, the distribution is heavy tailed, but falls off faster than would be expected in a scale-free distribution.}
\label{fig:week_node_strengthed_predicted_one_over_q}
\end{figure*}

\begin{figure*}[!htp]
\centering
\subfigure[]{\includegraphics[width=.45\textwidth]{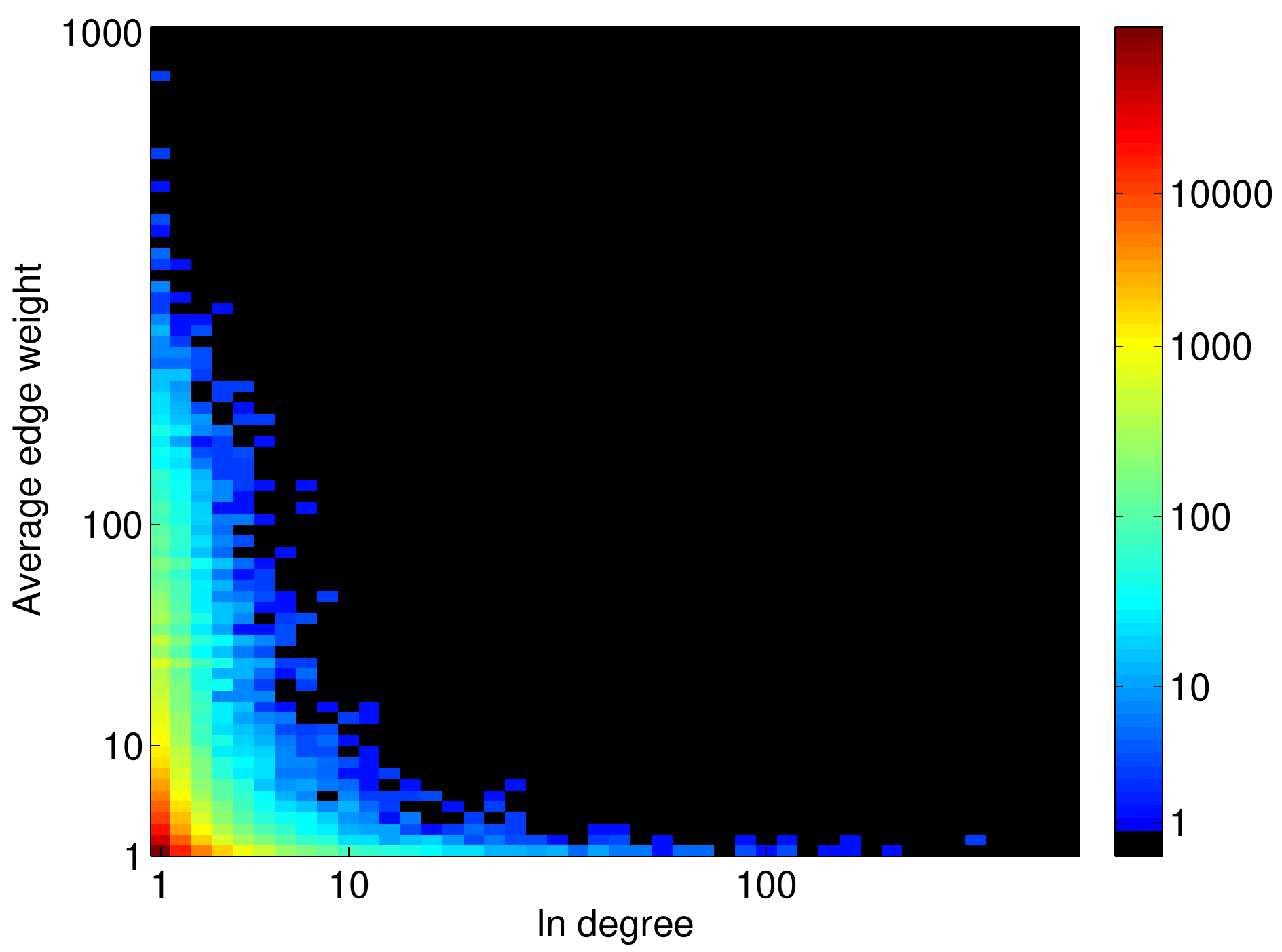}}
\subfigure[]{\includegraphics[width=.45\textwidth]{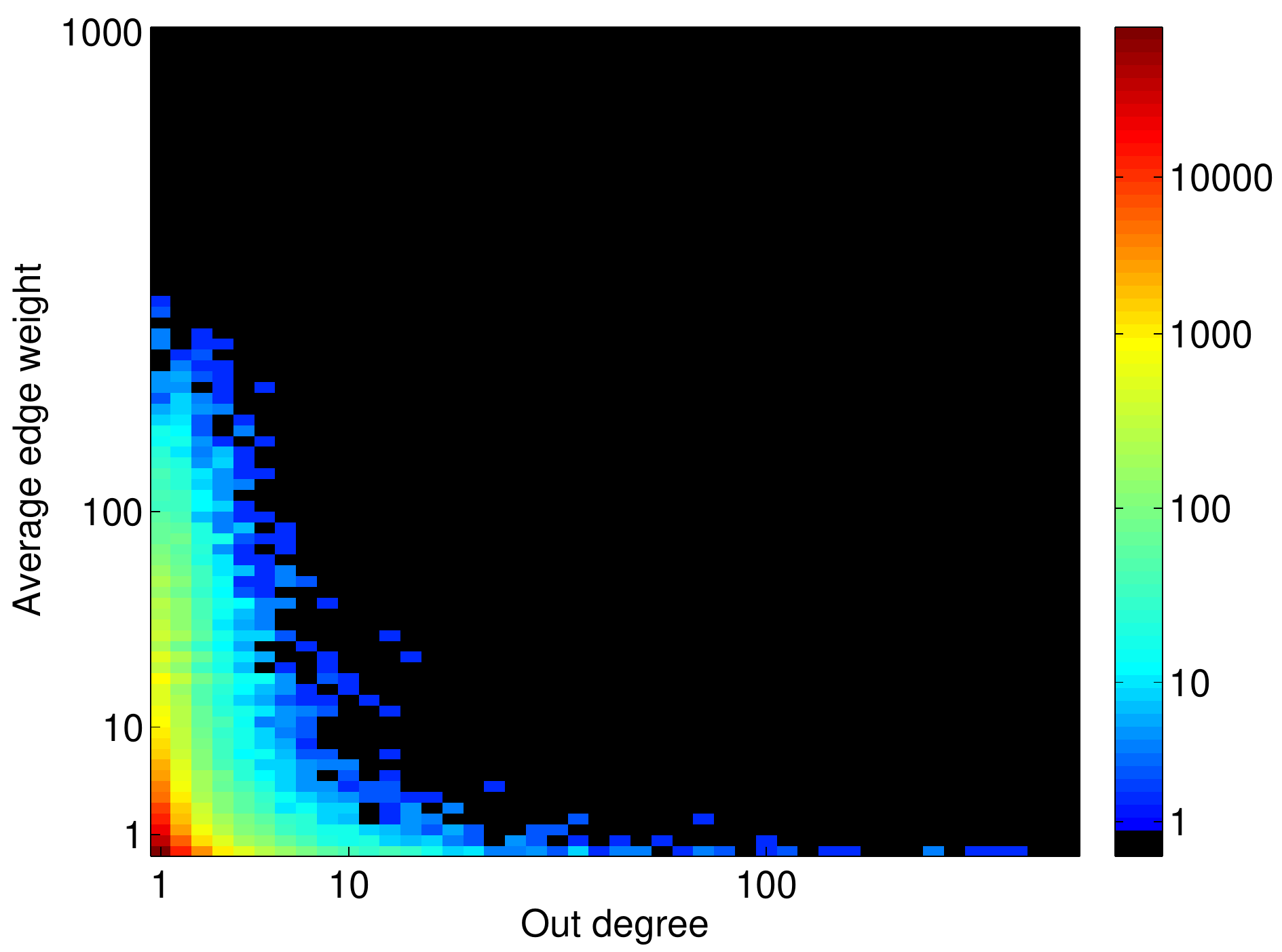}}\\
\subfigure[]{\includegraphics[width=.45\textwidth]{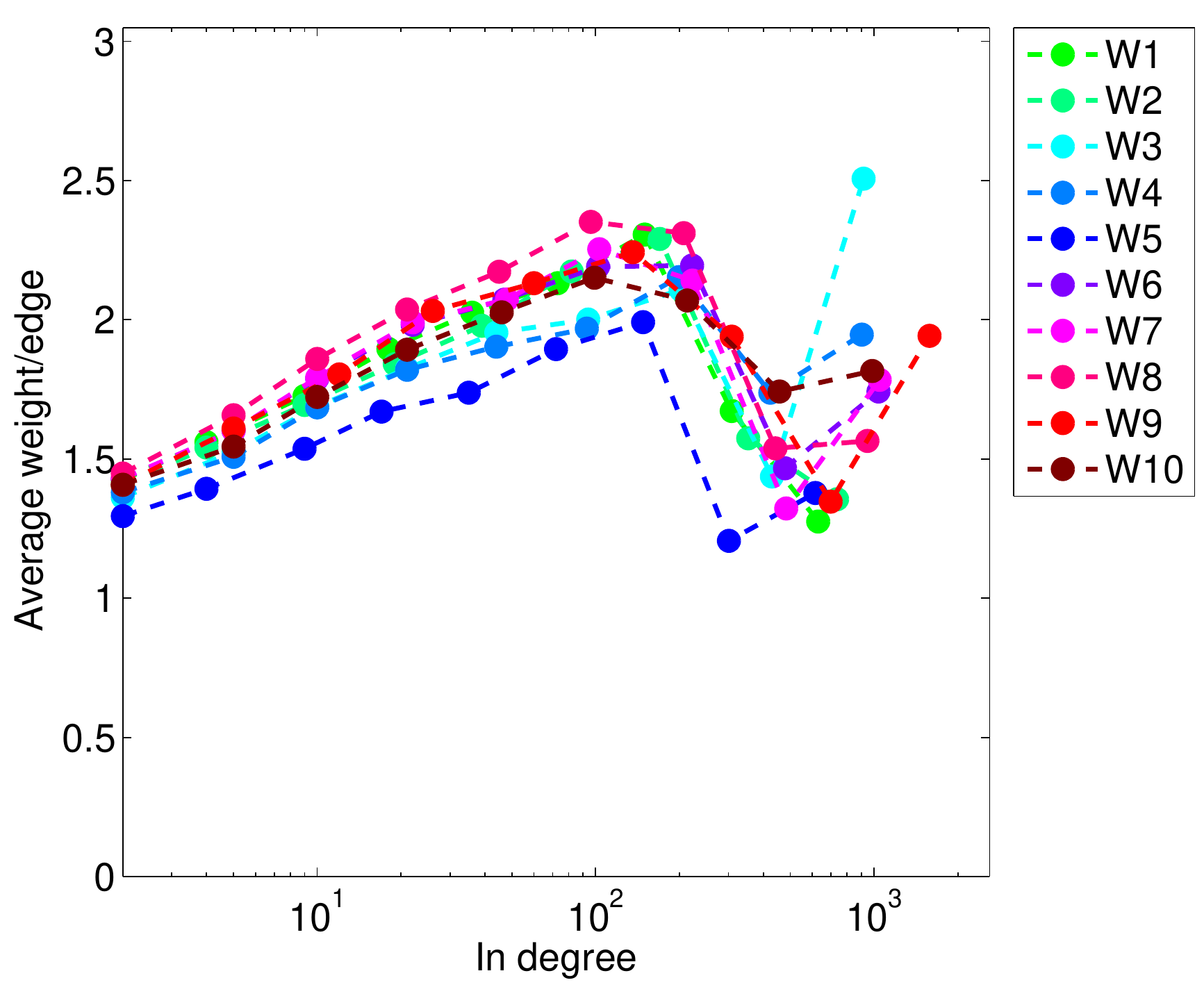}}
\subfigure[]{\includegraphics[width=.45\textwidth]{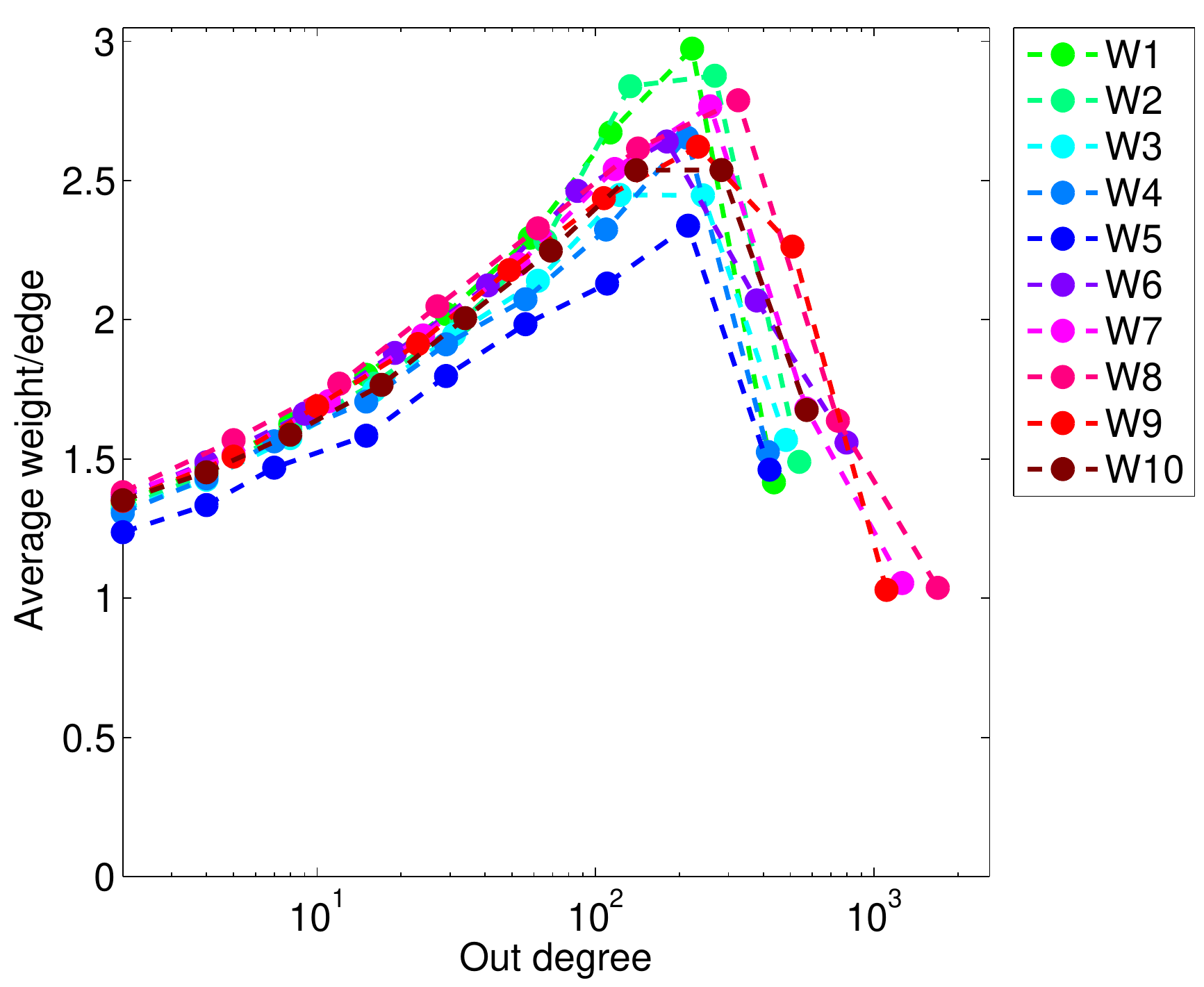}}
\caption[In, Out-degree vs. Average edge weight for Twitter reply networks]{In, Out-degree vs. Average edge weight for Twitter reply networks. (a.) The average in-coming edge weight for each node of degree $k$ is depicted in a logarithmically binned heatmap. (b.) The same as (a), except for out-going edges. (c.) The average weight per edge for in-coming edges as a function of $k_{\rm{in}}$ shows a gradual increase to $k_{\rm{in}} \approx 10^2$ with a peak of approximately 2.2 interactions per edge. (d.) The average weight per edge for out-going edges as a function of $k_{\rm{out}}$ shows a gradual increase to $k_{\rm{out}} \approx 10^2$ with a peak of between 2.5 and 3 interactions per edge.}
\label{fig:out_degree_vs_average_edge_weight}
\end{figure*}

Figure~\ref{fig:week_node_strengthed_predicted_one_over_q} depicts a log-log plot of the predicted node strength distribution. This plot reveals that there are fewer nodes in the high strength region than would be expected in a scale-free distribution. Figure~\ref{fig:out_degree_vs_average_edge_weight} reveals that low degree nodes dominate the dataset and that many of these low degree nodes often have low average edge weight ($w_{\text{avg}} \approx 1.5$). We find a peak in the average weight per edge as a function of degree around $k \approx 10^2$. This peak is more pronounced for out-going edges. Beyond this value, a limiting factor may prevent increases in the weight per edge, a result also noted by Gon{\c{c}}alves et al.~\cite{Goncalves2011}.

\subsection{Number of edges}
\begin{figure*}[!ht]
\centering
\subfigure[]{\includegraphics[width=.45\textwidth]{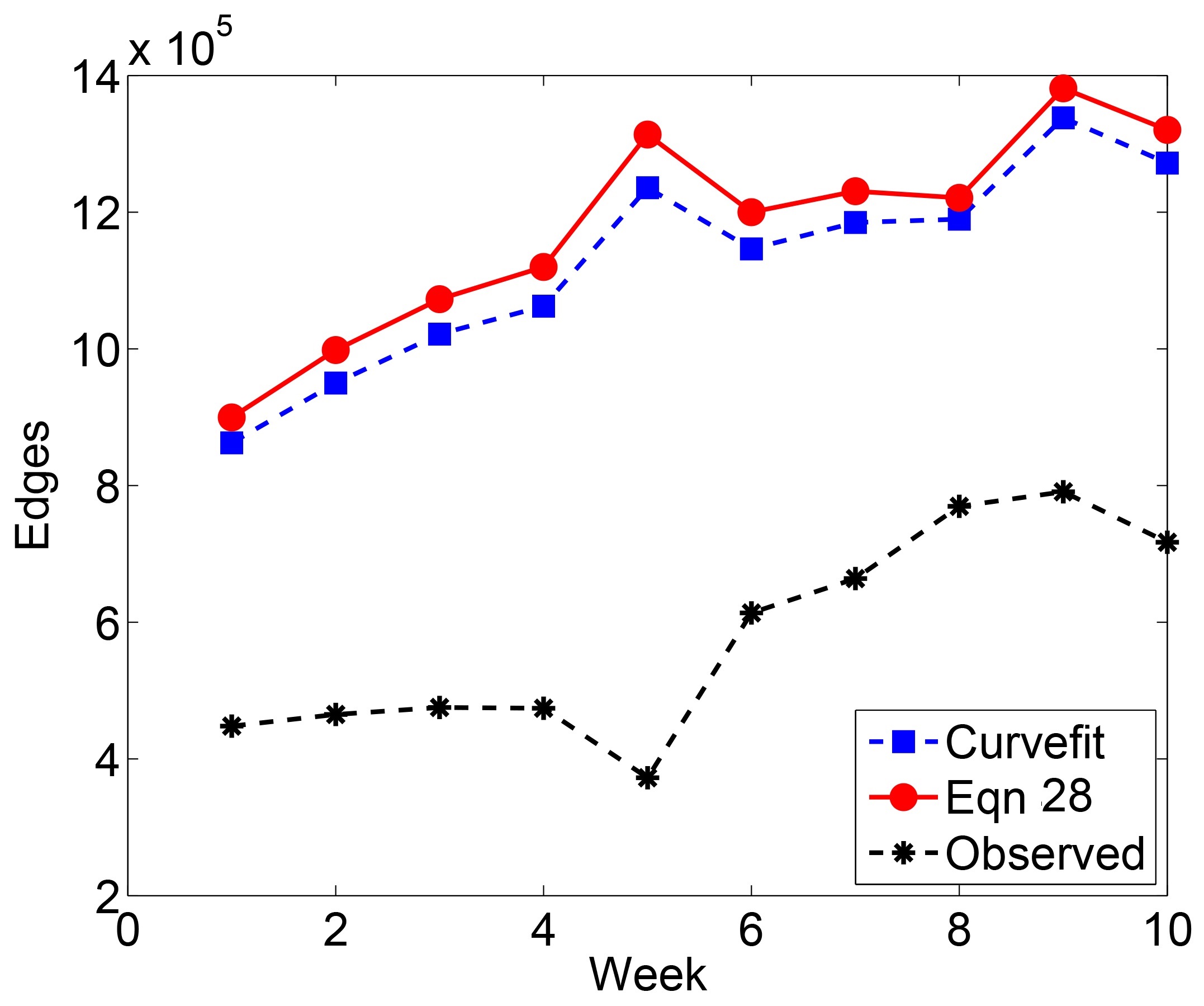}}
\subfigure[]{\includegraphics[width=.45\textwidth]{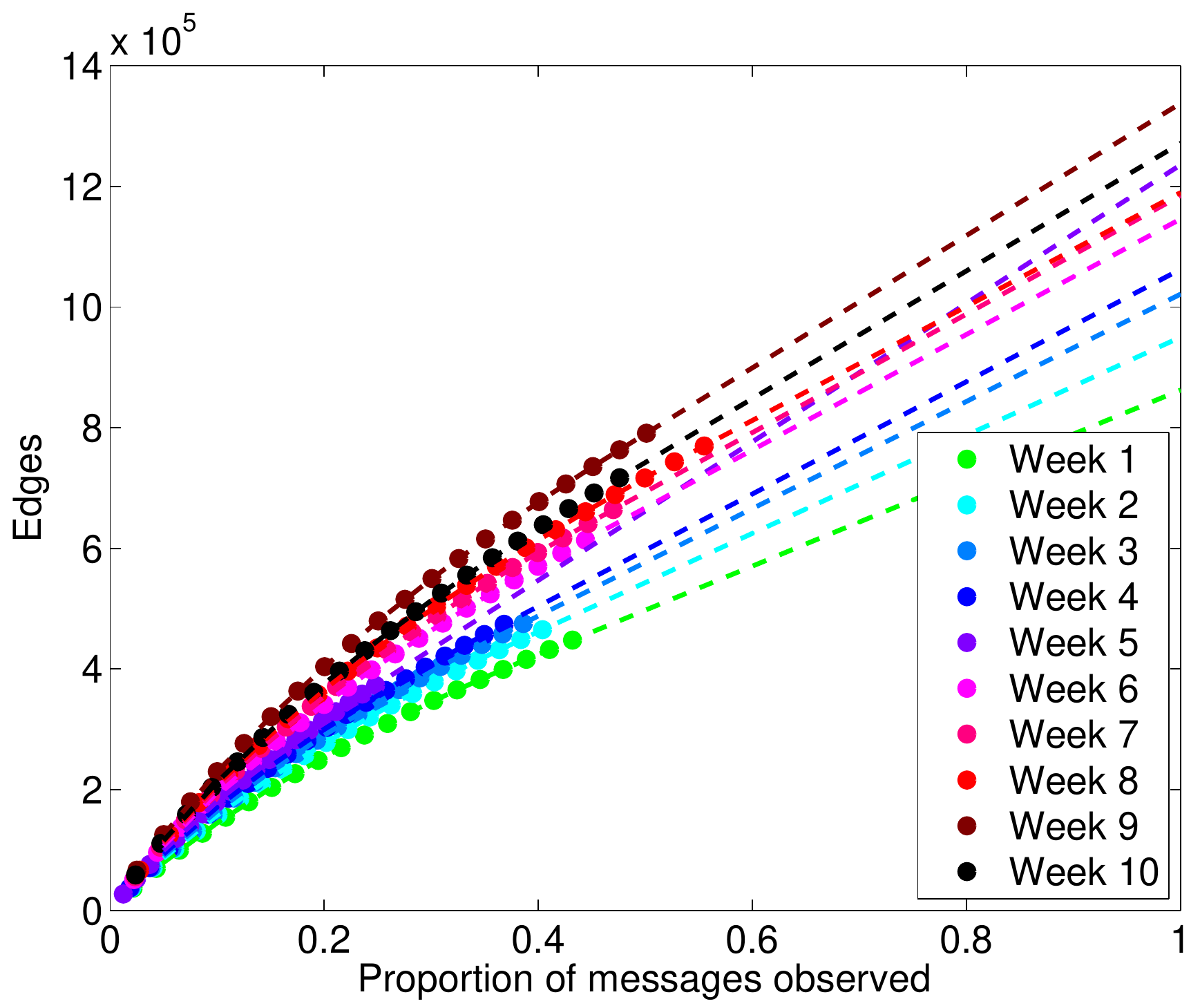}}
\caption[Predicted number of edges in Twitter reply networks]{Predicted number of edges in Twitter reply networks. (a.) A small proportion of observed messages for Week 5 ($<$ 25\%) may explain the spike in the estimated number of edges for that week. (b.) Each data point represents the number of directed edges observed, averaged over 100 simulated subsampling experiments. The dashed line extrapolates the predicted number of edges for greater proportions of sampled data.} 
\label{fig:predicted_edges_twitter}
\end{figure*}

\begin{figure*}[!ht]
\centering
\subfigure[Edge weights]{\includegraphics[width=.45\textwidth]{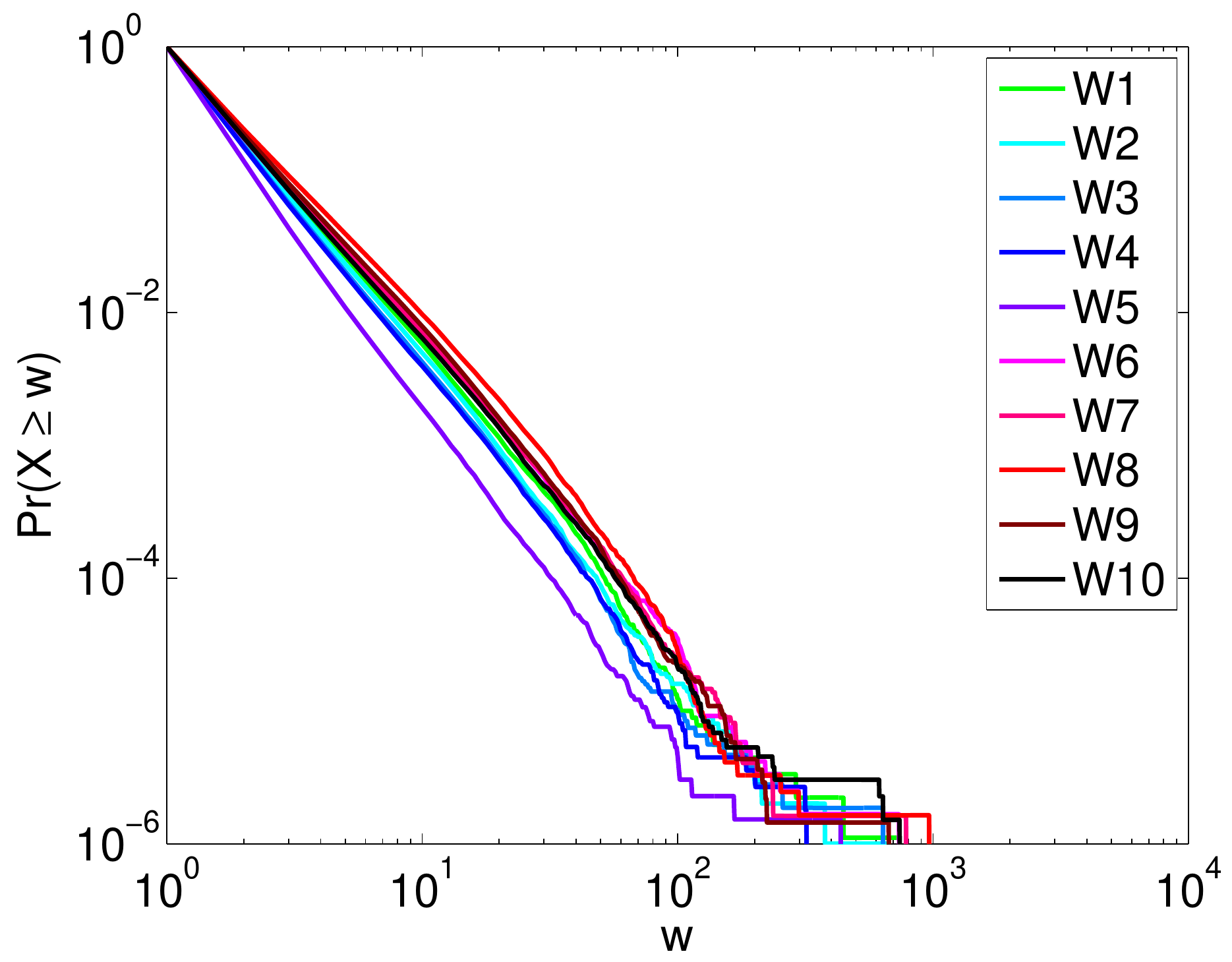}}\\
\subfigure[$Pr(k_{\rm{in}})$]{\includegraphics[width=.45\textwidth]{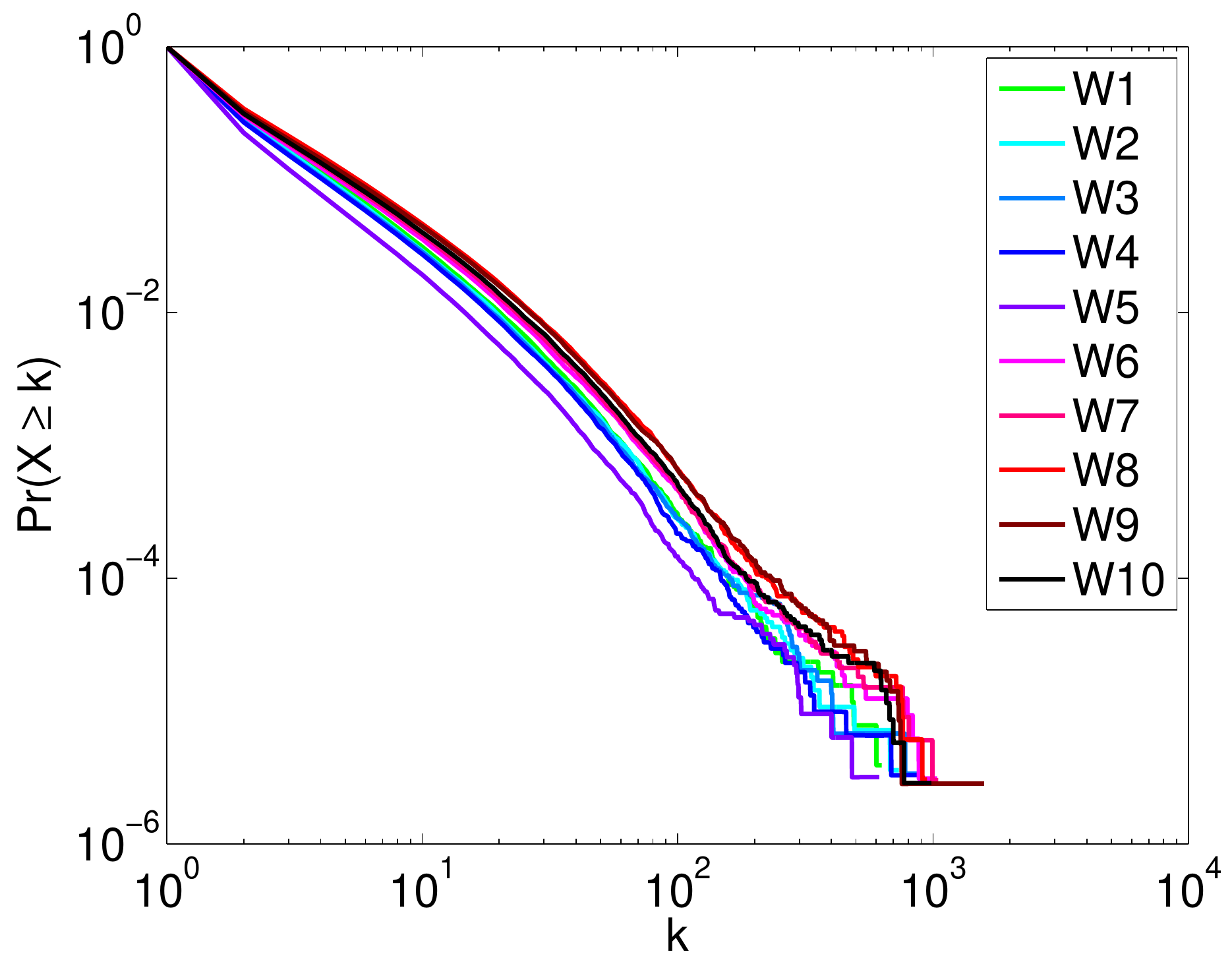}}
\subfigure[$Pr(k_{\rm{out}})$]{\includegraphics[width=.45\textwidth]{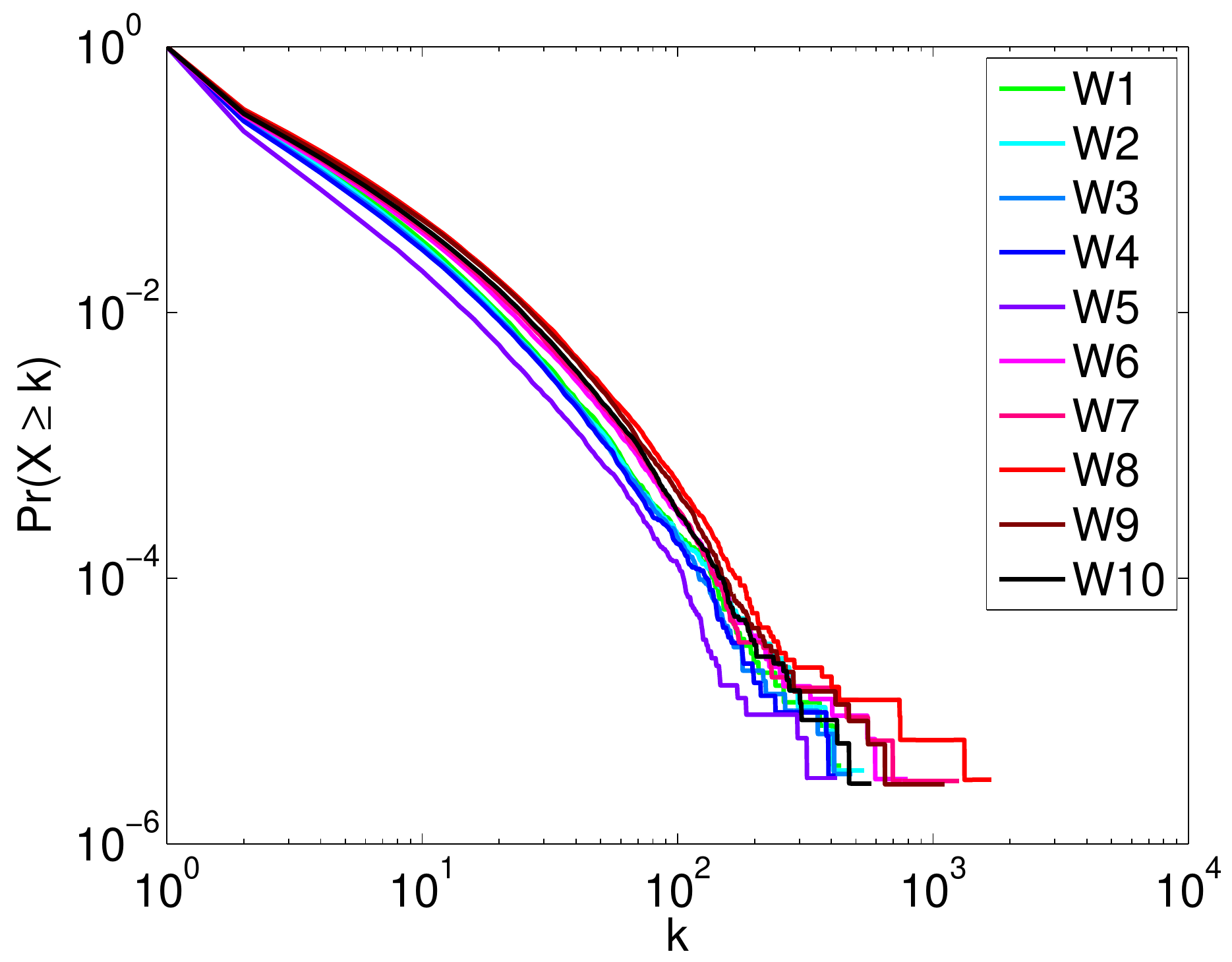}}
\caption[Predicted edge weight and degree distributions for Twitter reply networks]{Predicted edge weight and degree distributions for Twitter reply networks. (a.) The predicted edge weight distribution. (b.) Predicted $Pr(k_{\rm{in}})$ and (c.) $Pr(k_{\rm{out}})$ for Twitter reply networks.}
\label{fig:Twitter_edge_weight_distribution}
\end{figure*}

The number of edges can be predicted using Equations~\ref{eq:HT_estimator_sampled_interactions_edges} and~\ref{eq:approx_edges_sampled_interactions}. We present our results in Figure~\ref{fig:predicted_edges_twitter}. In all cases, the number of edges increases throughout the period of the study. Figure~\ref{fig:Twitter_edge_weight_distribution} depicts the predicted edge weight and degree distributions. The edge weight distribution shows that very few ($< .001\%$) edges have weight greater than $10^2$. The degree distribution of the observed subnetwork can be rescaled by reassigning nodes of degree $k$, to nodes of degree $\frac{\hat{M}}{m}k$. Figure~\ref{fig:Twitter_edge_weight_distribution} demonstrates a slightly heavier tail in the in-degree distribution as compared to the out-degree distribution. The degree distribution reveals that fewer than .01\% of the nodes have more than $10^2$ distinct neighbors. This value is approximately Dunbar's number, a value suggested to be the upper limit on the number of active social contacts for humans~\cite{Dunbar1992}.  

\subsection{Average degree}
\begin{figure*}[!ht]
\centering
\subfigure[$k_{\rm avg,in}$]{\includegraphics[width=.45\textwidth]{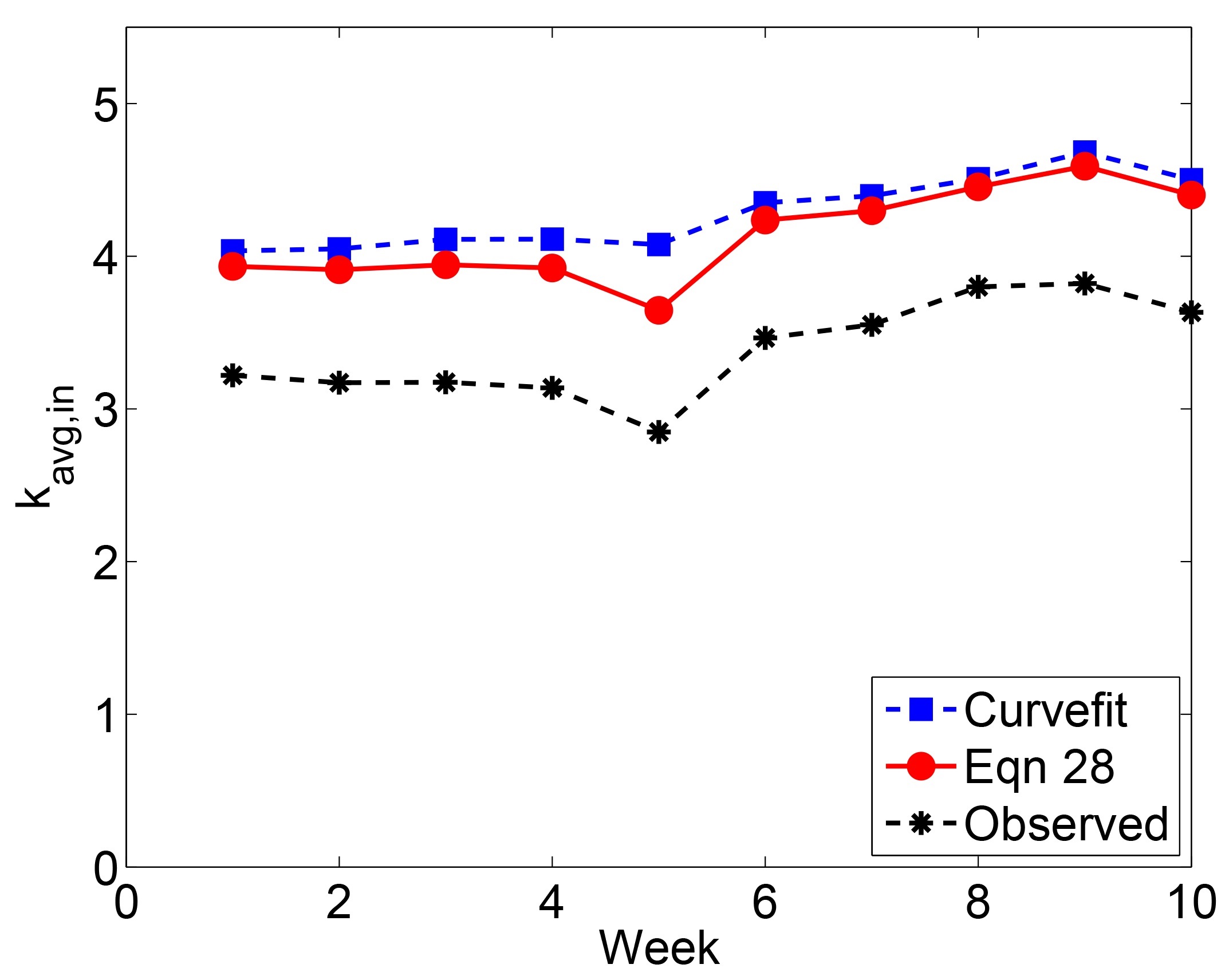}}
\subfigure[$k_{\rm avg,out}$]{\includegraphics[width=.45\textwidth]{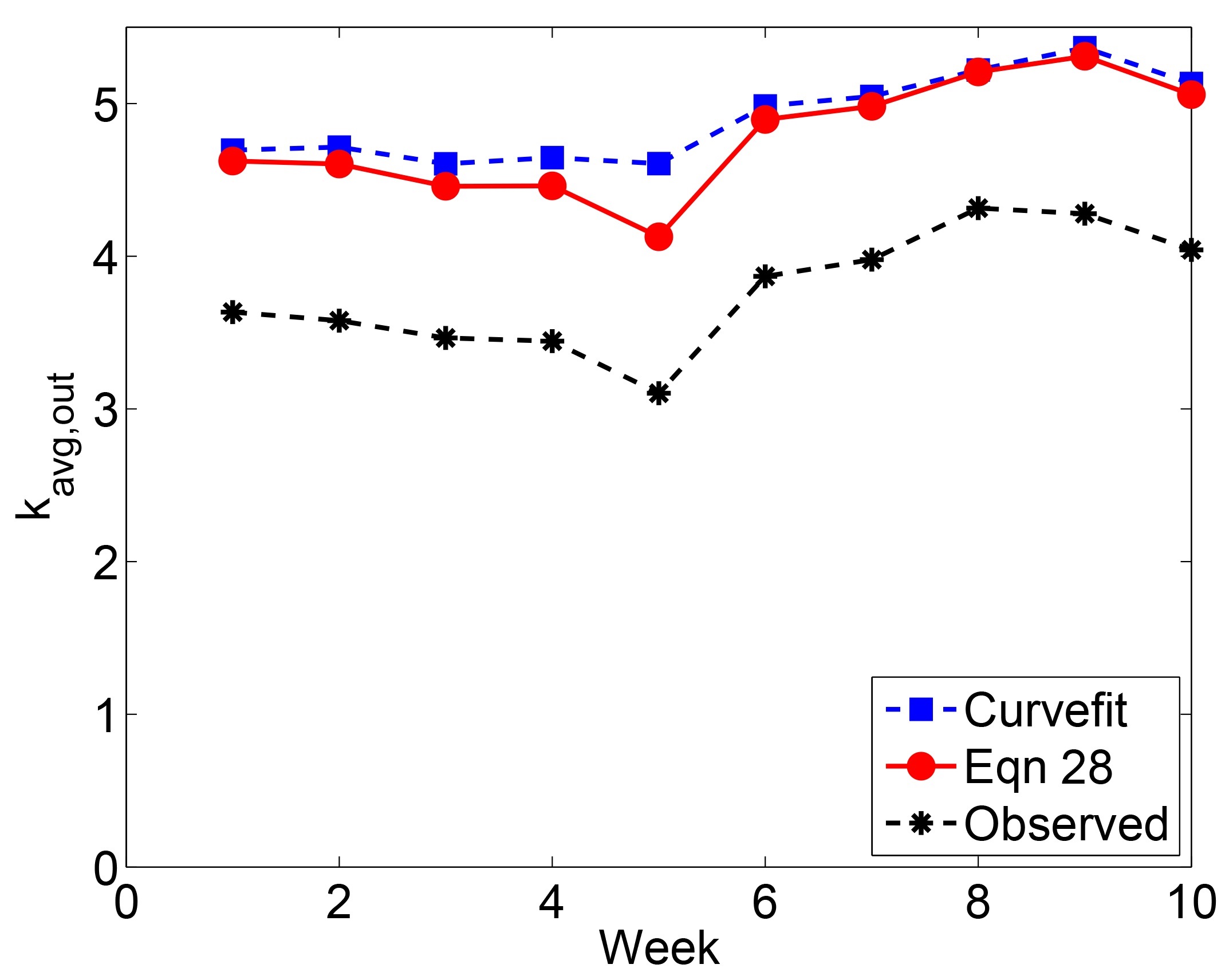}}
\caption[Predicted $k_{avg,in}$ and $k_{avg,out}$ in Twitter reply networks]{Predicted $k_{\rm avg,in}$ and $k_{\rm avg,out}$ in Twitter reply networks.}
\label{fig:predicted_avk_twitter}
\end{figure*}

\begin{figure*}[!ht]
\centering
\subfigure[$k_{\rm avg,in}$]{\includegraphics[width=.45\textwidth]{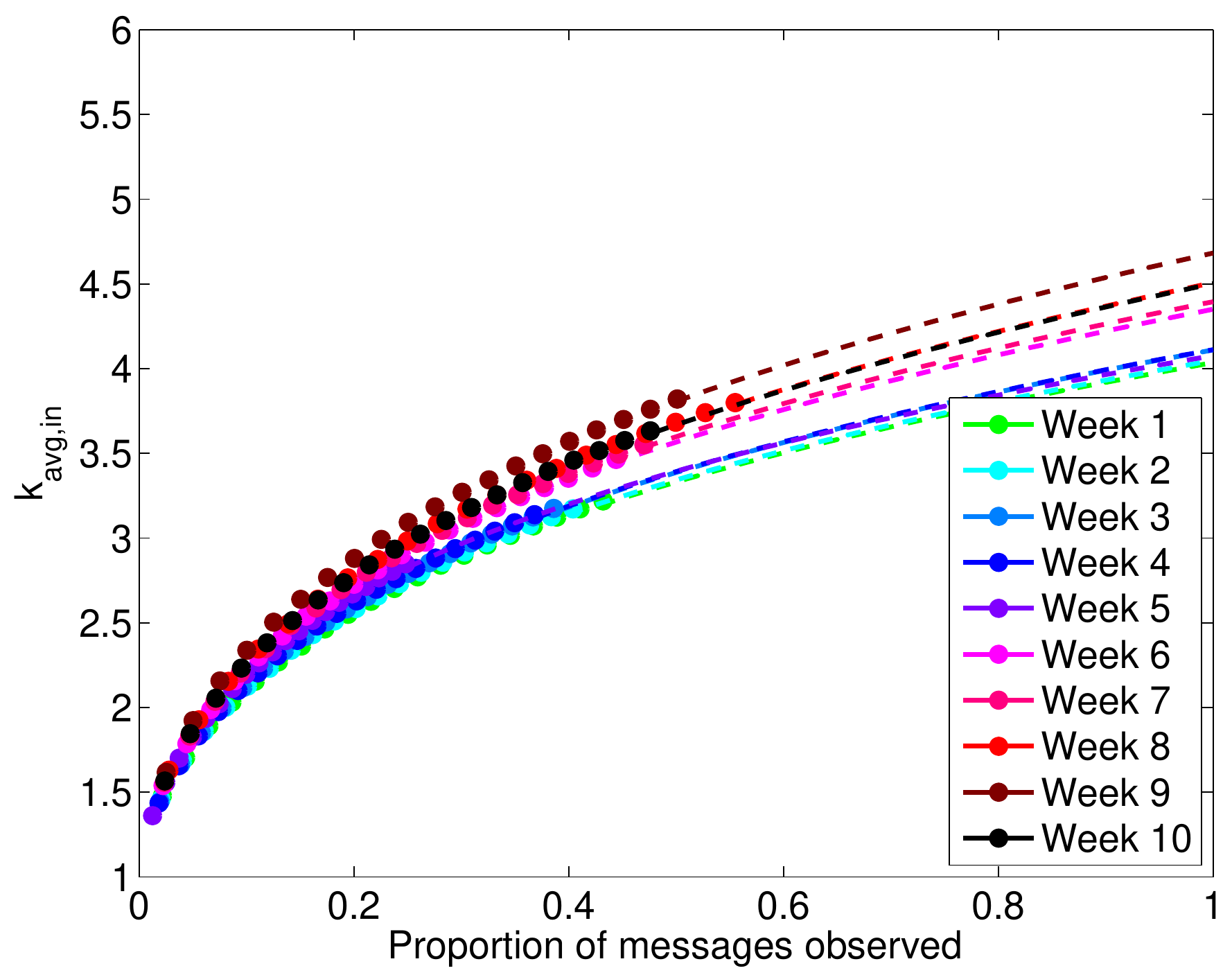}}
\subfigure[$k_{\rm avg,out}$]{\includegraphics[width=.45\textwidth]{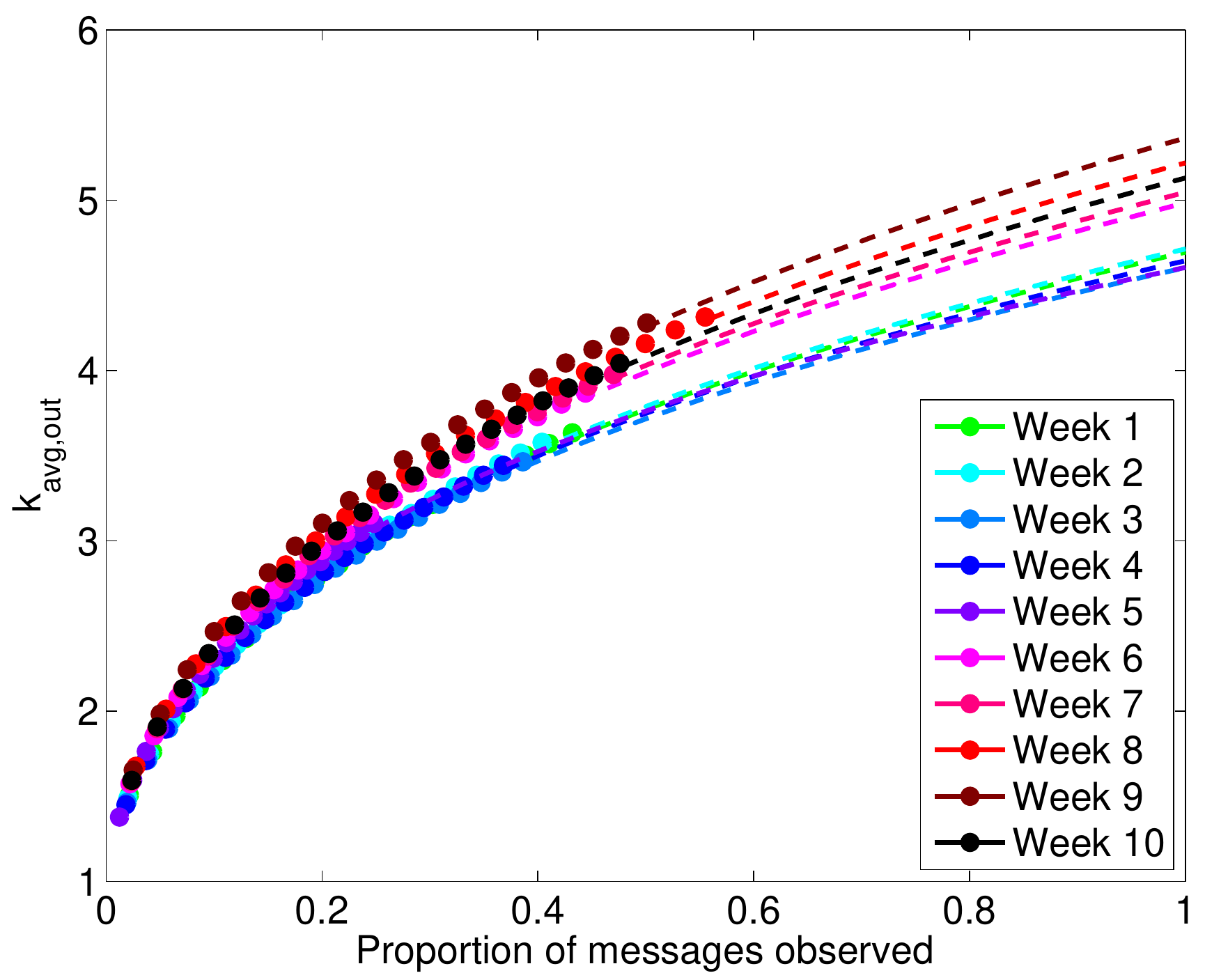}}\\
\caption[$k_{\rm avg, in}$ and $k_{\rm avg, in}$ for Twitter reply networks]{$k_{\rm avg, in}$ and $k_{\rm avg, in}$ for Twitter reply networks. Each data point represents the observed average in- and out-degree, averaged over 100 simulated subsampling experiments. The dashed line extrapolates the predicted number of edges for greater proportions of sampled data. }
\label{fig:Twitter_RRN_stats}
\end{figure*}

Once the number of nodes and edges have been predicted for the network, we may simply compute the average degree as $\hat{k}_{\rm{avg},\rm{in}}=\frac{\hat{M}}{\hat{N}_{\rm{receivers}}}$ and $\hat{k}_{\rm{avg},\rm{out}}=\frac{\hat{M}}{\hat{N}_{\rm{repliers}}}$. Upon doing so, we find that the average degree for Twitter reply networks is between 4 and 5 (Fig.~\ref{fig:predicted_avk_twitter}). We find that the average in-degree is less than the average out-degree (Fig.~\ref{fig:Twitter_RRN_stats}).

\subsection{Maximum degree}
\begin{figure*}[!ht]
\centering
\subfigure[$k_{\max,\rm{in}}$]{\includegraphics[width=.45\textwidth]{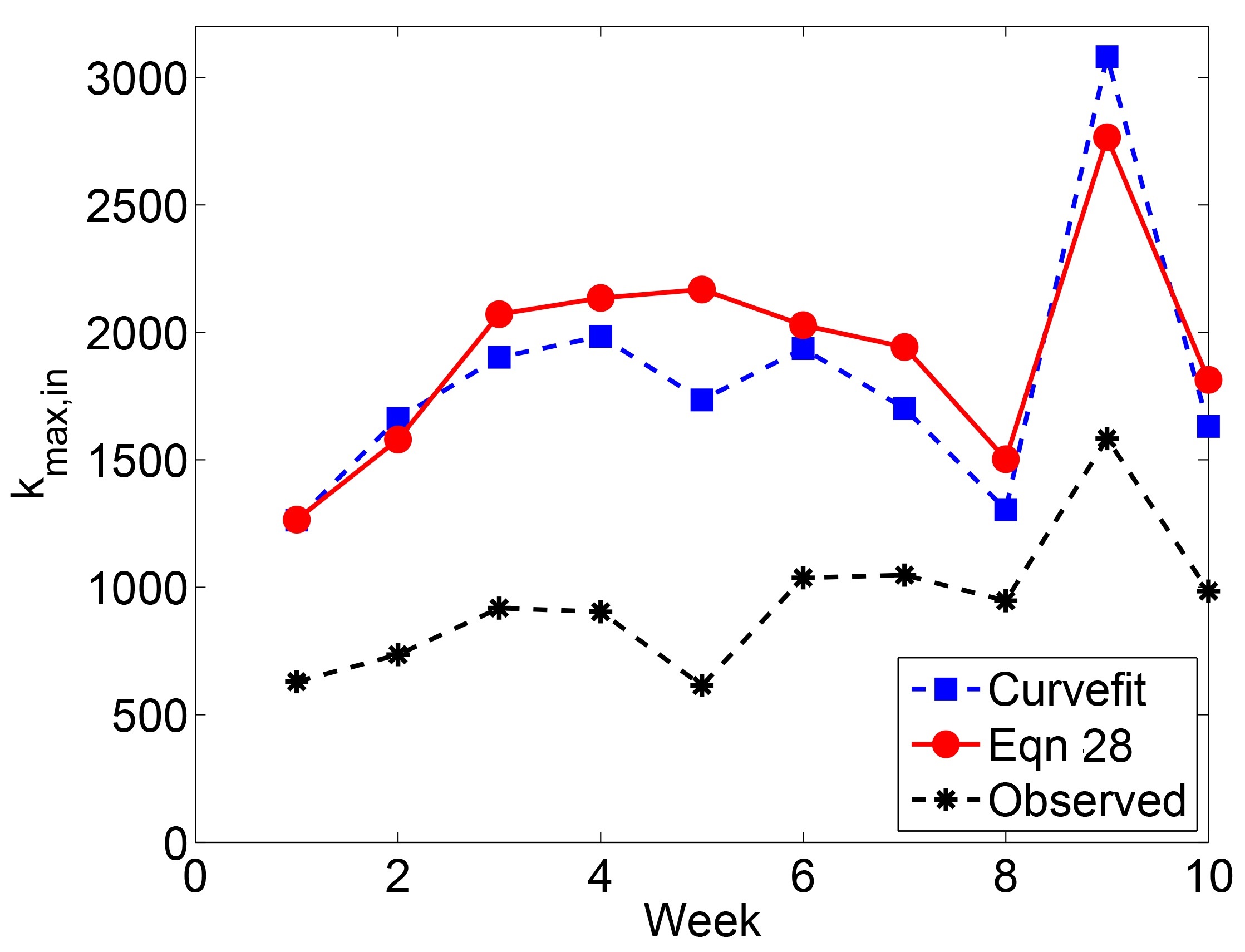}}
\subfigure[$k_{\max,\rm{out}}$]{\includegraphics[width=.45\textwidth]{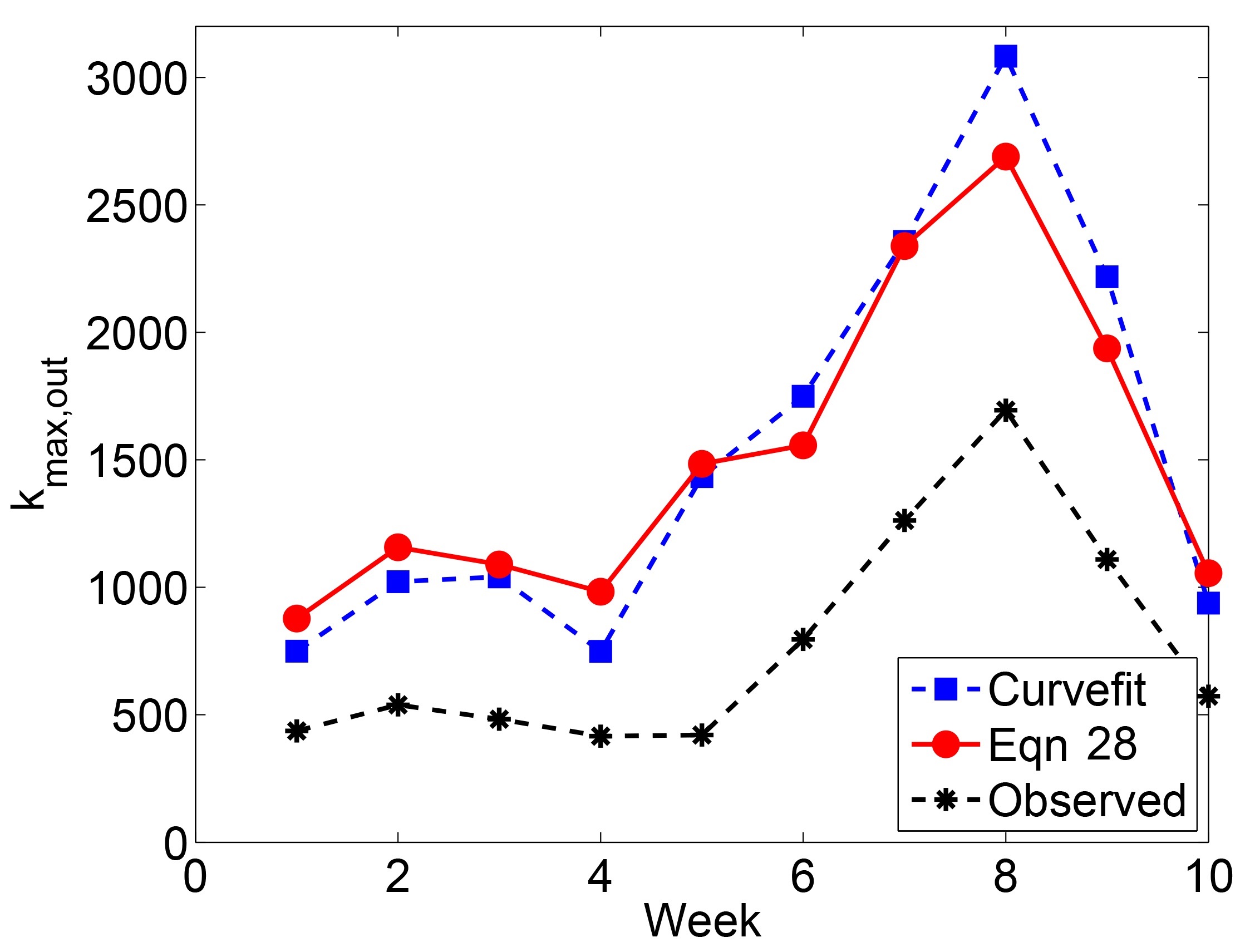}}
\caption[Predicted $k_{\max,\rm{in}}$ and $k_{\max,\rm{out}}$ in Twitter reply networks]{Predicted $k_{\max,\rm{in}}$ and $k_{\max,\rm{out}}$ in Twitter reply networks.}
\label{fig:predicted_maxk_twitter}
\end{figure*}

\begin{figure*}[!ht]
\centering
\subfigure[$k_{\rm max,\rm{in}}$]{\includegraphics[width=.45\textwidth]{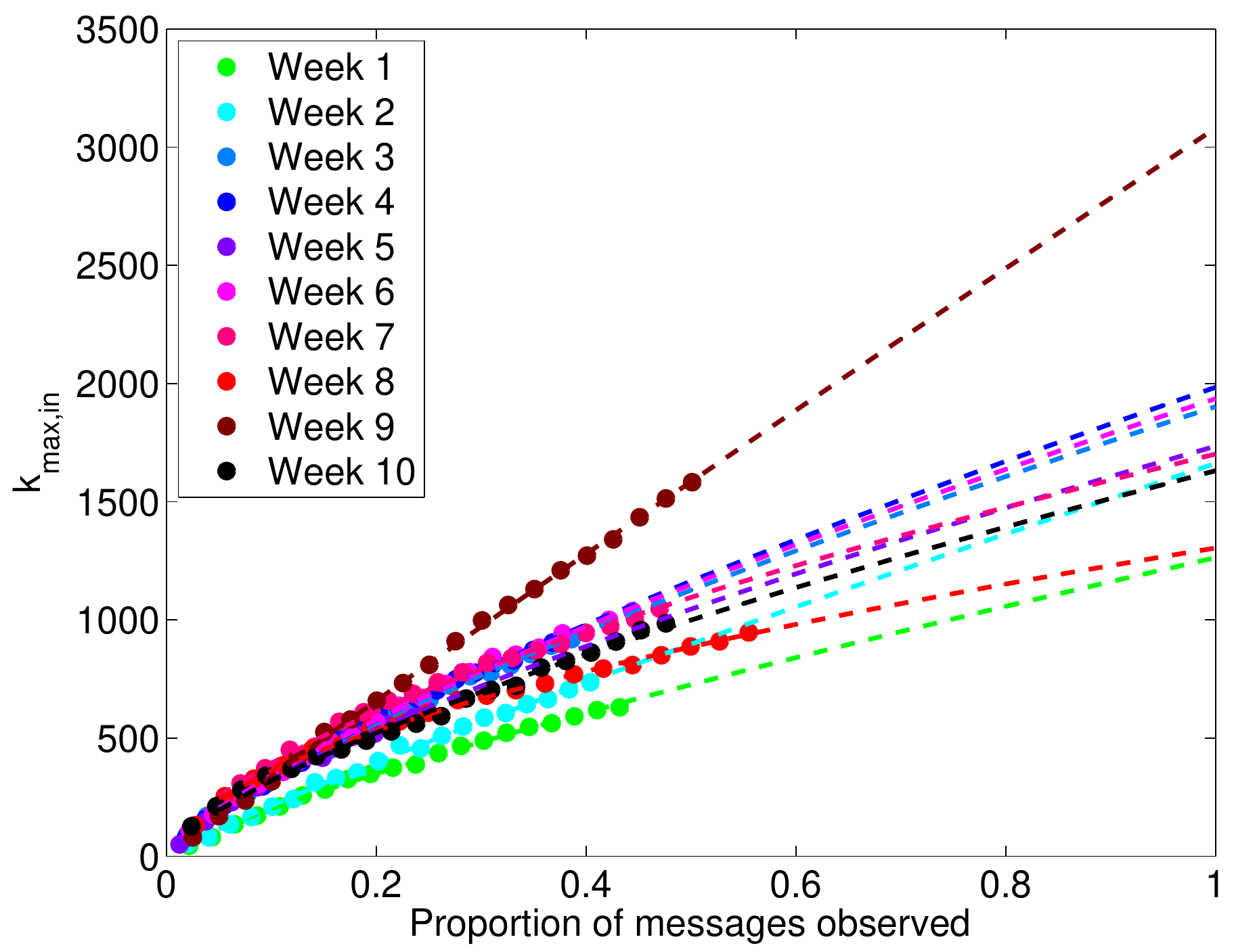}} 
\subfigure[$k_{\rm max,\rm{out}}$]{\includegraphics[width=.45\textwidth]{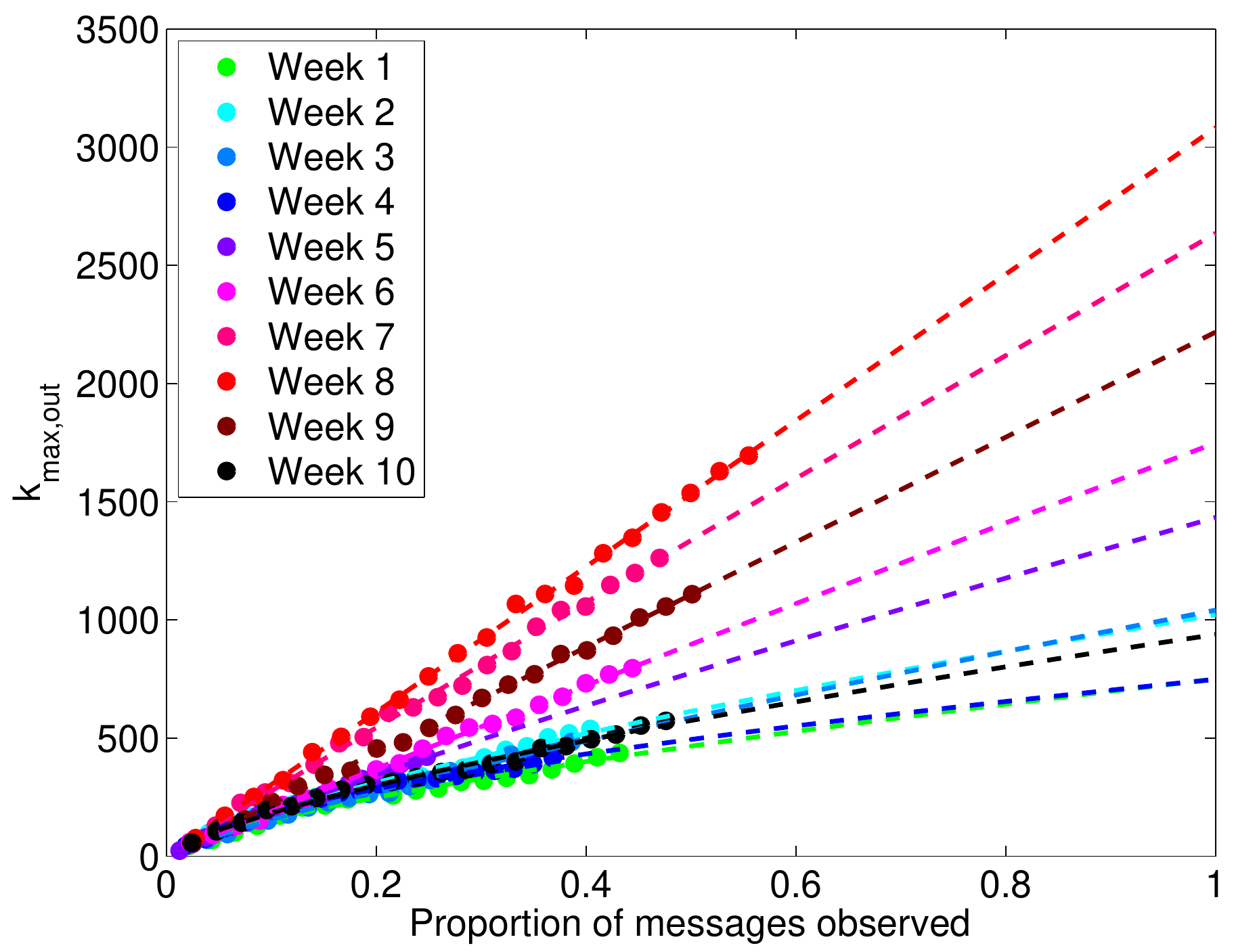}}\\
\caption[$k_{\max, \rm{in}}$ and $k_{\max, \rm{in}}$ for Twitter reply networks]{$k_{\max, \rm{in}}$ and $k_{\max, \rm{in}}$ for Twitter reply networks. Each data point represents the observed maximum in- and out-degree, averaged over 100 simulated subsampling experiments. The dashed line extrapolates the predicted number of edges for greater proportions of sampled data. }
\label{fig:Twitter_RRN_max_degree}
\end{figure*}

The maximum degree simply scales in proportion to the probability of edge inclusion. Since the probability of edge inclusion is no longer $q$, as in the case of sampling by links, we may approximate the probability of edge inclusion by $\frac{m}{\hat{M}}$ and thus $\hat{k}_{\max} = \frac{\hat{M}}{m} k^{\textnormal{obs}}_{\max}$. The predicted maximum degree for Twitter reply networks is shown in Figures~\ref{fig:predicted_maxk_twitter} and~\ref{fig:Twitter_RRN_max_degree}.

\section{Discussion}
Network measures derived from empirical observations will often be poor estimators of the true underlying network structure of the system. We have explored four sampling regimes: (1) subnetworks induced on randomly sampled nodes, (2) subnetworks obtained when all nodes are known and some links fail or are hidden, (3) subnetworks generated from randomly sampled links, and (4) weighted subnetworks generated by randomly sampled interactions. We have described how network statistics scale under these regimes via sampling experiments on simulated and empirical networks. Our paper advances an understanding of how network statistics scale, and more importantly how to correct for missing data when the proportion of missing nodes, links or interactions is known. 

A major obstacle to generating scaling techniques for subnetworks generated by sampled links or interactions has previously been the lack of a practical method for estimating the true degree distribution or node strength distribution. Problematically, the random selection of links creates a biased sample of nodes whereby hubs are more likely to be detected, and nodes of small degree are more likely to go undetected. Although scaling methods have been suggested, they are based on knowledge of (or a reasonable estimate of) the degree or node strength distribution~\cite{Kolaczyk2009}. In this paper, we have overcome this obstacle by our proposed scaling techniques for the degree distribution and apply this to several simulated and empirically derived networks with reasonably good results.

Very few studies have addressed the missing data problem in empirically studied networks, such as those constructed from tweets. An exception is work by Morstatter et al.~\cite{morstatter2013sample} who compared network statistics for the current Twitter's Spritzer ($\approx$ 1\% of all tweets) to the full Firehose (100\% of all tweets), however no methods for scaling from data collected via the API were suggested. 

We concluded our work by applying our derived scaling methods to Twitter reply networks. Our work supports Dunbar's hypothesis which suggests that individuals maintain an upper limit of roughly 100-150 contacts each week~\cite{Dunbar1992}. Further evidence for this hypothesis comes from previous work in link prediction. Bliss et al.~\cite{bliss2013evolutionary} detect the Resource Allocation index to often evolve a large, positive weight--thus contributing heavily (and positively) in the prediction of new links. This index considers the amount of time and attention one individual has as a ``social resource'' to spend in the social network and assumes that each node will distribute its resource equally among all neighbors. Although the presence of hubs is suggestive of preferential attachment, it is clear that the constraints of time and attention limit truly scale-free behavior in weekly Twitter reply networks. We find that the number of individuals who make replies is less than the number of individuals who receive replies.   

One limitation of our work is that our scaling methods are based upon the assumption that the sampling fraction, $q$ is known, while in practice this need not be the case. In cases where one may establish an upper and lower bound for $q$, our methods could be used to help establish bounds for the predicted network measures. In some cases, particularly when sampling by links or interactions, small changes in $q$ may have relatively little impact on the predicted statistics, especially for large $q$. Future work that seeks to classify subnetworks by network class based on signature subsampling properties may also prove to be fruitful. With some knowledge of network class or generative model, methods for estimating $q$ may be possible. Additionally, efforts to predict structural holes in networks from localized information may also greatly advance the field~\cite{bagrow2013shadow}.

To our knowledge, this is the first attempt provide scaling methods for $\kmax$. While our scaling techniques for predicting $\kmax$ perform well for several networks, they did not perform as well on simulated networks with a regularized structure.\footnote{Our rewiring probability for the simulated Small world networks was quite low, with $p=0.1$. Our methods perform well on other networks which are known to exhibit to Small world structure, such as our empirical networks Powergrid and \textit{C. elegans}.} Future work which detects and accounts for motif distributions may improve upon our efforts here.

With an increased interest in large, networked datasets, we hope that continued efforts will aid in the understanding of how subsampled network data can be used to infer properties of the true underlying system. Our methods advance the field in this direction, not only adding to the body of literature surrounding sampling issues and Twitter's API~\cite{morstatter2013sample}, but also to the growing body of literature on incomplete network data.
\FloatBarrier
\section{Acknowledgments}
The authors acknowledge the Vermont Advanced Computing
Core and support by NASA (NNX-08AO96G) at the University
of Vermont for Providing High Performance Computing resources
that have contributed to the research results reported within this
paper. CAB and PSD were funded by an NSF CAREER Award to PSD (\# 0846668). CMD and PSD were funded by a grant from the MITRE Corporation.
\newpage

\bibliographystyle{unsrt}
\bibliography{example4}
\newpage
\onecolumn
\section{Appendix}
  \renewcommand{\thefigure}{A\arabic{equation}}    
  \renewcommand{\thetable}{A\arabic{equation}}  

\noindent Derivation of Equation (2)
\noindent Restating Equation (1) for convenience, with $K=\kmax$ for ease of notation, we have 
\begin{align*}
\tilde{P_k} = \sum^{K}_{i=k} \binom{i}{k} q^k \left( 1-q \right)^{i-k} P_i. 
\end{align*}
\noindent Assume a finite network with maximum degree, $K$. Equation (1) may be rewritten as:
\footnotesize{
\begin{align*}
\left[ \begin{array}{c}\tilde{P}_1\\[1.3em]
\tilde{P}_2 \\[1.2em]
\tilde{P}_3 \\[1.2em]
\vdots \\[1.2em]
\tilde{P}_{K-2} \\[1.2em]
\tilde{P}_{K-1} \\[1.2em]
\tilde{P}_{K}\end{array} \right] &= \left[ \begin{array}{cccccc}
\binom{1}{1} q^1 \left( 1-q \right)^{1-1}  + &\binom{2}{1} q^1 \left( 1-q \right)^{2-1}  + &\binom{3}{1} q^1 \left( 1-q \right)^{3-1}  &+    \ \ \  \hdots \ \ \ +    & \binom{K-1}{1} q^1 \left( 1-q \right)^{K-2} +   &\binom{K}{1} q^1 \left( 1-q \right)^{K-1} \\[1em]
 &  \binom{2}{2} q^2 \left( 1-q \right)^{2-2} + & \binom{3}{2} q^2 \left( 1-q \right)^{3-2}  +  & + \ \ \  \hdots \ \ \  +  & \binom{K-1}{2} q^2 \left( 1-q \right)^{K-3}  + & \binom{K}{2} q^2 \left( 1-q \right)^{K-2}  \\[1em]
 & & \binom{3}{3} q^3 \left( 1-q \right)^{3-3} & + \ \ \  \hdots \ \ \ + & \binom{K-1}{3} q^3 \left( 1-q \right)^{K-4} +& \binom{K}{3} q^3 \left( 1-q \right)^{K-3} \\[1em]
 & & & &  & \vdots \\[1em]
 & &  & \ddots    &    \vdots   & \vdots\\[1em]
  &   & &      &  \binom{K-1}{K-1} q^{K-1} \left( 1-q \right)^{0}  + & \binom{K}{K-1} q^{K-1} \left( 1-q \right)^{1} \\[1em]
 & & & & & \binom{K}{K} q^K \left( 1-q \right)^{0}  \end{array} \right]\left[ \begin{array}{c} P_1 \\[1.2em] P_2 \\[1.2em] P_3 \\[1.2em] \vdots \\ [1.2em]
 P_{K-2} \\[1.2em] P_{K-1} \\[1.2em] P_{K} \end{array}\right]\end{align*}
}
\noindent Back solving for $P_K$ yields:
\begin{align*}
P_k &= \frac{1}{q^K }\tilde{P}_K 
\end{align*}
\noindent Continuing, we solve for $P_{K-1}$. 
\begin{align*}
\tilde{P}_{K-1} &=  \binom{K-1}{K-1} q^{K-1} \left( 1-q \right)^{0} P_{K-1} + \binom{K}{K-1} q^{K-1} \left( 1-q \right)^{1} P_K\\
 P_{K-1} &= \frac{\tilde{P}_{K-1} - \binom{K}{K-1}q^{K-1}(1-q)P_K}{q^{K-1} } \\
P_{K-1} &= \frac{\tilde{P}+{K-1} - \binom{K}{K-1}q^{K-1}(1-q) \left( \frac{\tilde{P}_K}{q^K} \right) }{q^{K-1} } \\
P_{K-1} &= \frac{\tilde{P}_{K-1}}{q^{K-1}} - \frac{K (1-q)}{q^K} \tilde{P}_K 
\end{align*}
\noindent Solving for $P_{K-2}$ yields,
\begin{align*}
\tilde{P}_{K-2} &=  \binom{K-2}{K-2} q^{K-2} \left( 1-q \right)^{0} P_{K-2} + \binom{K-1}{K-2} q^{K-2} \left( 1-q \right)^{1} P_{K-1} + \binom{K}{K-2} q^{K-2} \left( 1-q \right)^{2} P_K\\
 P_{K-2} &= \frac{ \tilde{P}_{K-2} - \binom{K-1}{K-2} q^{K-2} \left( 1-q \right)^{1} P_{K-1} - \binom{K}{K-2} q^{K-2} \left( 1-q \right)^{2} P_K}{q^{K-2}} \\
P_{K-2} &= \frac{ \tilde{P}_{K-2}}{q^{K-2}} - (K-1)( 1-q ) P_{K-1} - \frac{K(K-1)}{2} ( 1-q )^{2} P_K 
\end{align*}

\noindent Now, using our previous results, we have
\begin{align}
P_{K-2} &= \frac{ \tilde{P}_{K-2}}{q^{K-2}} - (K-1)( 1-q ) \left( \frac{\tilde{P}_{K-1}}{q^{K-1}} - \frac{K (1-q)}{q^K} \tilde{P}_{K} \right) - \frac{K(K-1)}{2} ( 1-q )^{2} \left( \frac{\tilde{P}_{K}}{q^K} \right)\\
P_{K-2} &= \frac{ \tilde{P}_{K-2}}{q^{K-2}} - \frac{(K-1)(1-q)}{q^{K-1}} \tilde{P}_{K-1} + \frac{K(K-1)(1-q)^2}{q^K} \tilde{P}_{K}  - \frac{K(K-1)( 1-q )^{2}}{2q^K} \tilde{P}_{K}\\
P_{K-2} &= \frac{ \tilde{P}_{K-2}}{q^{K-2}} - \frac{(K-1)(1-q)}{q^{K-1}} \tilde{P}_{K-1} + \frac{K(K-1)(1-q)^2}{2q^K} \tilde{P}_{K} 
\end{align}

\noindent Proceeding, we may generalize our result as 
\begin{align}
\hat{P}_{K-n} = \sum^{K}_{i=K-n} \frac{(-1)^{i-(K-n)} \binom{i}{K-n} (1-q)^{i-(K-n)}}{q^i} \tilde{P}_i,
\end{align}
\noindent which is
\begin{align}
\hat{P}_{k} = \sum^{K}_{i=k} \frac{(-1)^{i-k} \binom{i}{k} (1-q)^{i-k}}{q^i} \tilde{P}_i,
\end{align}
\noindent with $k-K-n$.

\begin{proof}[Verification of Equation 2]
\noindent We will demonstrate that our derivation for the predicted degree distribution, $\hat{P}_{k},$ agrees with the true degree distribution $P_{k}$.  Observe that
{\footnotesize
\begin{align*}
\hat{P}_{k}&=\sum^{K}_{i=k} \frac{(-1)^{i-k} \binom{i}{k} \left(1-q\right)^{i-k}}{q^i} \tilde{P}_i,\\
&=\underbrace{ \tilde{P}_{k}}_{i=k} - \underbrace{\frac{ \binom{k+1}{k} \left(1-q\right)^1}{q^{k+1}} \tilde{P}_{k+1}}_{i=k+1} + \hdots \underbrace{\frac{(-1)^{K-k} \binom{K}{k} \left(1-q\right)^{K-k}}{q^K} \tilde{P}_K}_{i=K}\\
&=\underbrace{\sum^{K}_{j=k} \binom{j}{k} q^{k} \left(1-q\right)^{j-k} P_j }_{i=k} - \underbrace{\frac{ \binom{k+1}{k} \left(1-q\right)^1}{q^{k+1}} \sum^{K}_{j=k+1} \binom{j}{k+1} q^{k+1} \left(1-q\right)^{j-(k+1)} P_j}_{i=k+1} \\
& \hspace{2cm}+  \hdots \underbrace{\frac{(-1)^{K-k} \binom{K}{k} \left(1-q\right)^{K-k}}{q^K} \sum^{K}_{j=K} \binom{j}{K} q^K \left(1-q\right)^{j-K} P_j}_{i=K}\\
&=\underbrace{\sum^{K}_{j=k} \binom{j}{k} q^{k} \left(1-q\right)^{j-k} P_j }_{i=k} - \underbrace{ \binom{k+1}{k} \left(1-q\right)^1 \sum^{K}_{j=k+1} \binom{j}{k+1} \left(1-q\right)^{j-(k+1)} P_j}_{i=k+1} \\
& \hspace{2cm}+  \hdots \underbrace{(-1)^{K-k} \binom{K}{k} \left(1-q\right)^{K-k} \sum^{K}_{j=K} \binom{j}{K}  \left(1-q\right)^{j-K} P_j}_{i=K}
\end{align*}}
\noindent Expanding the sums in all terms and collecting powers of $(1-q)$ yields
{\footnotesize
\begin{align*}
\hat{P}_{k}= P_{k} +\binom{k+1}{k}(1-q)P_{k+1} +\binom{k+2}{k}(1-q)^2P_{k+2} + \hdots &+ \binom{K}{k}(1-q)^{K-k}P_{K}\\
        -\binom{k+1}{k}(1-q)P_{k+1} -\binom{k+1}{k}\binom{k+2}{k+1}(1-q)^2P_{k+2} - \hdots &- \binom{k+1}{k}\binom{K}{k}(1-q)^{K-k}P_K\\
        &\vdots\\
        &(-1)^{{K-k}} \binom{K}{k}(1-q)^{K-k} P_K
\end{align*}}
\noindent For each power of $(1-q)$, we have 
\begin{align*}
\sum^{a}_{i=0} (-1)^{a} \binom{k+i}{k} \binom{k+b}{k+i} (1-q)^a P_{k+a} &= \sum^{a}_{i=0} (-1)^{a} \binom{k+a}{k} \binom{a}{i} (1-q)^a P_{k+a}\\
&=\binom{k+a}{k} \left(\sum^{a}_{i=0} (-1)^{a}  \binom{a}{i} \right)(1-q)^a P_{k+a} \\
&=\begin{cases}
\binom{k+a}{k} \left(0 \right)(1-q)^a P_{k+a}, & \text{if }a \geq 1\\
 \binom{k+a}{k} \left(1 \right)(1-q)^a P_{k+a}, & \text{if }a=0\\
 \end{cases}\\
&=\begin{cases}
0, & \text{if }a \geq 1\\
  P_{k}, & \text{if }a=0\\
 \end{cases}
\end{align*}
\noindent whereby we have made use of the binomial theorem $(x+y)^a=\sum^a_{i=0} \binom{a}{i}(x^{a-i}y^i$ for $x=1, y=-1$ which implies that $0=\sum^{a}_{i=0} (-1)^{a}  \binom{a}{i},$ for $a \geq 1$. Thus, all terms of $\hat{P}_{k}$ cancel, except for the $P_{k}$ and so $\hat{P}_{k}=P_{k}$.
\end{proof}
\setcounter{equation}{1}
\begin{figure*}[!ht]
\centering
\subfigure[Nodes]{\includegraphics[width=.28\textwidth]{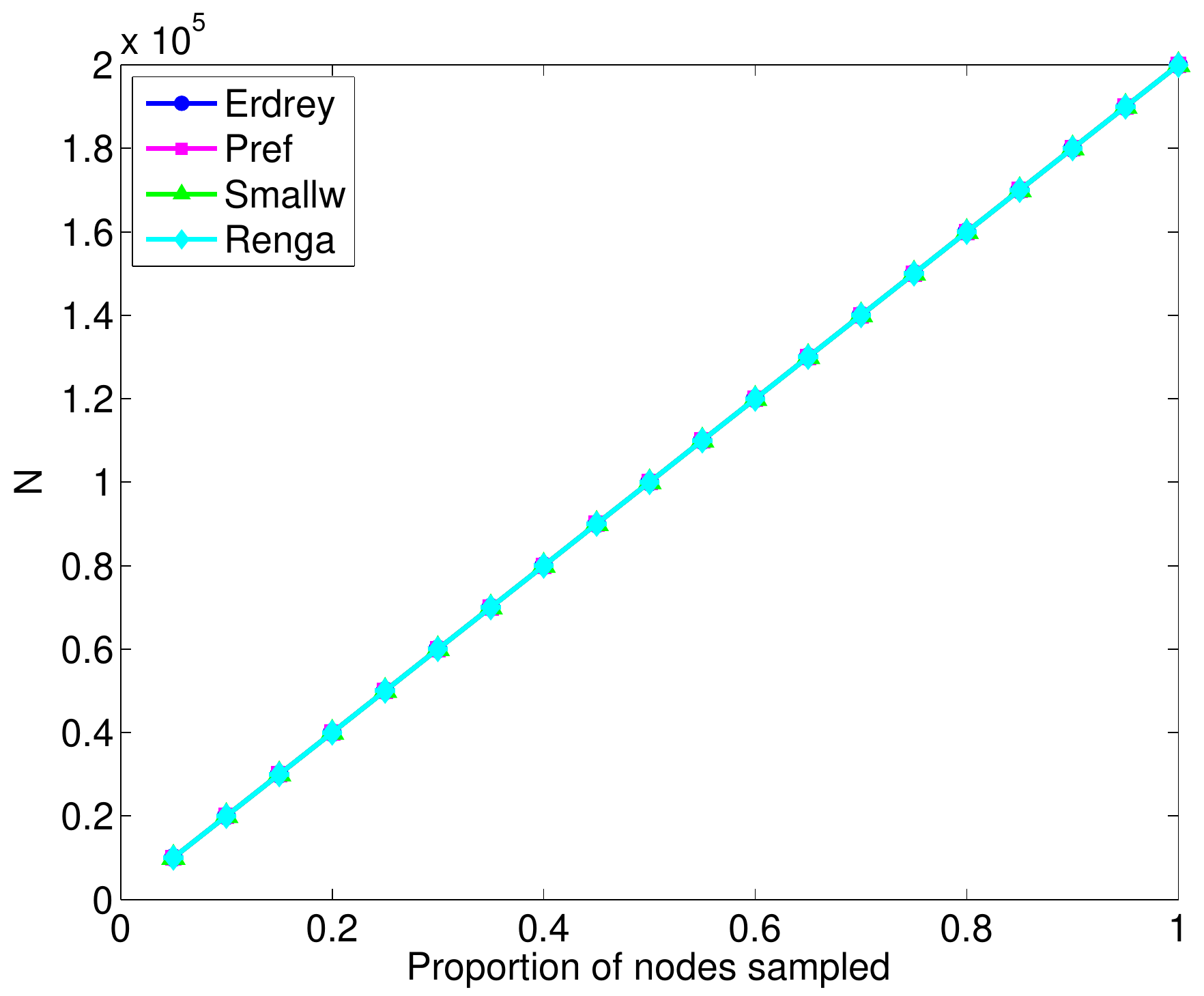}}
\subfigure[Edges]{\includegraphics[width=.28\textwidth]{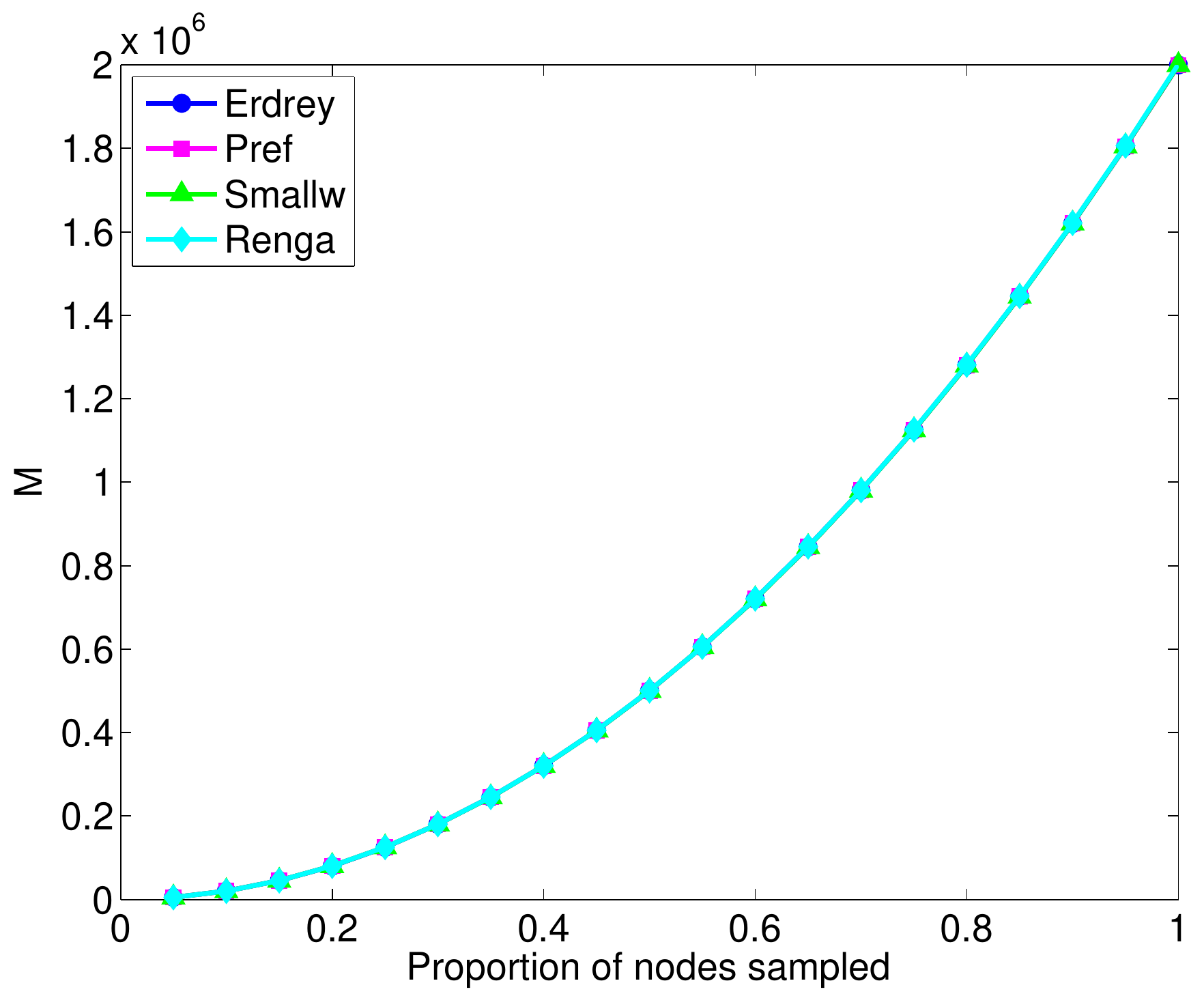}}
\subfigure[Average degree]{\includegraphics[width=.28\textwidth]{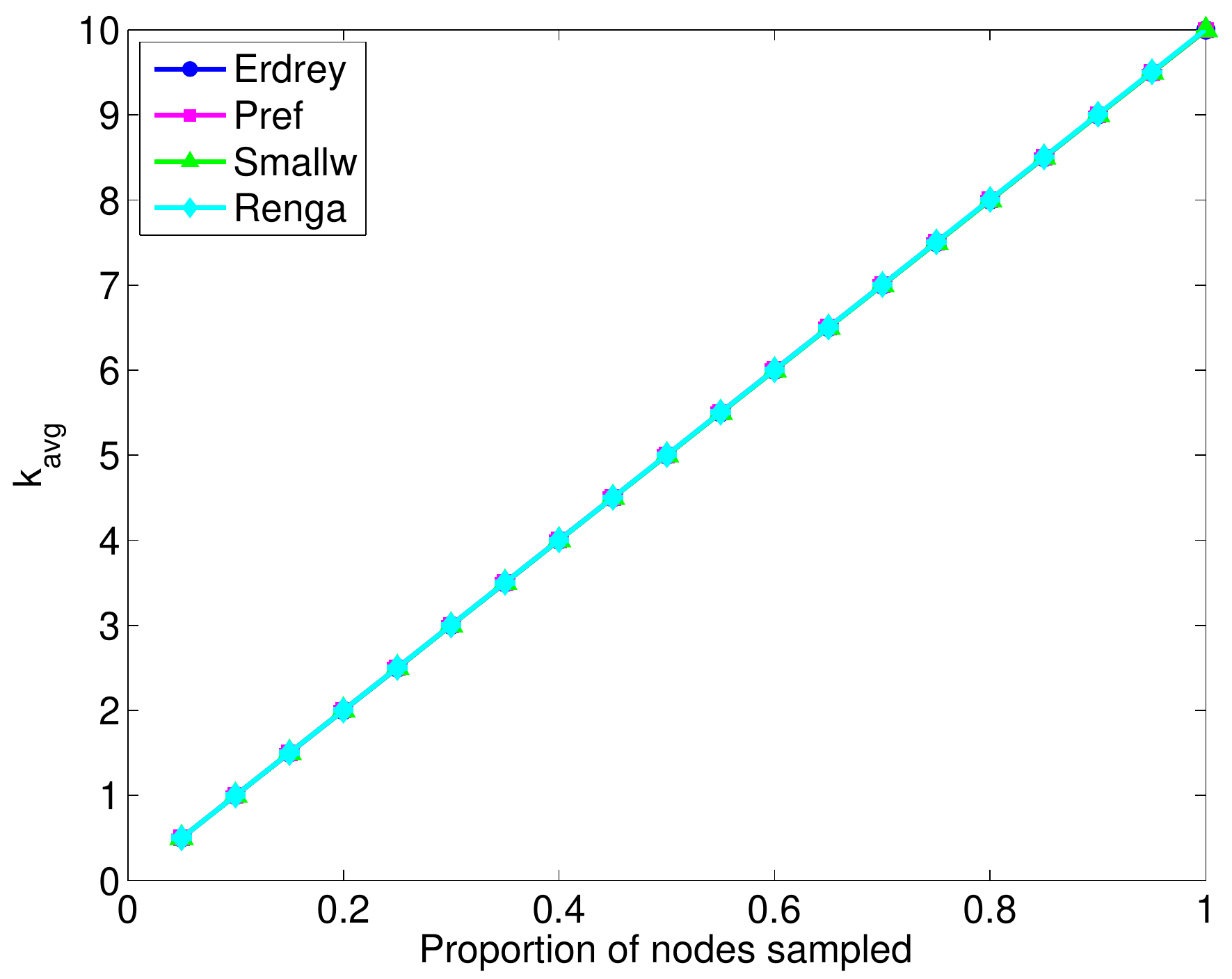}}\\
\subfigure[Max degree]{\includegraphics[width=.28\textwidth]{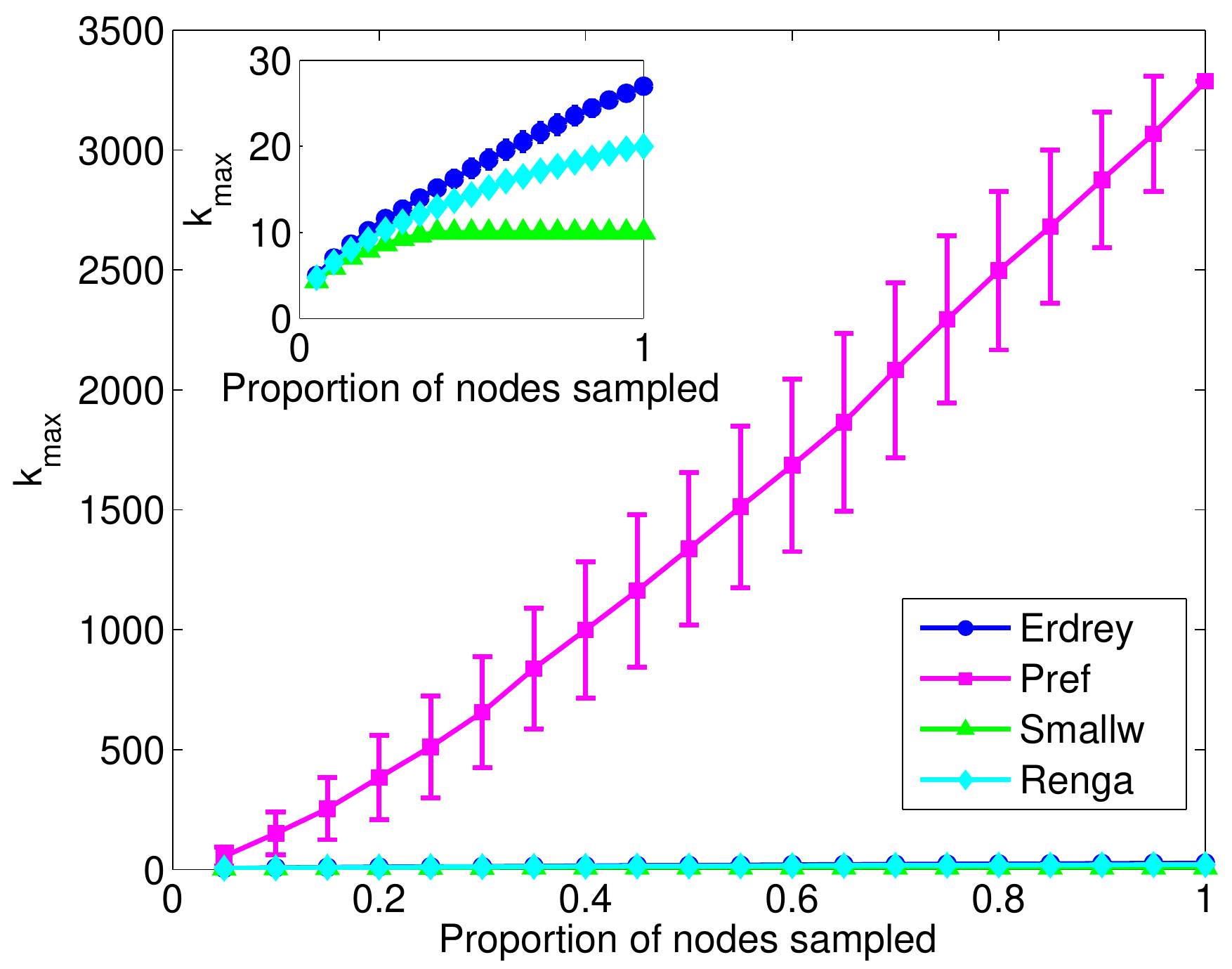}}
\subfigure[Clustering]{\includegraphics[width=.28\textwidth]{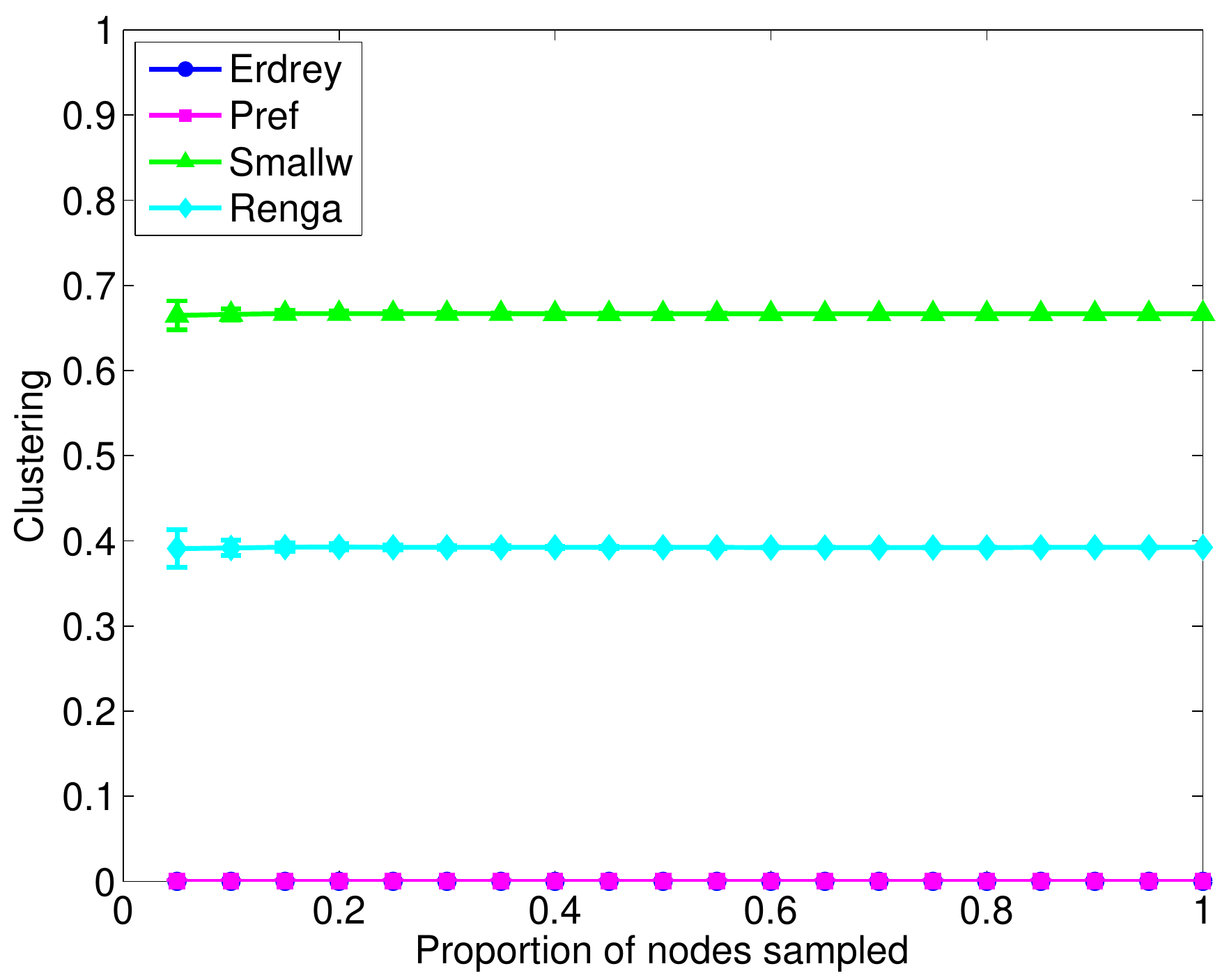}}
\subfigure[Prop. of nodes in Giant Component]{\includegraphics[width=.28\textwidth]{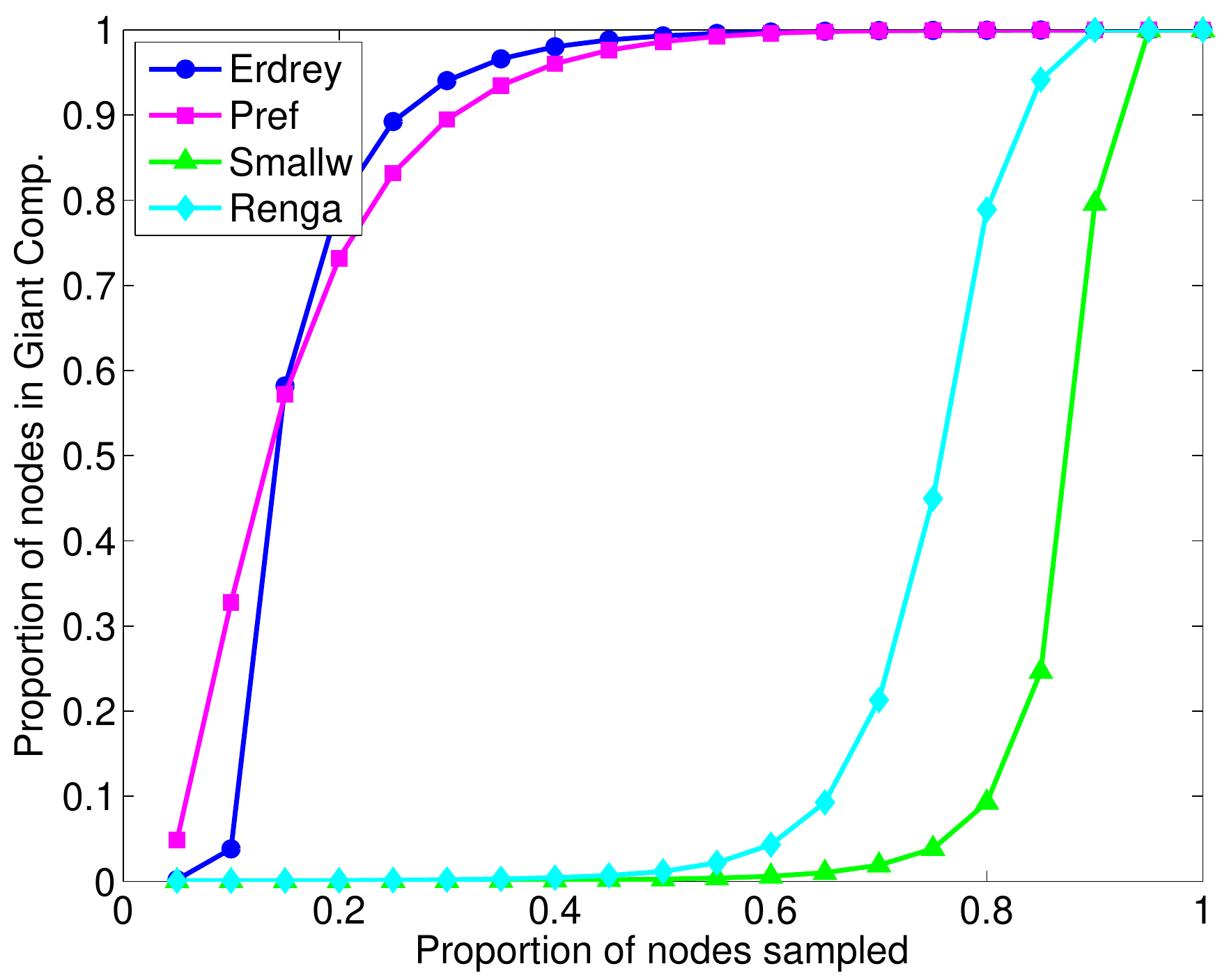}}\\
\caption[Scaling of statistics for simulated subnetworks induced on sampled nodes]{Scaling of statistics for simulated subnetworks induced on sampled nodes. (a.) The number of nodes in a subnetwork sampled by nodes scales as $n = qN$ precisely because only $qN$ nodes are selecting during subsampling. (b.) The number of edges scales as $m \approx M   \frac{n(n-1)}{N(N-1)} \approx Mq^2$, for $n \gg 1$ and $N \gg 1$. (c.) The average degree scales linearly with the proportion of nodes subsampled. (d.) The scaling of the max degree is dependent on network type. For networks with few large hubs, $k^{\textnormal{obs}}_{\kmax} \approx q \kmax$. For networks exhibiting a nontrivial number of nodes with degrees relatively close to $\kmax$, the max. degree scales nonlinearly. (e.) The clustering coefficient~\cite{newman2001clustering} shows little variation with respect to $q$ as suggested by the analytical result from Frank~\cite{Frank1978}. This suggests that $\hat{C} \approx C_{\text{obs}}$. (f.) The proportion of nodes in the giant component increases with the proportion of nodes sampled. For the random graphs (Erdrey and Pref) there is a critical point corresponding to the approximate sampling level  corresponding to when $k^{\textnormal{obs}}_{\rm avg}>1$. The thresholds for Small World and Range dependent networks are much higher due to the uniformity of the motif distribution in these networks. Markers indicates the mean over 100 simulations. Error bars showing one standard deviation are too small to see, except for (d.).}
\label{fig:sampling_by_nodes_scaling}
\end{figure*}
\setcounter{equation}{2}
\begin{figure*}[!ht]
\centering
\subfigure[Nodes]{\includegraphics[width=.28\textwidth]{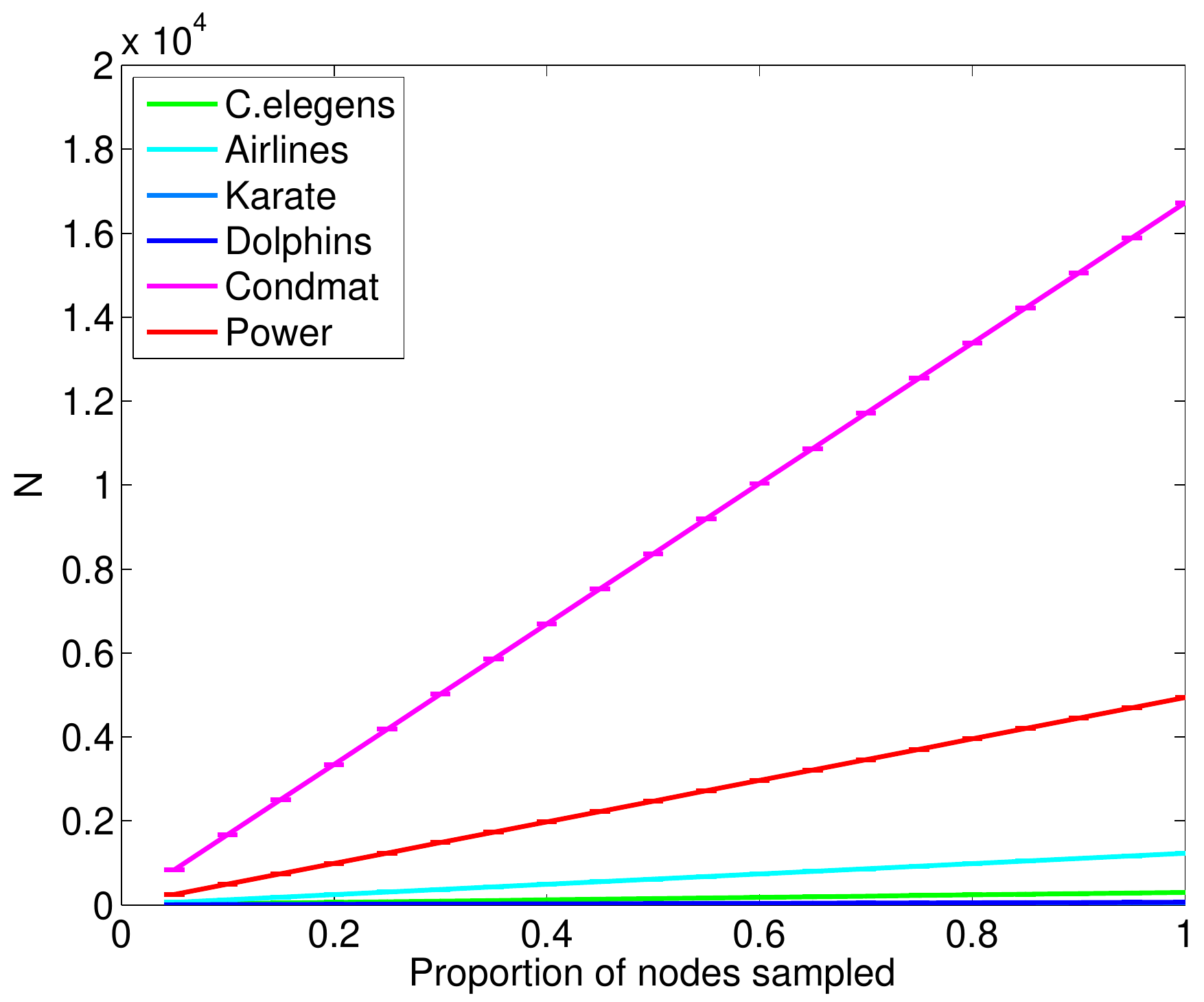}}
\subfigure[Edges]{\includegraphics[width=.28\textwidth]{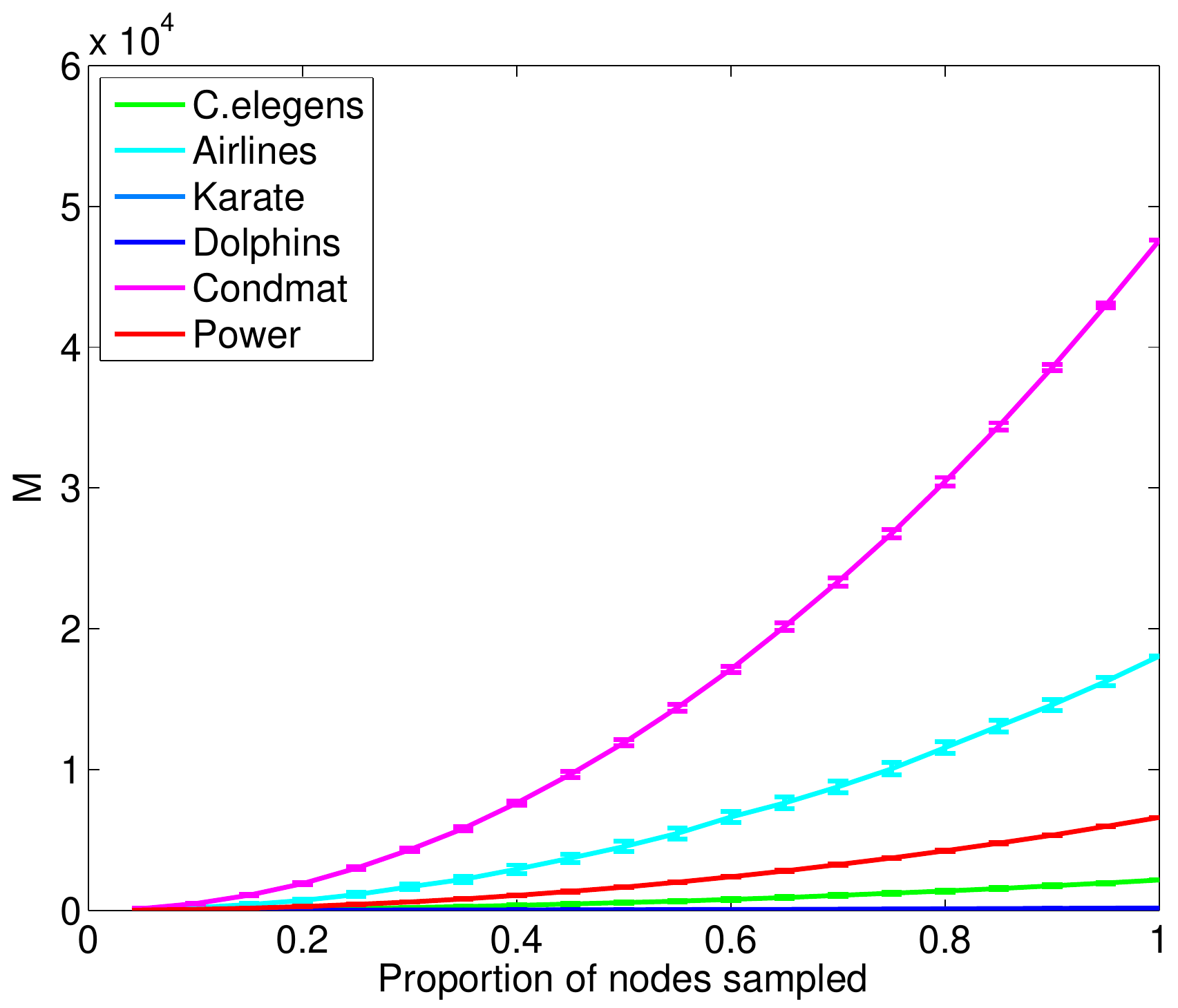}}
\subfigure[Average degree]{\includegraphics[width=.28\textwidth]{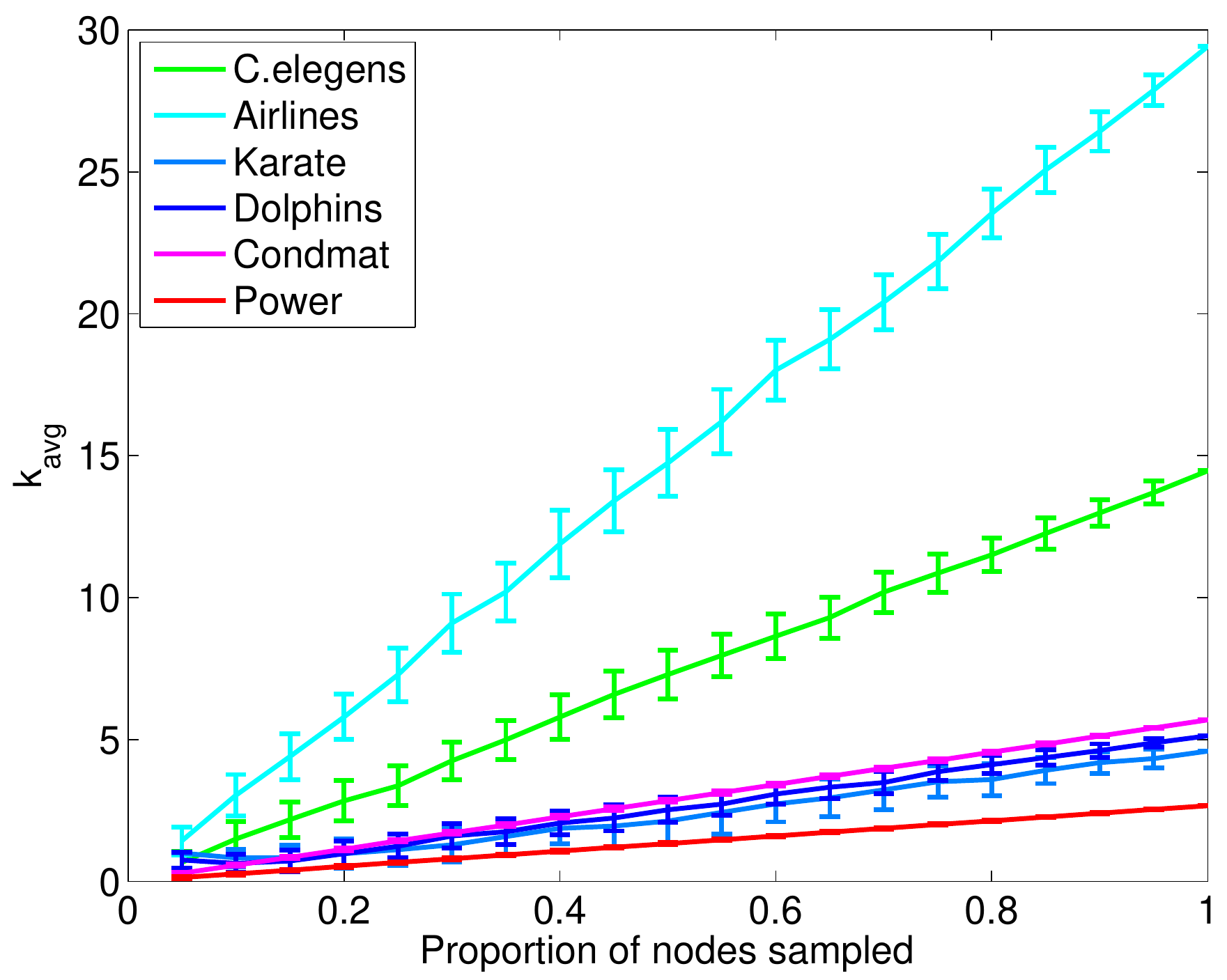}}\\
\subfigure[Max degree]{\includegraphics[width=.28\textwidth]{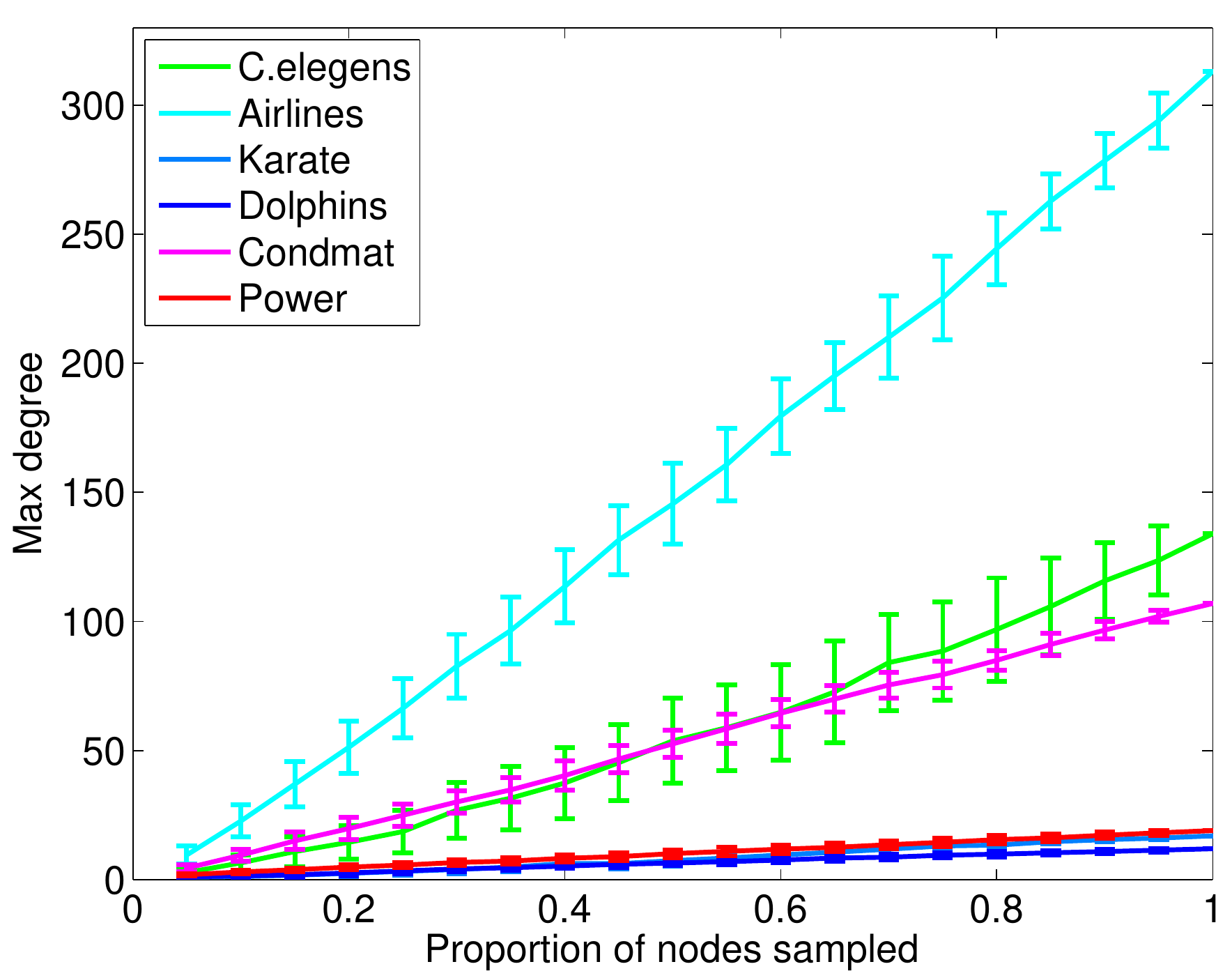}}
\subfigure[Clustering]{\includegraphics[width=.28\textwidth]{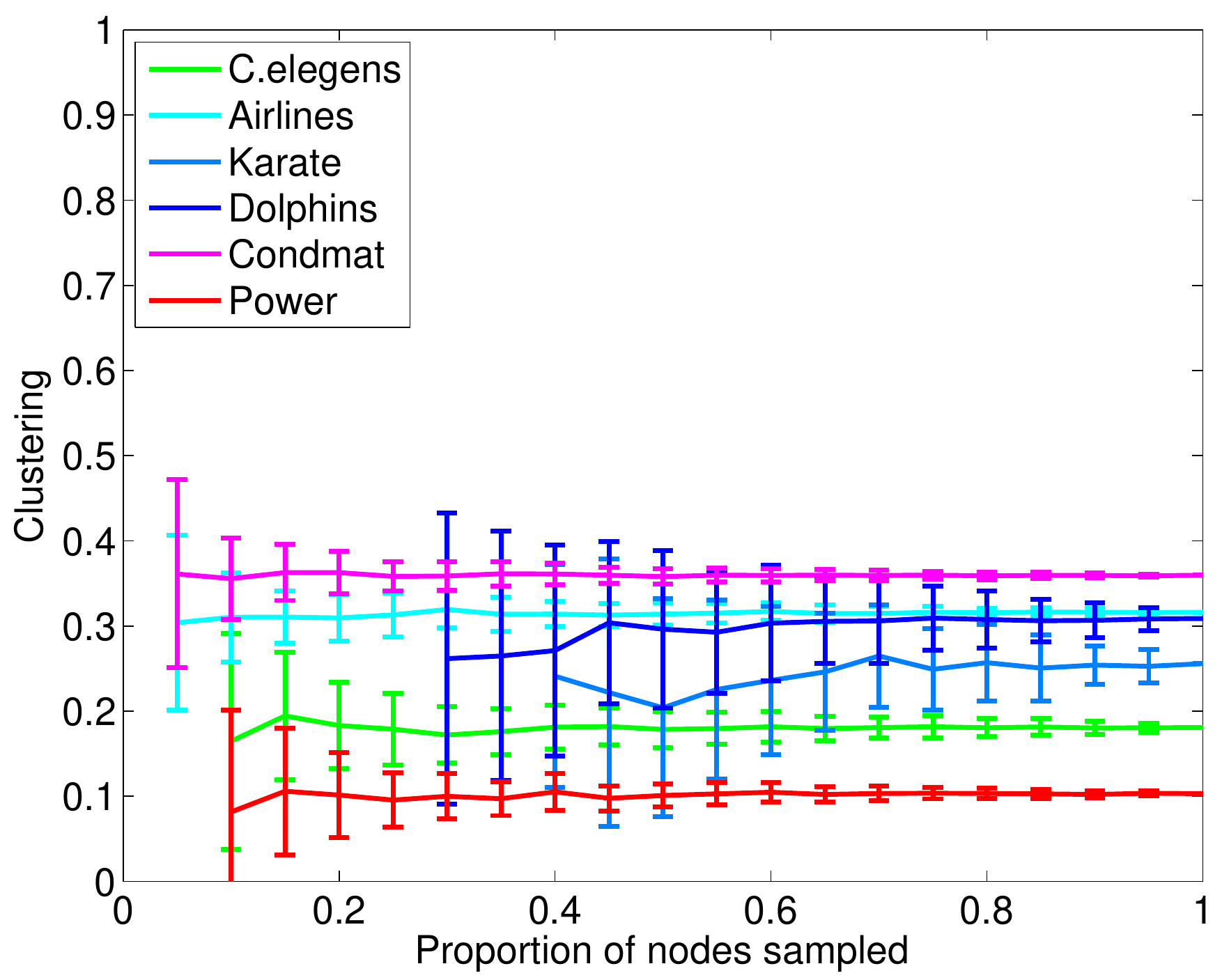}}
\subfigure[Prop. of nodes in Giant Component]{\includegraphics[width=.28\textwidth]{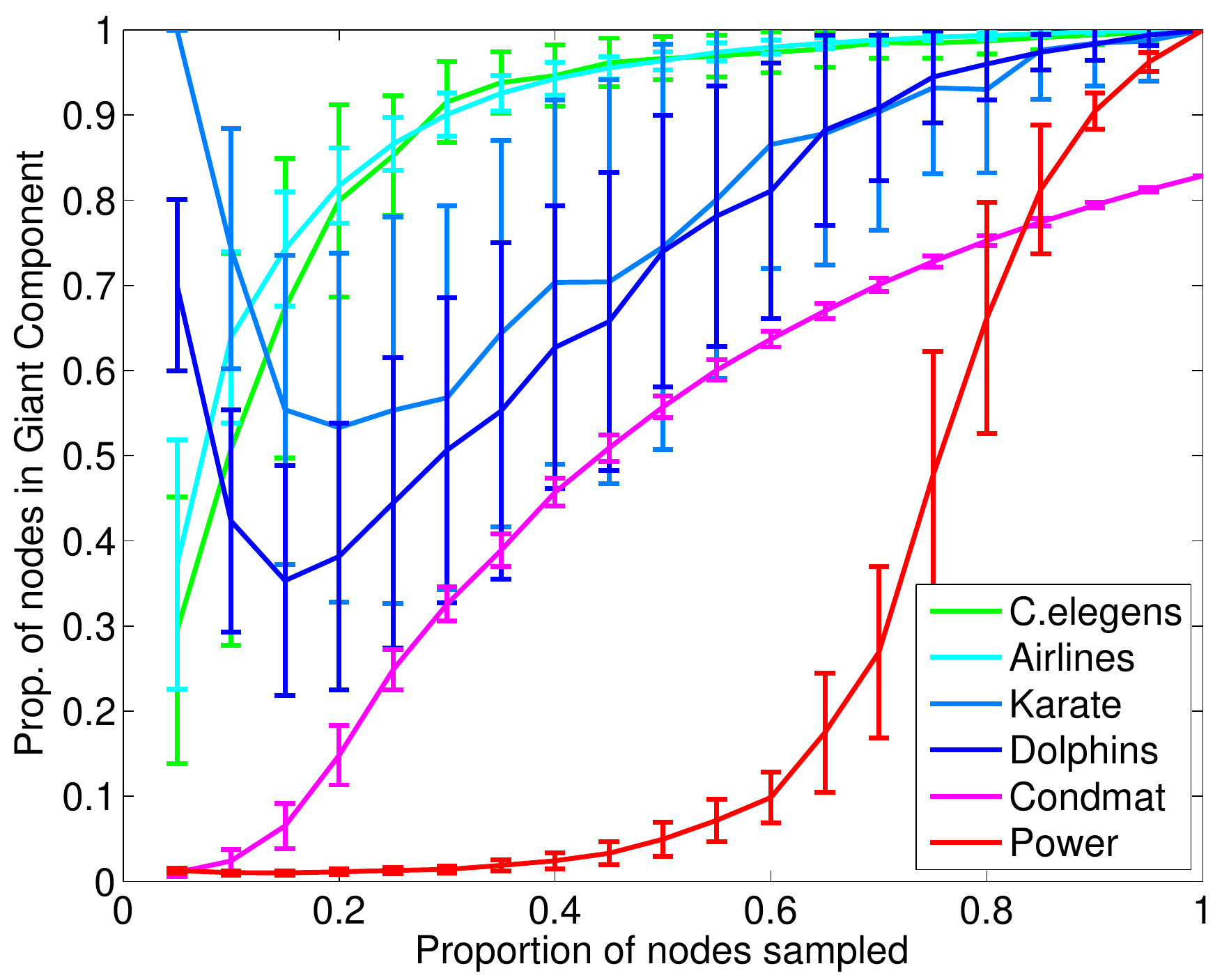}}\\
\caption[Scaling of statistics for empirical subnetworks induced on sampled nodes]{Scaling of statistics for empirical subnetworks induced on sampled nodes. (a.) The number of nodes scales as $n = qN$ precisely because only $qN$ nodes are selecting during subsampling. (b.) The number of edges scales as $m \approx M   \frac{n(n-1)}{N(N-1)} \approx Mq^2$, where $q$ is the proportion of nodes subsampled. (c.) The average degree scales as $k^{\text{obs}}_{\rm avg} \approx q k^{\text{true}}_{\rm avg}$. (d.) The max degree scales roughly linearly as $k^{\text{obs}}_{\max} \approx q k^{\text{true}}_{\max}$. (e.) The clustering coefficient~\cite{newman2001clustering} shows little variation with respect to $q$ as suggested by the analytical result from Frank~\cite{Frank1978},  $\hat{C} \approx C_{\text{obs}}$. (f.) Large networks, such as the Powergrid and Condensed Matter author collaboration networks show the expected transition to the giant component as $q$ increases corresponding to when $k^{\textnormal{obs}}_{\rm avg}>1$. Smaller networks, such as the Karate club and Dolphin network show a high proportion of nodes in the giant component, for low $q$ because the subnetwork generated for these levels of $q$ contains fewer than 10 nodes (i.e., the network is degenerate).}
\label{fig:sampling_by_nodes_empirical_scaling}
\end{figure*}

\setcounter{equation}{3}
\begin{figure*}[!ht]
\centering
\subfigure[Erdrey]{\includegraphics[width=.28\textwidth]{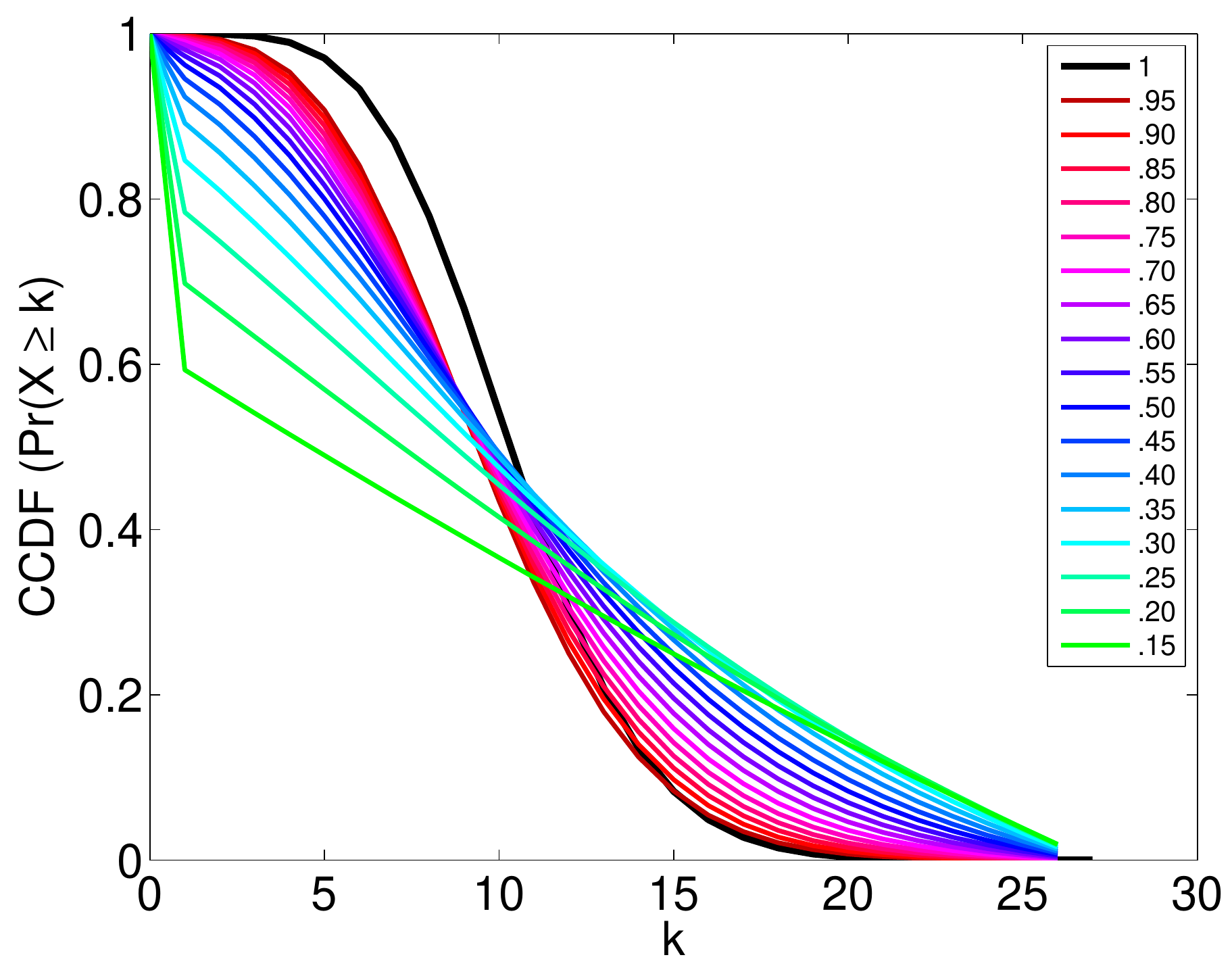}}
\subfigure[Pref]{\includegraphics[width=.28\textwidth]{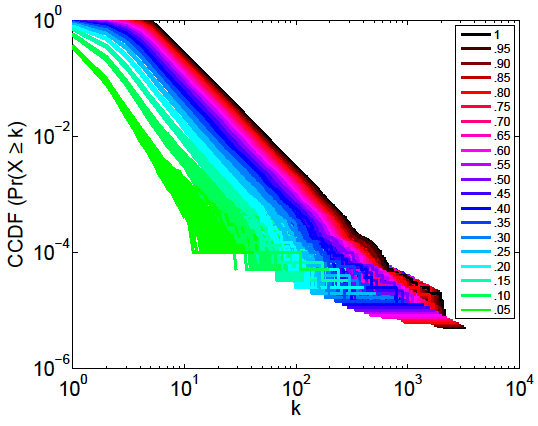}}
\subfigure[Smallworld]{\includegraphics[width=.28\textwidth]{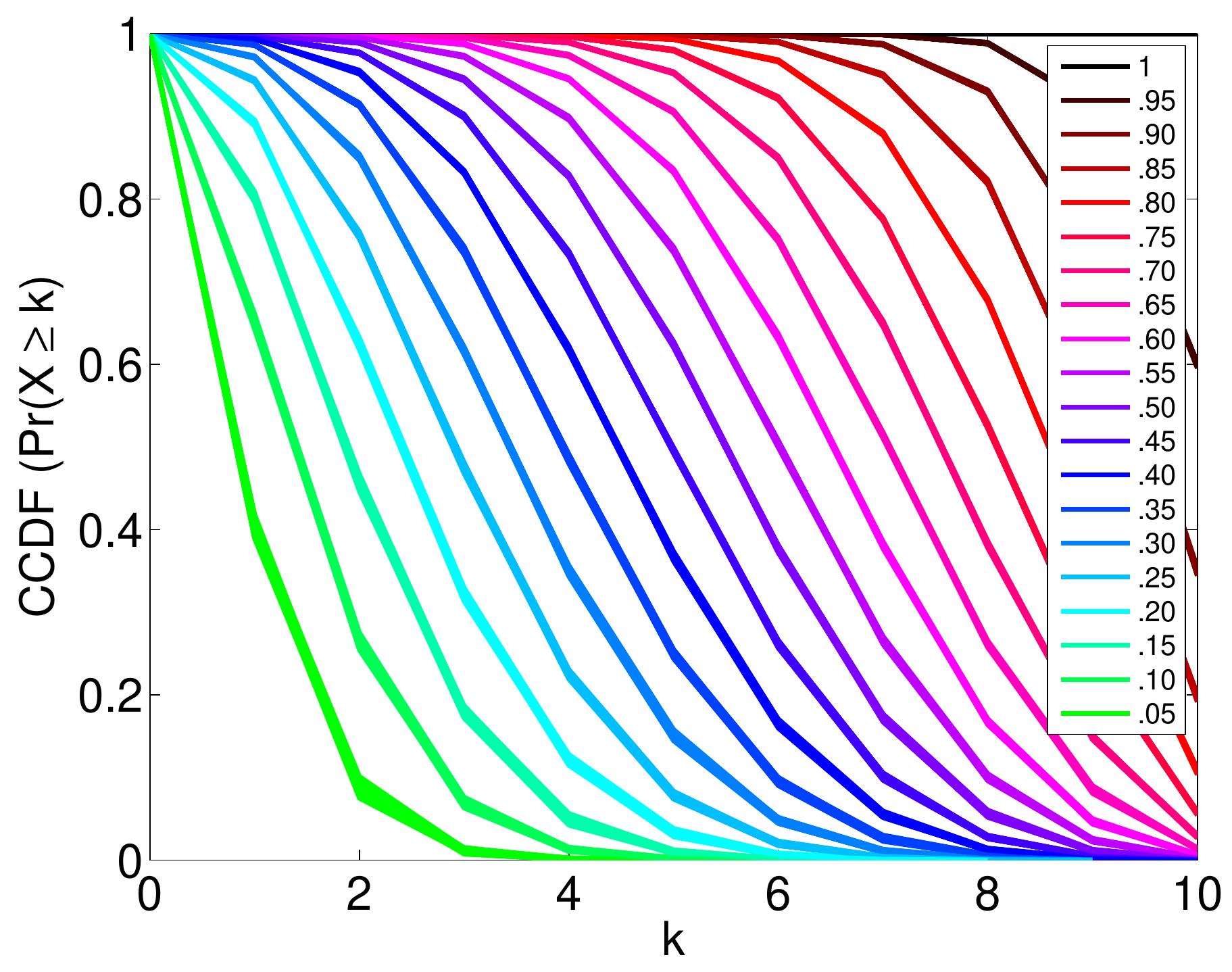}}\\
\subfigure[Renga]{\includegraphics[width=.28\textwidth]{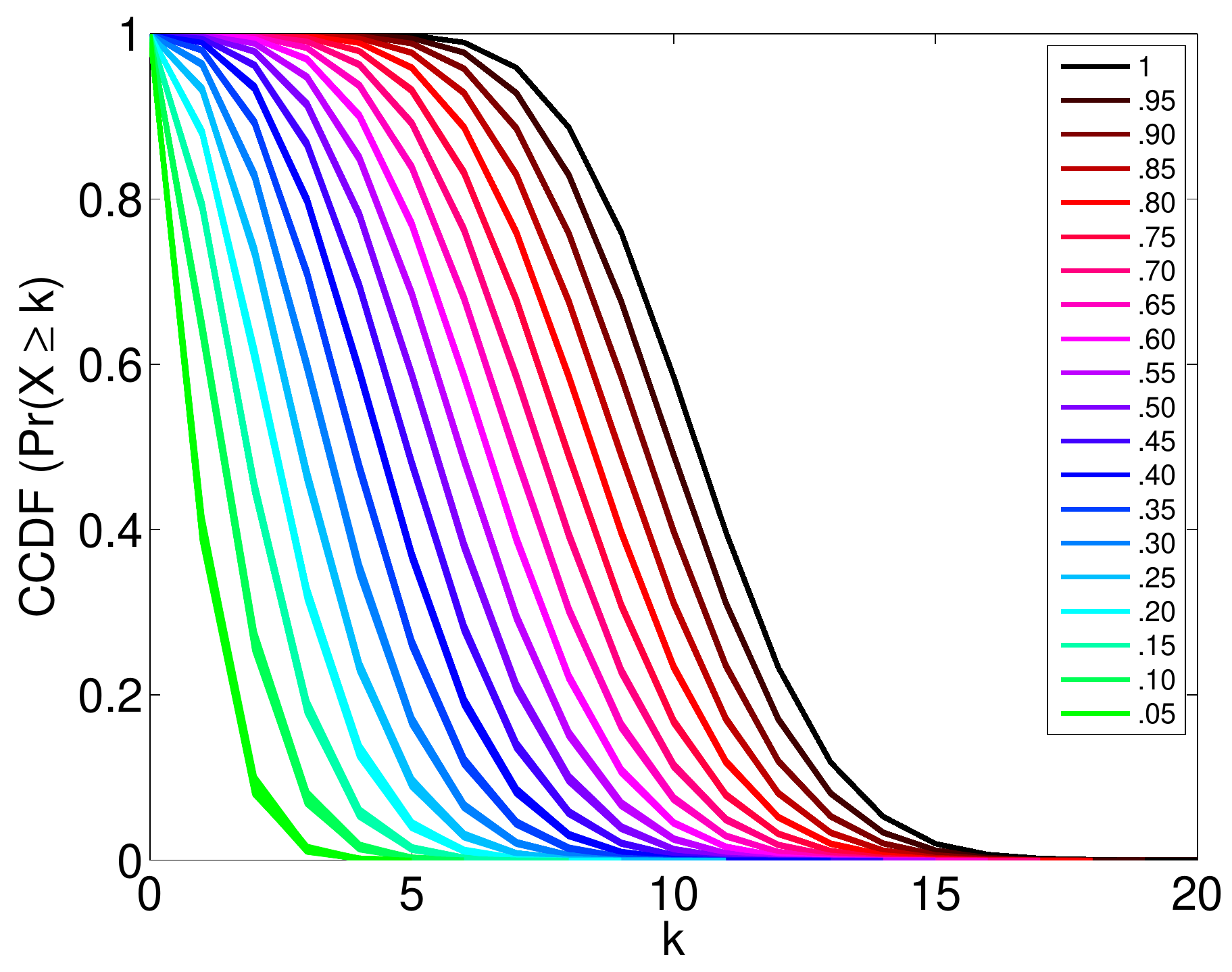}}
\subfigure[C. elegans]{\includegraphics[width=.28\textwidth]{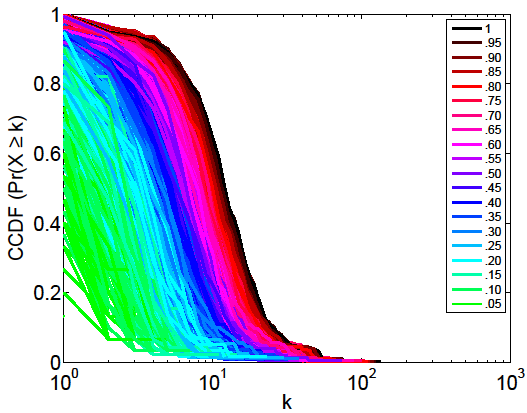}}
\subfigure[Airlines]{\includegraphics[width=.28\textwidth]{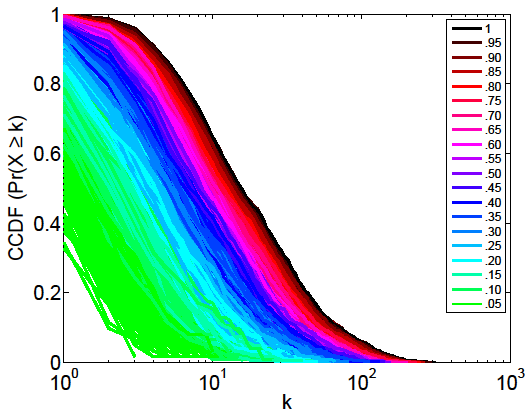}}\\
\subfigure[Karate]{\includegraphics[width=.28\textwidth]{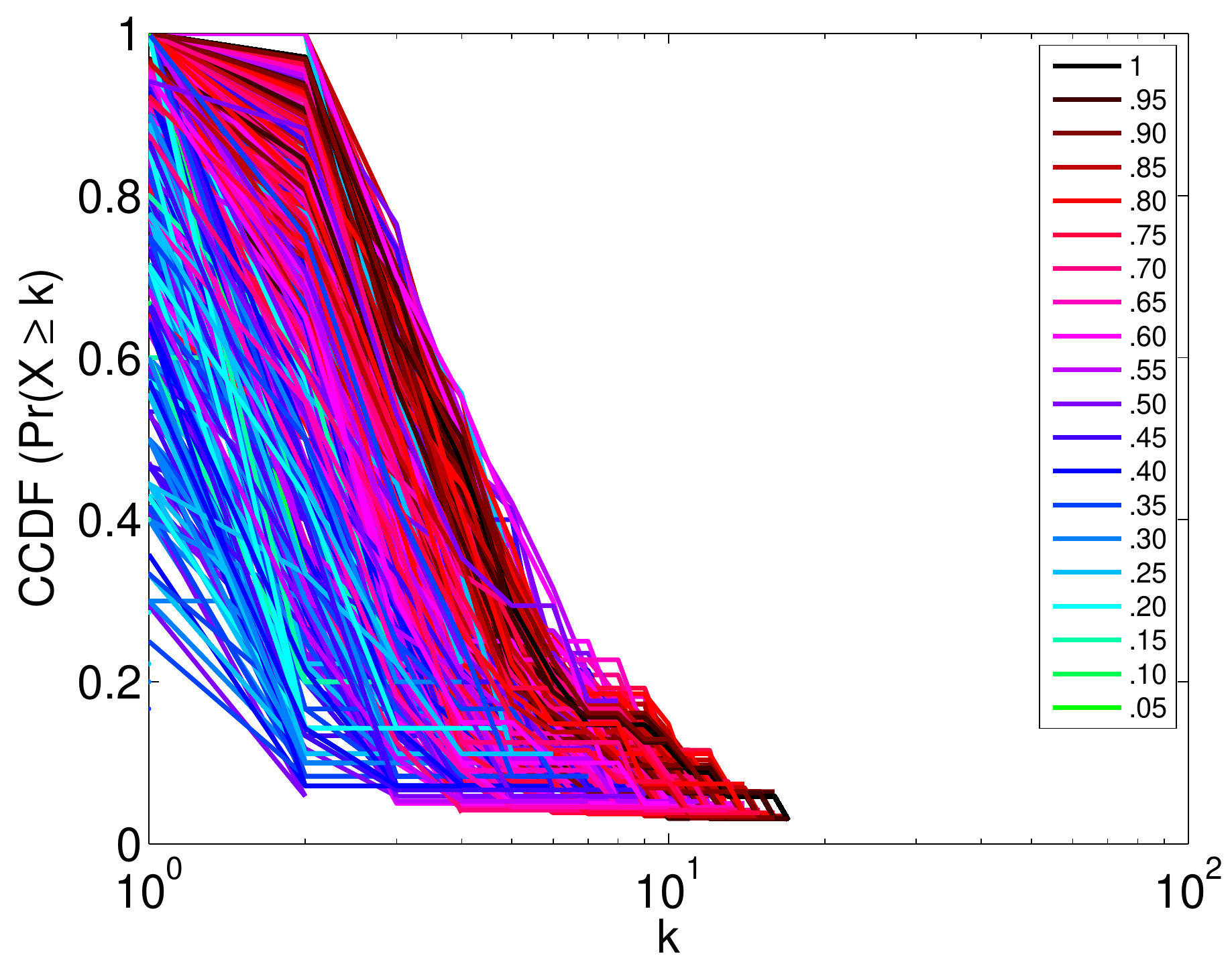}}
\subfigure[Dolphins]{\includegraphics[width=.28\textwidth]{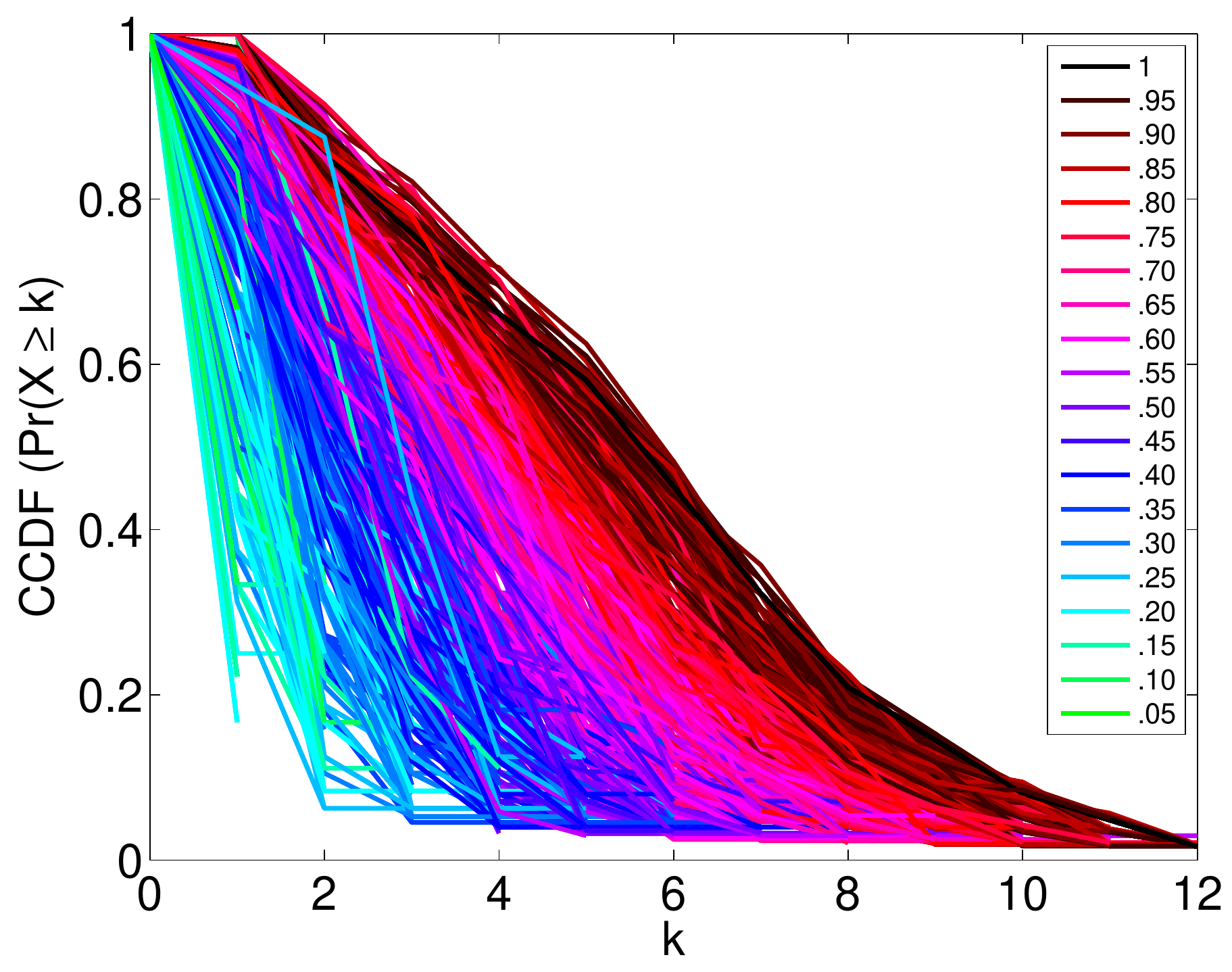}}
\subfigure[Condmat]{\includegraphics[width=.28\textwidth]{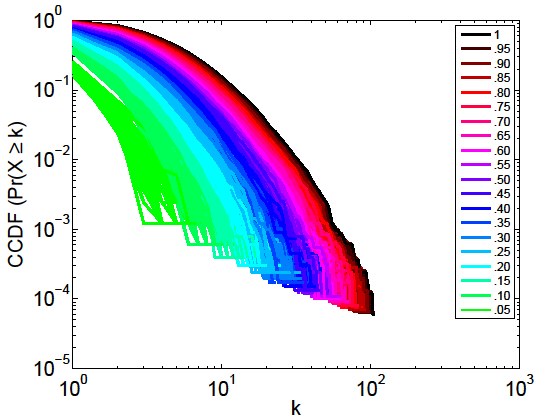}}\\
\subfigure[Powergrid]{\includegraphics[width=.28\textwidth]{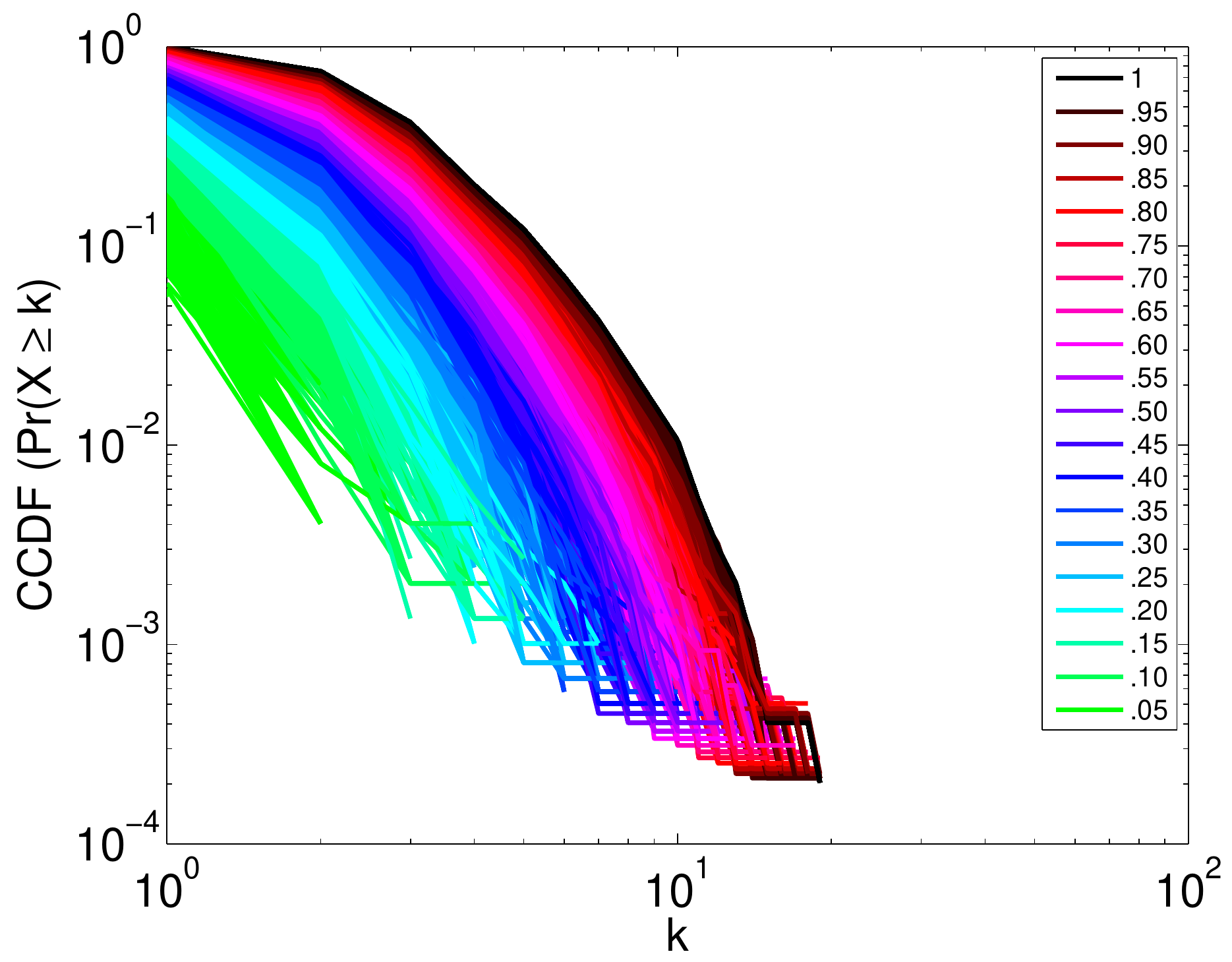}}
\caption[CCDF distortion for subnetworks induced on sampled nodes]{CCDF distortion for subnetworks induced on sampled nodes. Subnetwork degree distributions do not capture the true degree distribution, especially for small $q$.}
\label{fig:sampling_by_nodes_empirical_ccpdf}
\end{figure*}

\setcounter{equation}{4}
\begin{figure*}[!ht]
\centering
\subfigure[$\text{Erdrey}^+$]{\includegraphics[width=.28\textwidth]{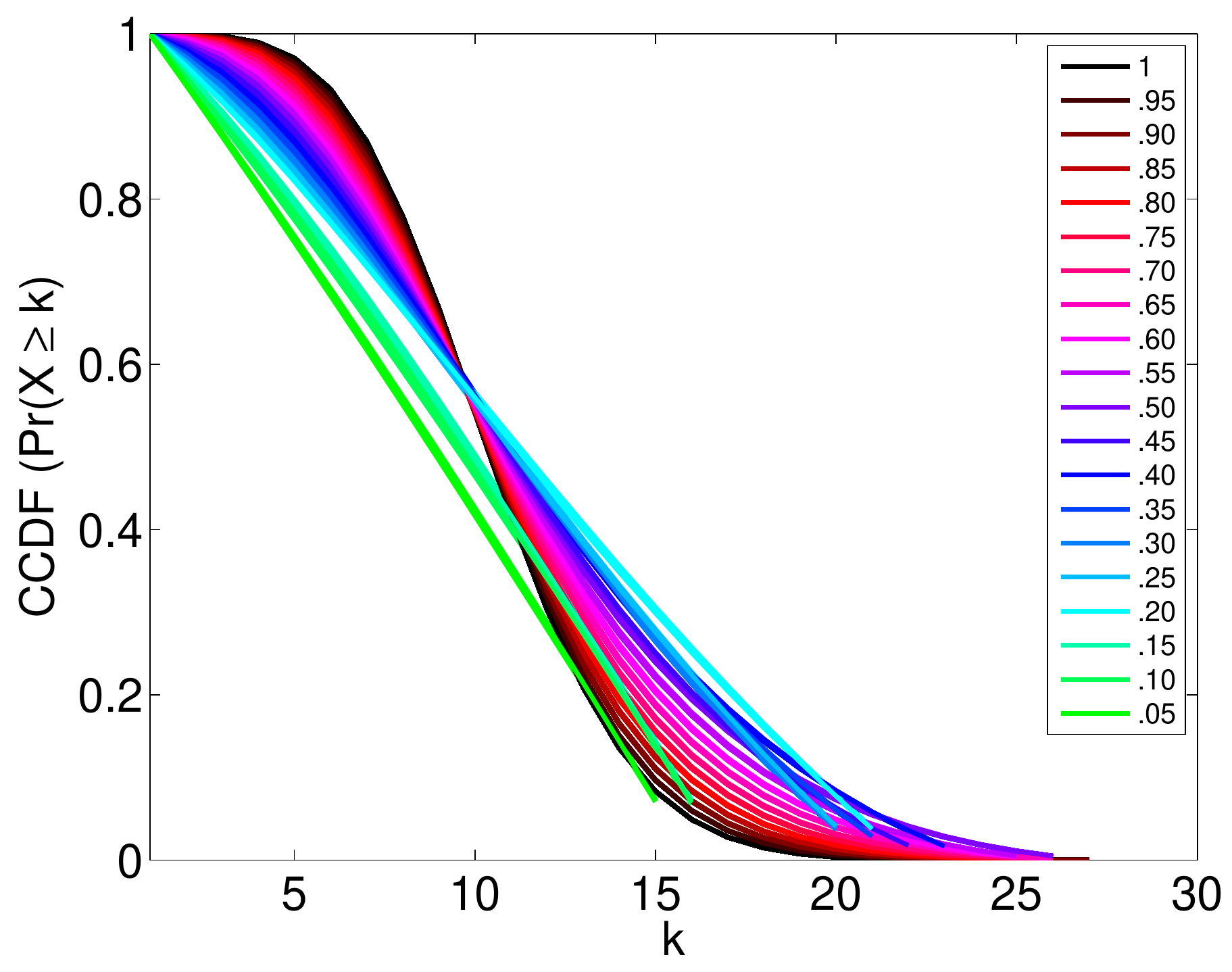}}
\subfigure[$\text{Pref}^*$]{\includegraphics[width=.28\textwidth]{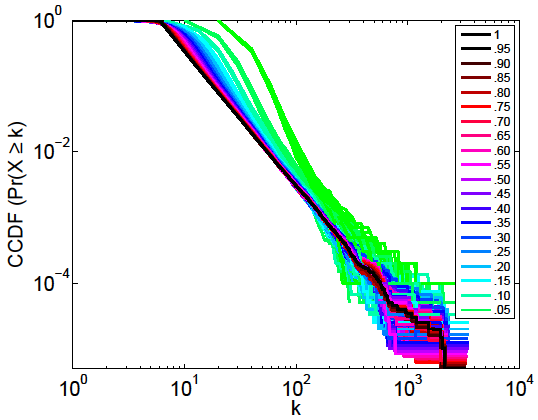}}
\subfigure[$\text{Smallworld}^+$]{\includegraphics[width=.28\textwidth]{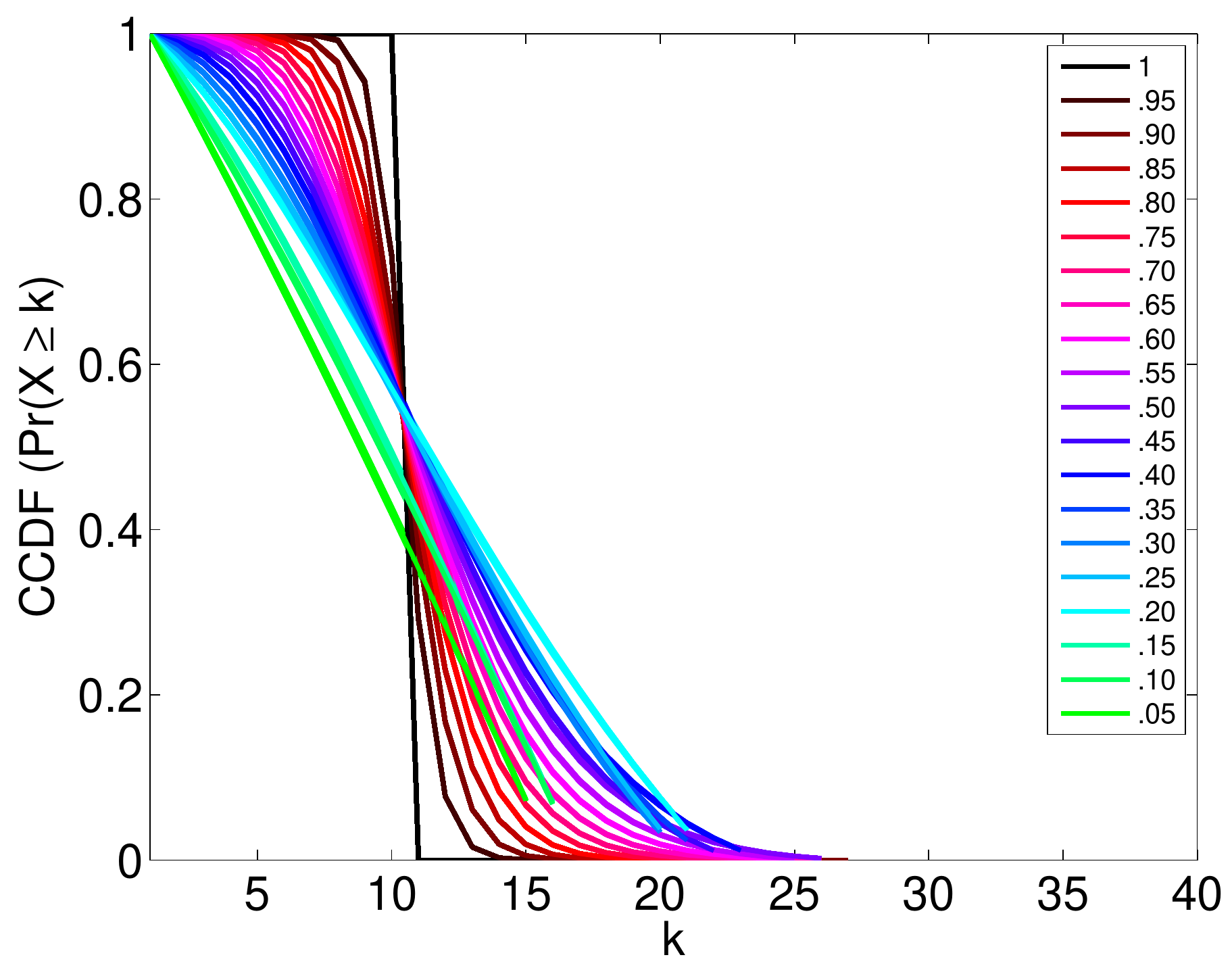}}\\
\subfigure[$\text{Renga}^+$]{\includegraphics[width=.28\textwidth]{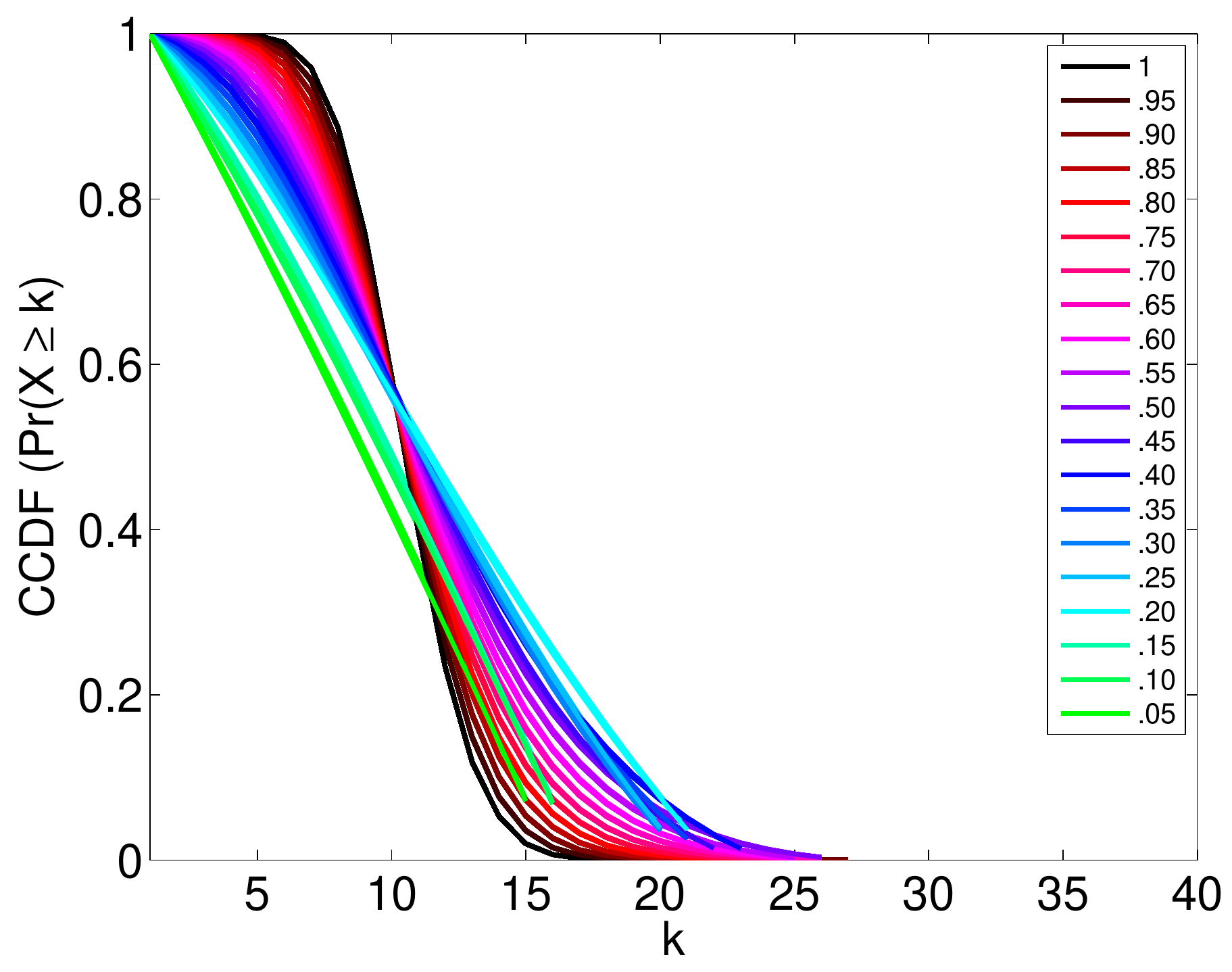}}
\subfigure[$\text{C. elegans}^*$]{\includegraphics[width=.28\textwidth]{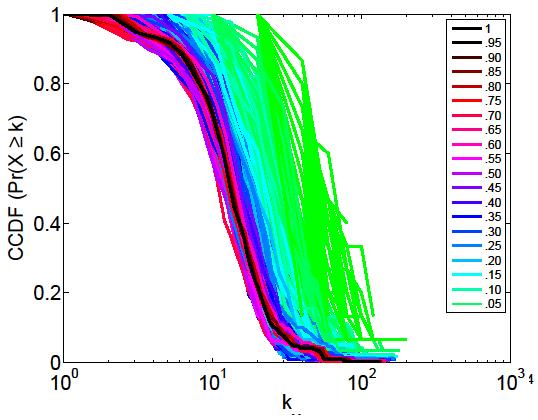}}
\subfigure[$\text{Airlines}^*$]{\includegraphics[width=.28\textwidth]{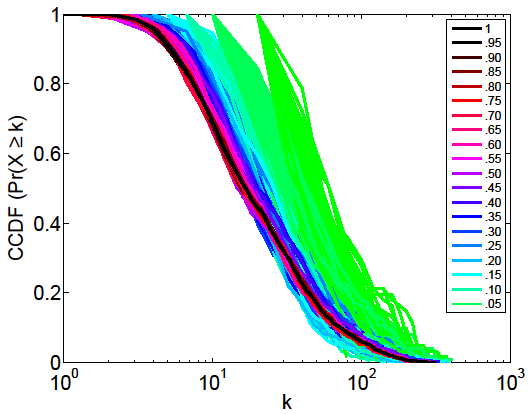}}\\
\subfigure[$\text{Karate}^+$]{\includegraphics[width=.28\textwidth]{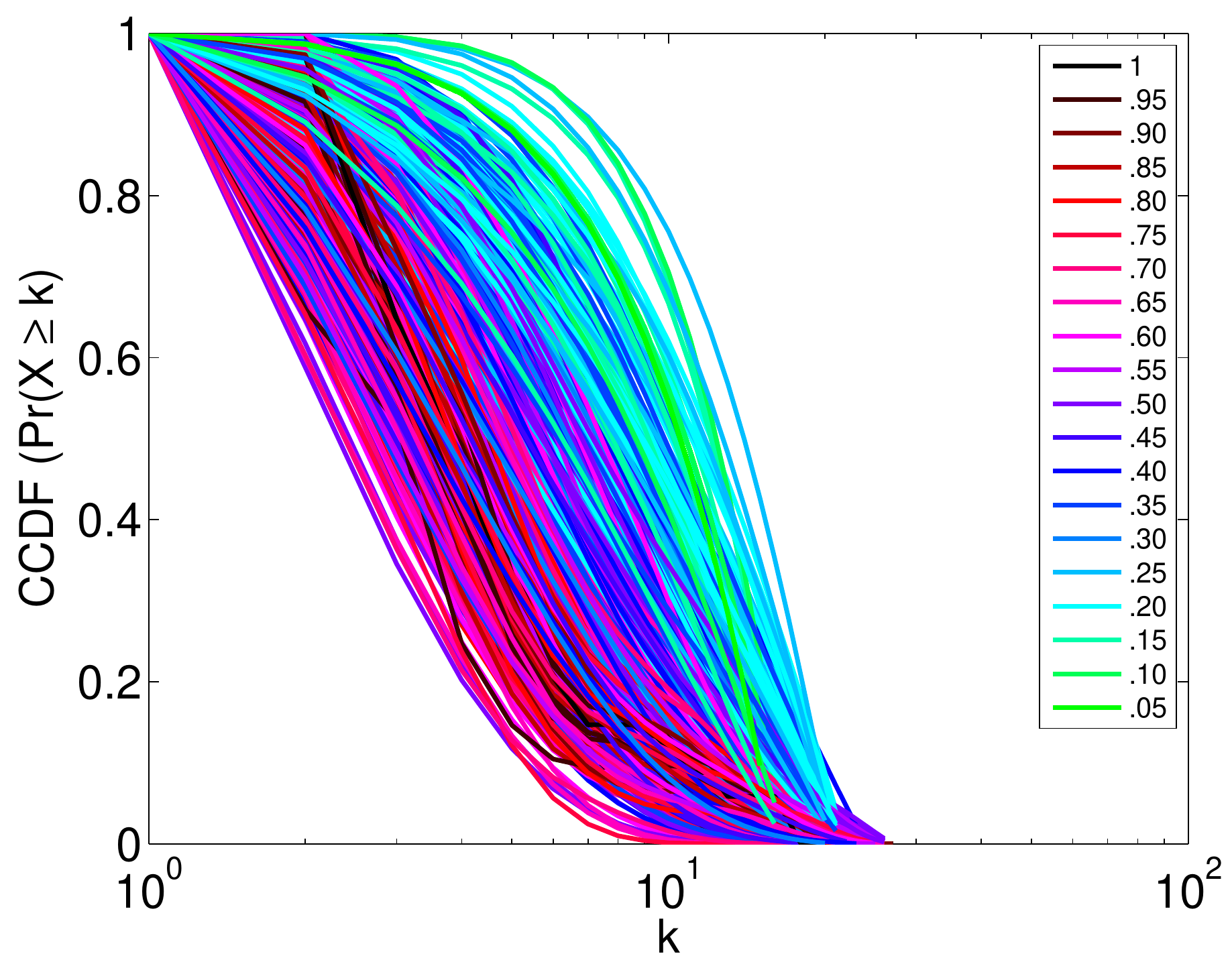}}
\subfigure[$\text{Dolphins}^+$]{\includegraphics[width=.28\textwidth]{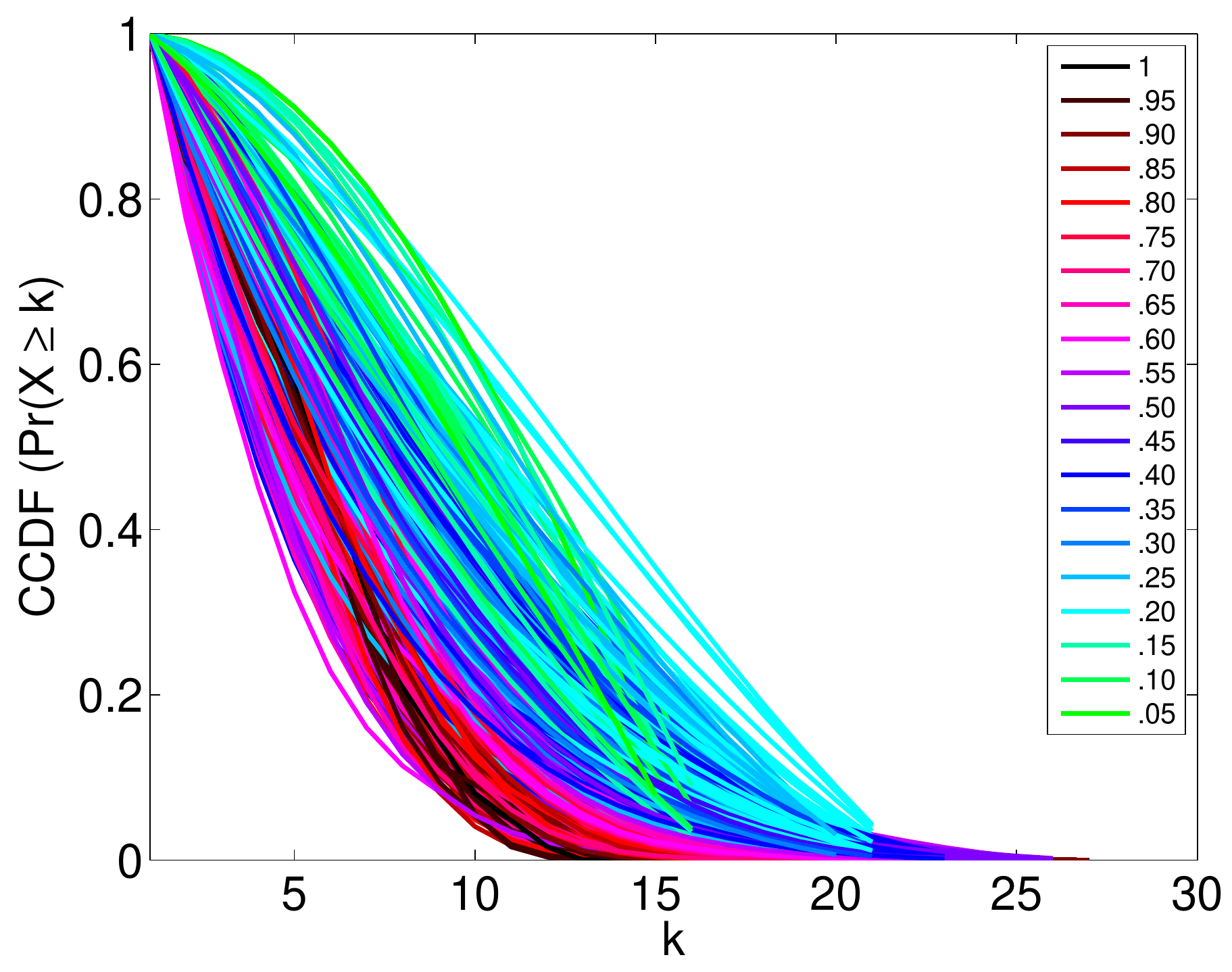}}
\subfigure[$\text{Condmat}^*$]{\includegraphics[width=.28\textwidth]{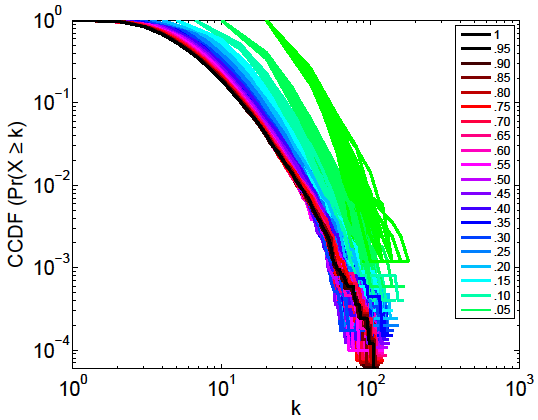}}\\
\subfigure[$\text{Powergrid}^*$]{\includegraphics[width=.28\textwidth]{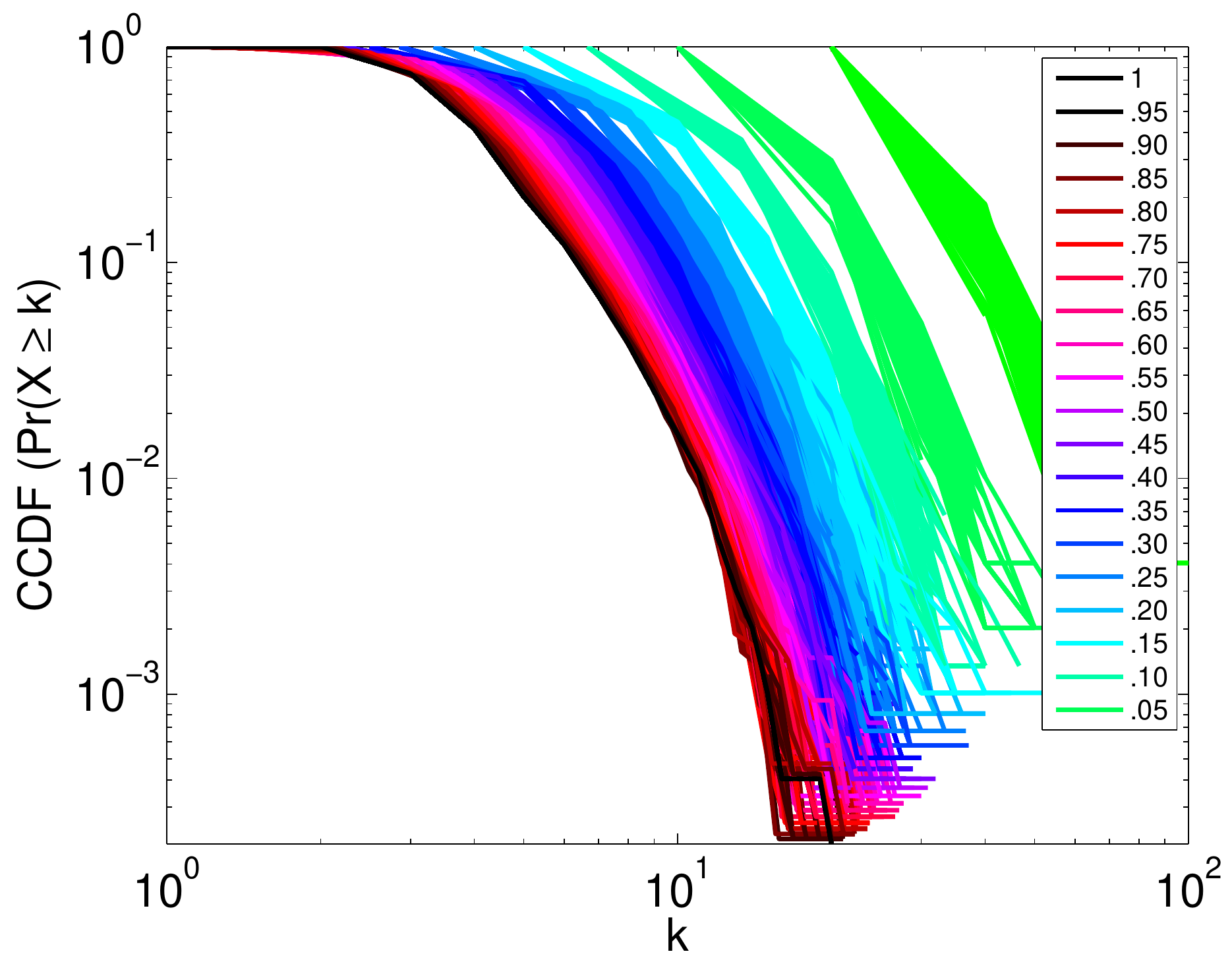}}
\caption[Predicted CCDF from subnetworks induced on sampled nodes]{Predicted CCDF from subnetworks induced on sampled nodes. The predicted CCDF shows relatively good agreement with the true CCDF for most networks. Karate club and Dolphins exhibit significant deviation, possible due to the small number of nodes in these networks. Networks designated $\text{with }^+$ utilized Equation~\ref{eq:dist_rollback} and those designated with $\text{with }^*$ utilized Equation~\ref{eq:my_dist_shorter}.  }
\label{fig:predicting_by_nodes_combined_ccpdf}
\end{figure*}

\setcounter{equation}{5}
\begin{figure*}[!ht]  
\centering
\subfigure[Nodes]{\includegraphics[width=.3\textwidth]{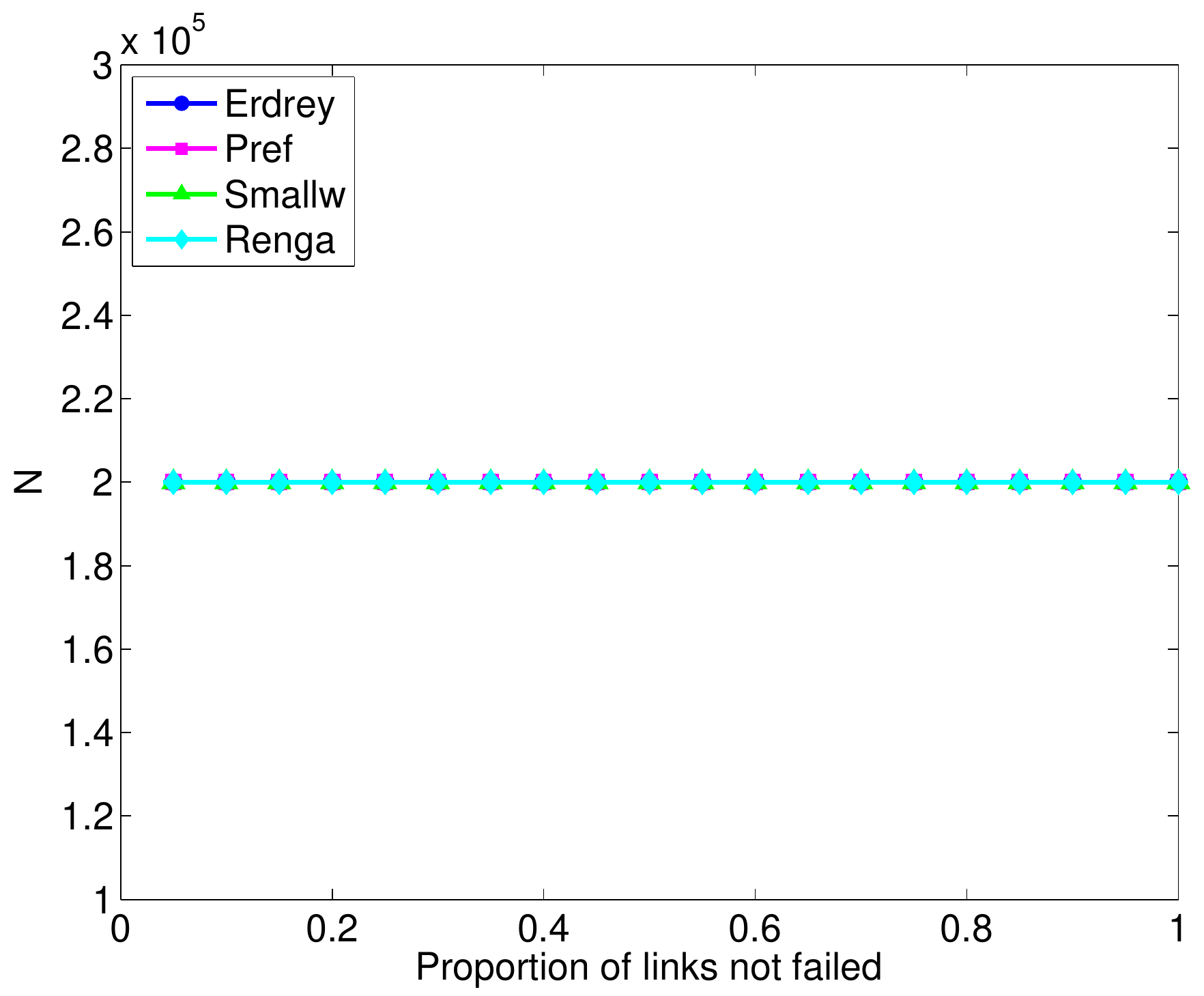}}
\subfigure[Edges]{\includegraphics[width=.3\textwidth]{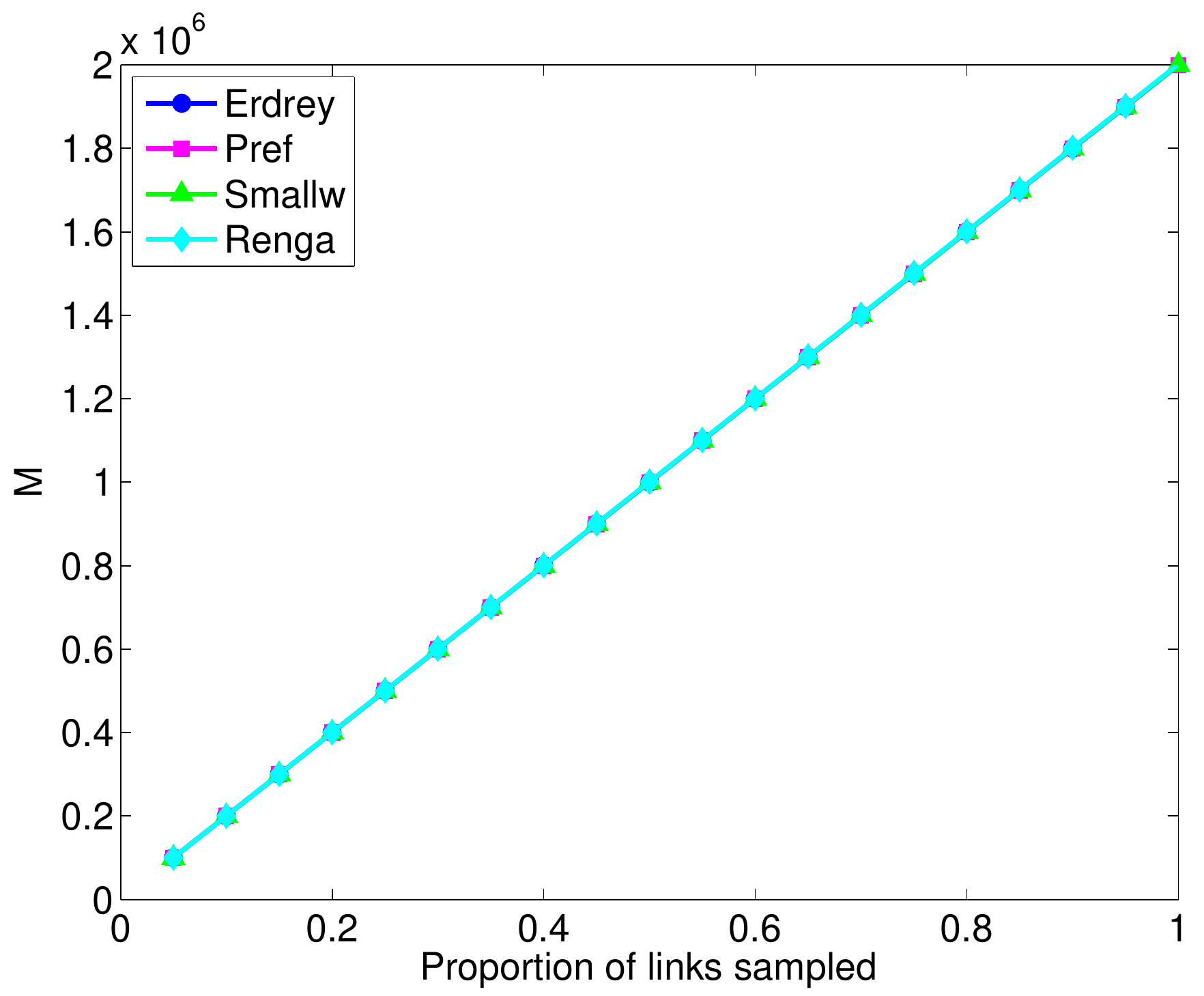}}
\subfigure[Average degree]{\includegraphics[width=.3\textwidth]{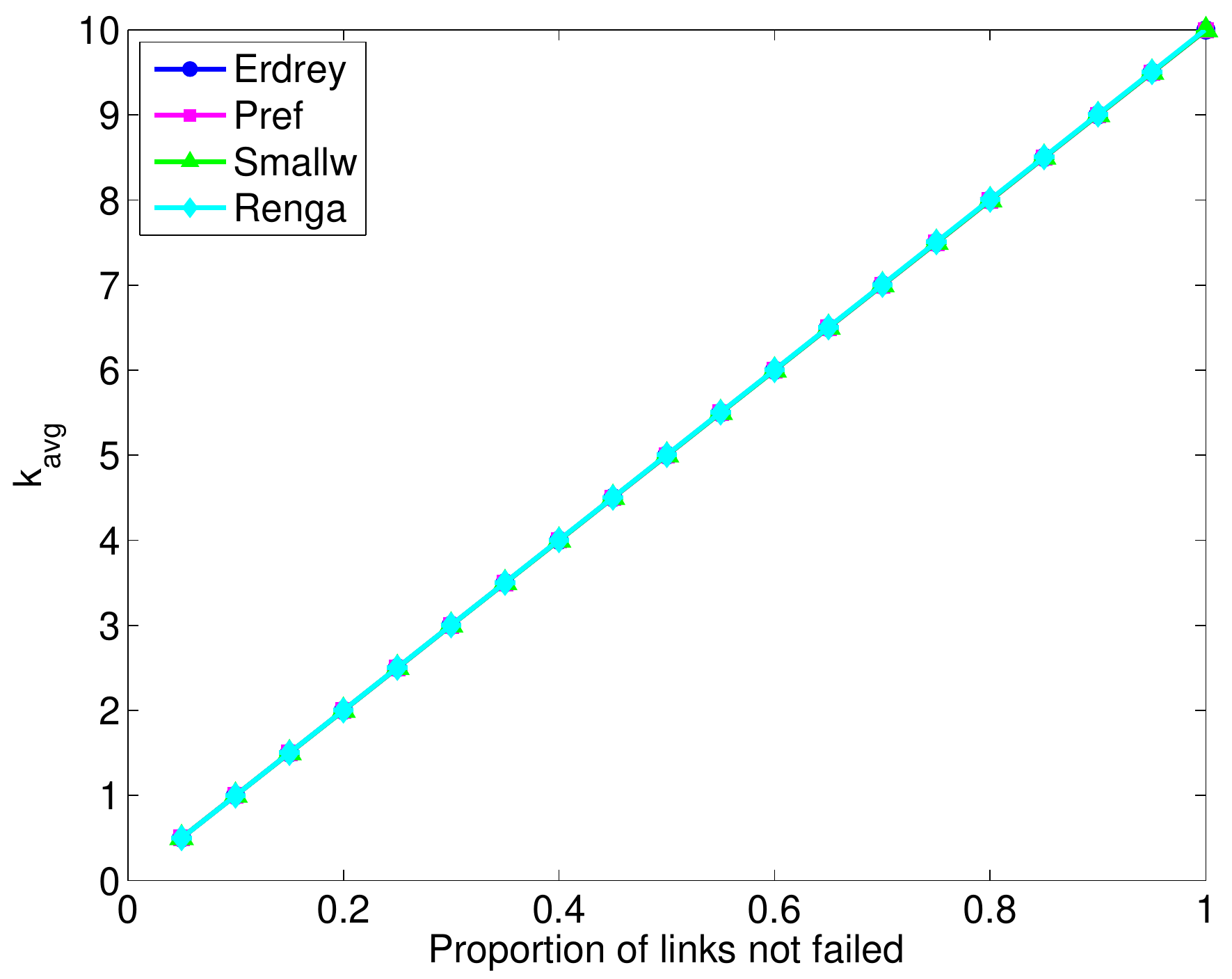}}\\
\subfigure[Max degree]{\includegraphics[width=.3\textwidth]{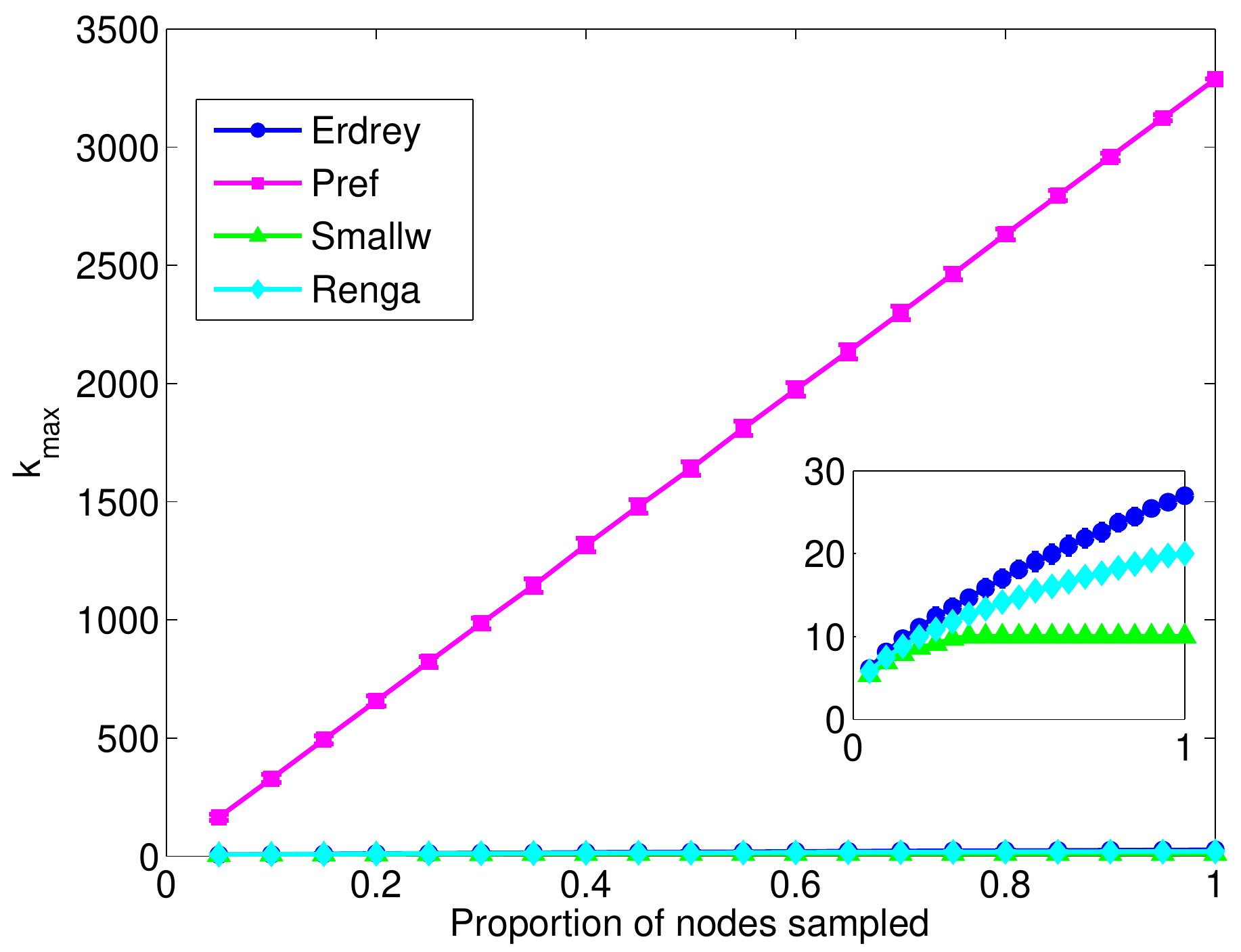}}
\subfigure[Clustering]{\includegraphics[width=.3\textwidth]{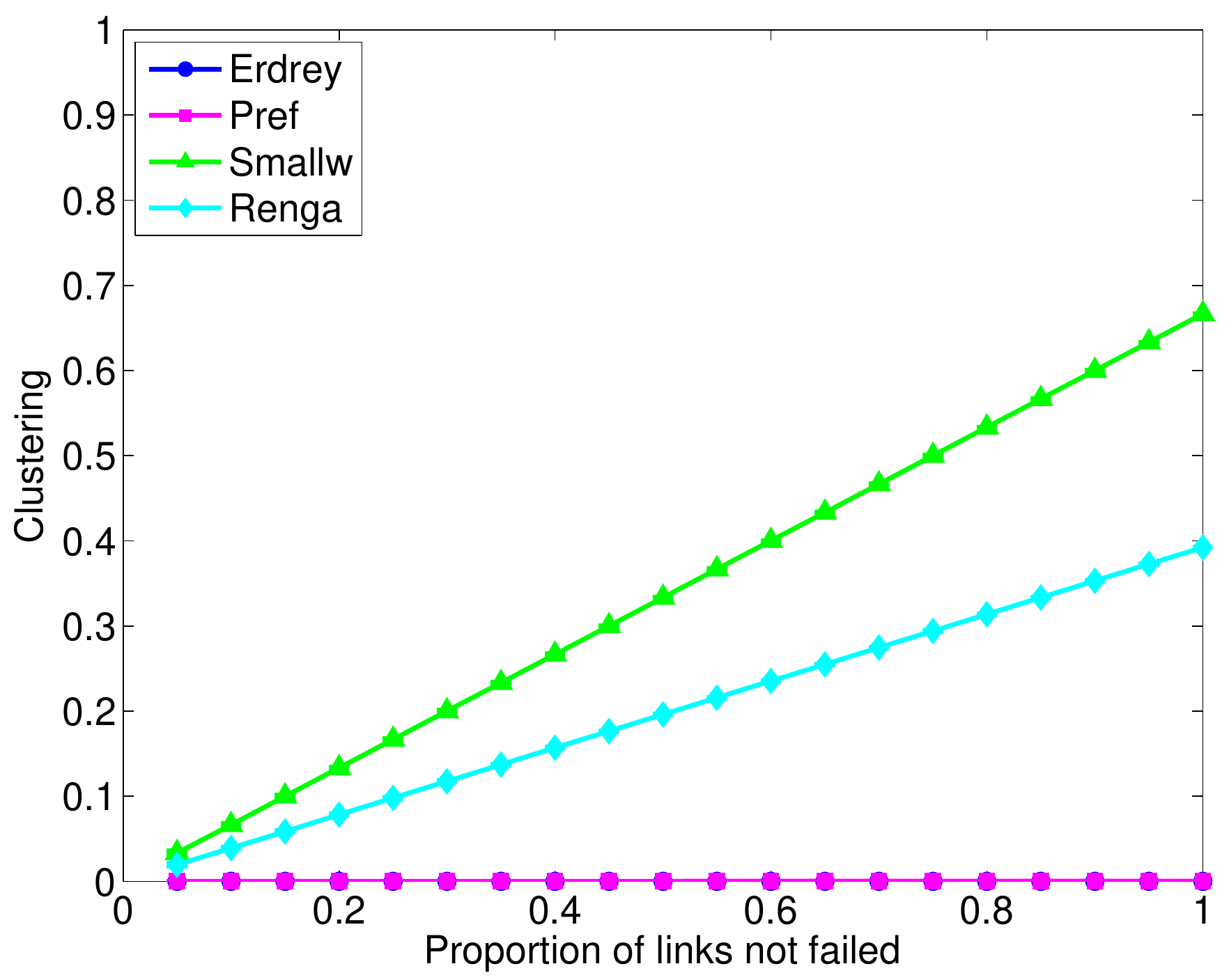}}
\subfigure[Prop. of nodes in Giant Component]{\includegraphics[width=.3\textwidth]{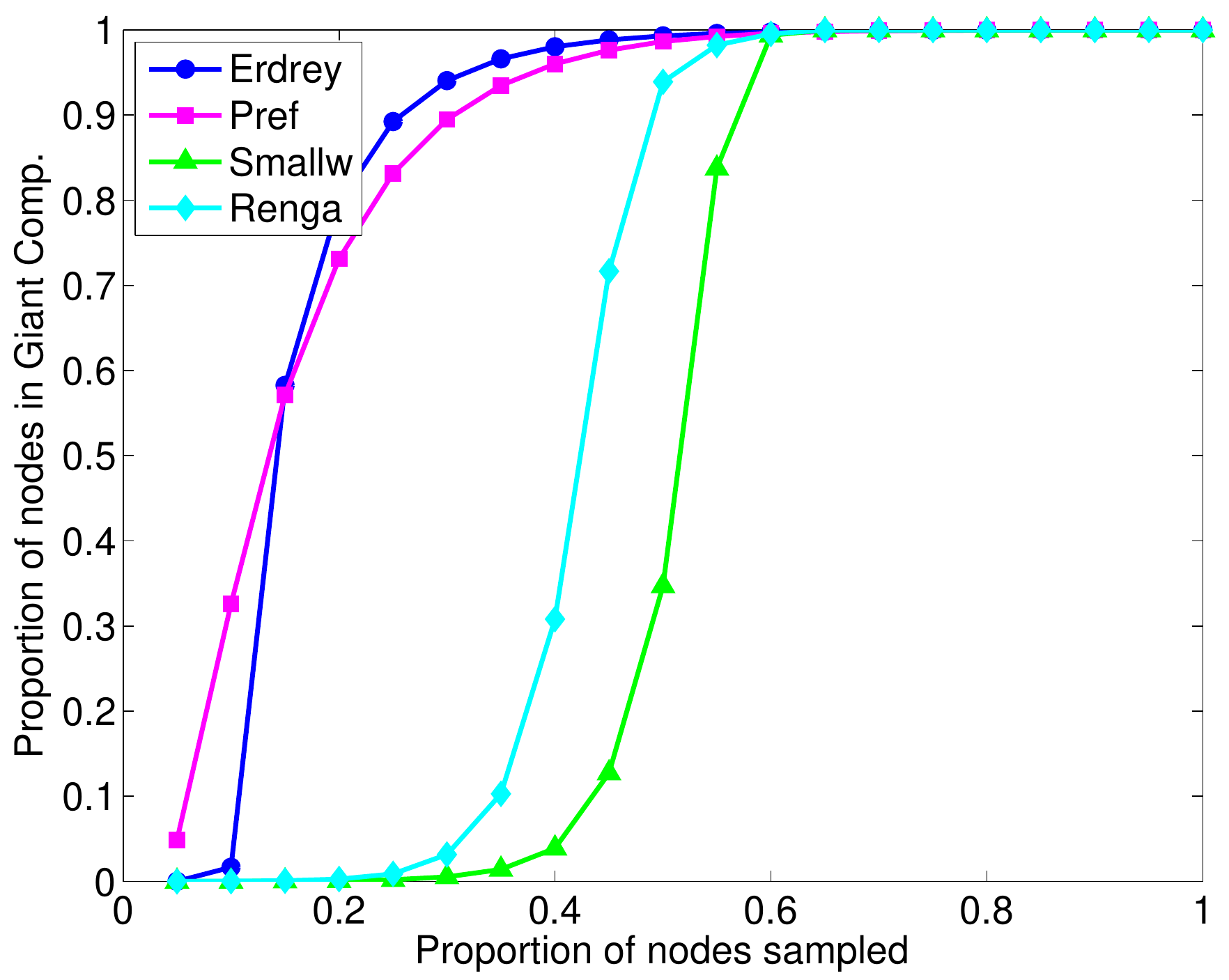}}\\
\caption[Scaling of subnetwork statistics for simulated networks obtained by failing links]{Scaling of subnetwork statistics for simulated networks obtained by failing links. (a.) When all nodes are known $q$ links are observed through sampling, the sample statistic for the number of nodes $n$ equals the true number of nodes $N$. It should be noted, though, that some nodes of degree 0 may be observed and these are counted as nodes (not discarded). (b.) The number of edges scales linearly as $M_{\text{obs}}=qM$. (c.) The average degree scales linearly as $k^{\textnormal{obs}}_{\rm avg}=\frac{k^{\textnormal{true}}_{\rm avg}}{q}$. (d.) The max degree scales linearly for Pref, but nonlinearly for other networks which have several nodes with degree similar to $\kmax$. (e.) Clustering scales roughly linearly with $q$. (f.) The percolation threshold for random graphs (Erd{\"o}s-R\'{e}nyi and Preferential attachment) roughly corresponds to the $q$ for which $k_{\rm avg} \geq 1$. Smallworld and Renga show more fragility and have a threshold which is closer to $q \approx 0.4$.}
\label{fig:sampling_by_failing_links_scaling}
\end{figure*}
\setcounter{equation}{6}
\begin{figure*}[!ht]  
\centering
\subfigure[Nodes]{\includegraphics[width=.3\textwidth]{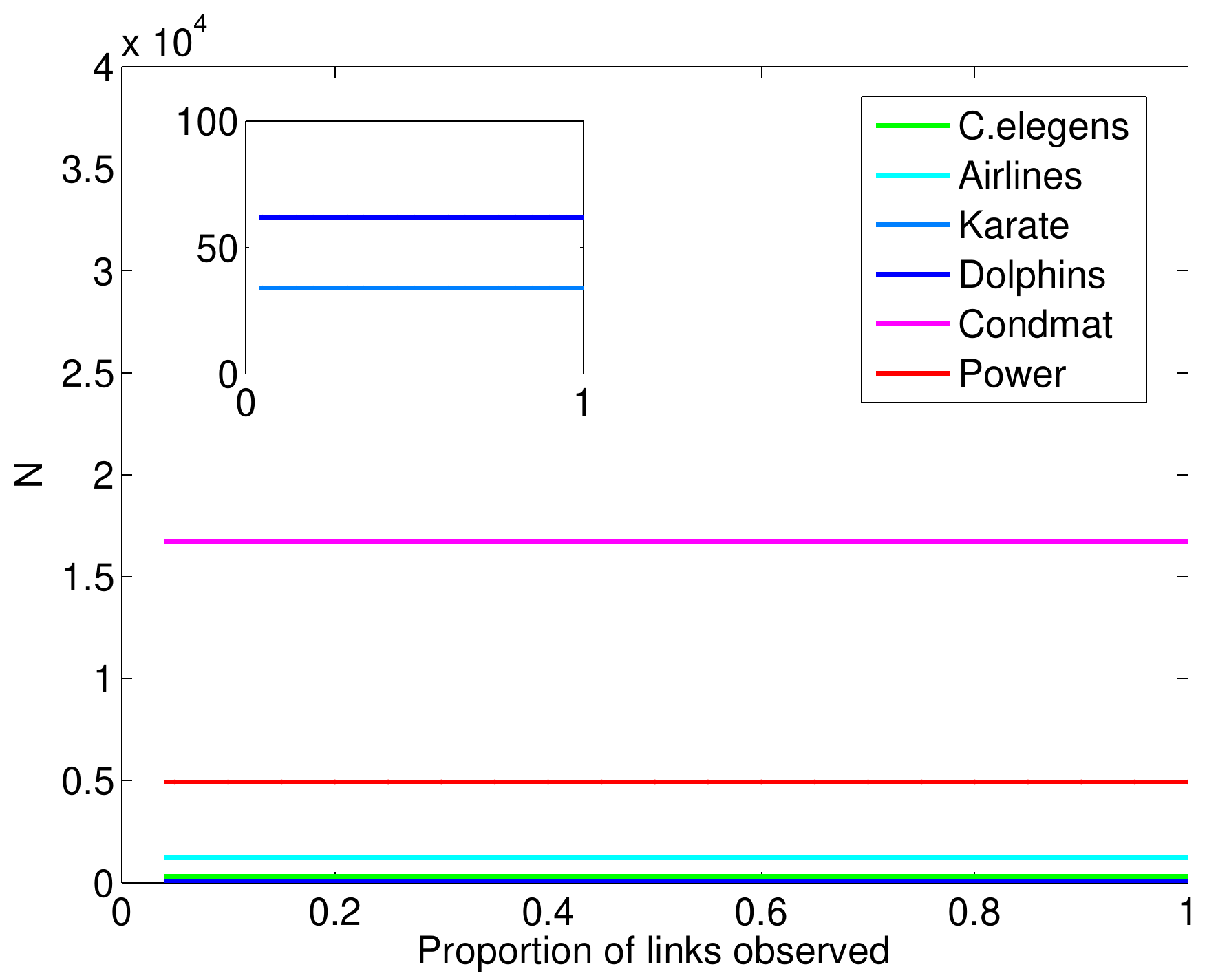}}
\subfigure[Edges]{\includegraphics[width=.3\textwidth]{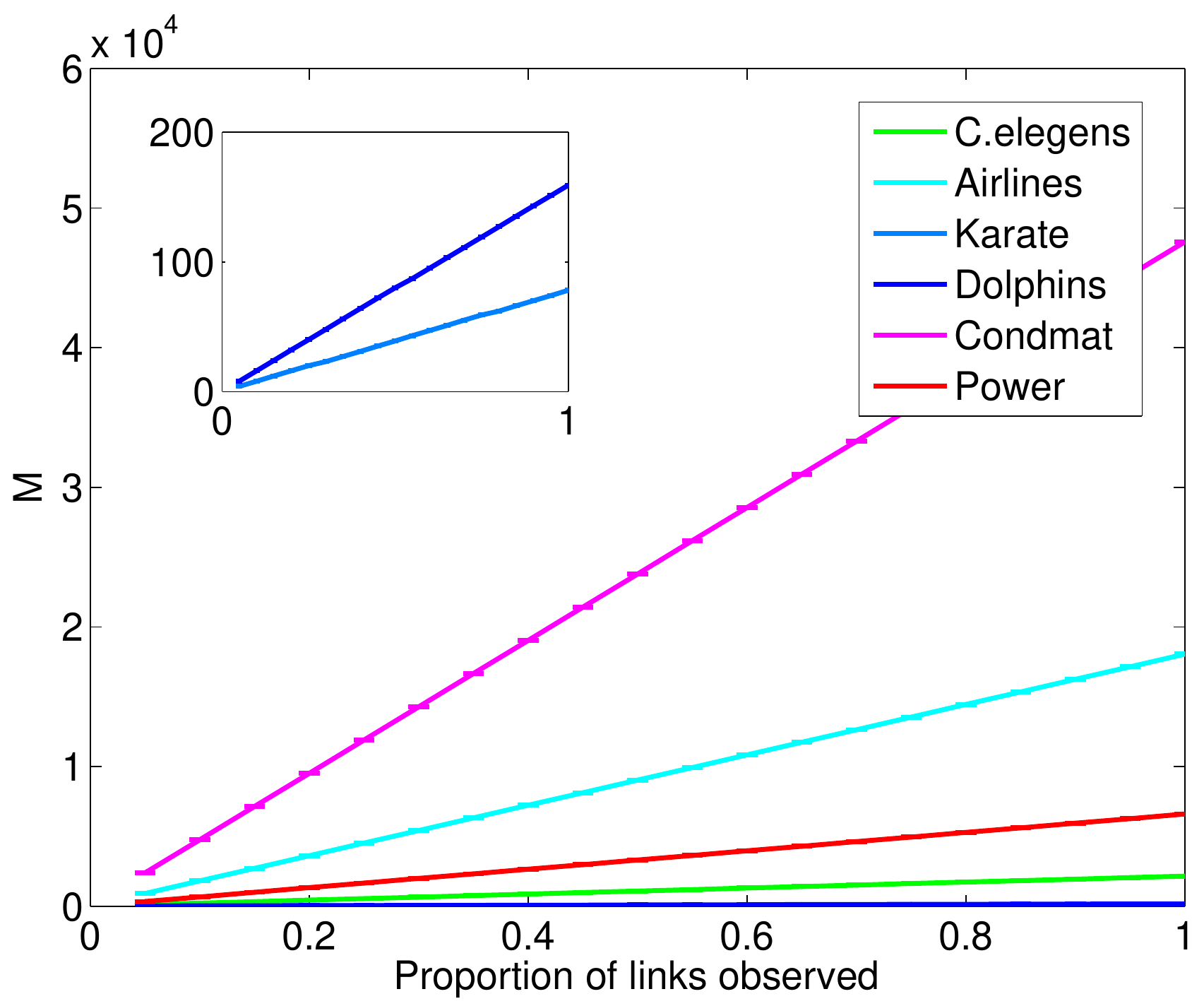}}
\subfigure[Average degree]{\includegraphics[width=.3\textwidth]{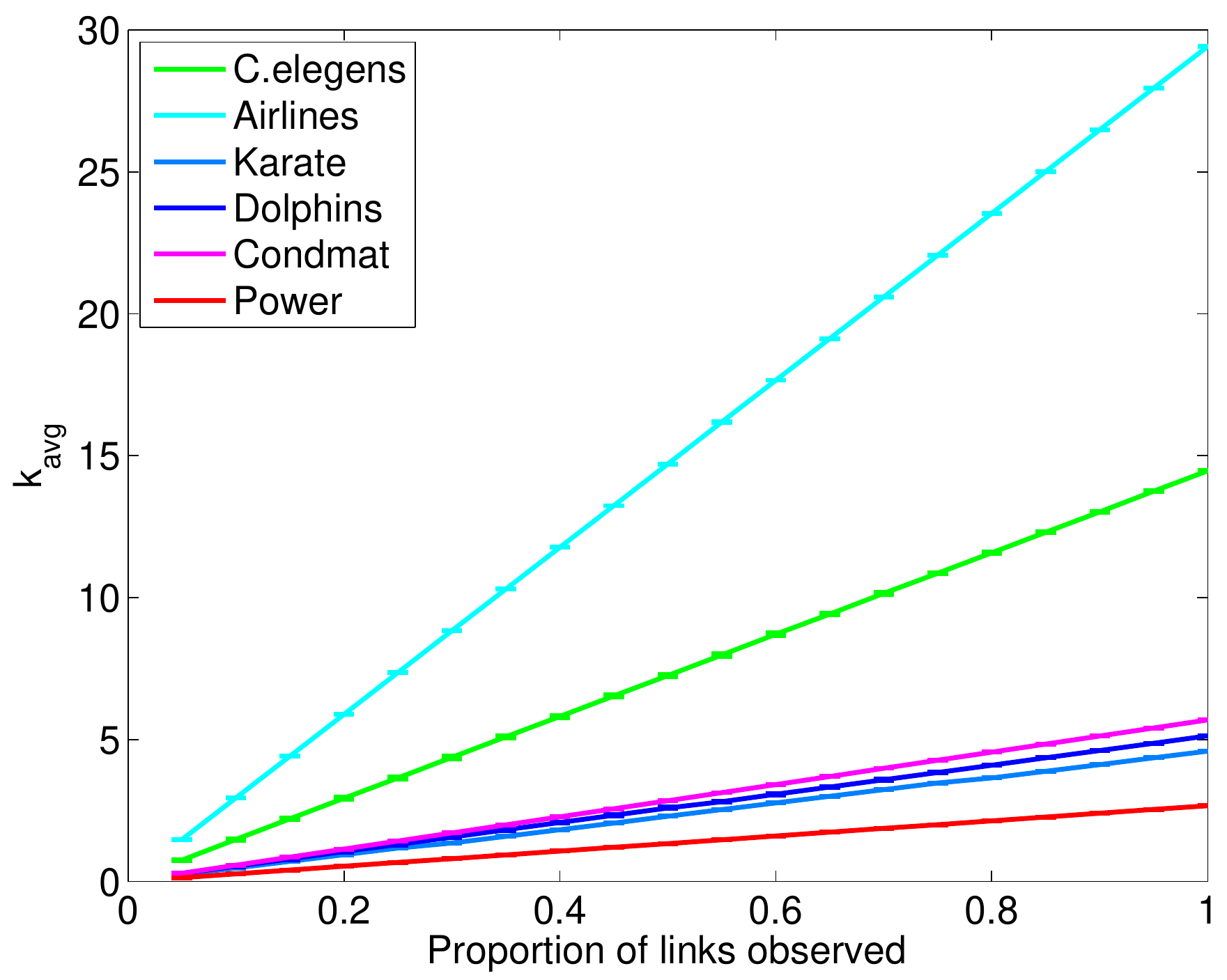}}\\
\subfigure[Max degree]{\includegraphics[width=.3\textwidth]{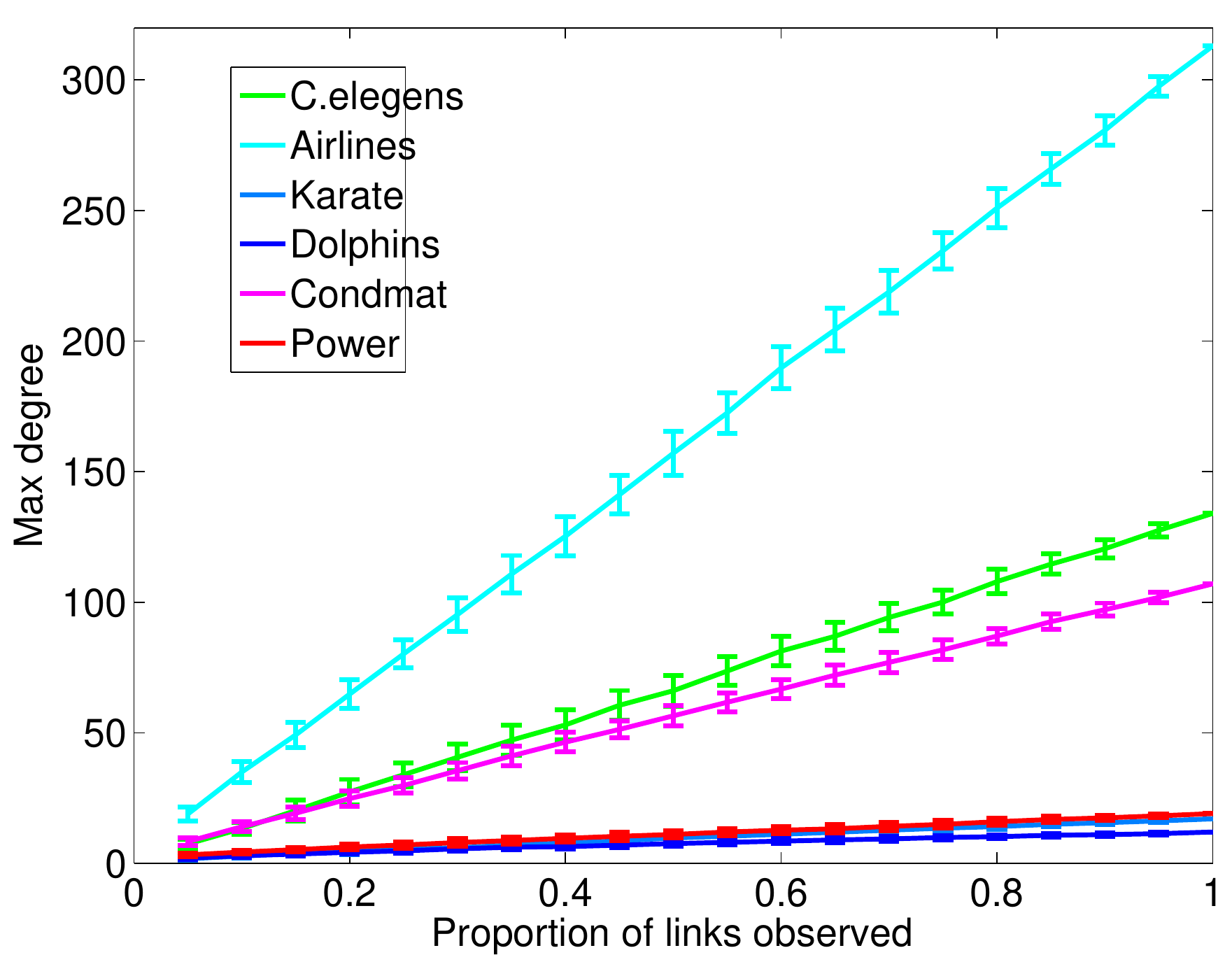}}
\subfigure[Clustering]{\includegraphics[width=.3\textwidth]{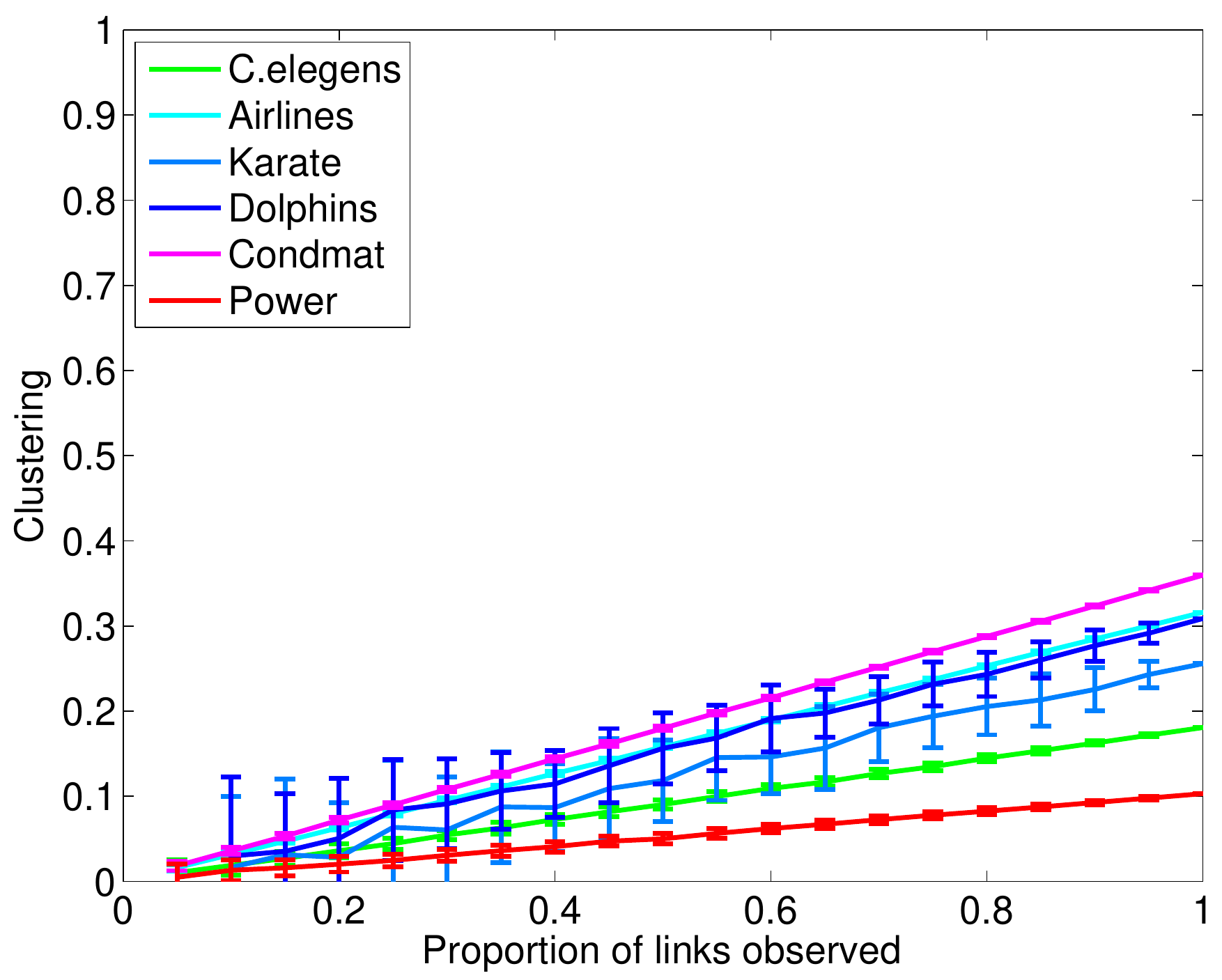}}
\subfigure[Prop. of nodes in Giant Component]{\includegraphics[width=.3\textwidth]{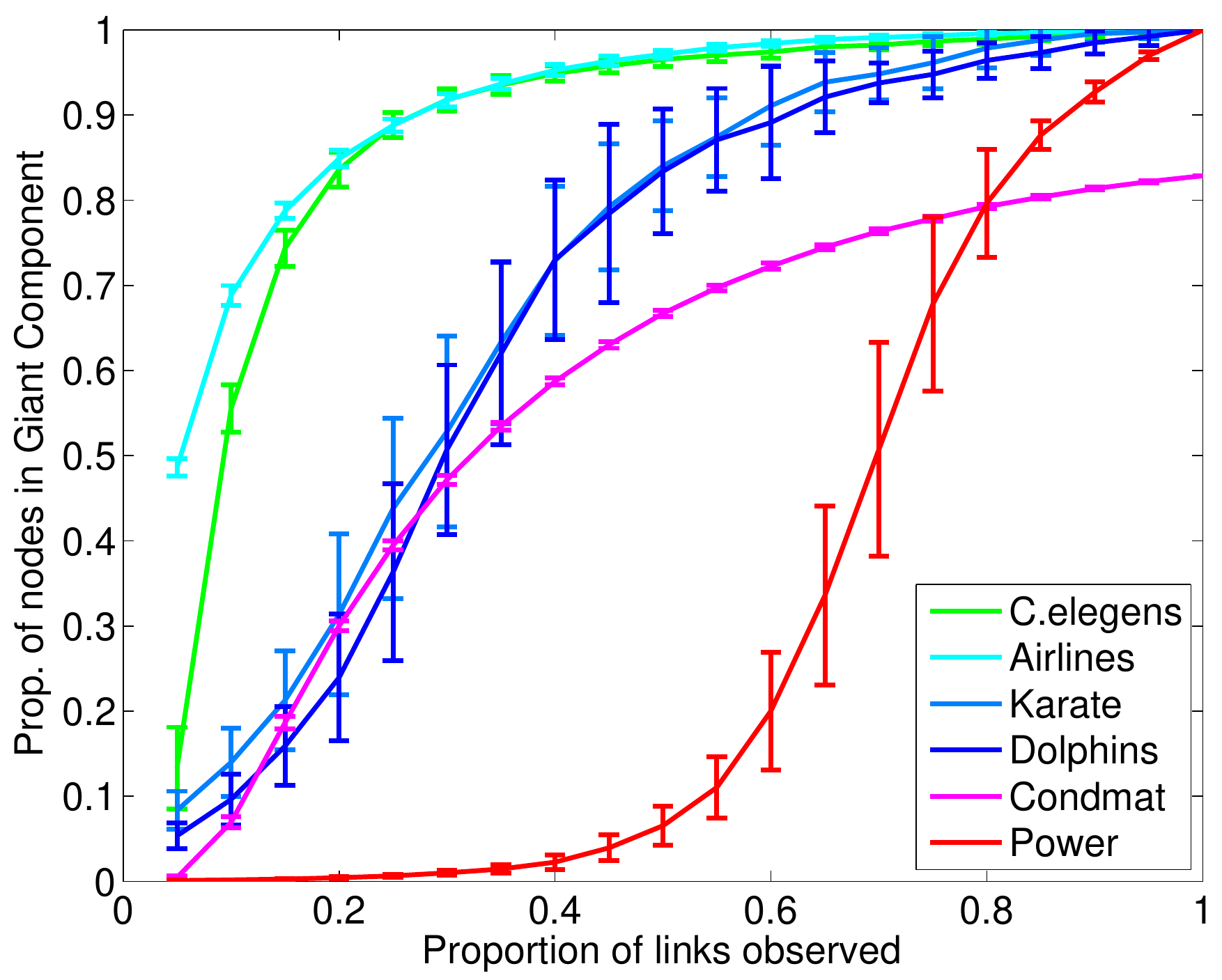}}\\
\caption[Scaling of subnetwork statistics for empirical networks obtained by failing links]{Scaling of subnetwork statistics for empirical networks obtained by failing links. (a.) When all nodes are known $q$ links are observed through sampling, the sample statistic for the number of nodes $n$ equals the true number of nodes $N$. It should be noted, though, that some nodes of degree 0 may be observed and these are counted as nodes (not discarded). (b.) The number of edges scales linearly as $M_{\text{obs}}=qM$. (c.) The average degree scales linearly as $k^{\textnormal{obs}}_{\rm avg}=\frac{k^{\textnormal{true}}_{\rm avg}}{q}$. (d.) The max degree scales linearly  (e.) Clustering scales roughly linearly with $q$. (f.) The percolation threshold roughly corresponds to the $q$ for which $k_{\rm avg} \geq 1$.}
\label{fig:sampling_by_failing_links_scaling}
\end{figure*}

\setcounter{equation}{7}
\begin{figure*}[!ht]
\centering
\subfigure[Erdrey]{\includegraphics[width=.28\textwidth]{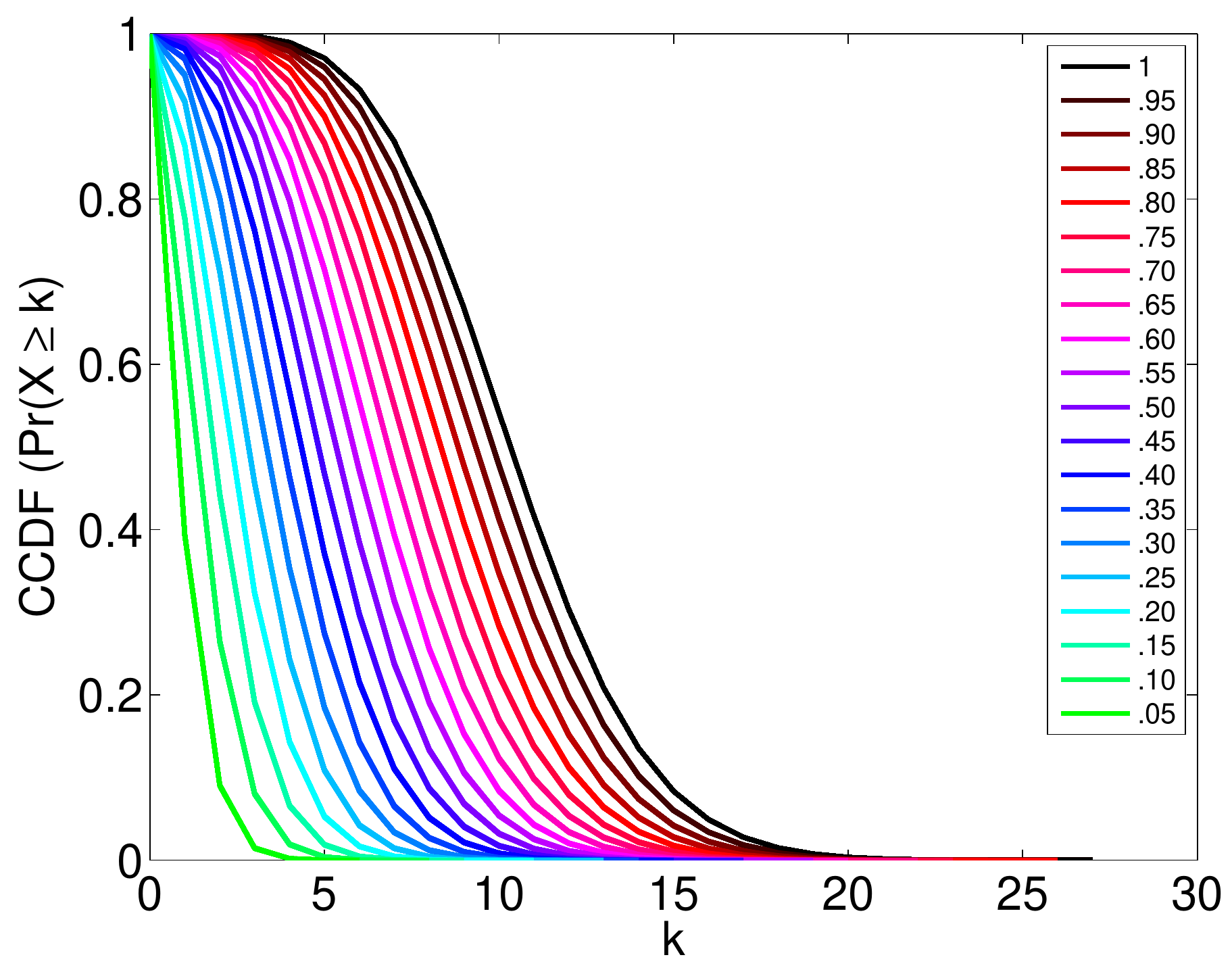}}
\subfigure[Pref]{\includegraphics[width=.28\textwidth]{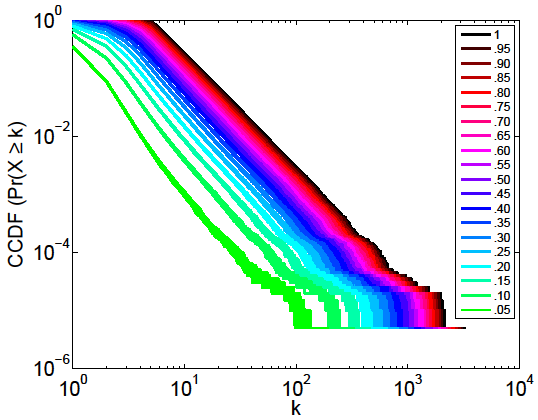}}
\subfigure[Smallworld]{\includegraphics[width=.28\textwidth]{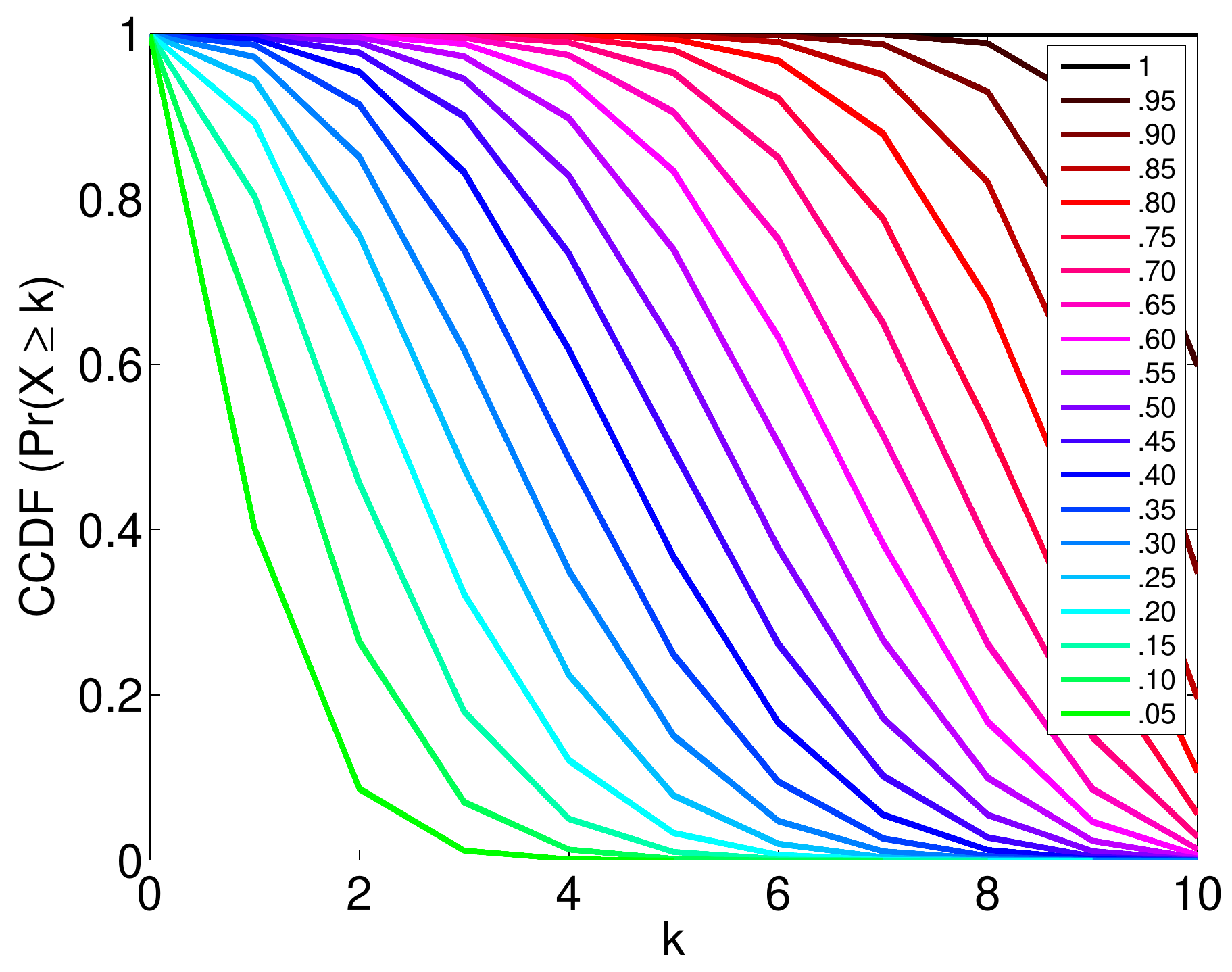}}\\
\subfigure[Renga]{\includegraphics[width=.28\textwidth]{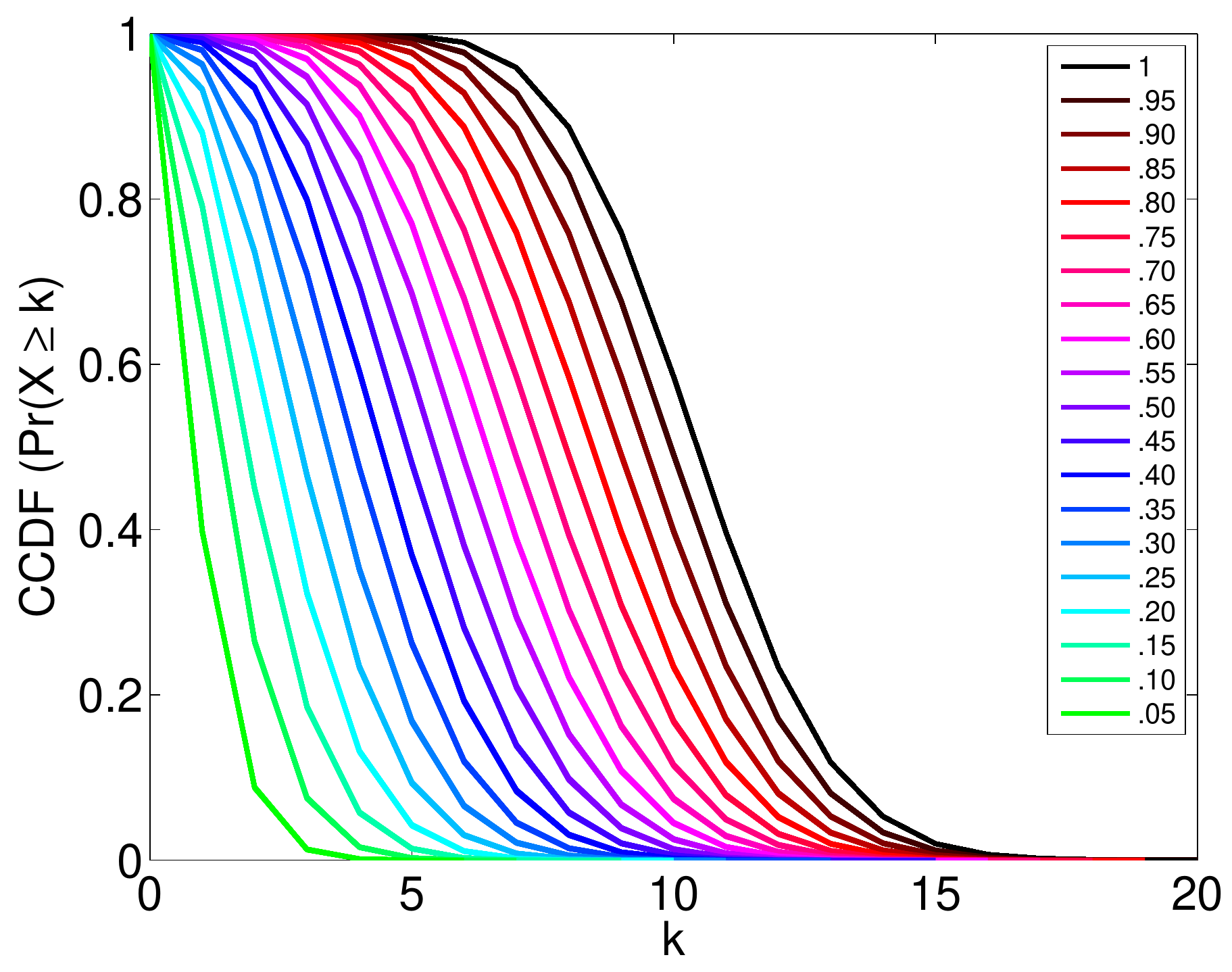}}
\subfigure[C. elgegans]{\includegraphics[width=.28\textwidth]{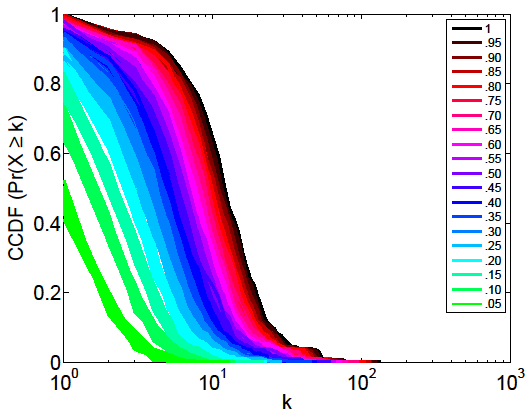}}
\subfigure[Airlines]{\includegraphics[width=.28\textwidth]{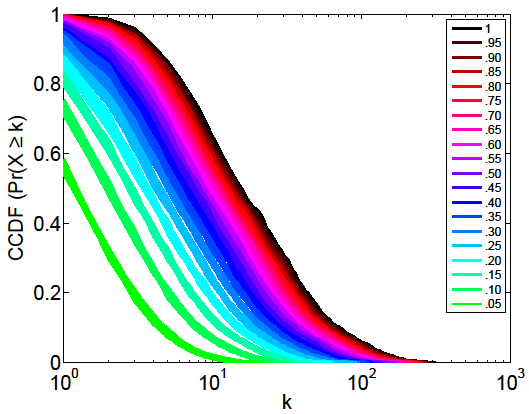}}\\
\subfigure[Karate]{\includegraphics[width=.28\textwidth]{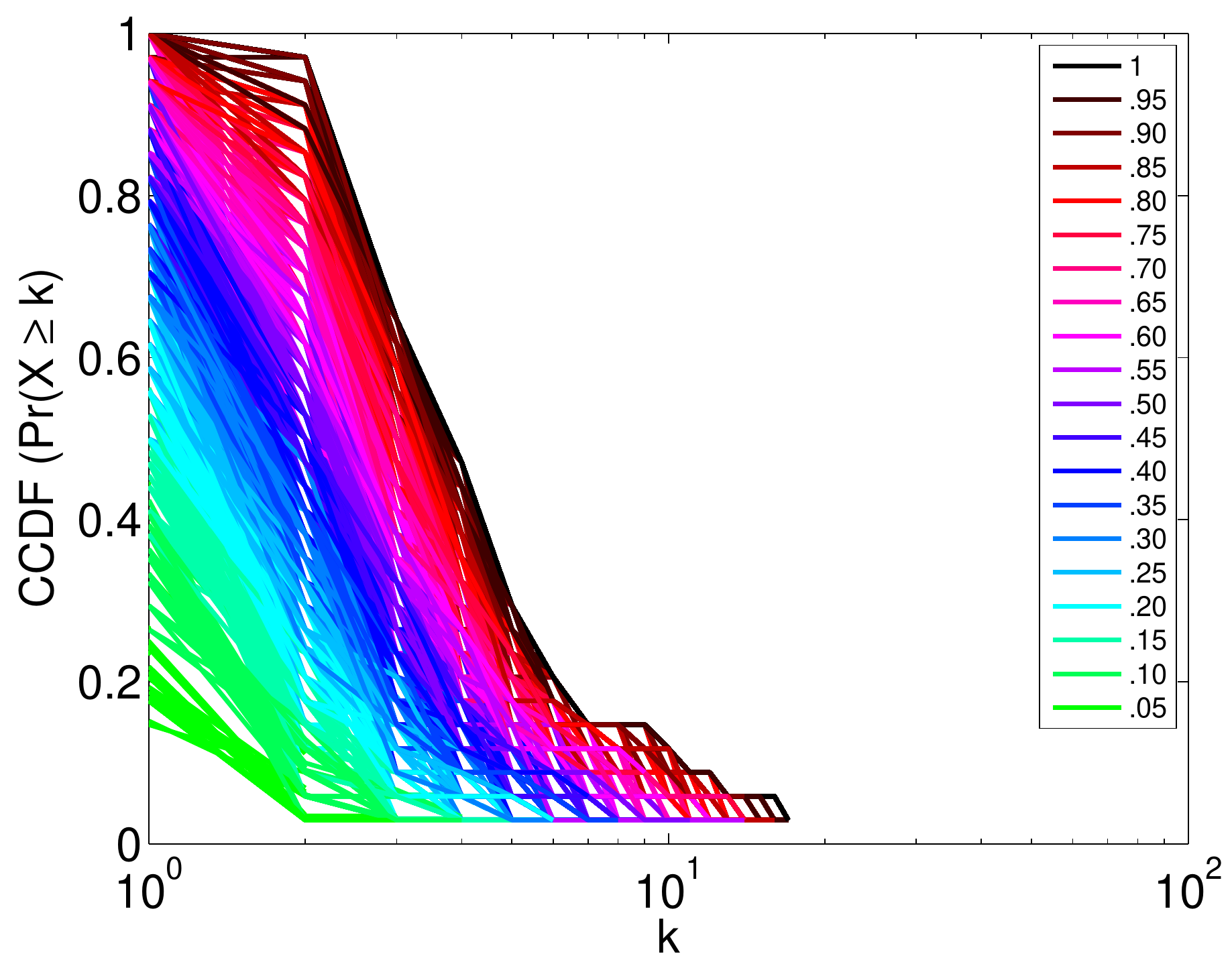}}
\subfigure[Dolphins]{\includegraphics[width=.28\textwidth]{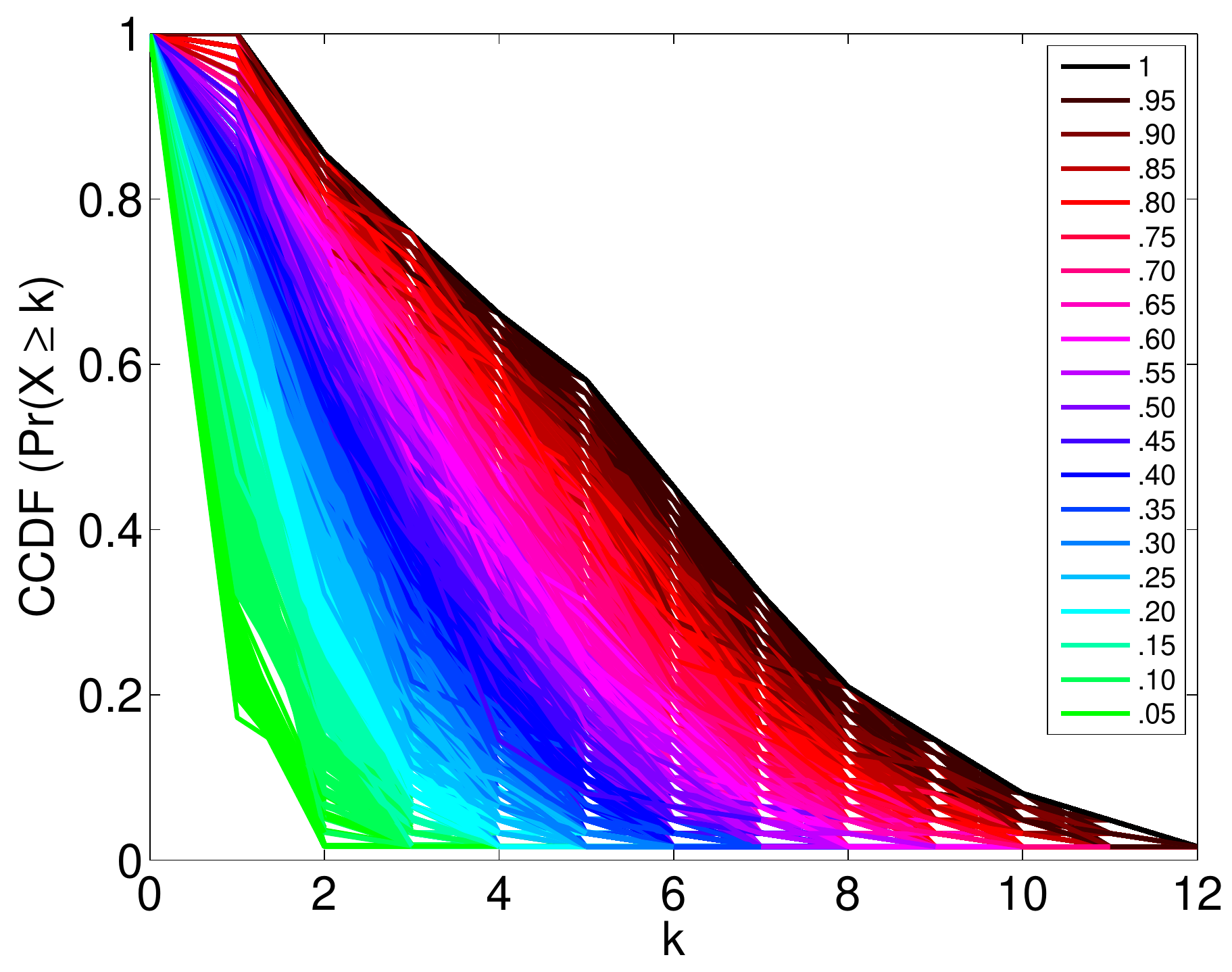}}
\subfigure[Condmat]{\includegraphics[width=.28\textwidth]{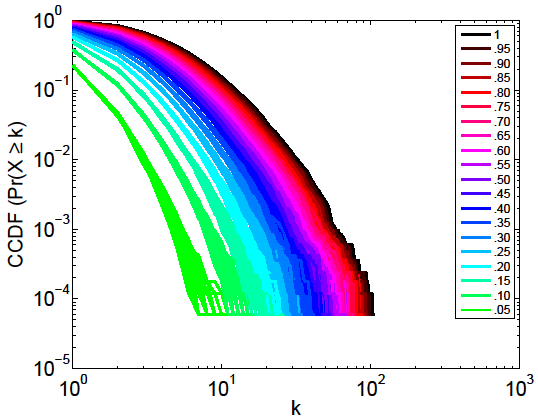}}\\
\subfigure[Powergrid]{\includegraphics[width=.28\textwidth]{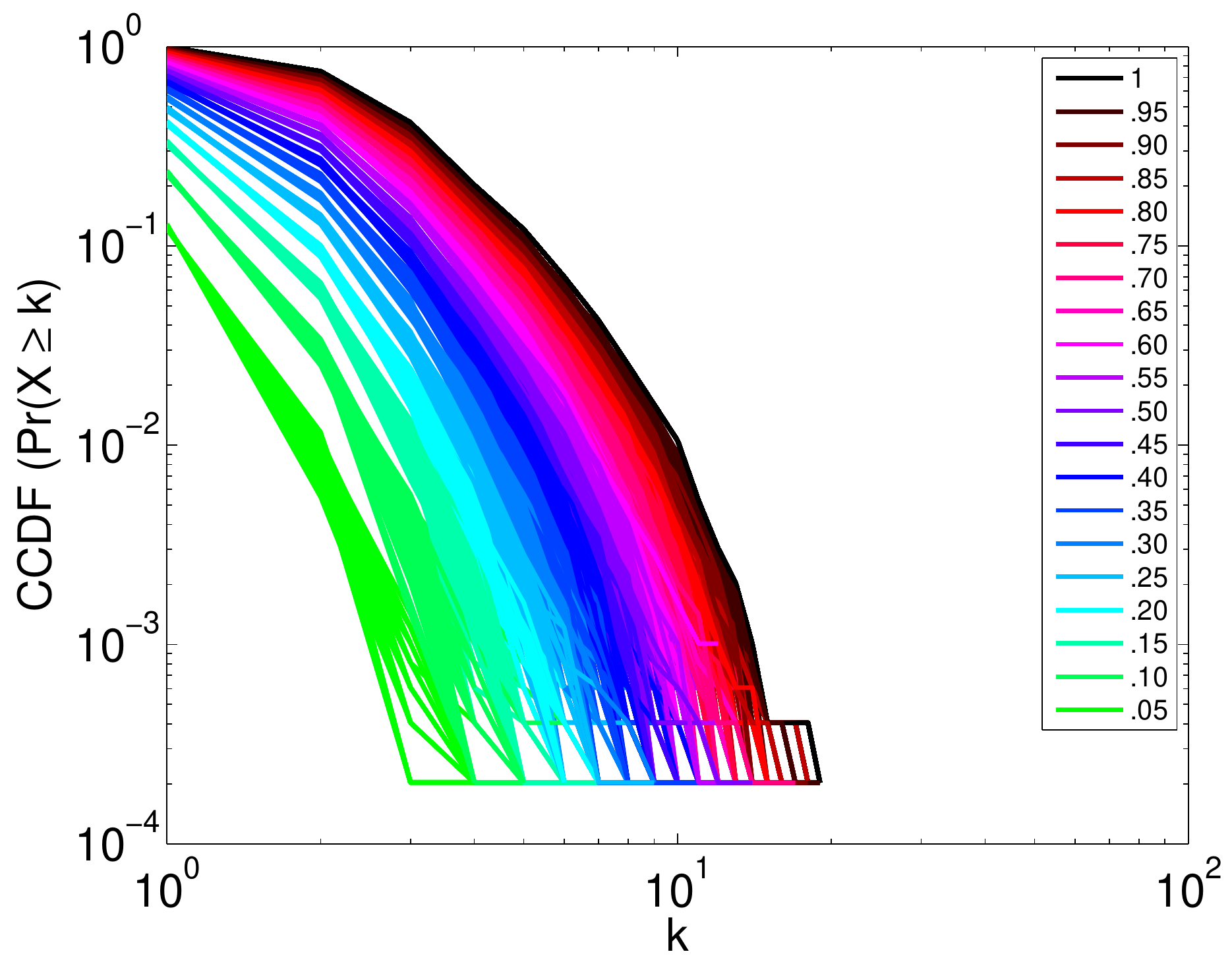}}
\caption[CCDF distortion for subnetworks obtained by failing links]{CCDF distortion for subnetworks obtained by failing links. Subnetwork degree distributions do not capture the true degree distribution, especially for small $q$.}
\label{fig:failed_links_ccpdf}
\end{figure*}
\setcounter{equation}{8}
\begin{figure*}[!ht]
\centering
\subfigure[Erdrey$^+$]{\includegraphics[width=.28\textwidth]{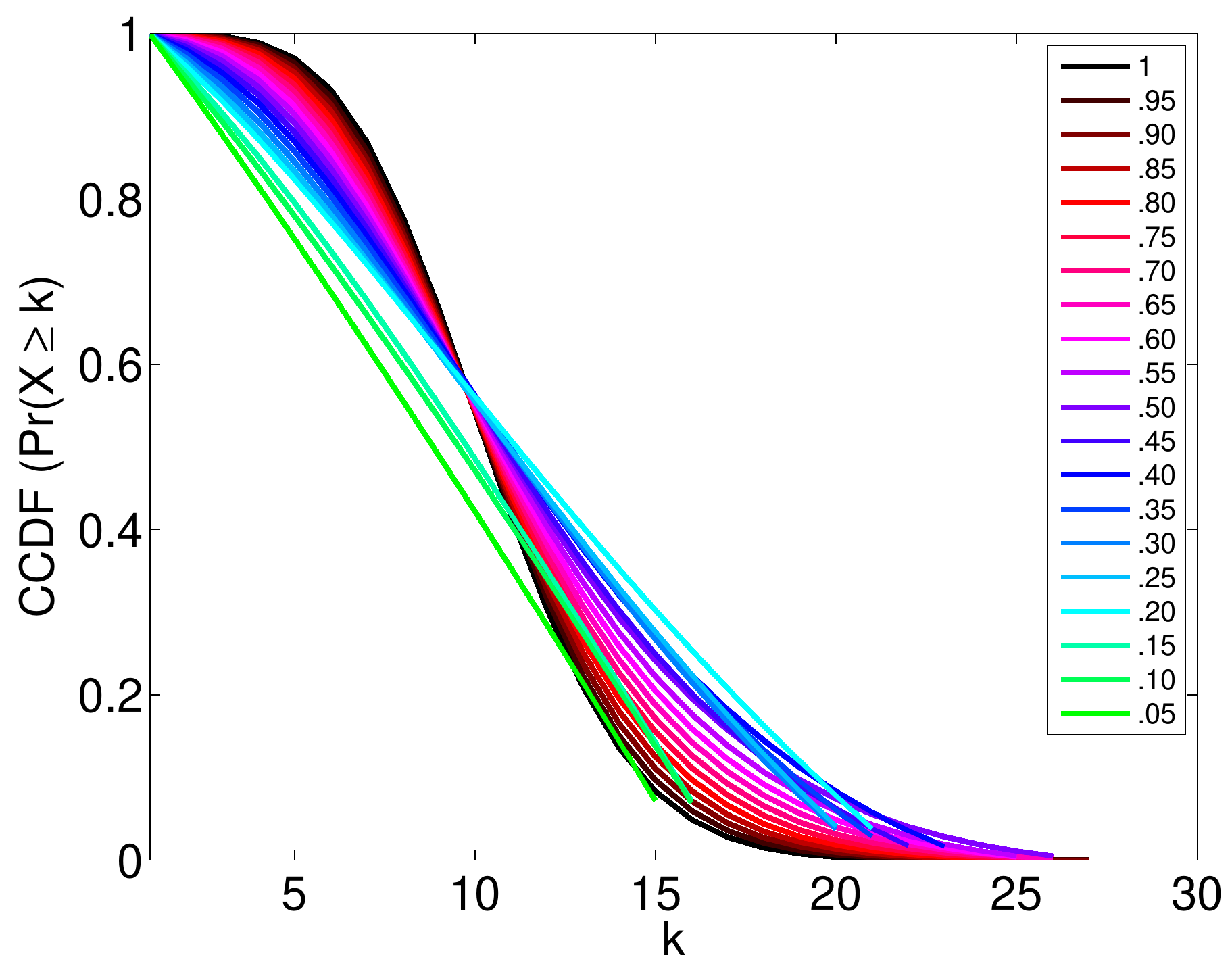}}
\subfigure[Pref$^*$]{\includegraphics[width=.28\textwidth]{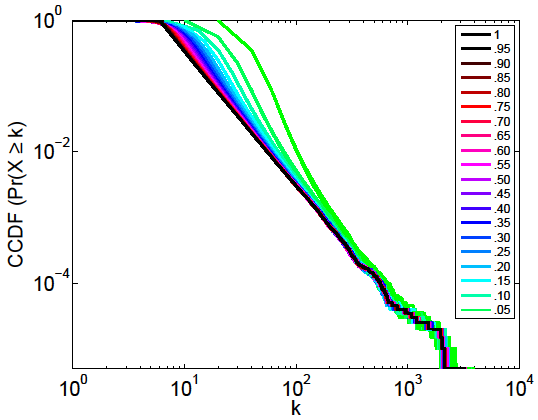}}
\subfigure[Smallworld$^+$]{\includegraphics[width=.28\textwidth]{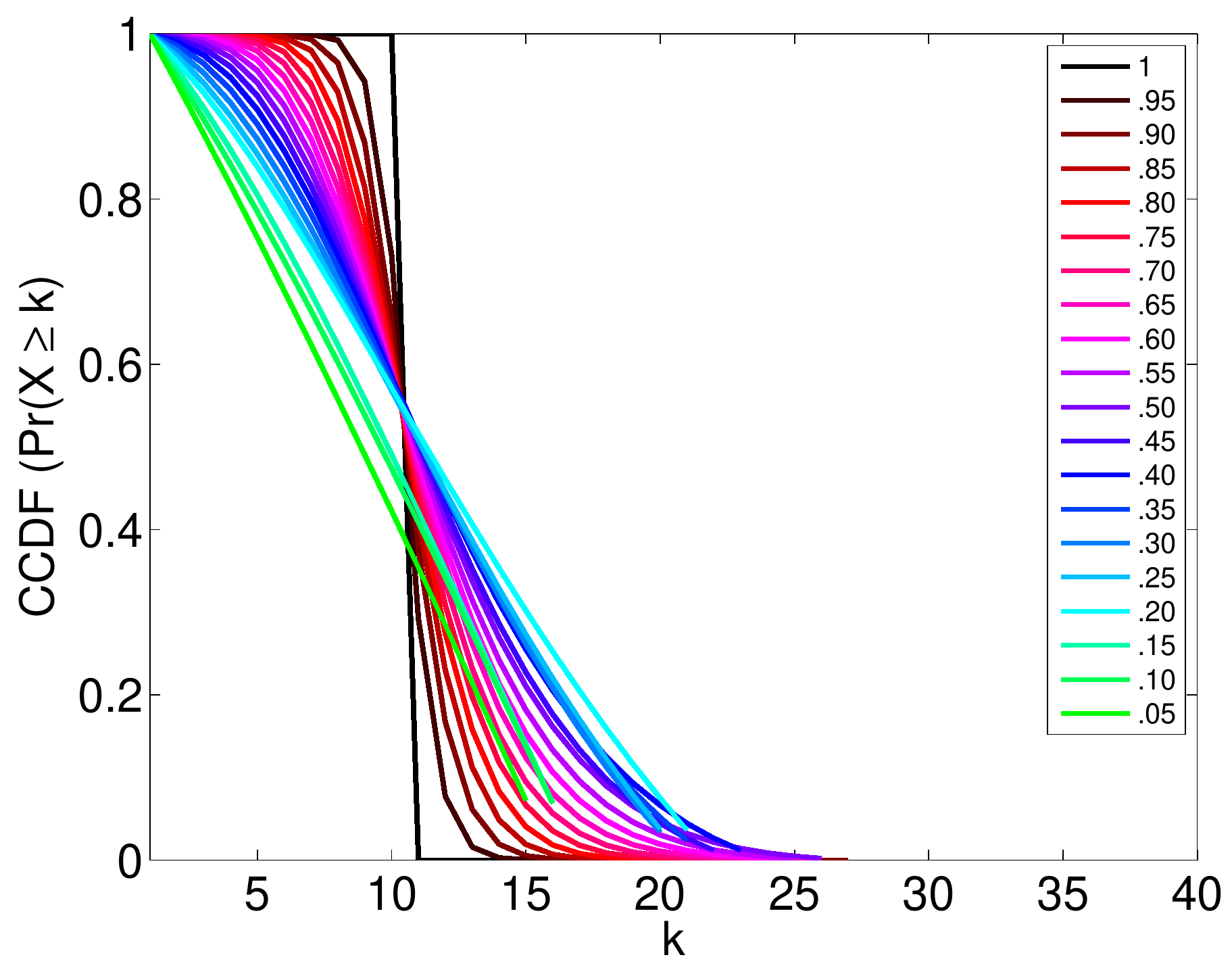}}\\
\subfigure[Renga$^+$]{\includegraphics[width=.28\textwidth]{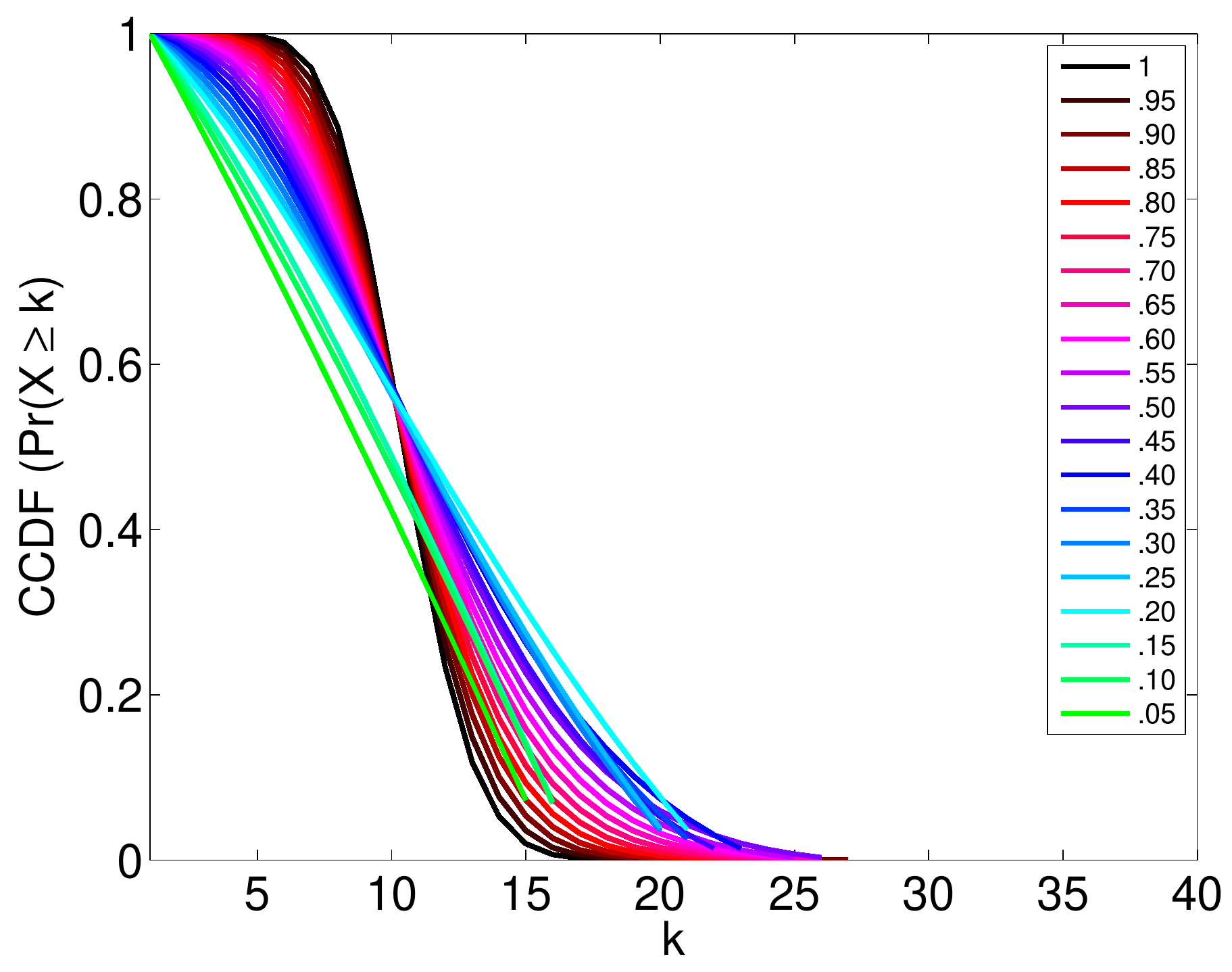}}
\subfigure[C. elgegans$^*$]{\includegraphics[width=.28\textwidth]{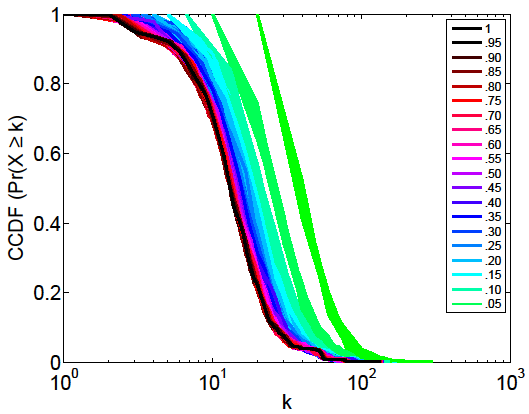}}
\subfigure[Airlines$^*$]{\includegraphics[width=.28\textwidth]{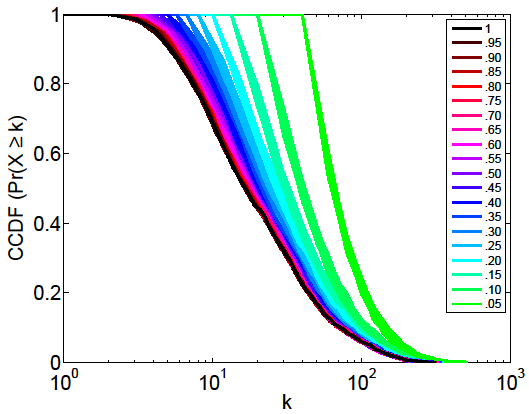}}\\
\subfigure[Karate$^+$]{\includegraphics[width=.28\textwidth]{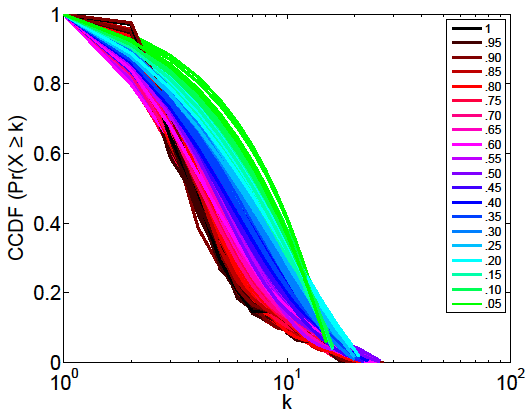}}
\subfigure[Dolphins$^+$]{\includegraphics[width=.28\textwidth]{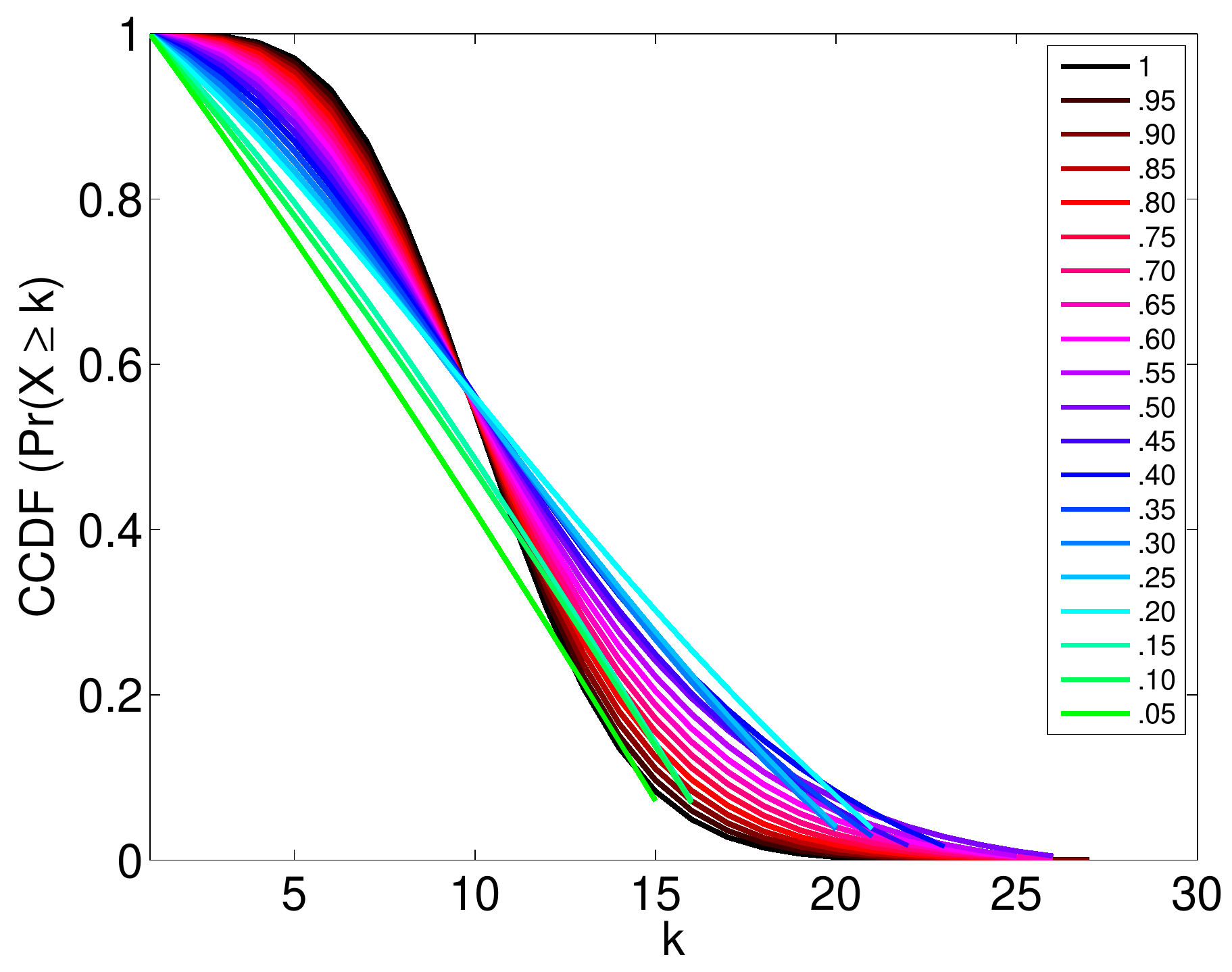}}
\subfigure[Condmat$^*$]{\includegraphics[width=.28\textwidth]{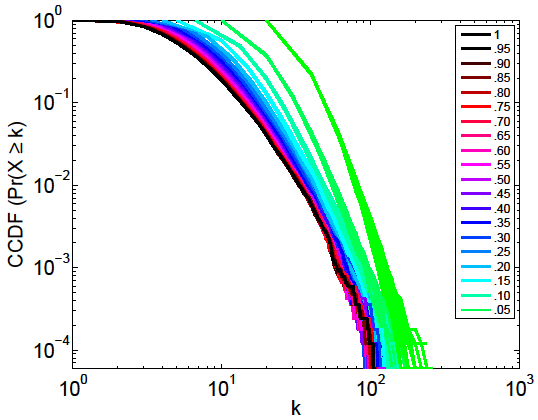}}\\
\subfigure[Powergrid$^*$]{\includegraphics[width=.28\textwidth]{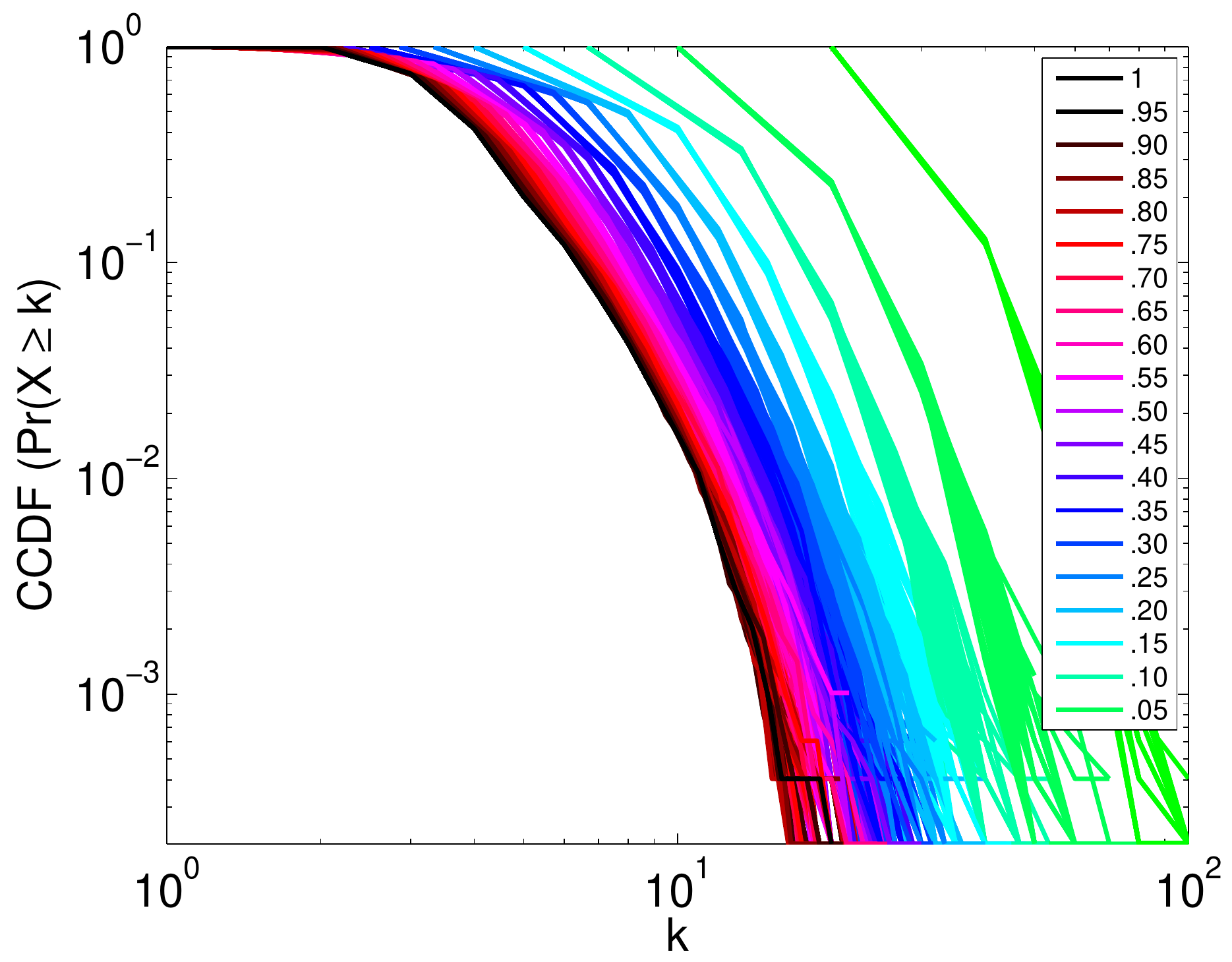}}
\caption[Predicted CCDF from subnetworks obtained by failing links]{Predicted CCDF from subnetworks obtained by failing links. The predicted CCDF shows relatively good agreement with the CCDF for most networks. Karate club and Dolphins exhibit significant deviations, possibly due to the small number of nodes in these networks. Networks designated $\text{with }^+$ utilized Equation~\ref{eq:dist_rollback} and those designated with $\text{with }^*$ utilized Equation~\ref{eq:my_dist_shorter}. }
\label{fig:predicting_failed_links_empirical_ccpdf}
\end{figure*}

\setcounter{equation}{9}
\begin{figure*}[!ht]
\centering
\subfigure[Nodes]{\includegraphics[width=.3\textwidth]{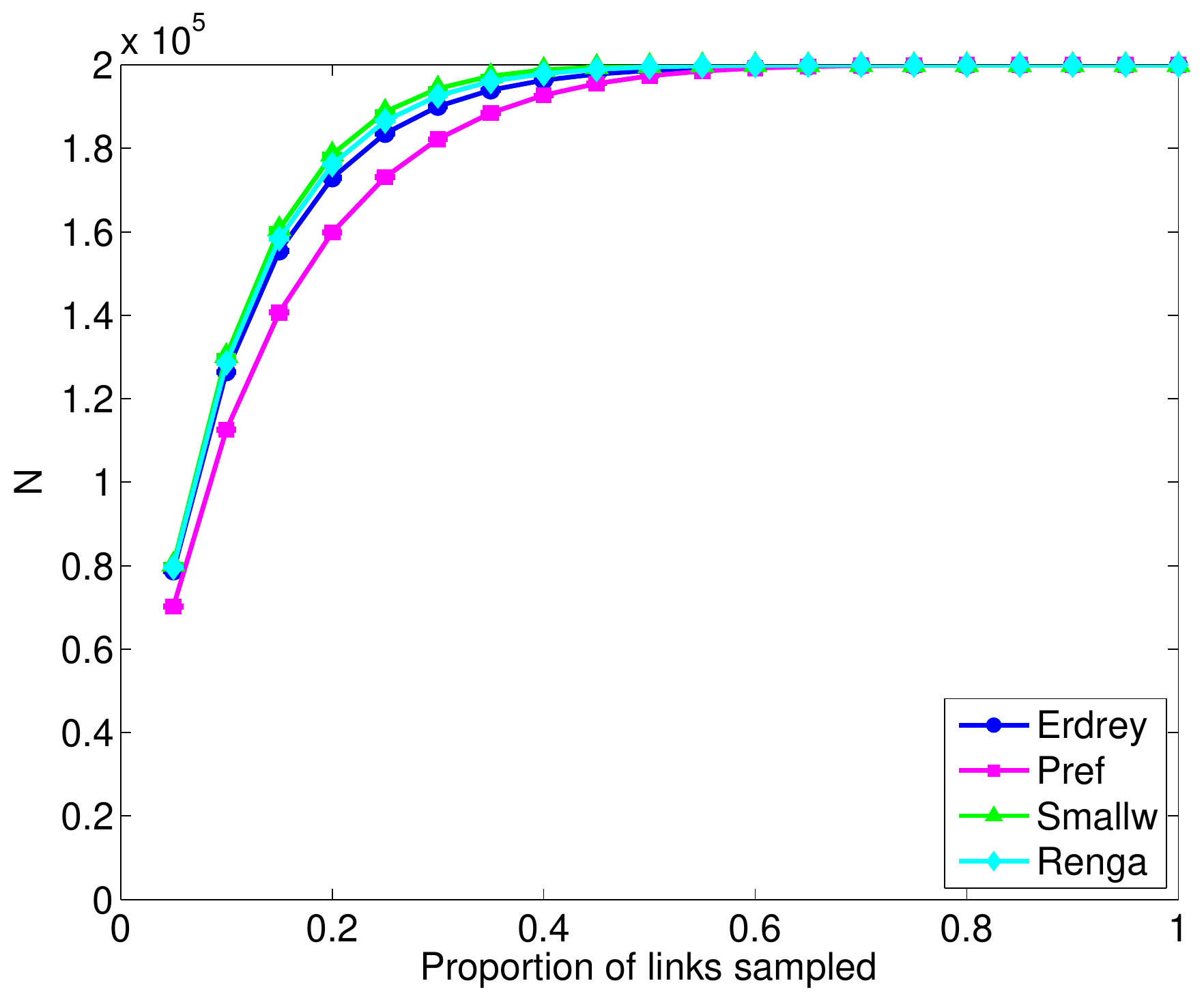}}
\subfigure[Edges]{\includegraphics[width=.3\textwidth]{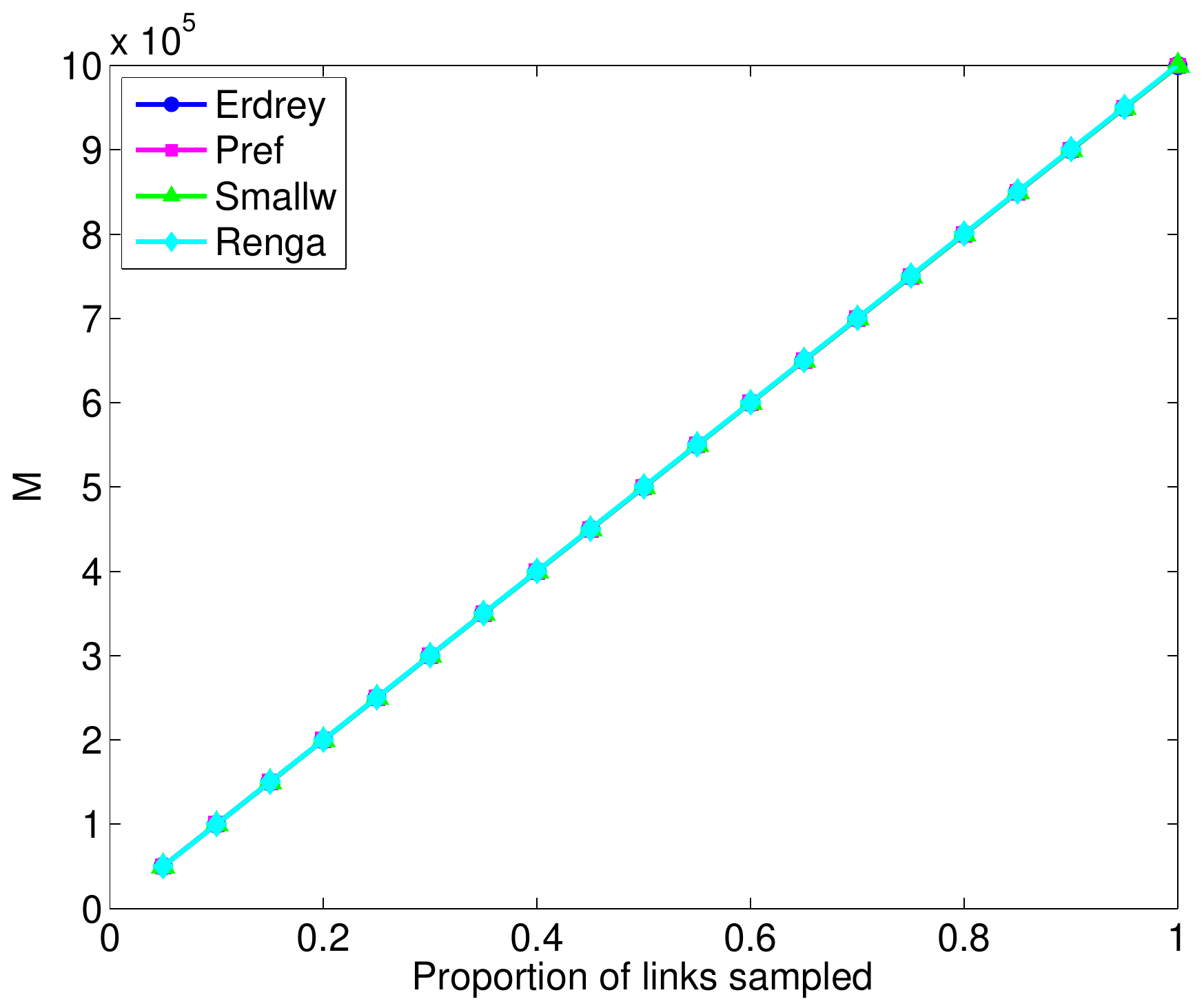}}
\subfigure[Average degree]{\includegraphics[width=.3\textwidth]{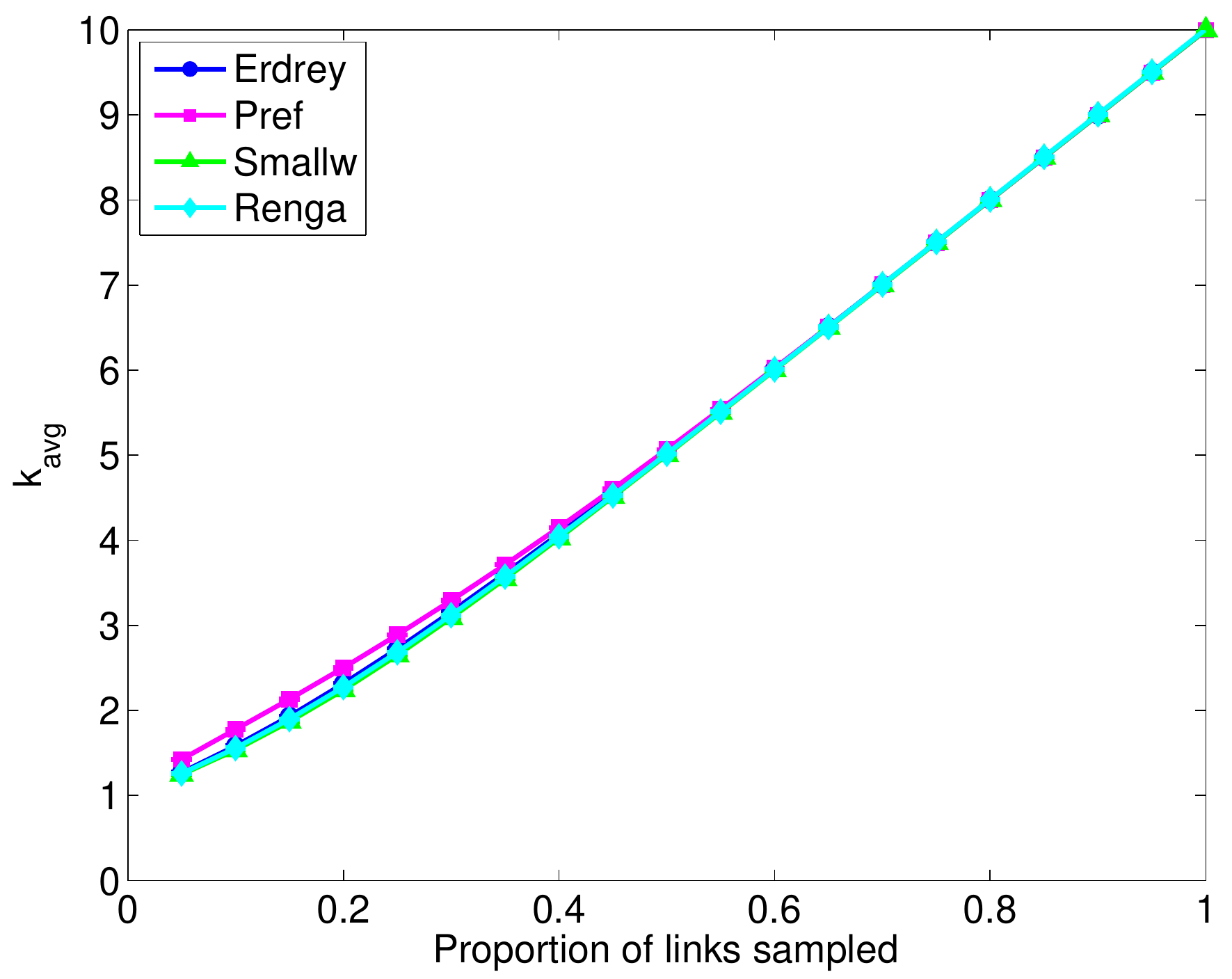}}\\
\subfigure[Max degree]{\includegraphics[width=.3\textwidth]{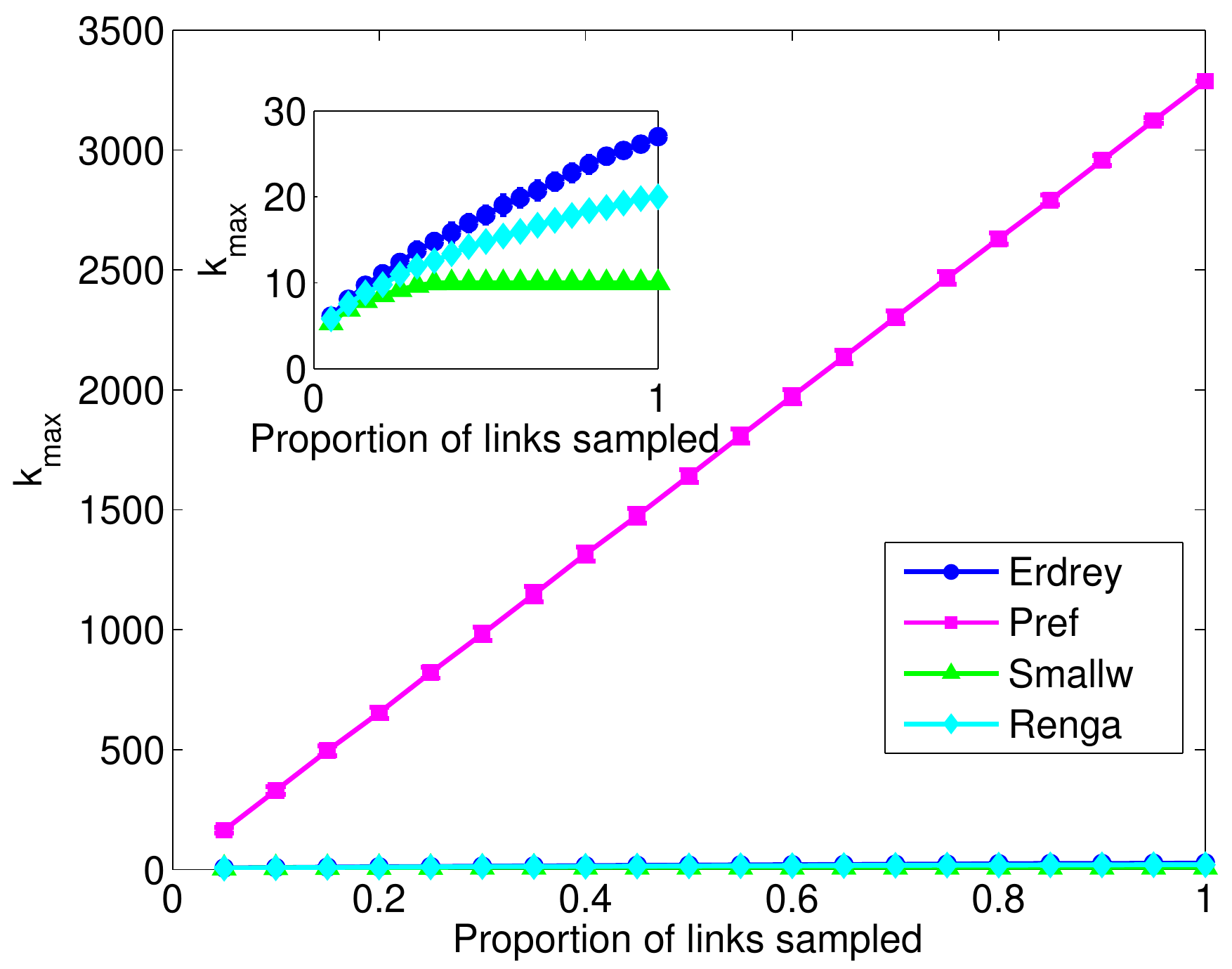}}
\subfigure[Clustering]{\includegraphics[width=.3\textwidth]{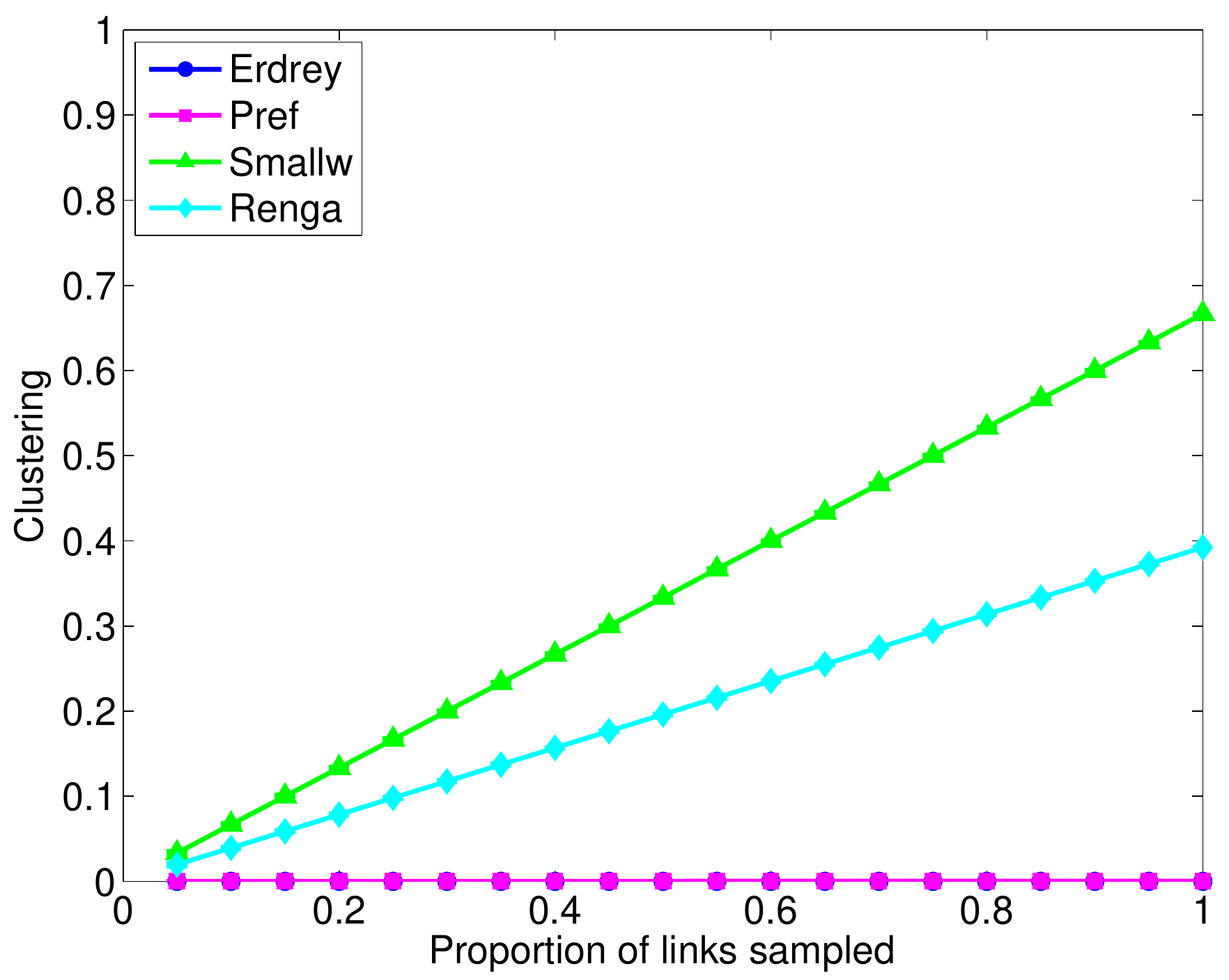}}
\subfigure[Prop. of links in Giant Component]{\includegraphics[width=.3\textwidth]{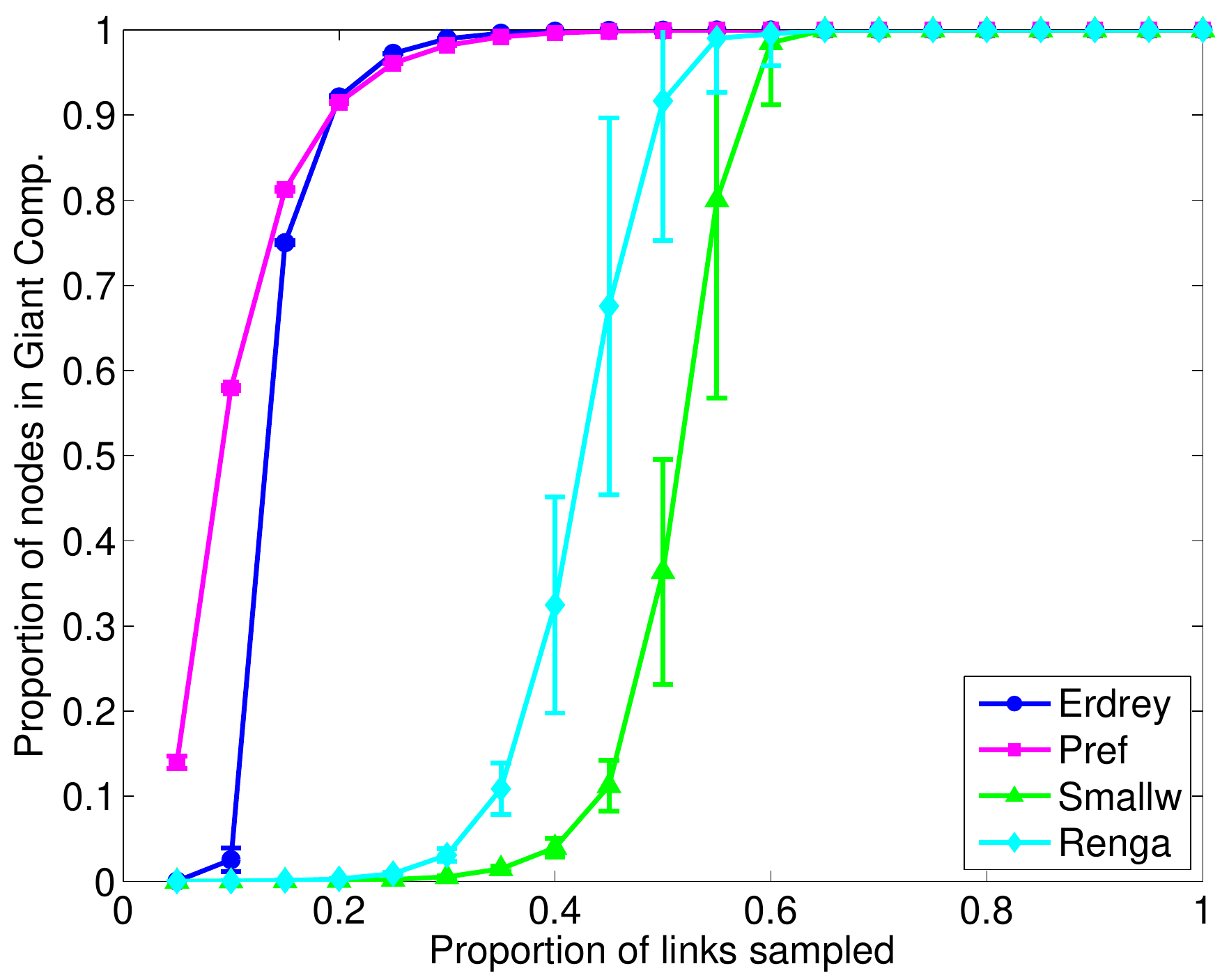}}\\
\caption[Scaling of subnetwork statistics for simulated networks induced on sampled links]{Scaling of subnetwork statistics for simulated networks induced on sampled links. (a.) The number of nodes in a subnetwork sampled by links scales nonlinearly with $q$. (b.) The number of edges scales as $m \approx qM$. (c.) The average degree scales roughly linearly with the proportion of nodes subsampled $k^{\textnormal{sub}}_{\rm avg} \approx qk_{\rm avg}$ . (d.) The max degree scales roughly linearly for networks with few large hubs (e.g., Pref) and nonlinearly when there are several nodes with degrees roughly similar to $\kmax$. (e.) The clustering coefficient scales roughly linearly $C^{\textnormal{sub}} \approx q C$. (f.) The proportion of nodes in the giant component increases with the proportion of nodes sampled. For the random graphs (Erdrey and Pref) there is a critical point corresponding to the approximate sampling level when $k_{\rm avg} > 1$ (which corresponds to $q=0.1$). The thresholds for Small World and Range dependent networks are much higher due to the uniformity of the motif distribution in these networks. Markers indicates the mean over 100 simulations. Error bars showing one standard deviation are too small to see.}
\label{fig:sampling_by_links_incident_scaling}
\end{figure*}

\setcounter{equation}{10}
\begin{figure*}[!ht]
\centering
\subfigure[Nodes]{\includegraphics[width=.3\textwidth]{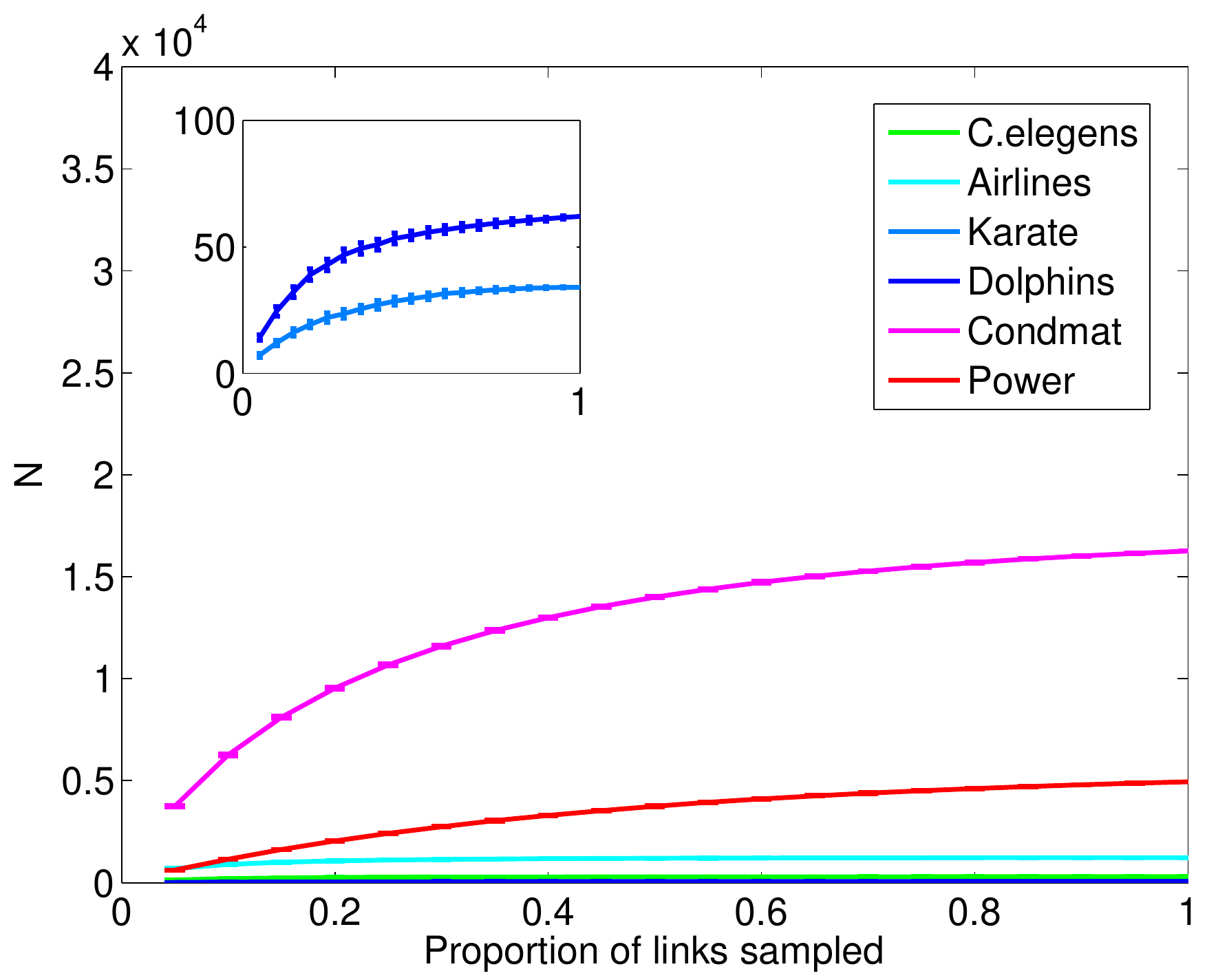}}
\subfigure[Edges]{\includegraphics[width=.3\textwidth]{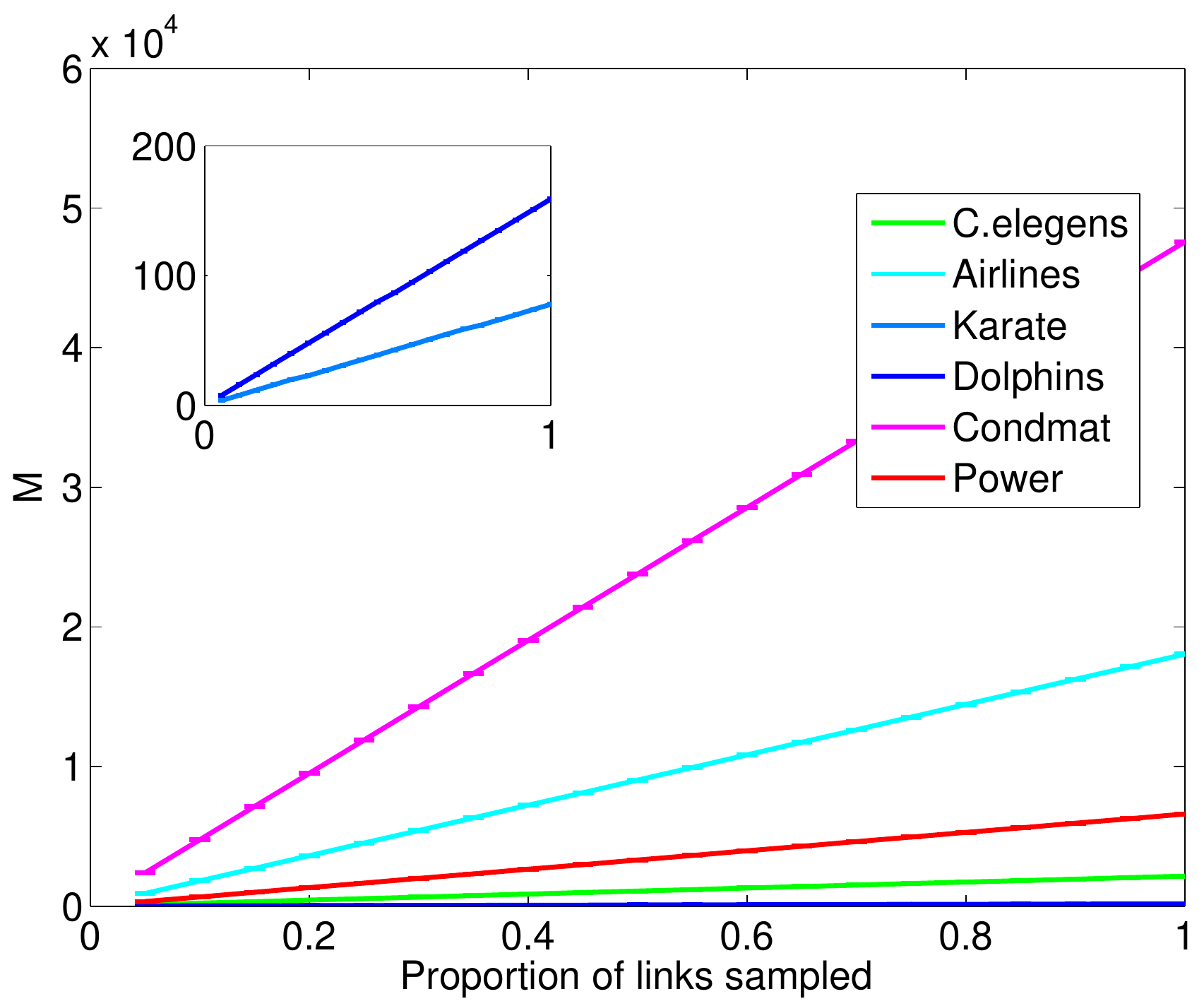}}
\subfigure[Average degree]{\includegraphics[width=.3\textwidth]{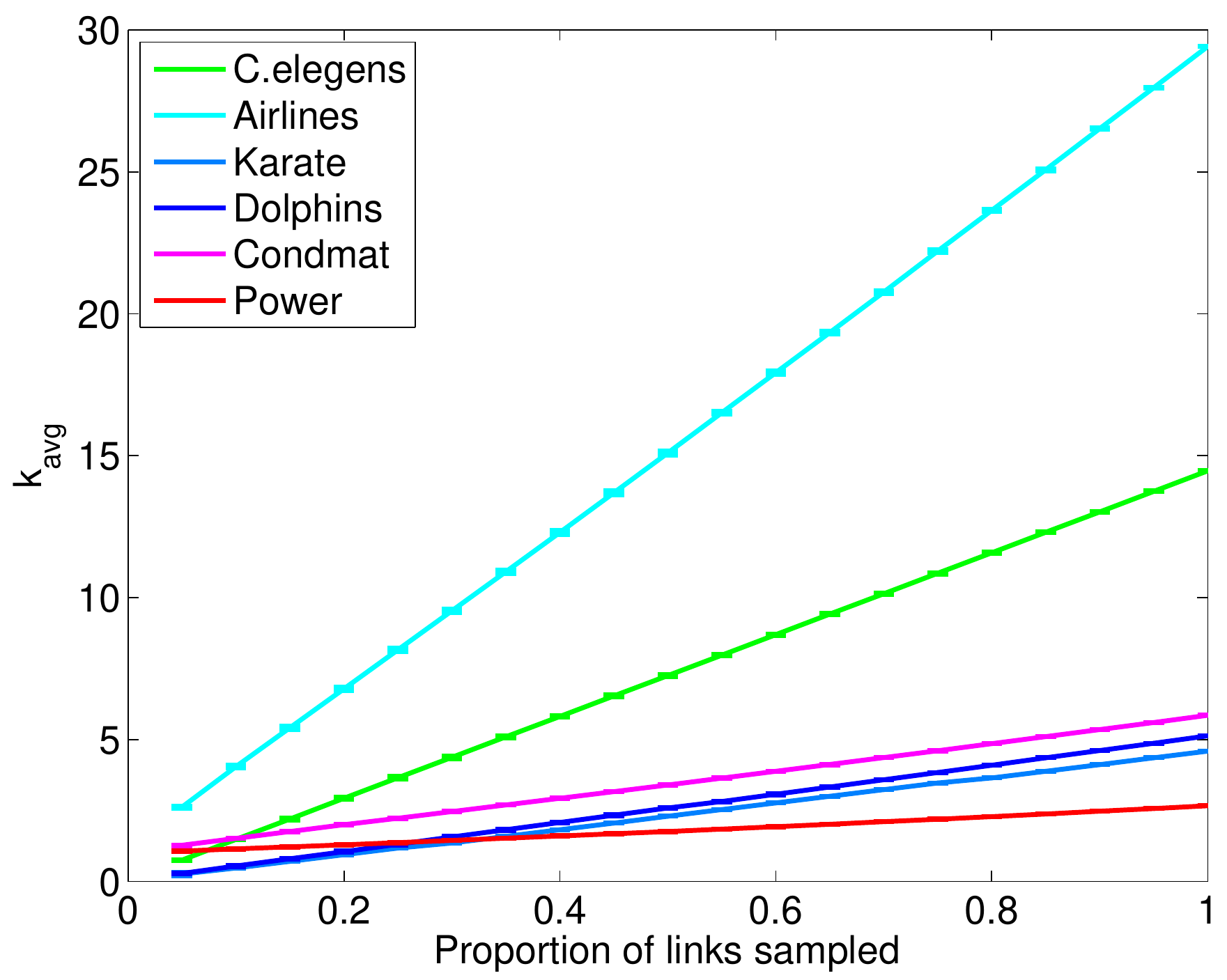}}\\
\subfigure[Max degree]{\includegraphics[width=.3\textwidth]{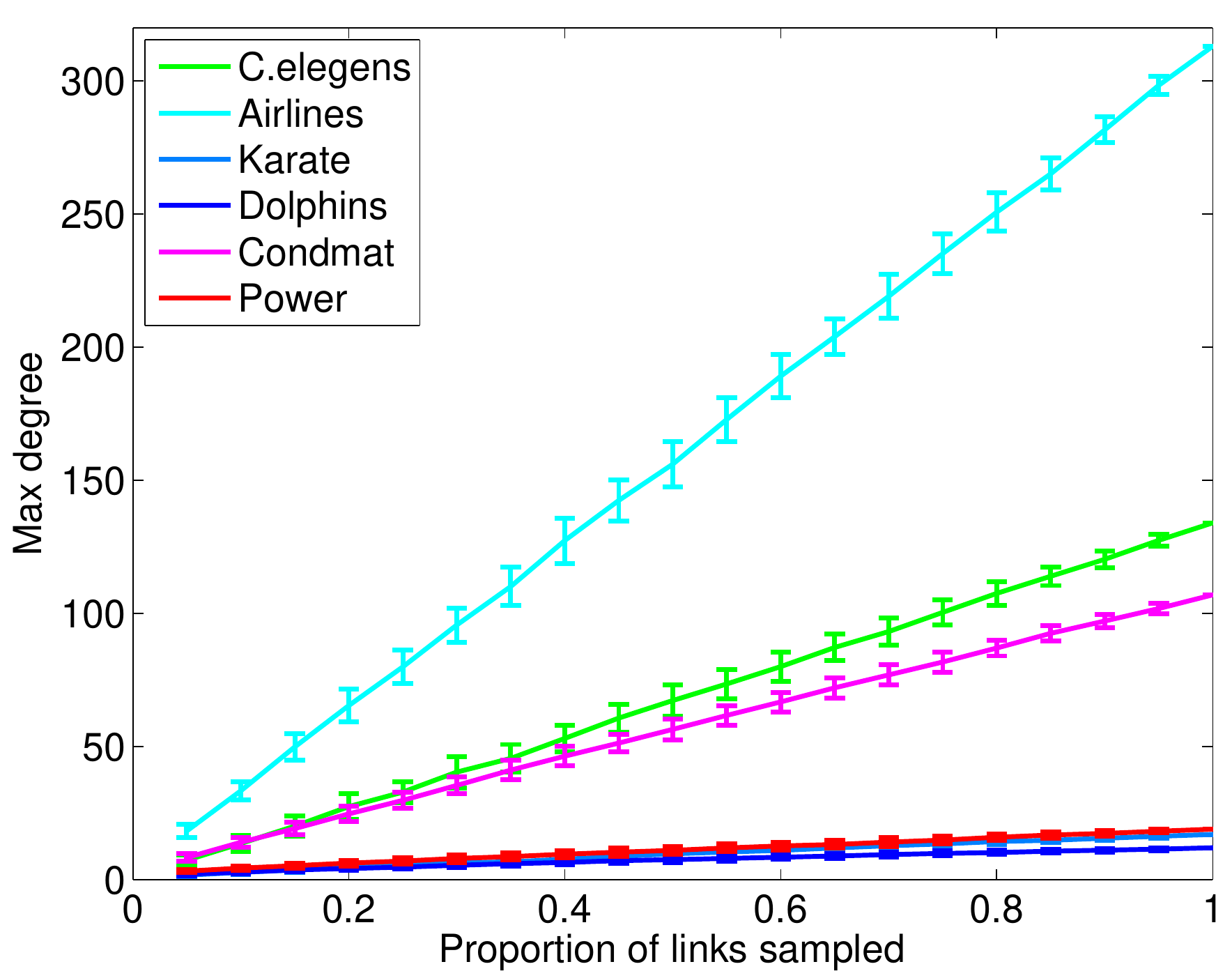}}
\subfigure[Clustering]{\includegraphics[width=.3\textwidth]{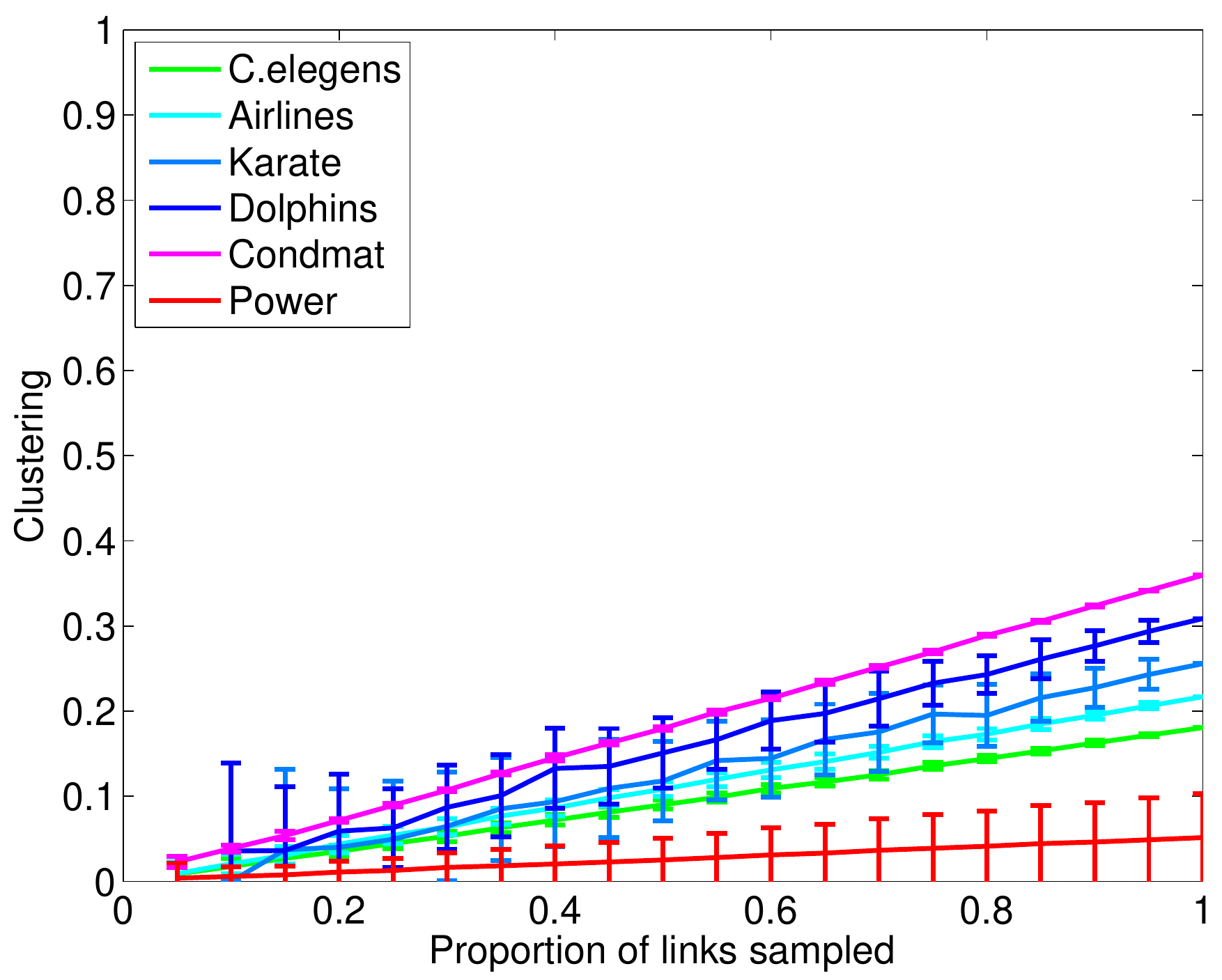}}
\subfigure[Giant Component]{\includegraphics[width=.3\textwidth]{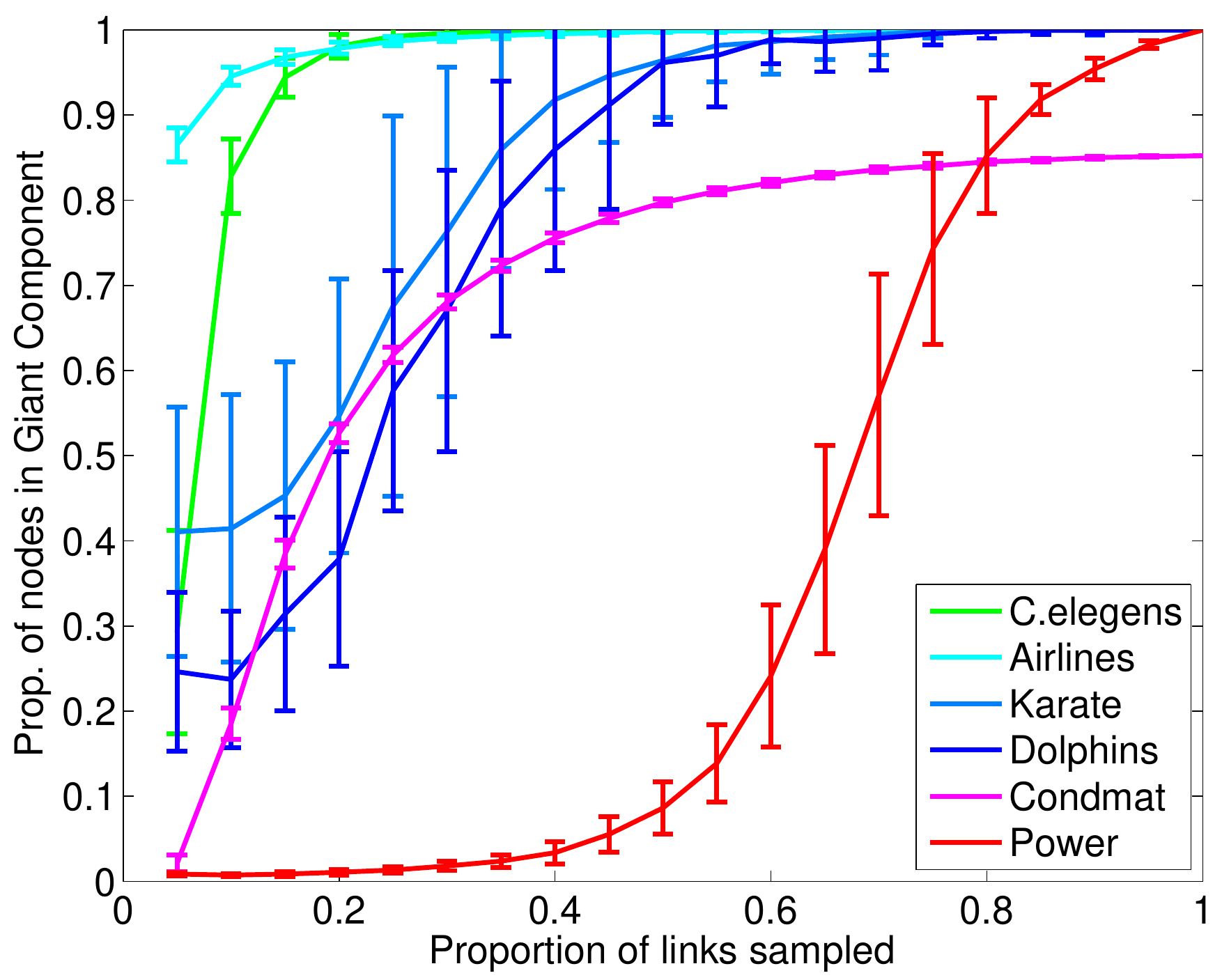}}\\
\caption[Scaling of subnetwork statistics for empirical networks induced on sampled links]{Scaling of subnetwork statistics for empirical networks induced on sampled links. (a.) The number of nodes in a subnetwork sampled by nodes scales nonlinearly with $q$. (b.) The number of edges scales as $m \approx qM$. (c.) The average degree scales roughly linearly with the proportion of nodes subsampled $k^{\textnormal{sub}}_{\rm avg} \approx qk_{\rm avg}$. (d.) The max degree scales roughly linearly for networks with few large hubs. (e.) The clustering coefficient scales roughly linearly $C^{\textnormal{sub}} \approx q C$. (f.) The proportion of nodes in the giant component increases with the proportion of links sampled. \textit{C. elegans} and airlines maintain a large proportion of nodes in the giant component, most likely because these networks have high average degree. Karate club and dolphins show considerable variability (as shown by error bars $\pm$ s.d.) because these are relatively small networks. Powergrid is fragile to sampling by links, meaning the a high proportion of sampled links must be obtained to reach a fully connected network.}
\label{fig:sampling_by_links_incident_empirical_scaling}
\end{figure*}

\FloatBarrier

\setcounter{equation}{11}
\begin{figure*}[!ht]
\centering
\subfigure[Erdrey]{\includegraphics[width=.28\textwidth]{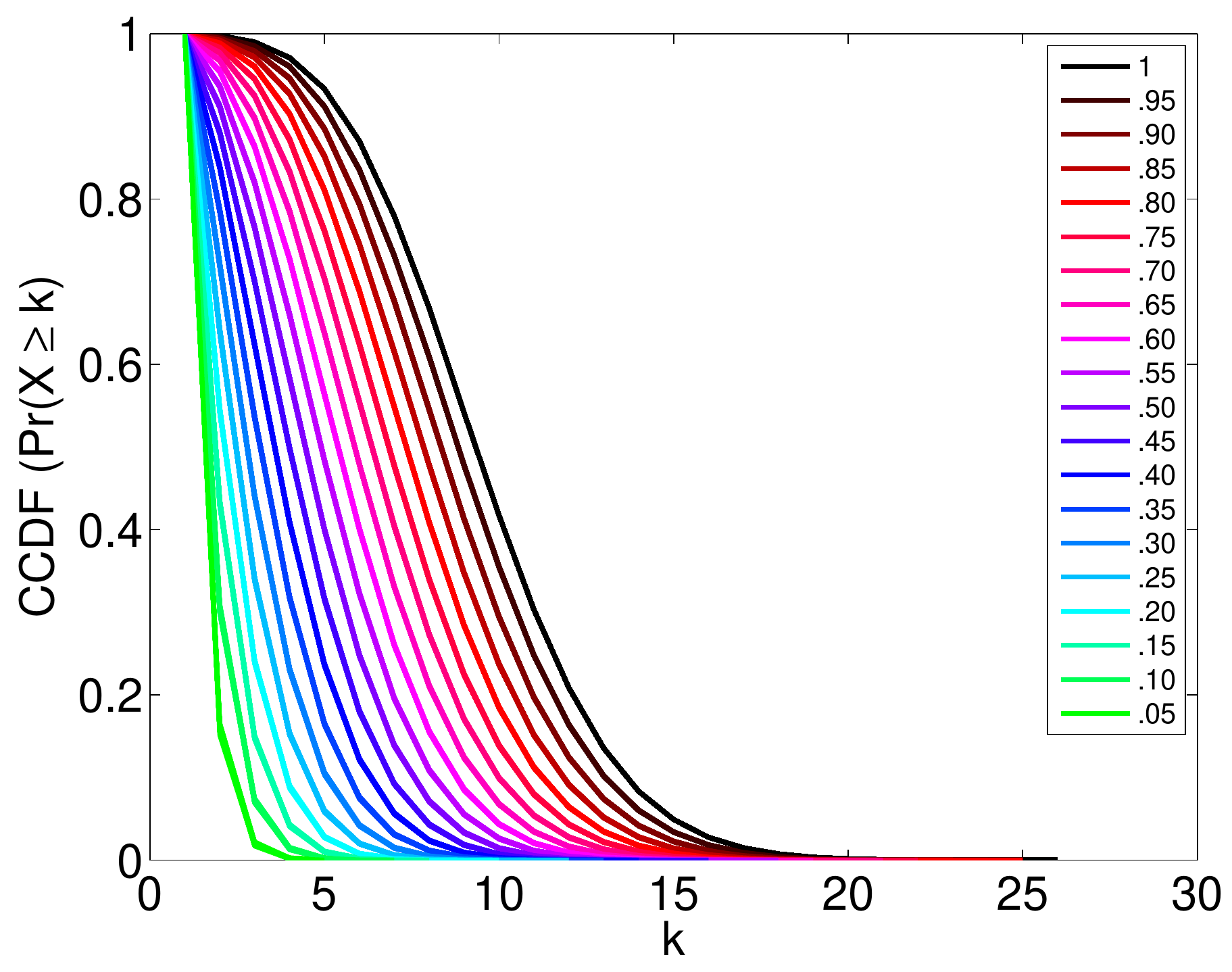}}
\subfigure[Pref]{\includegraphics[width=.28\textwidth]{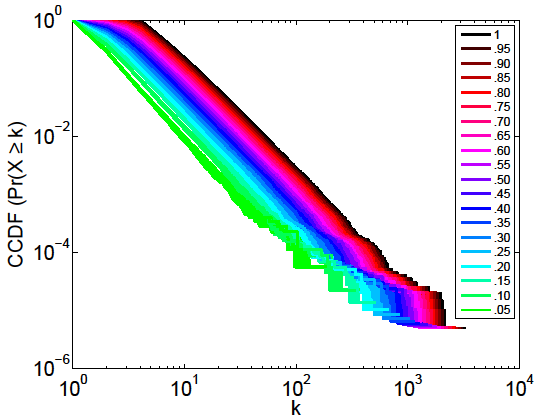}}
\subfigure[Smallworld]{\includegraphics[width=.28\textwidth]{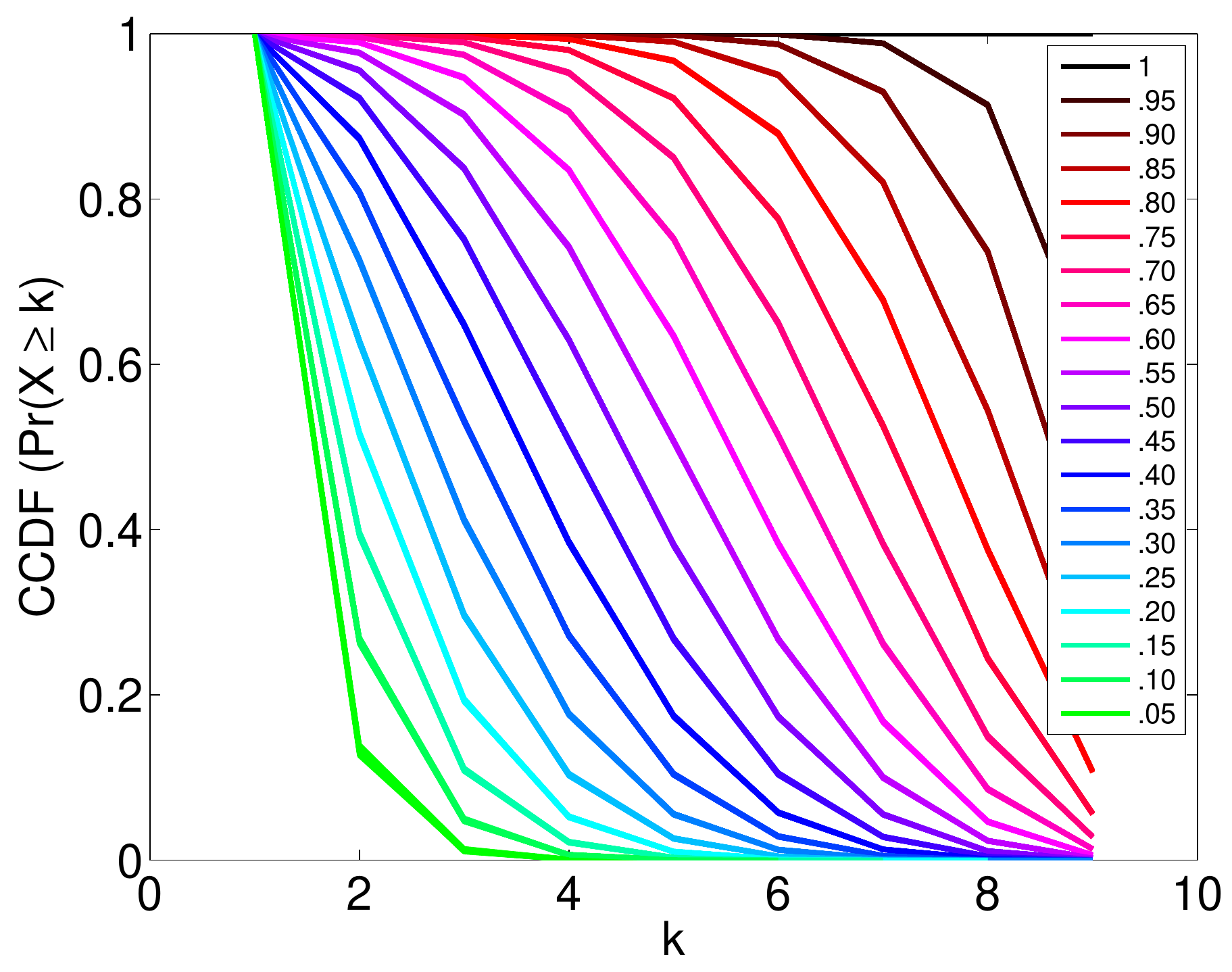}}\\
\subfigure[Renga]{\includegraphics[width=.28\textwidth]{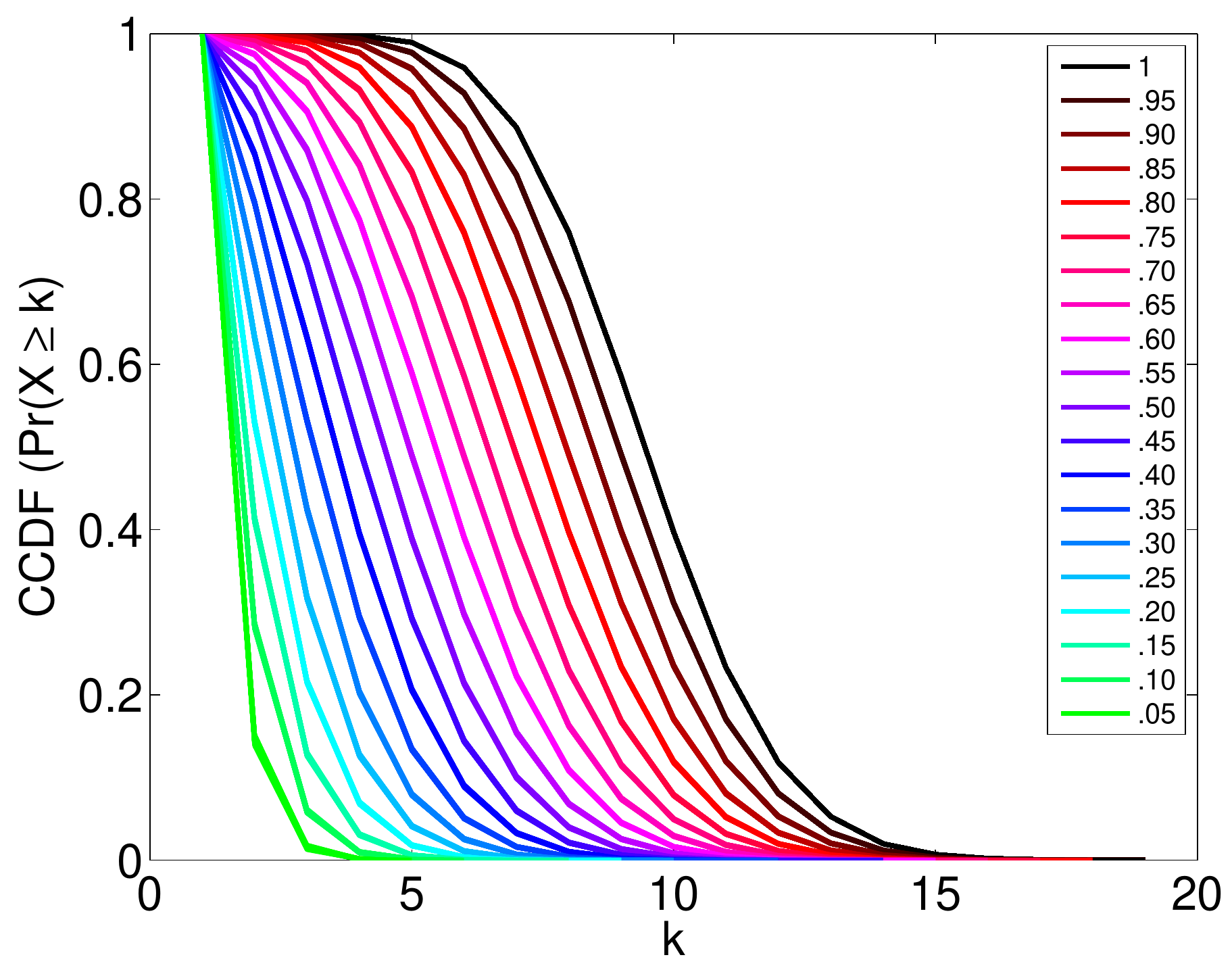}}
\subfigure[C. elgegans]{\includegraphics[width=.28\textwidth]{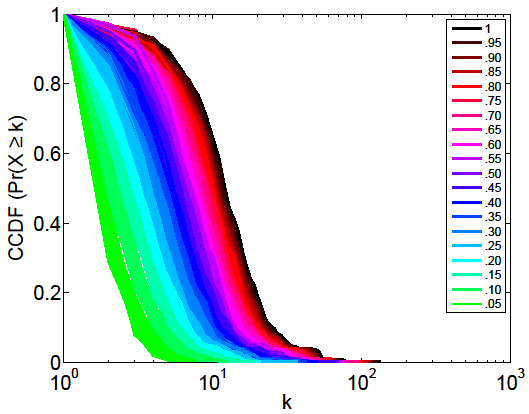}}
\subfigure[Airlines]{\includegraphics[width=.28\textwidth]{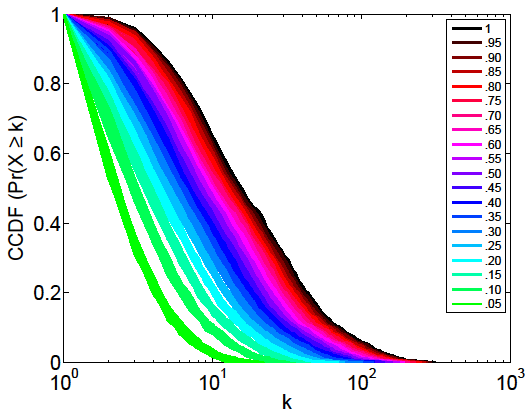}}\\
\subfigure[Karate]{\includegraphics[width=.28\textwidth]{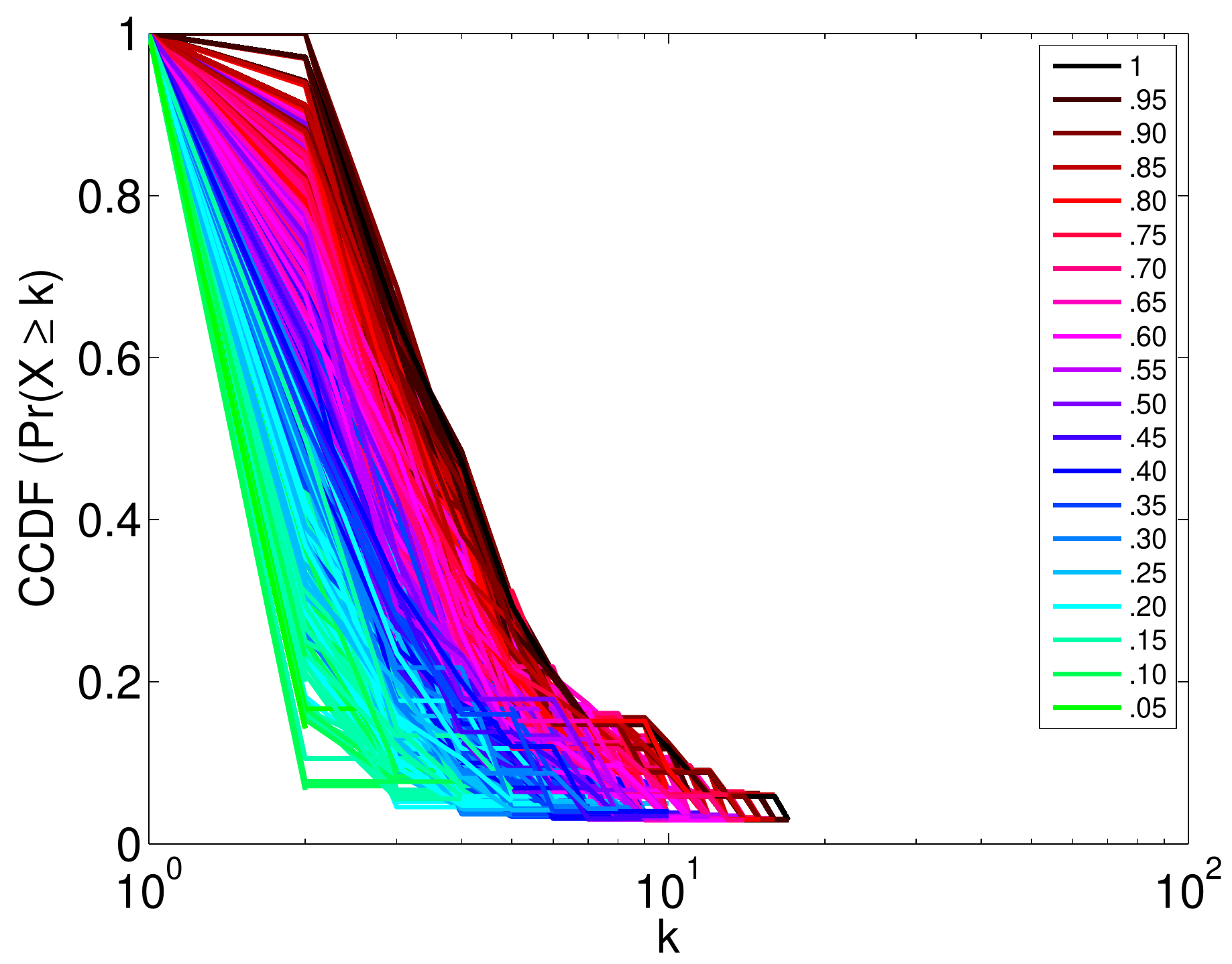}}
\subfigure[Dolphins]{\includegraphics[width=.28\textwidth]{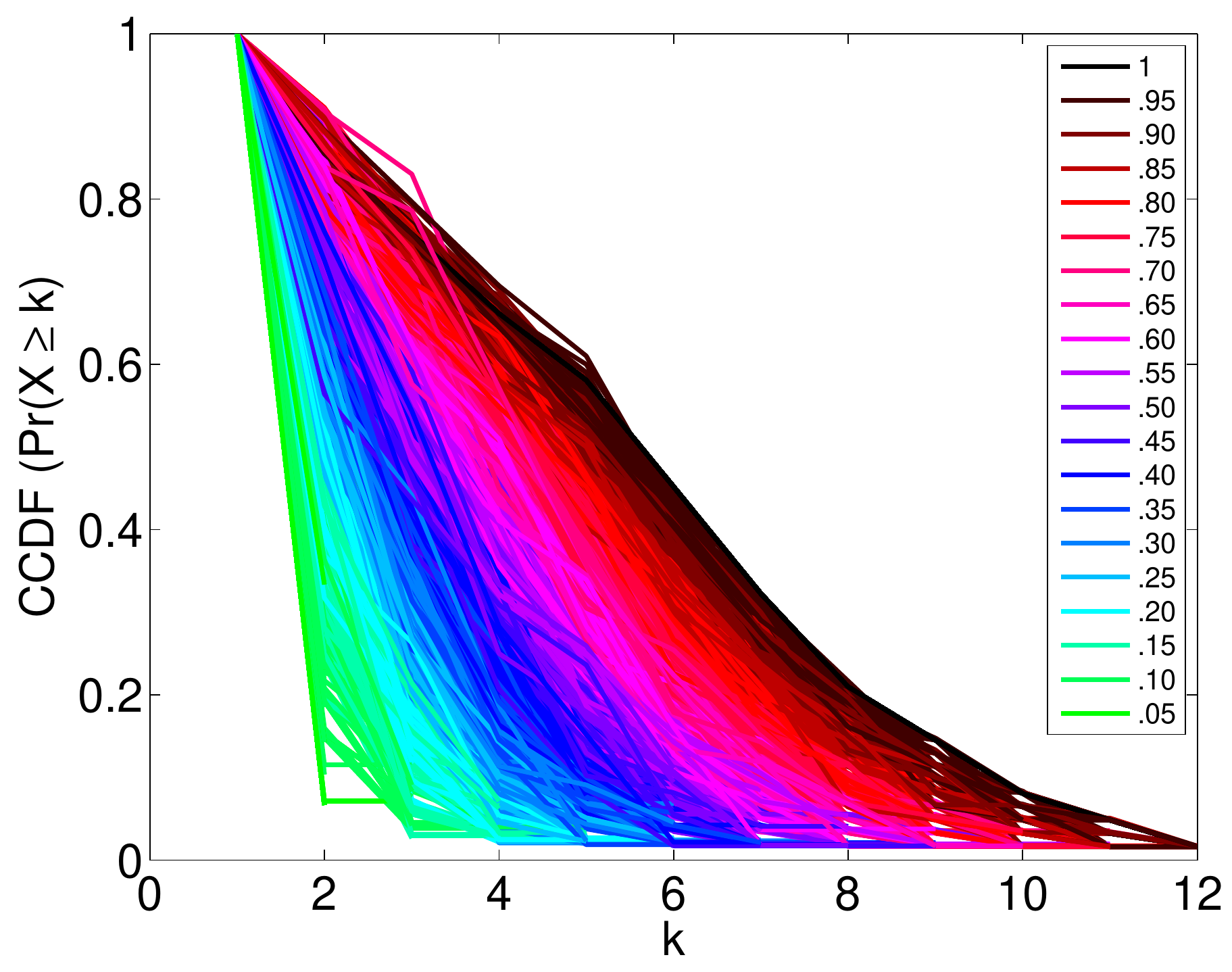}}
\subfigure[Condmat]{\includegraphics[width=.28\textwidth]{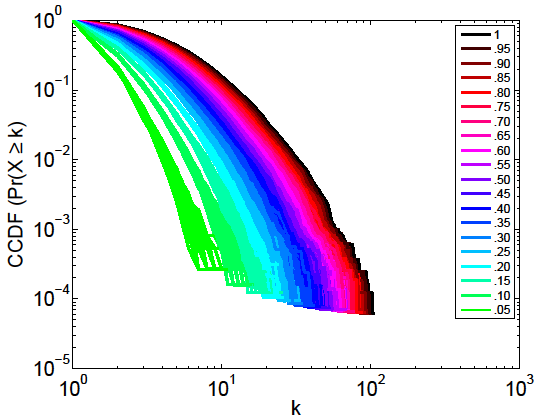}}\\
\subfigure[Powergrid]{\includegraphics[width=.28\textwidth]{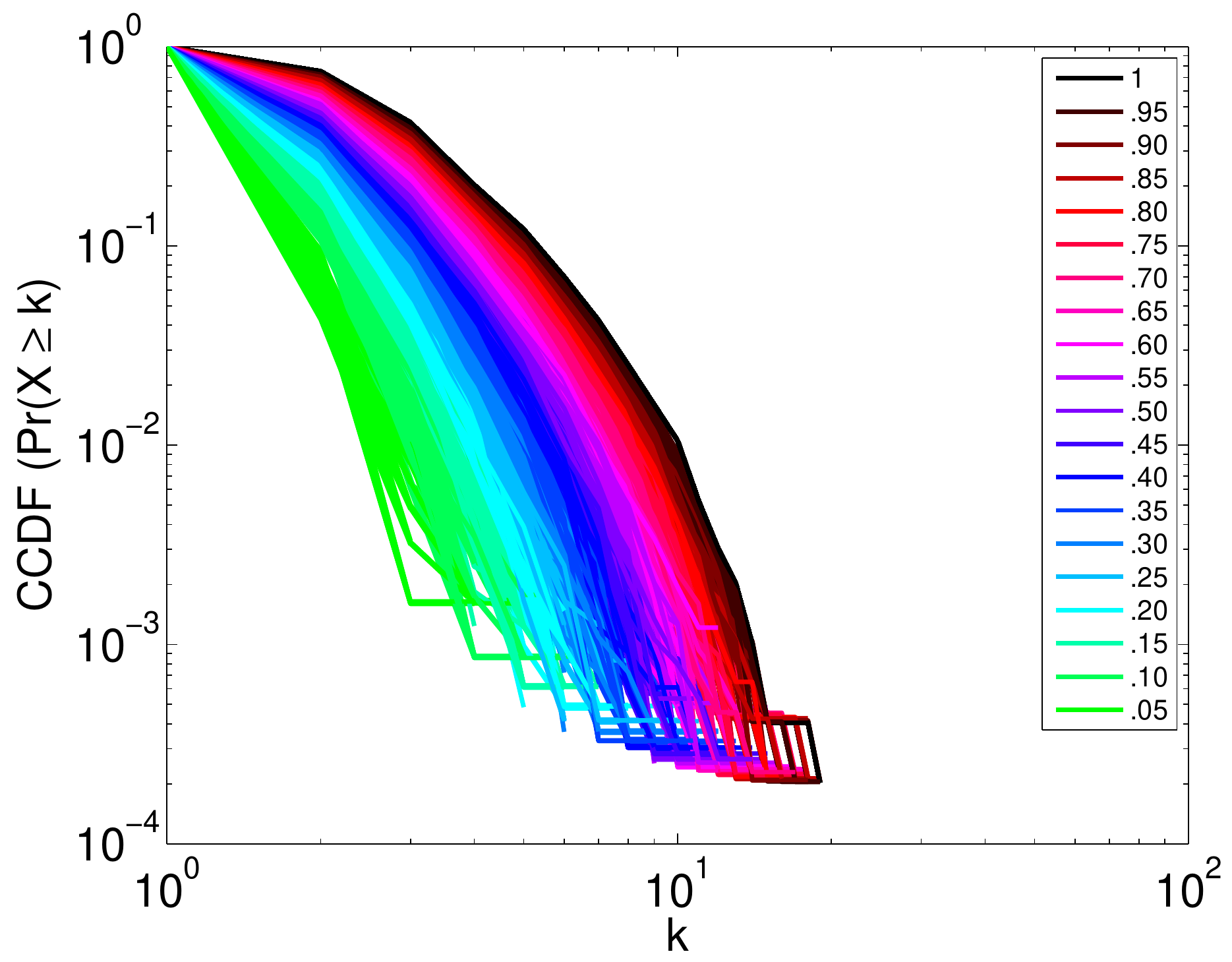}}
\caption[CCDF distortion for subnetworks induced on sampled links]{CCDF distortion for subnetworks induced on sampled links. Subnetwork degree distributions do not capture the true degree distribution, especially for small $q$.}
\label{fig:links_incident_ccpdf}
\end{figure*}
\setcounter{equation}{12}
\begin{figure*}[!ht]
\centering
\subfigure[Erdrey$^+$]{\includegraphics[width=.28\textwidth]{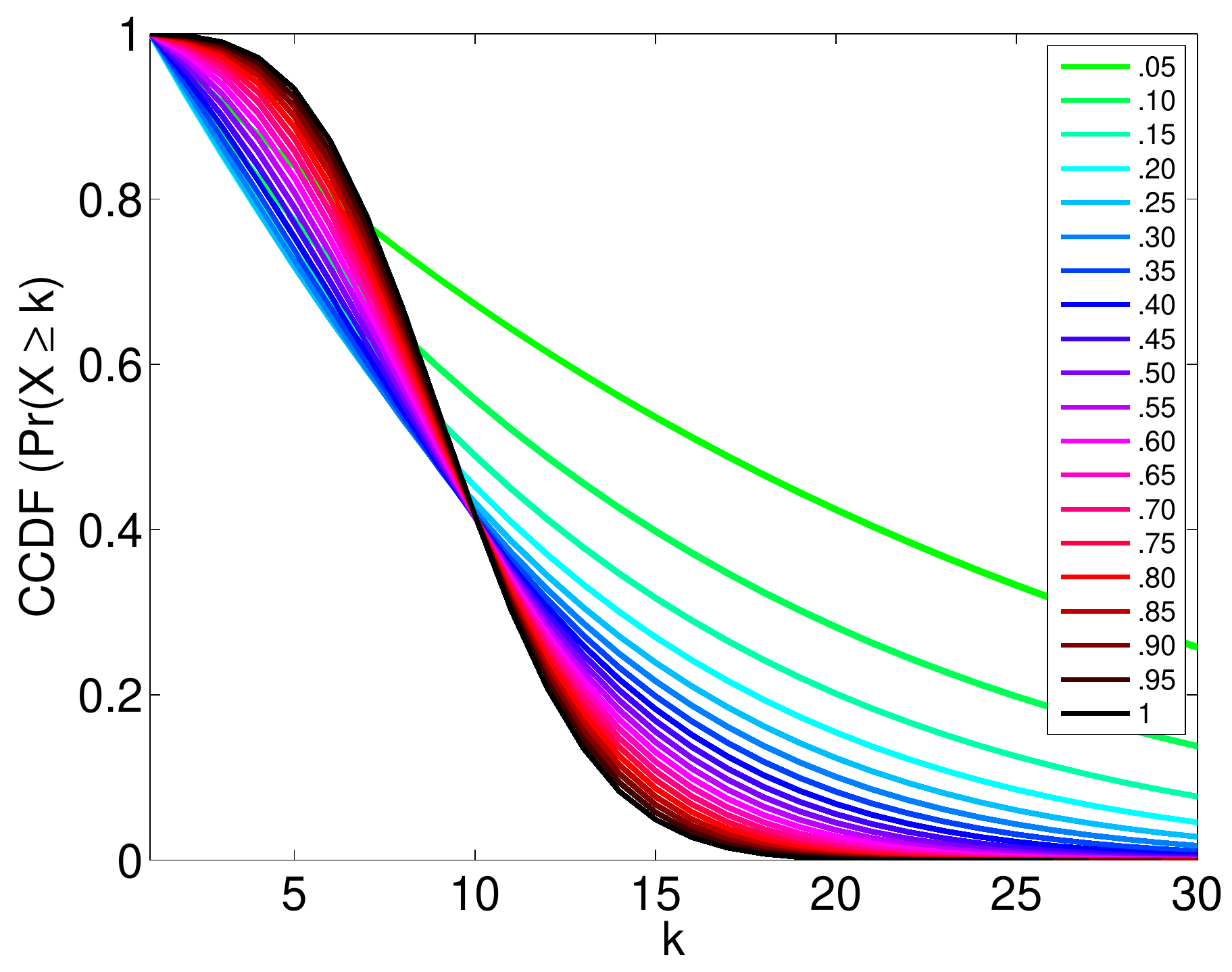}}
\subfigure[Pref$^*$]{\includegraphics[width=.28\textwidth]{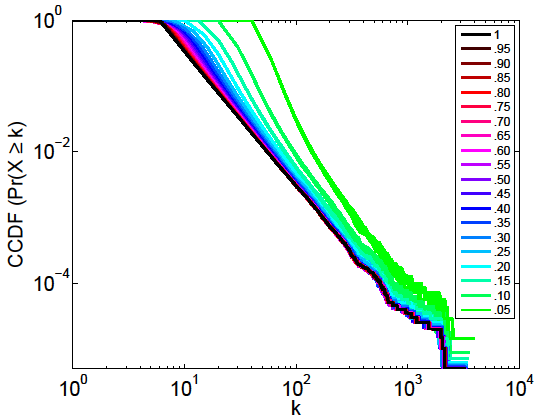}}
\subfigure[Smallworld$^+$]{\includegraphics[width=.28\textwidth]{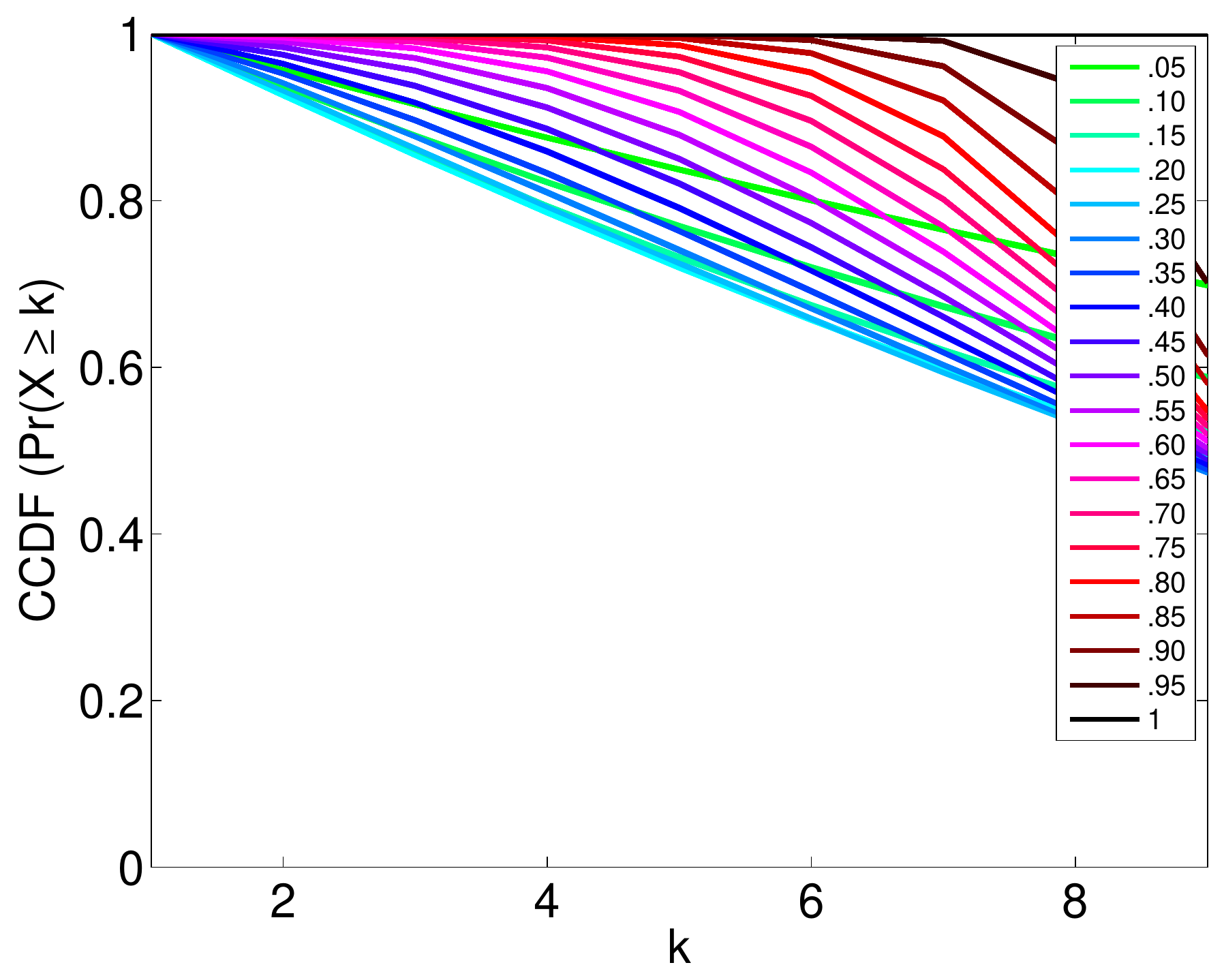}}\\
\subfigure[Renga$^+$]{\includegraphics[width=.28\textwidth]{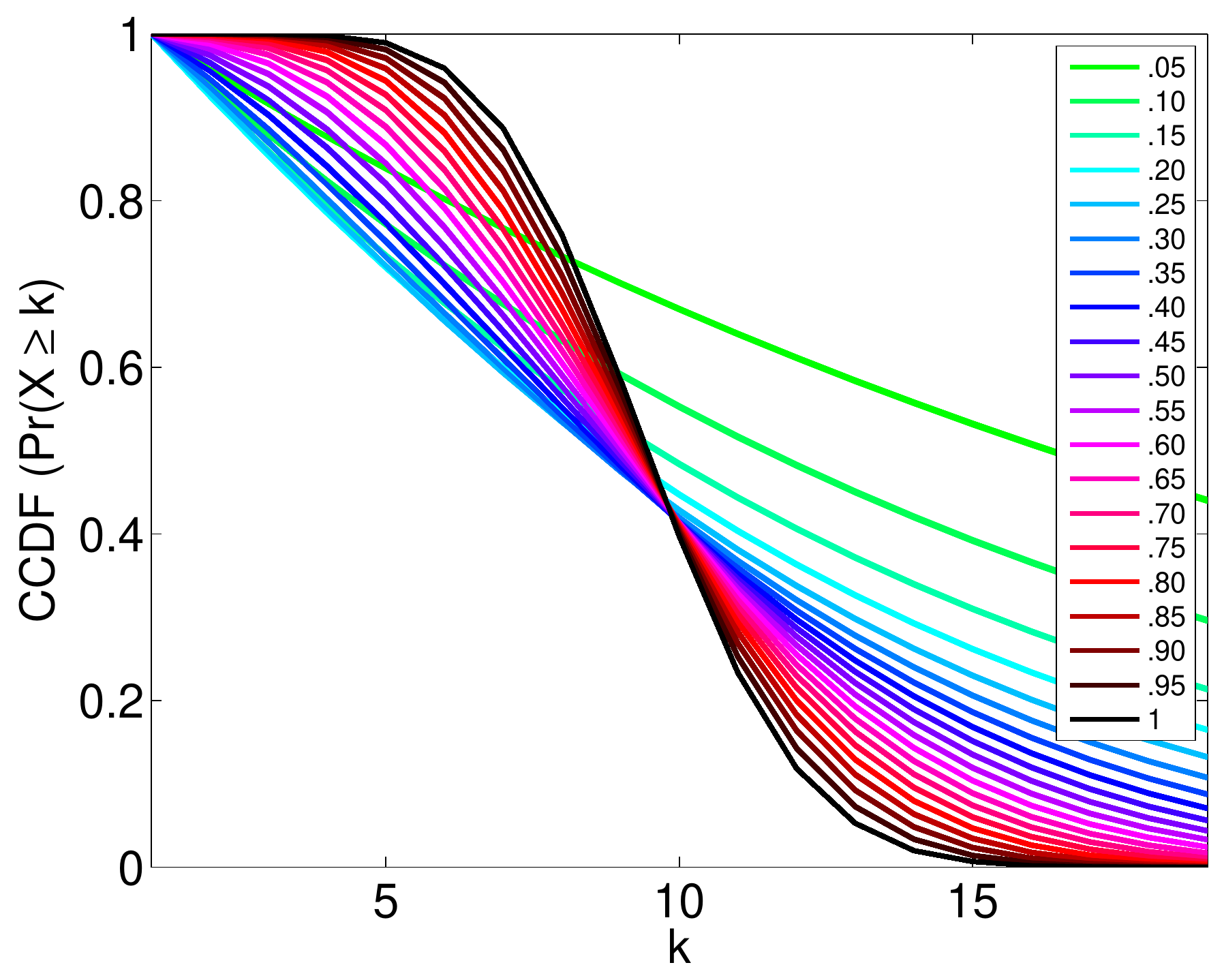}}
\subfigure[C. elgegans$^+$]{\includegraphics[width=.28\textwidth]{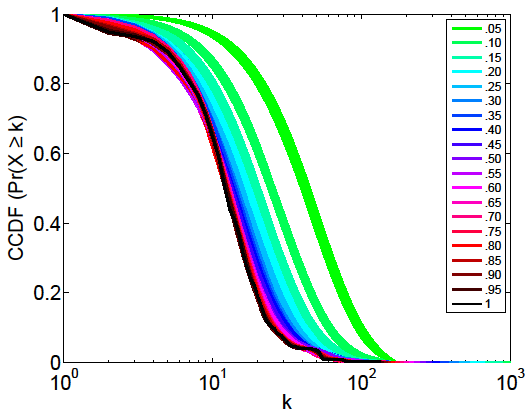}}
\subfigure[Airlines$^*$]{\includegraphics[width=.28\textwidth]{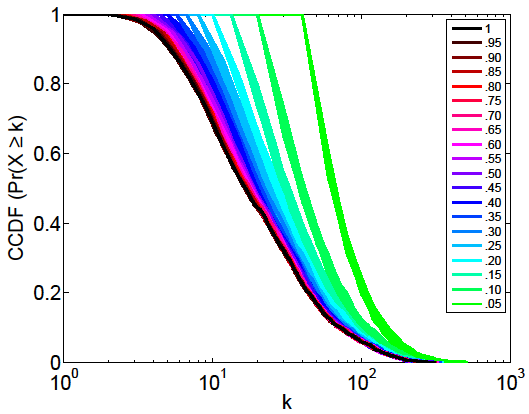}}\\
\subfigure[Karate$^+$]{\includegraphics[width=.28\textwidth]{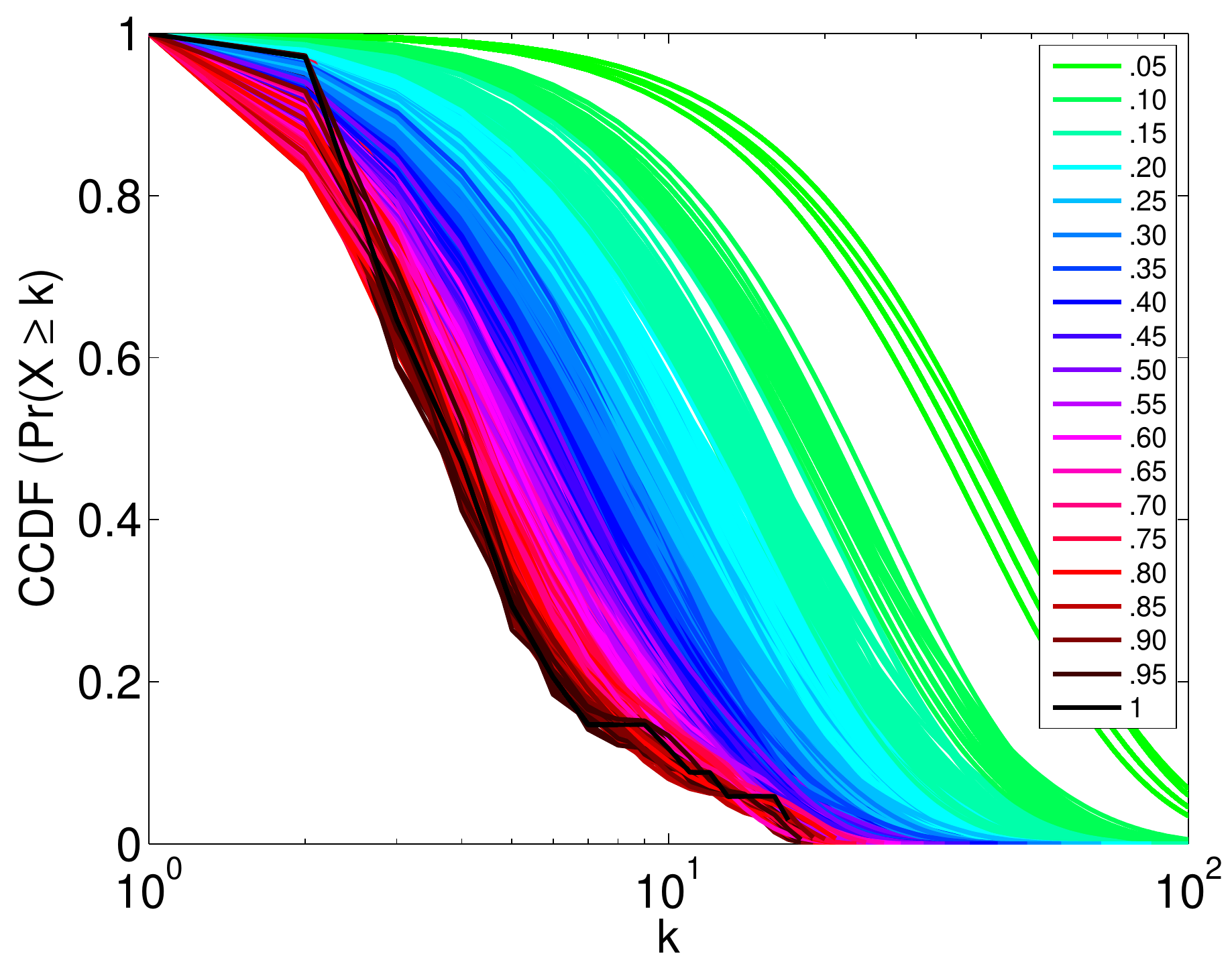}}
\subfigure[Dolphins$^+$]{\includegraphics[width=.28\textwidth]{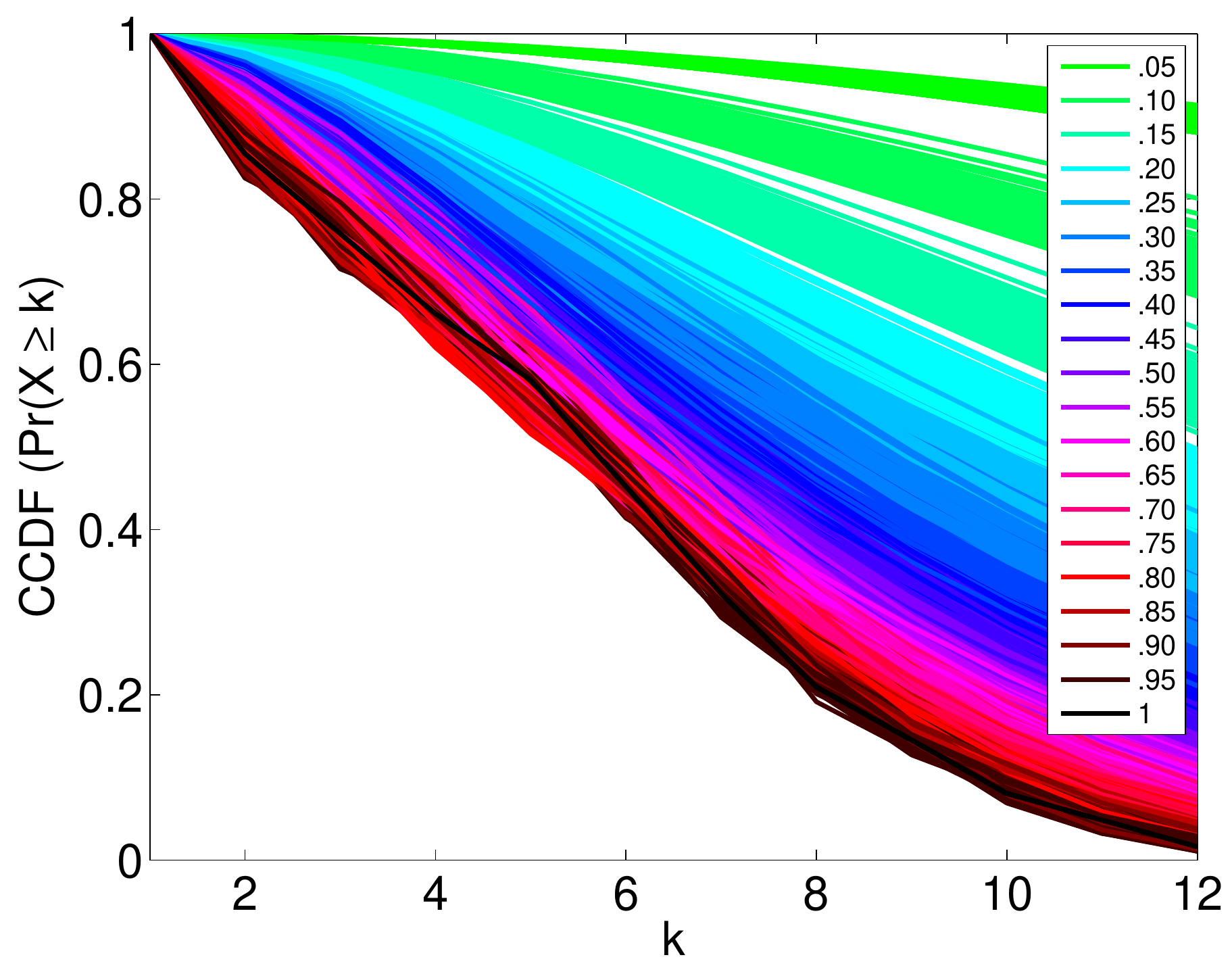}}
\subfigure[Condmat$^*$]{\includegraphics[width=.28\textwidth]{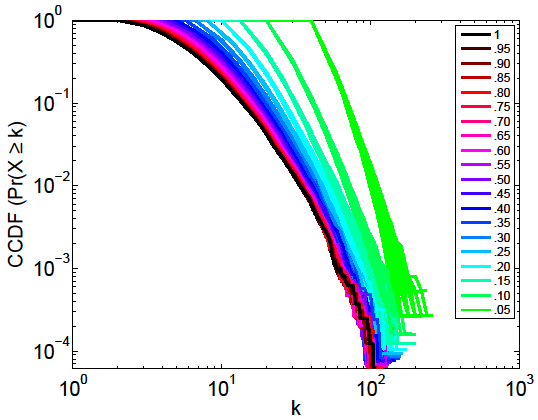}}\\
\subfigure[Powergrid$^*$]{\includegraphics[width=.28\textwidth]{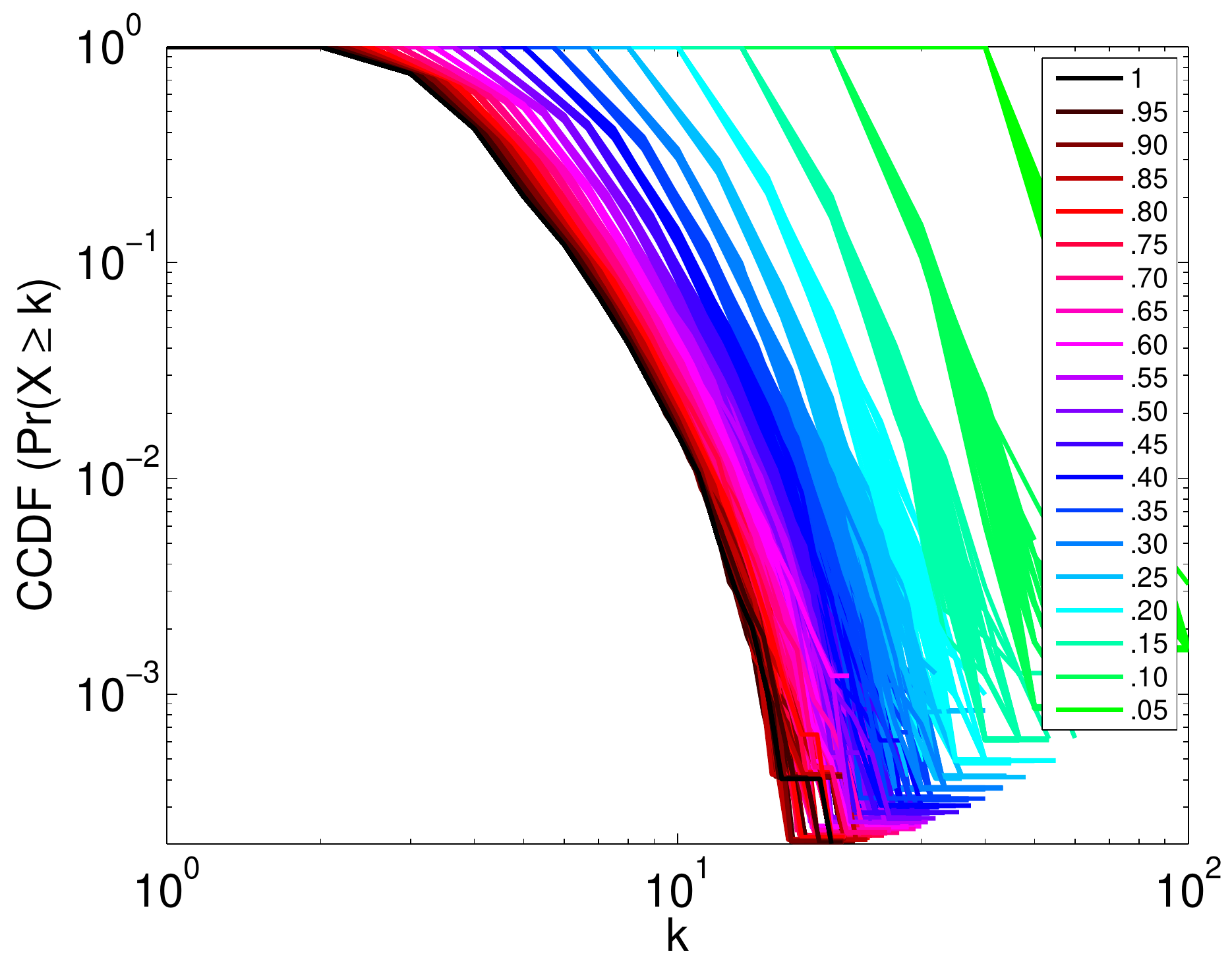}}
\caption[Predicted CCDF from subnetworks induced on sampled links]{Predicted CCDF from subnetworks induced on sampled links. The predicted CCDF shows relatively good agreement with the CCDF for most networks. Karate club and Dolphins exhibit significant deviations, possibly due to the small number of nodes in these networks. Networks designated $\text{with }^+$ utilized Equation~\ref{eq:dist_rollback} and those designated with $\text{with }^*$ utilized Equation~\ref{eq:my_dist_shorter}. }
\label{fig:predicting_by_links_incident_empirical_ccpdf}
\end{figure*}

\setcounter{equation}{13}
\begin{figure*}[!ht]
\centering
\subfigure[Erdrey, $N$]{\includegraphics[width=.24\textwidth]{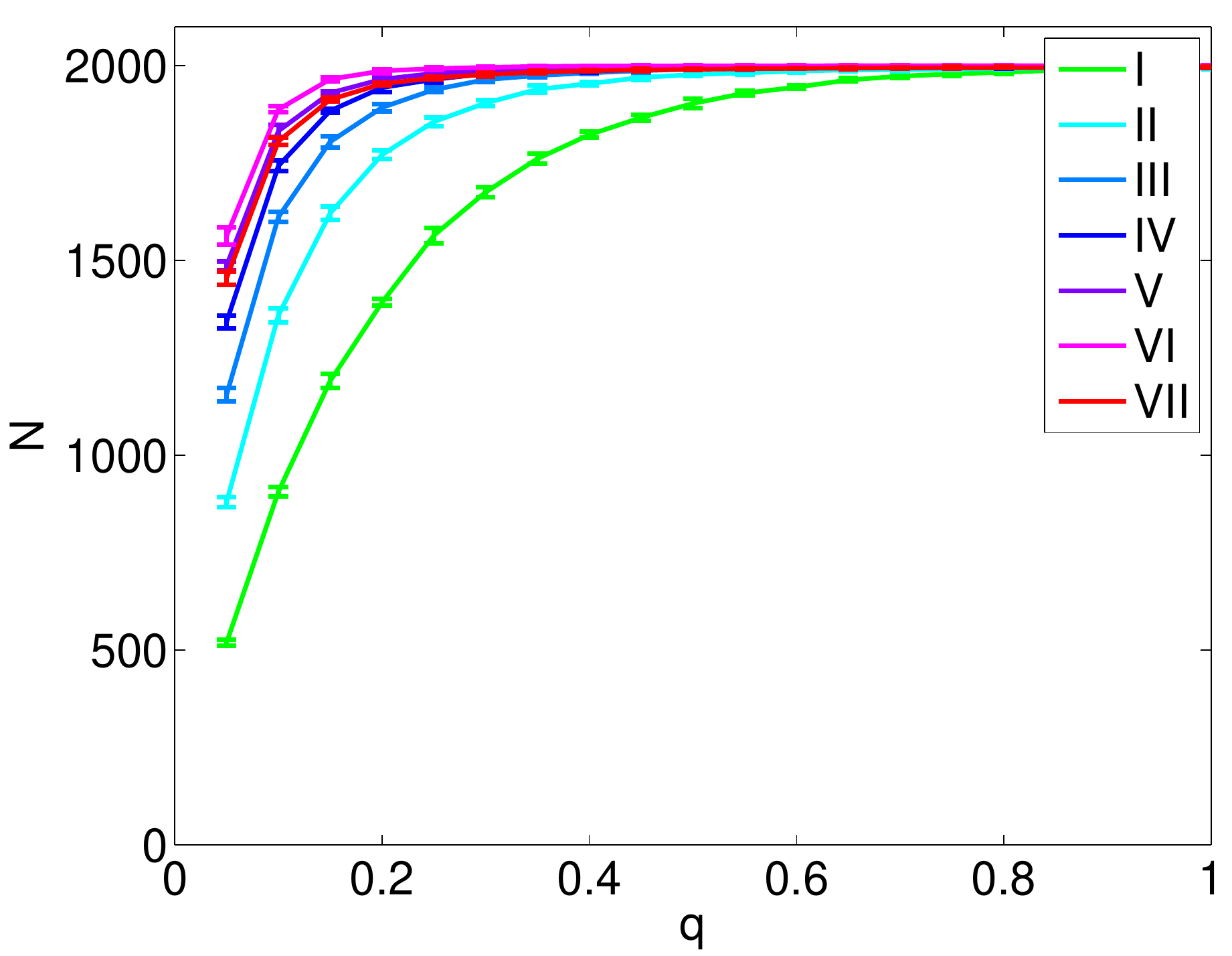}}
\subfigure[Erdrey, $M$]{\includegraphics[width=.24\textwidth]{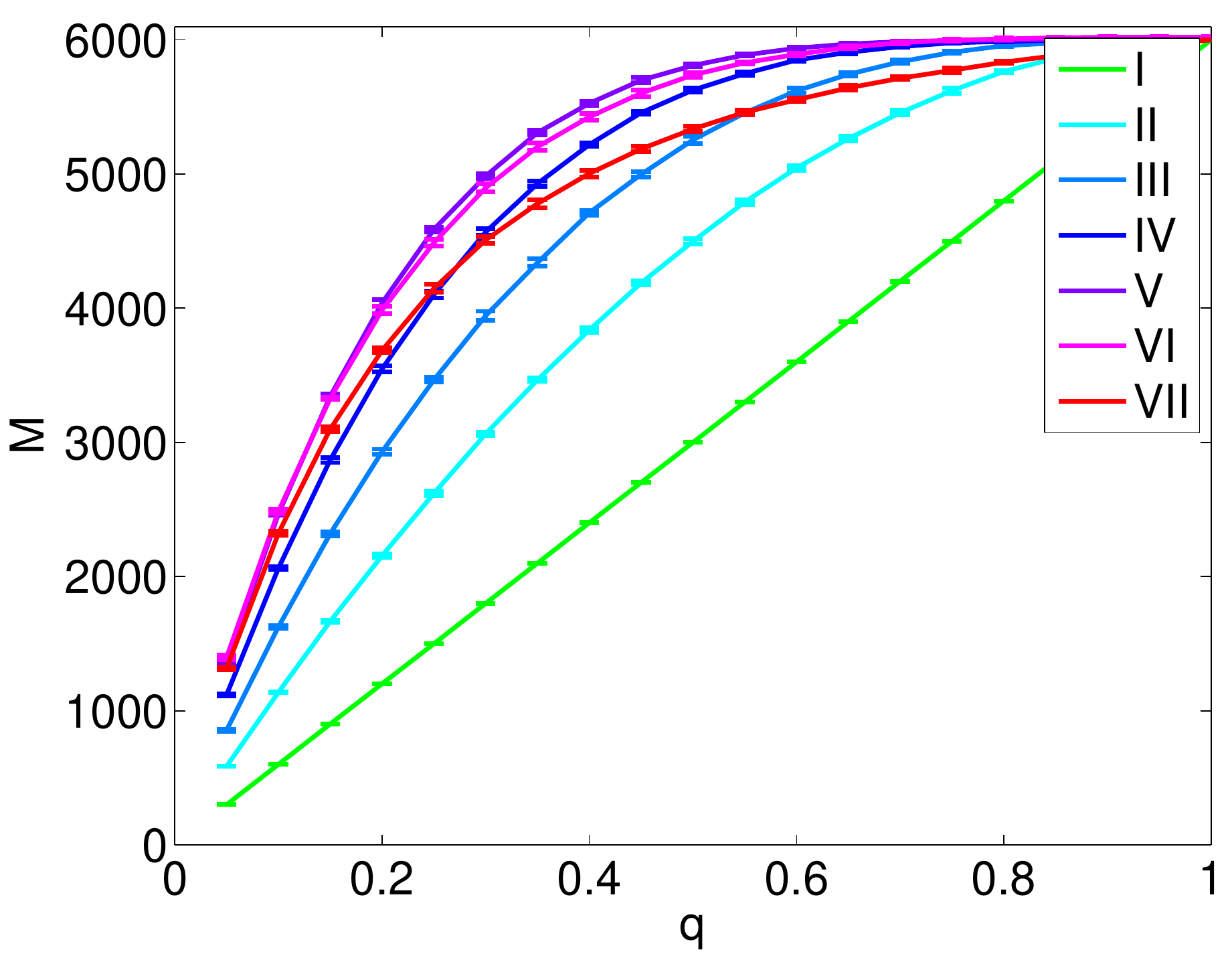}}
\subfigure[Erdrey, $k_{\rm avg}$]{\includegraphics[width=.24\textwidth]{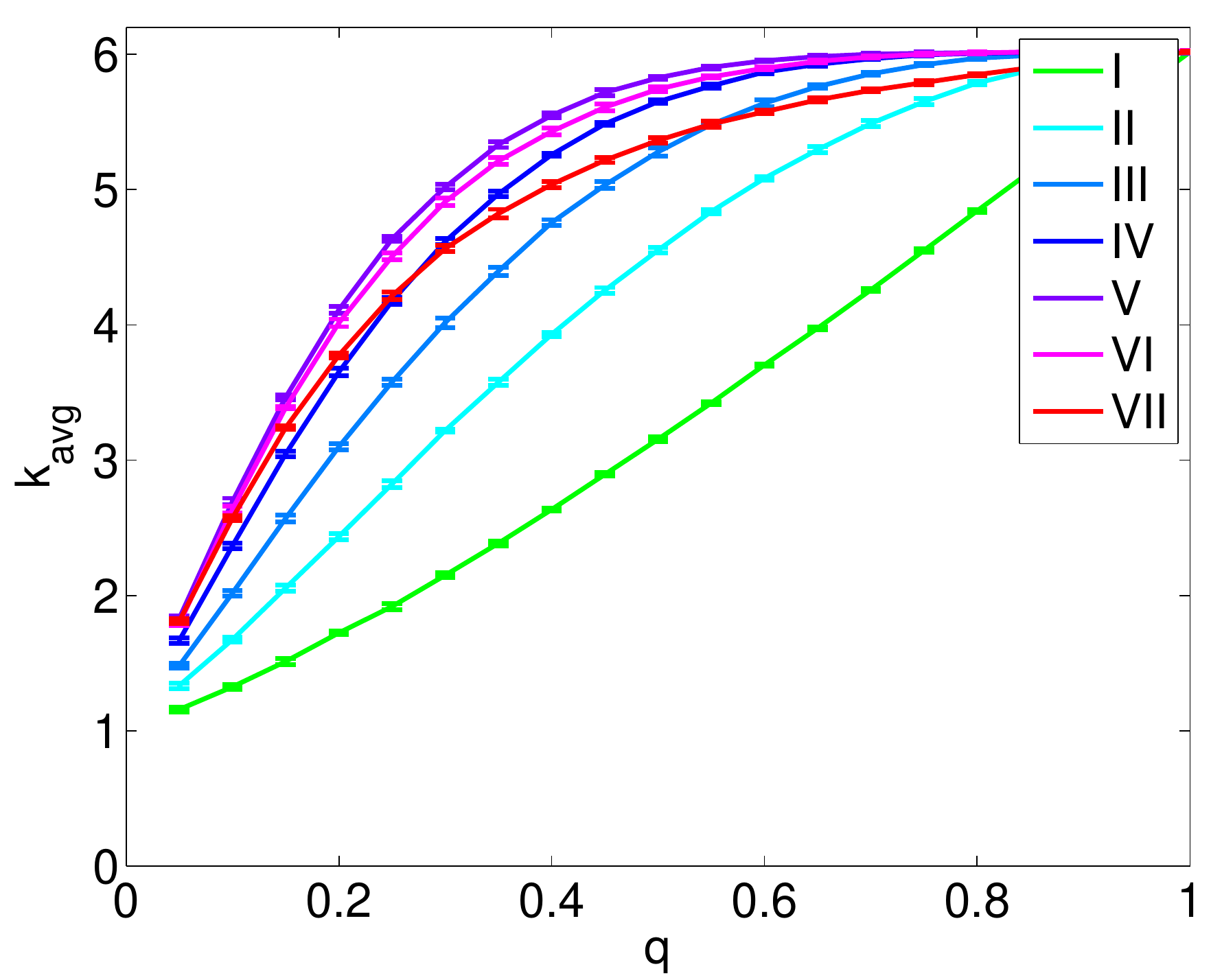}}
\subfigure[Erdrey, $\kmax$]{\includegraphics[width=.24\textwidth]{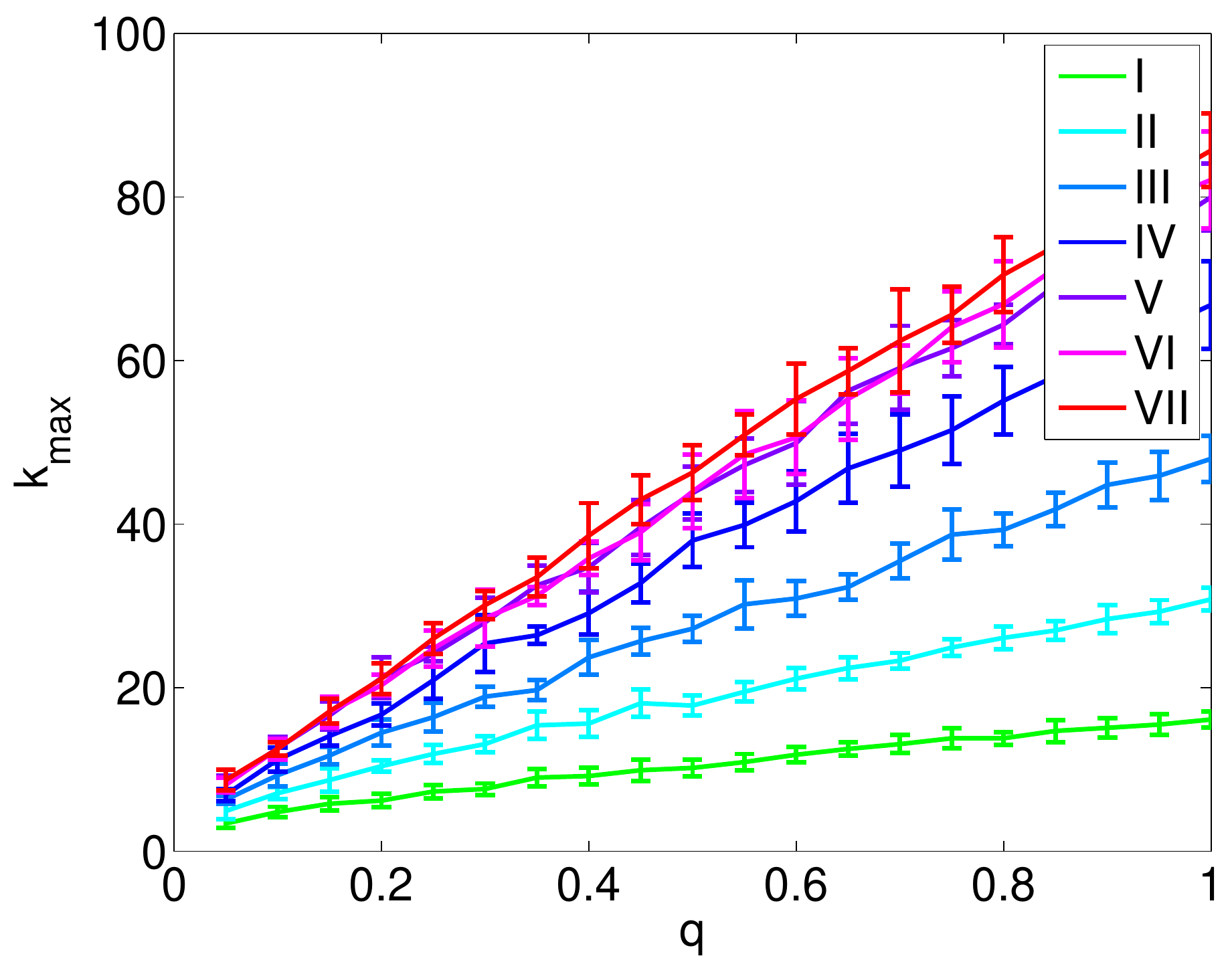}}\\
\subfigure[Pref, $N$]{\includegraphics[width=.24\textwidth]{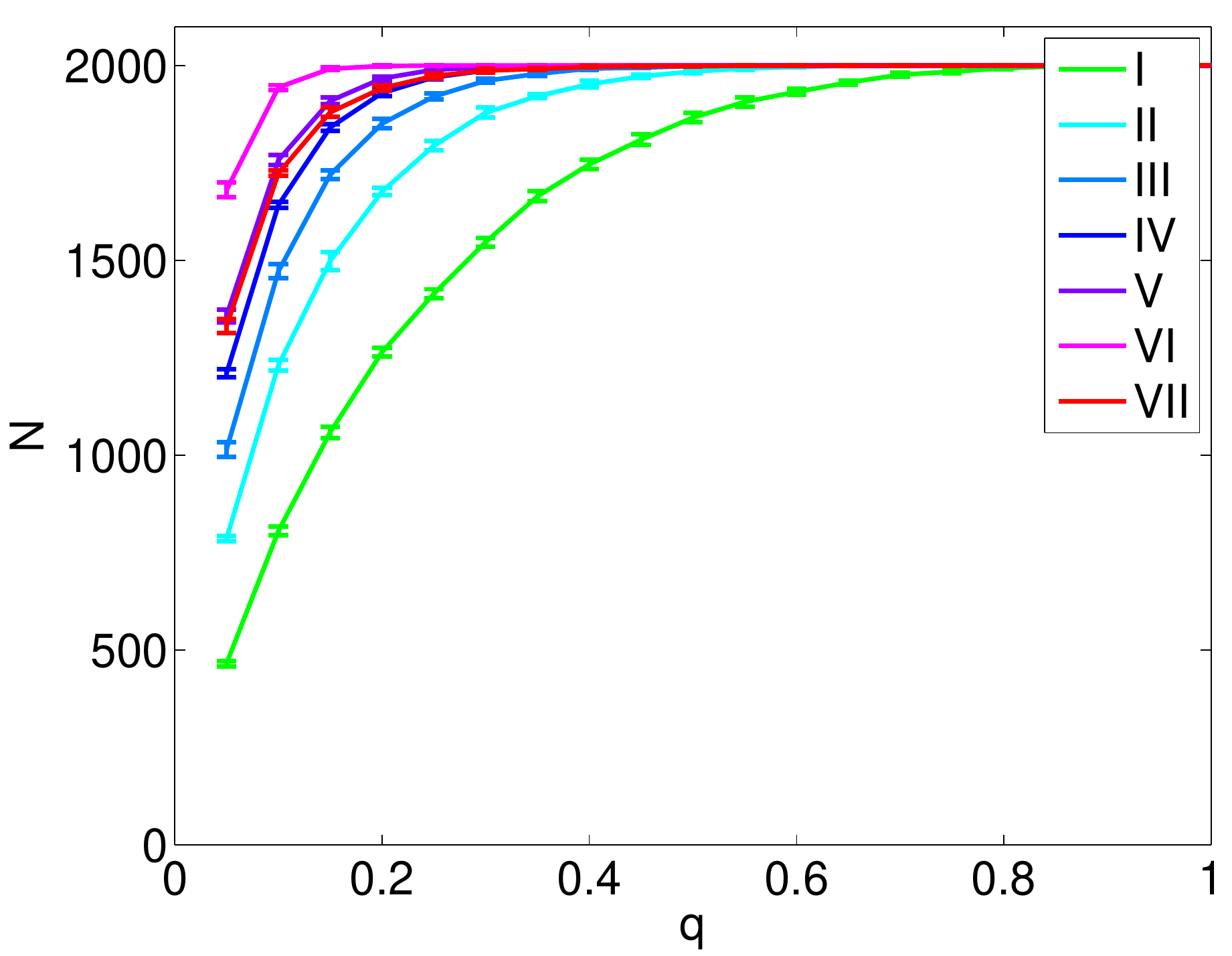}}
\subfigure[Pref, $M$]{\includegraphics[width=.24\textwidth]{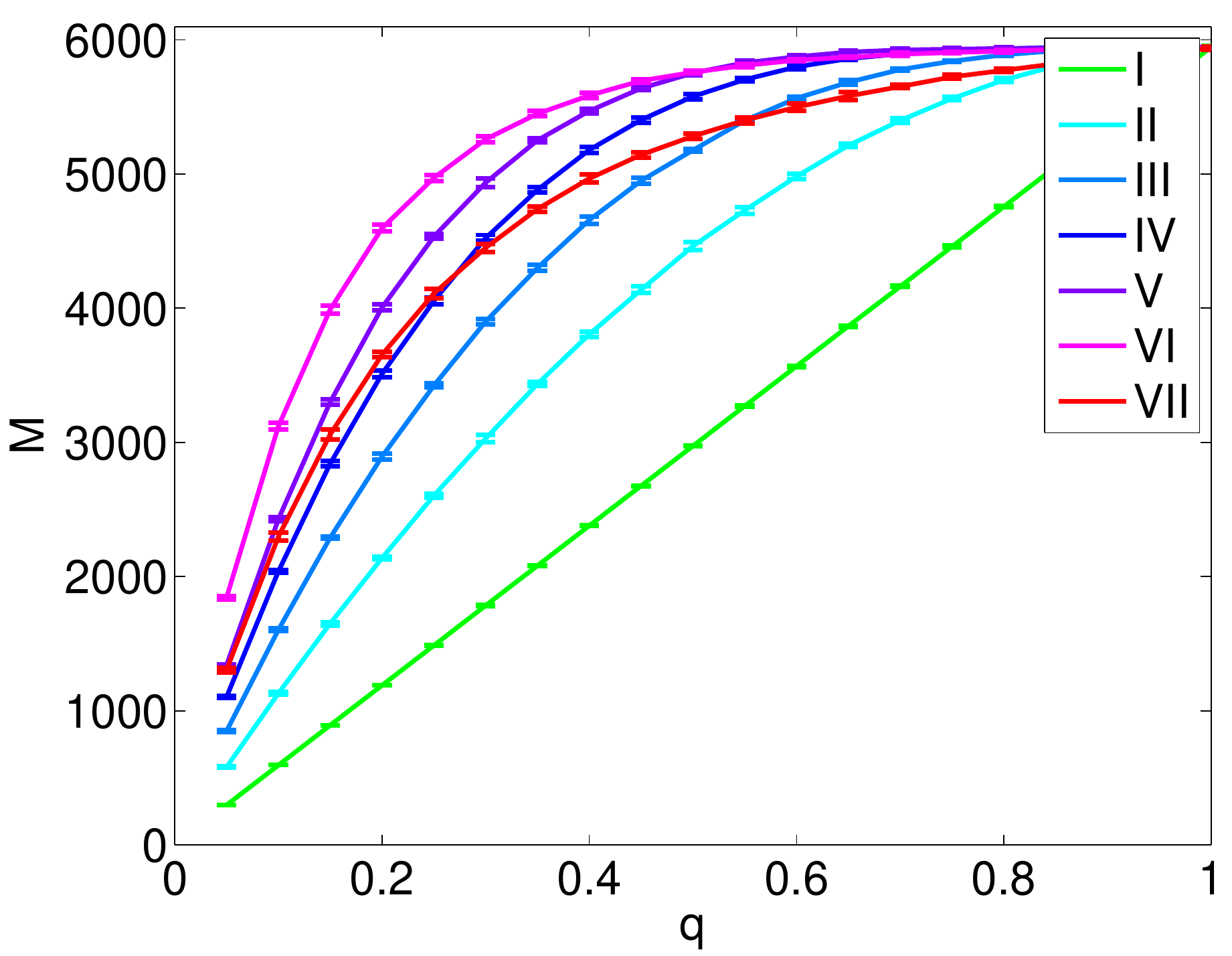}}
\subfigure[Pref, $k_{\rm avg}$]{\includegraphics[width=.24\textwidth]{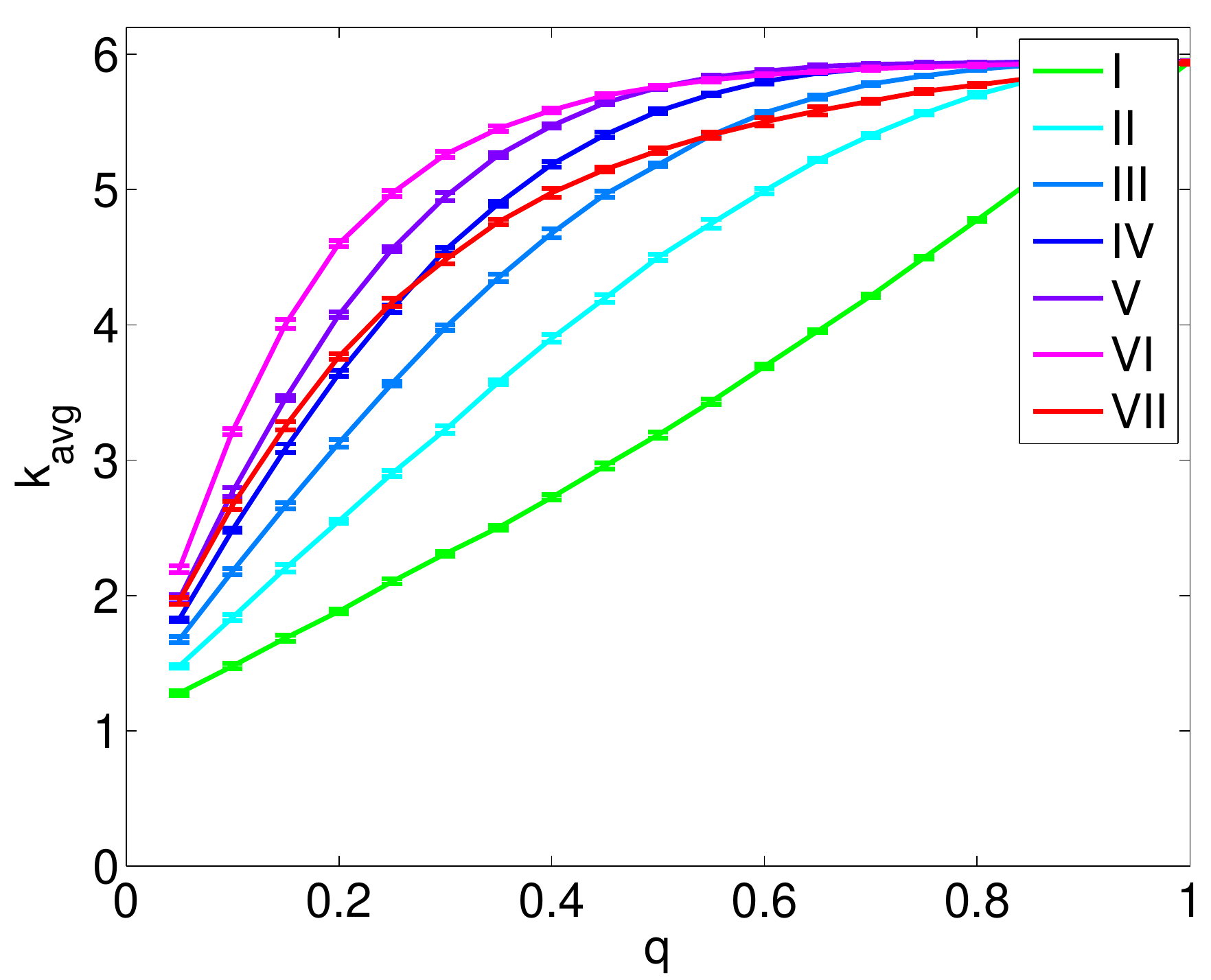}}
\subfigure[Pref, $\kmax$]{\includegraphics[width=.24\textwidth]{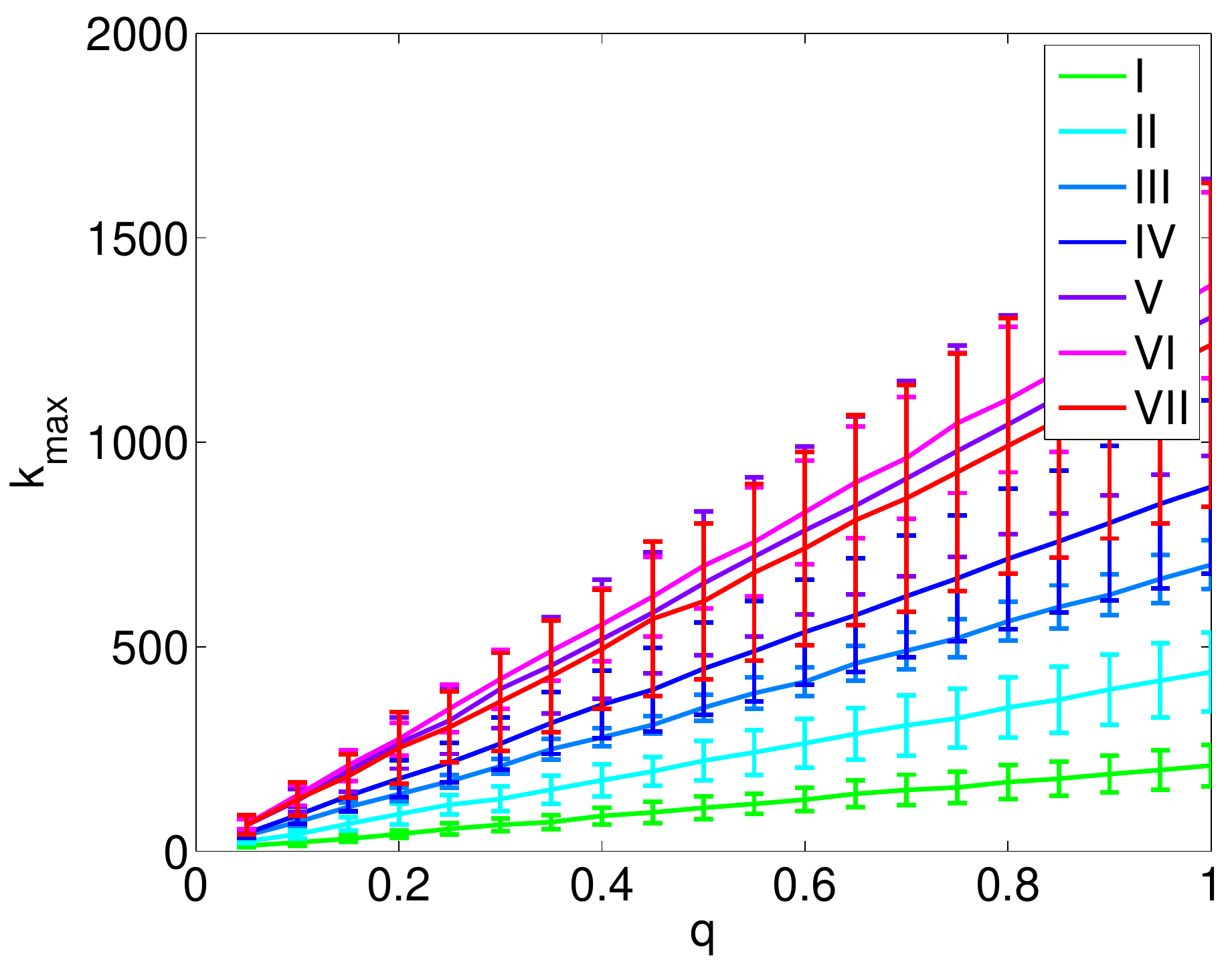}}\\
\caption[Scaling of subnetwork statistics for simulated networks induced on sampled interactions]{Scaling of subnetwork statistics for simulated networks induced on sampled interactions.}
\label{fig:weighted_subsampling}
\end{figure*}

\newpage
\setcounter{equation}{14}
\begin{figure*}[!ht]
\centering
\subfigure[Erdrey, Case 1]{\includegraphics[width=.19\textwidth]{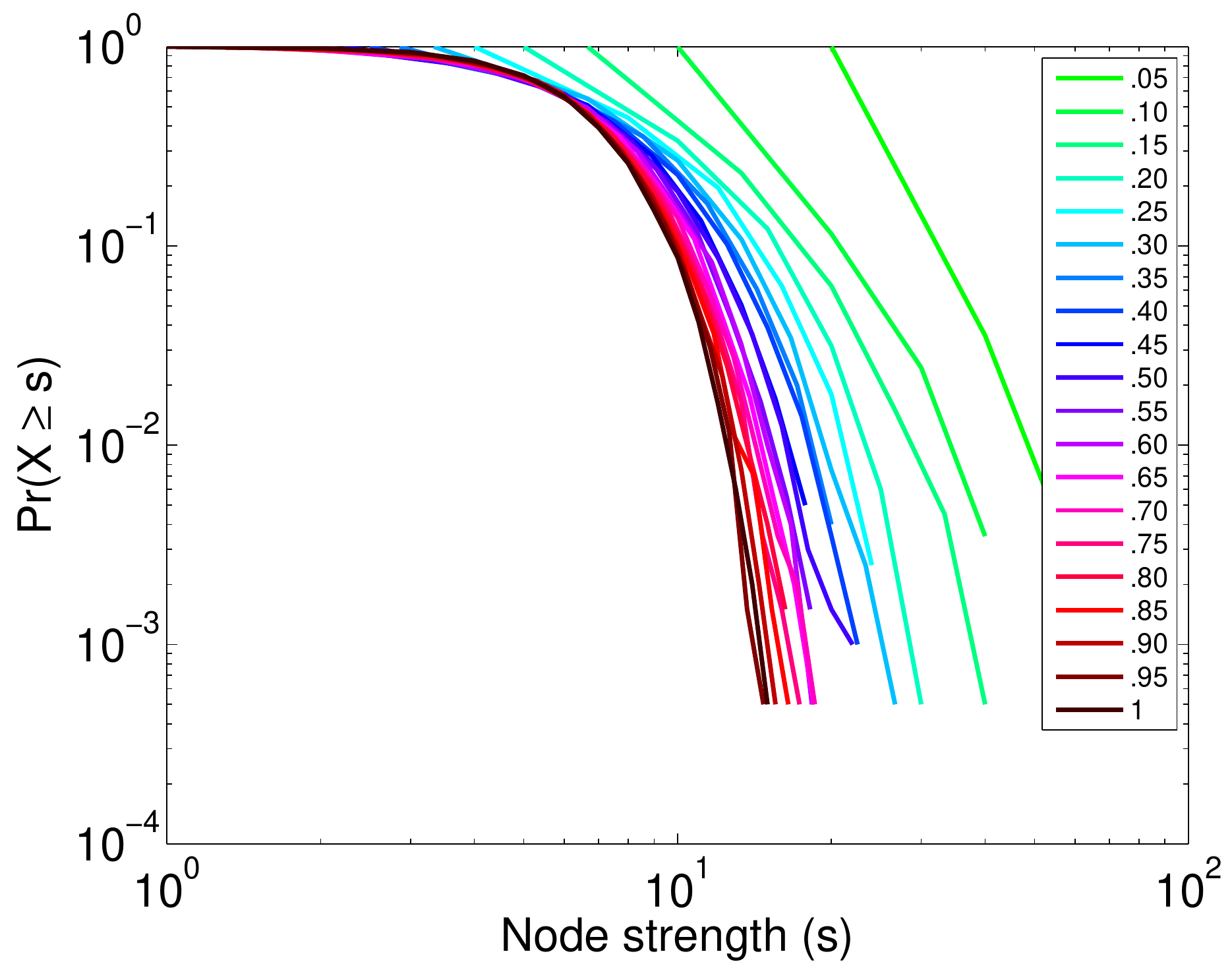}}
\subfigure[Erdrey, Case 2]{\includegraphics[width=.19\textwidth]{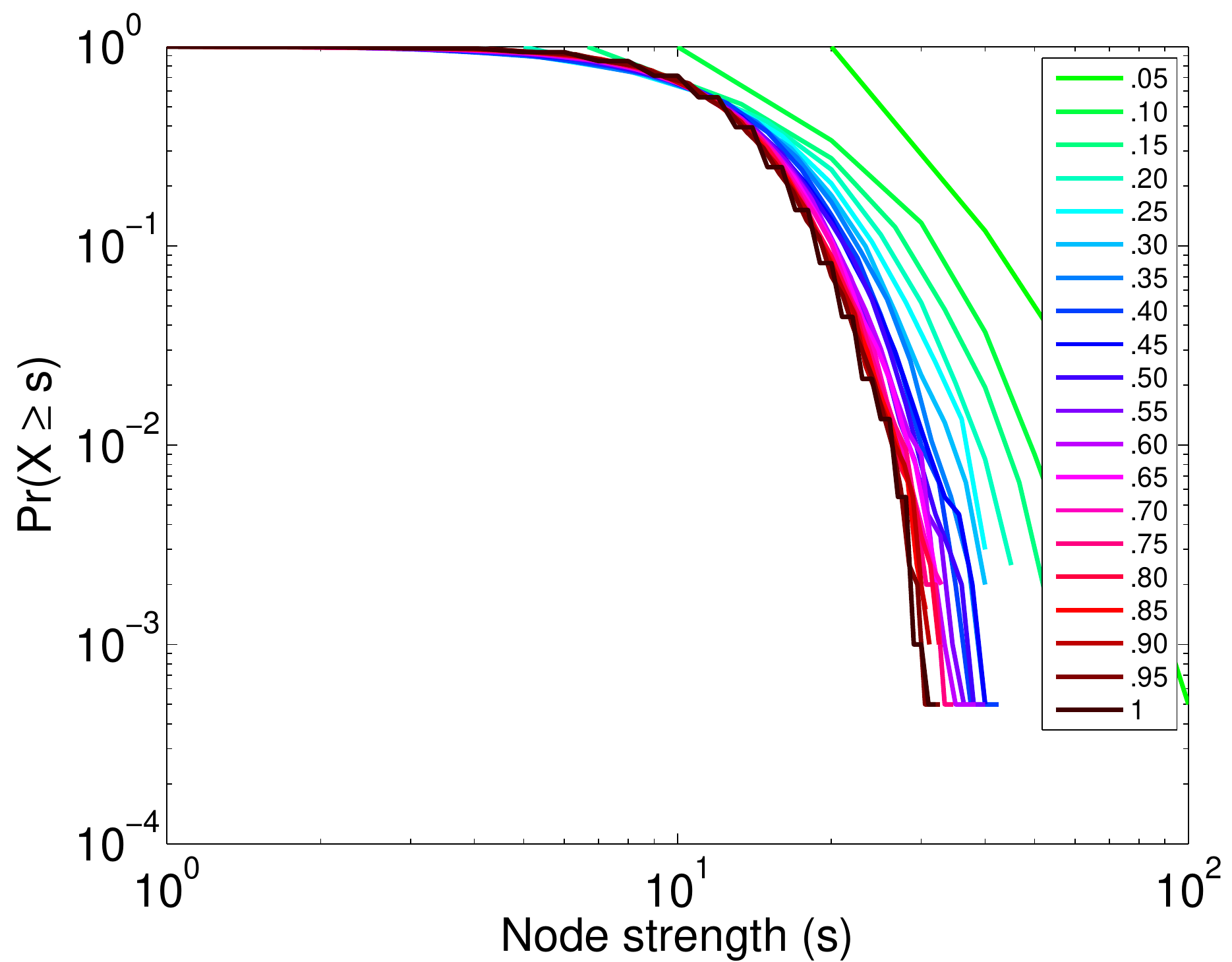}}
\subfigure[Erdrey, Case 3]{\includegraphics[width=.19\textwidth]{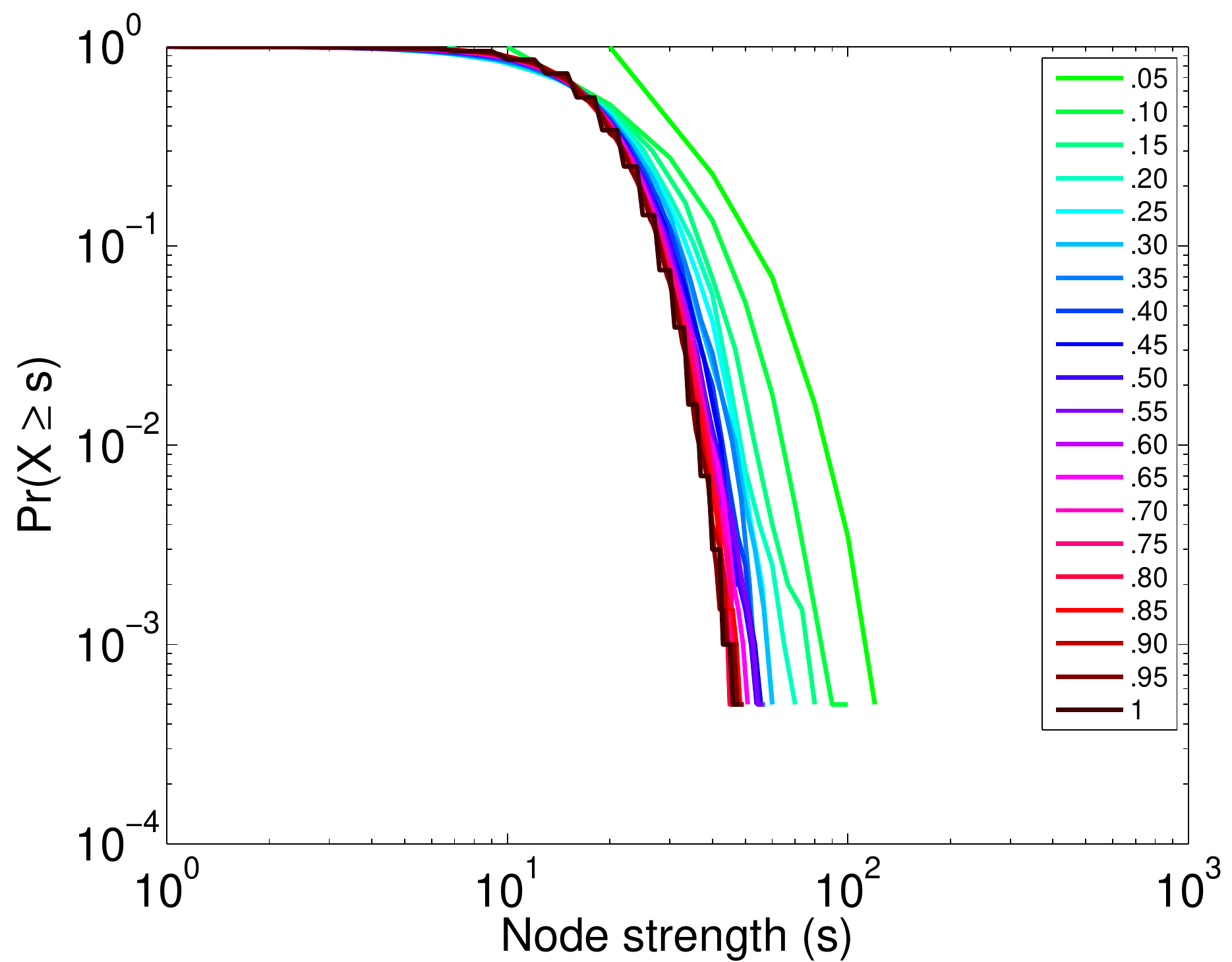}}
\subfigure[Erdrey, Case 4]{\includegraphics[width=.19\textwidth]{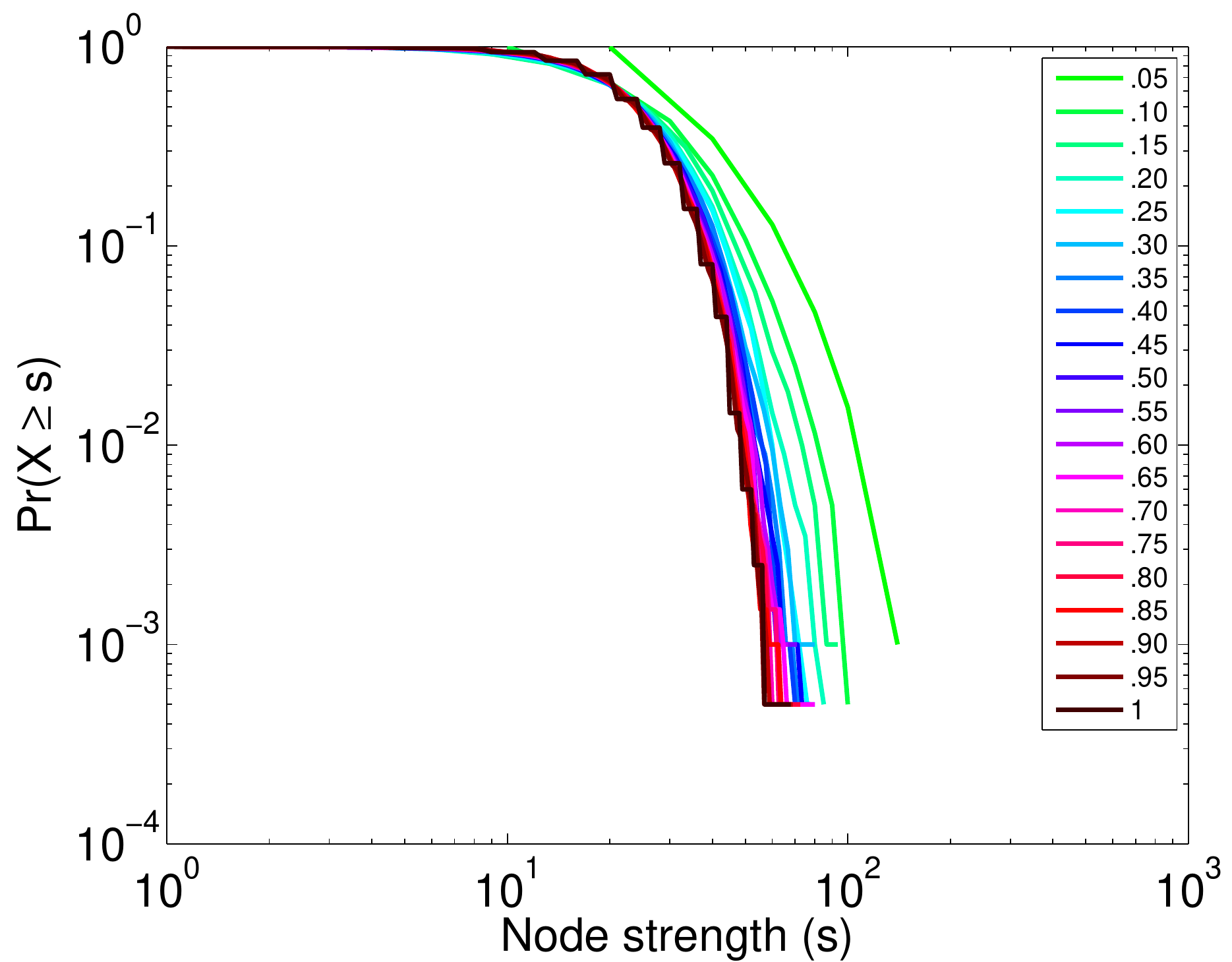}}
\subfigure[Erdrey, Case 5]{\includegraphics[width=.19\textwidth]{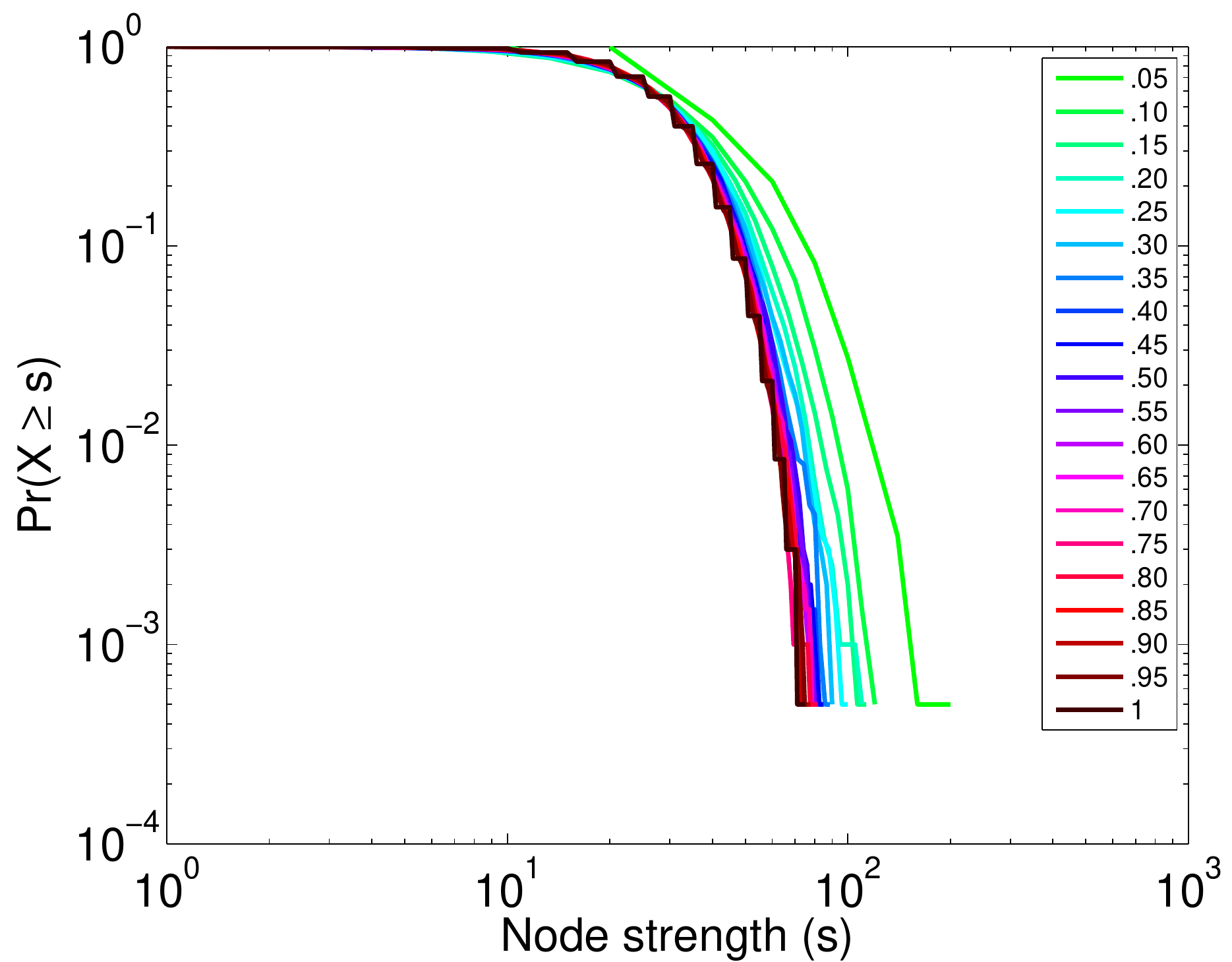}}\\
\subfigure[Erdrey, Case 6]{\includegraphics[width=.19\textwidth]{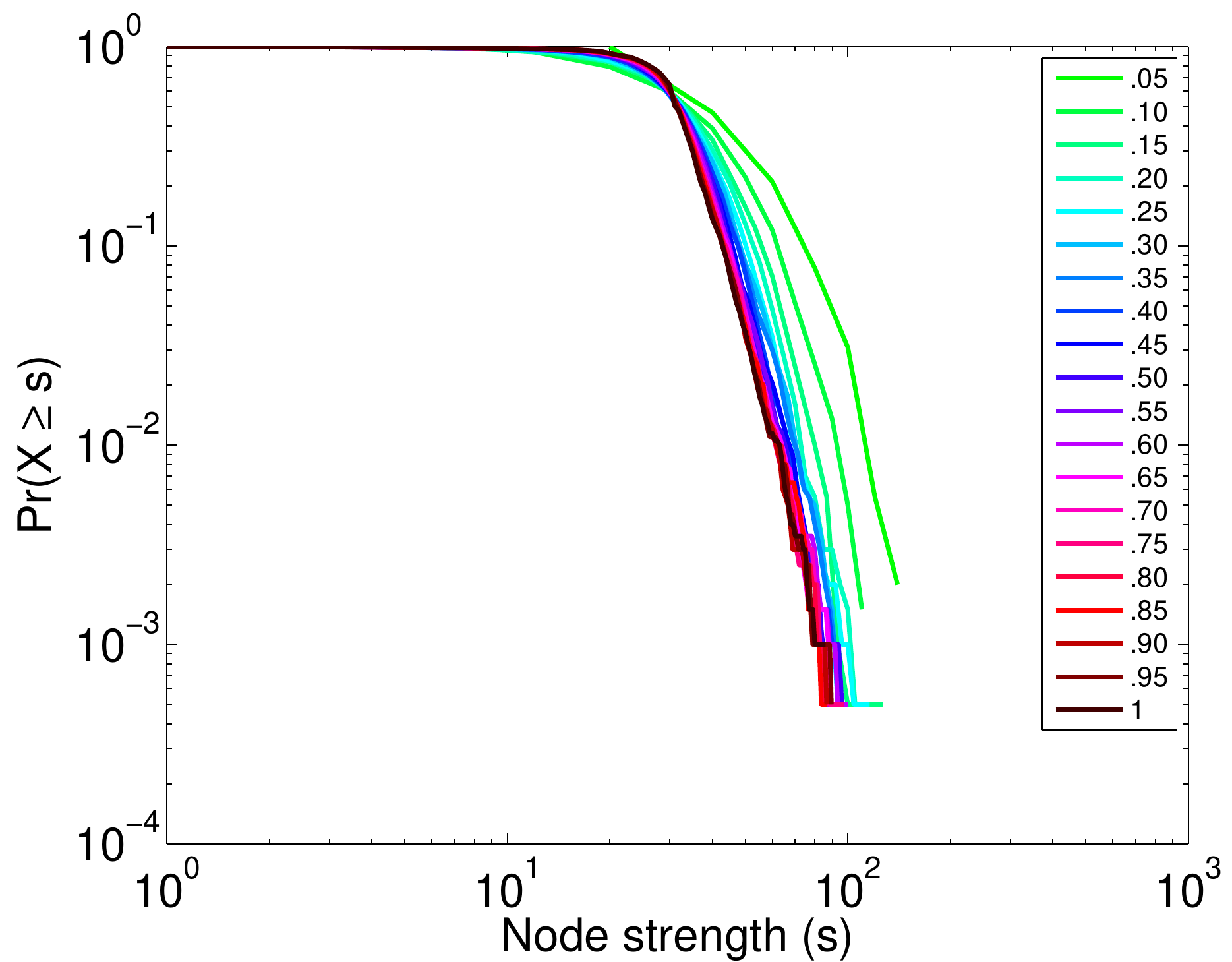}}
\subfigure[Erdrey, Case 7]{\includegraphics[width=.19\textwidth]{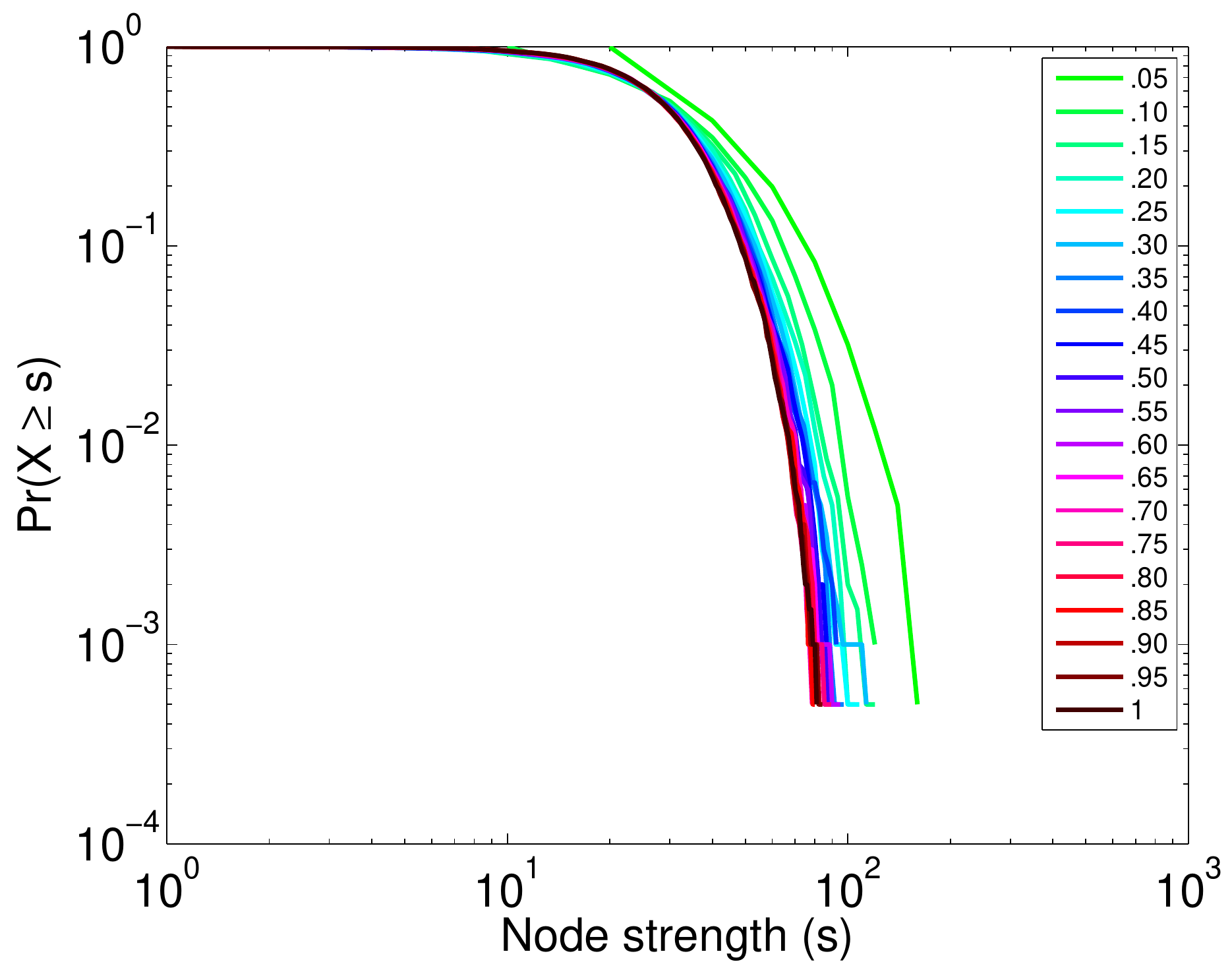}}\\
\subfigure[Pref, Case 1]{\includegraphics[width=.19\textwidth]{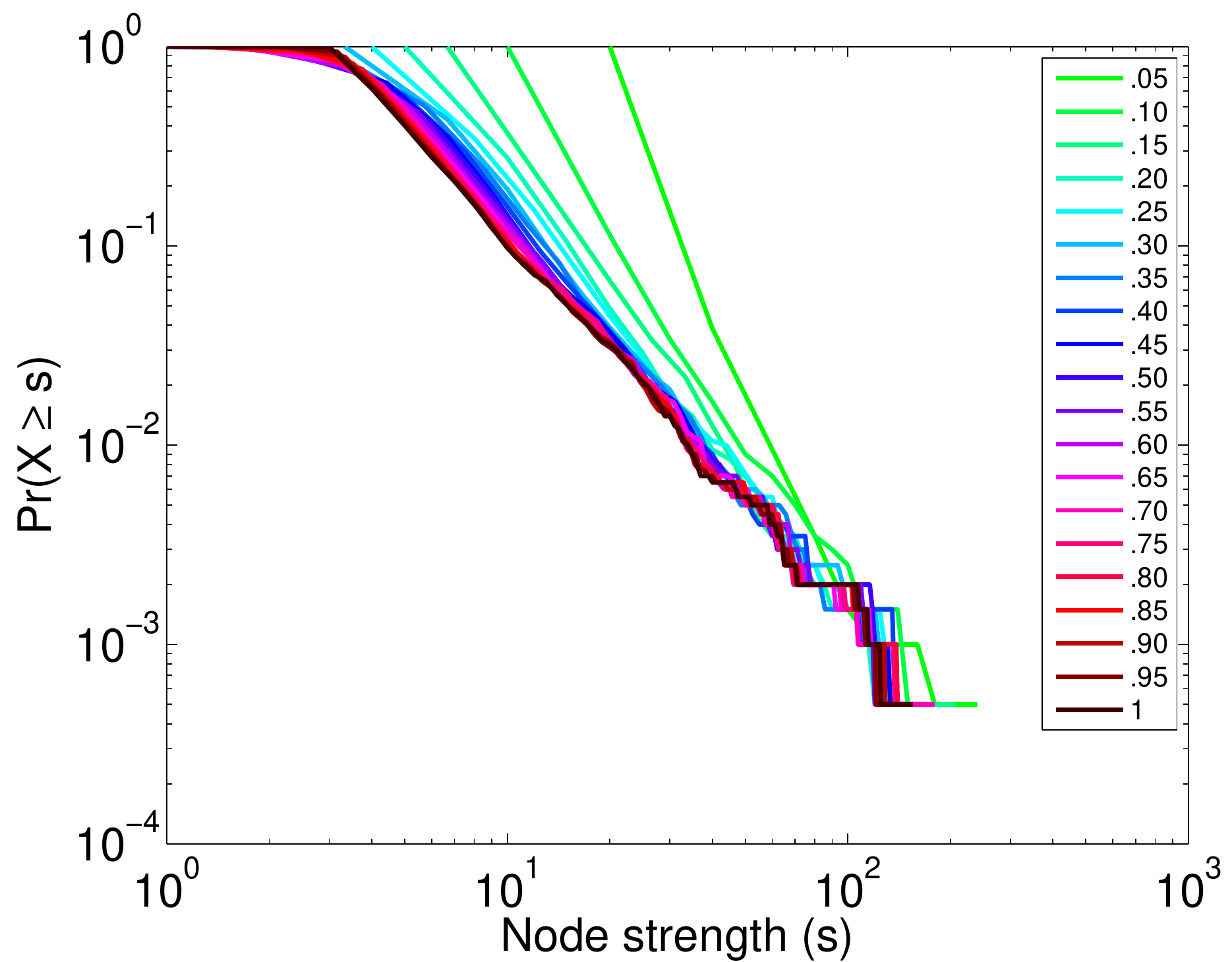}}
\subfigure[Pref, Case 2]{\includegraphics[width=.19\textwidth]{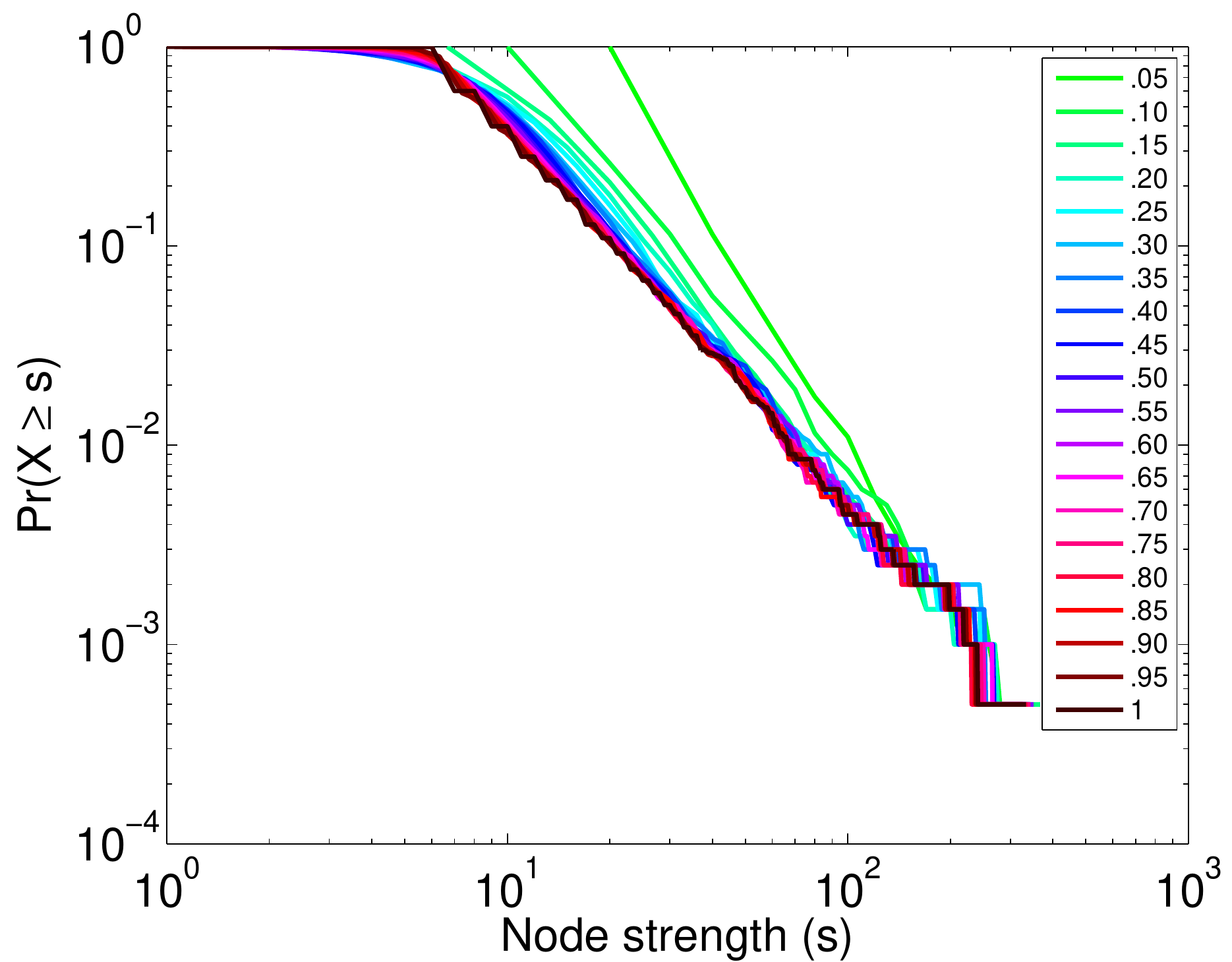}}
\subfigure[Pref, Case 3]{\includegraphics[width=.19\textwidth]{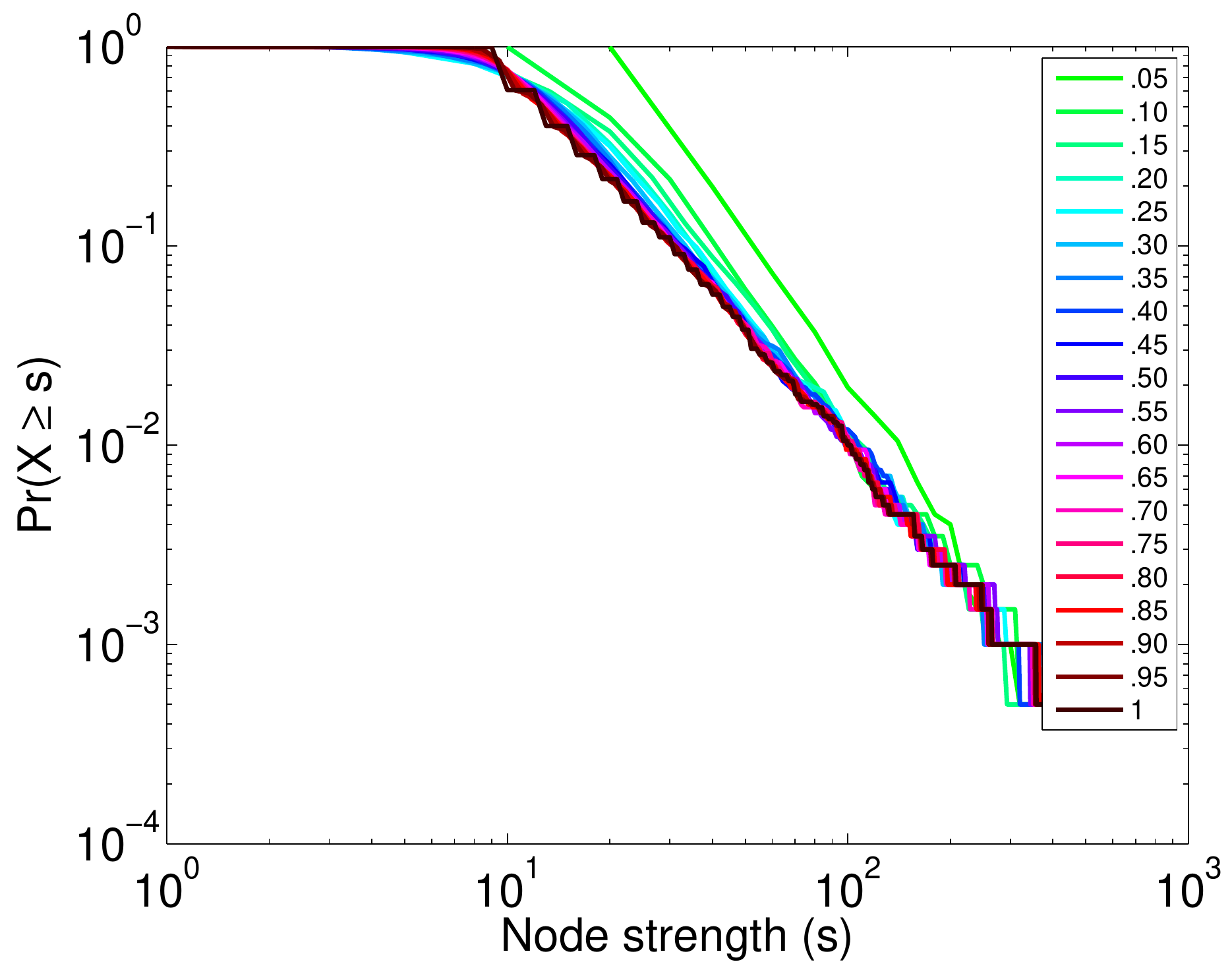}}
\subfigure[Pref, Case 4]{\includegraphics[width=.19\textwidth]{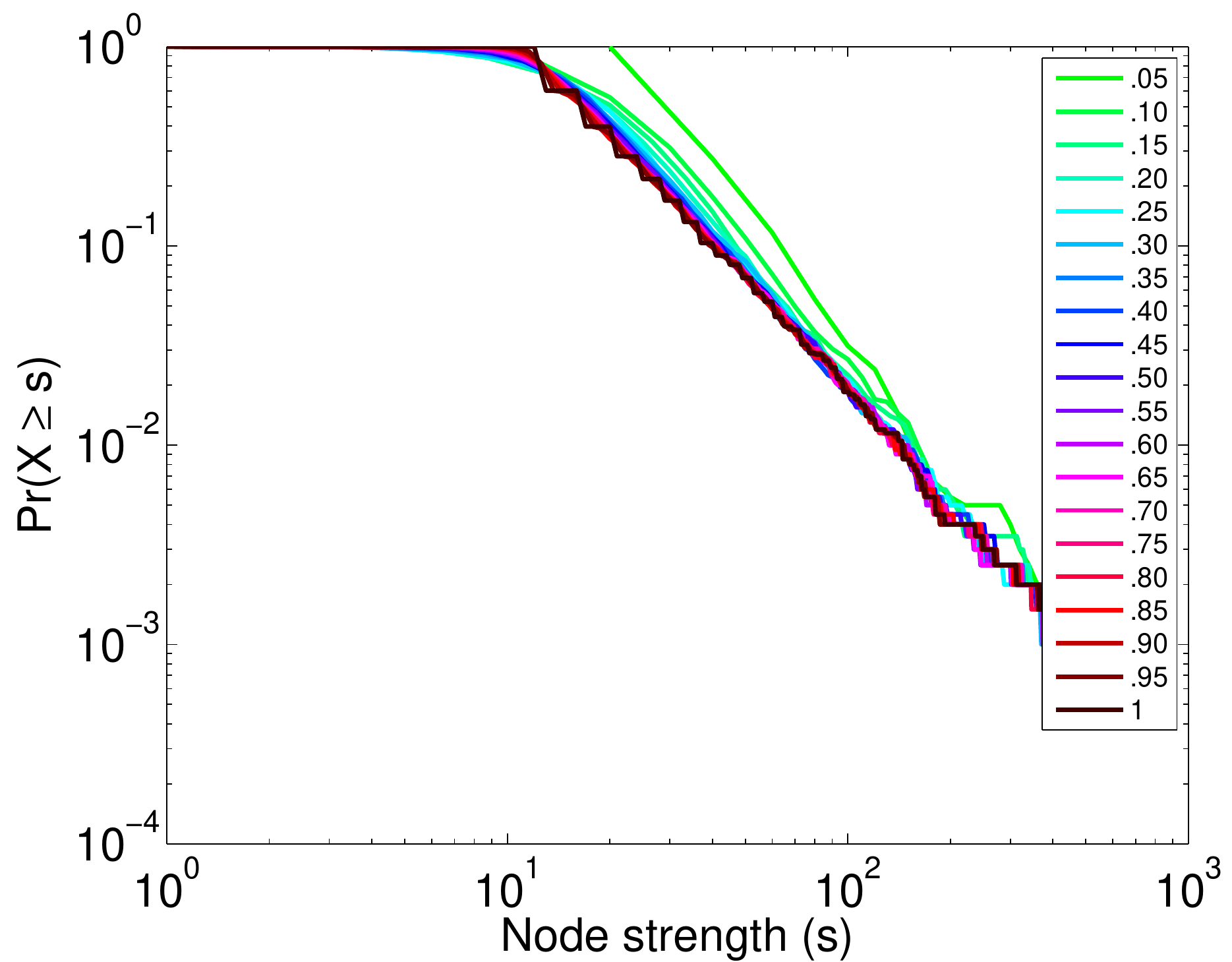}}
\subfigure[Pref, Case 5]{\includegraphics[width=.19\textwidth]{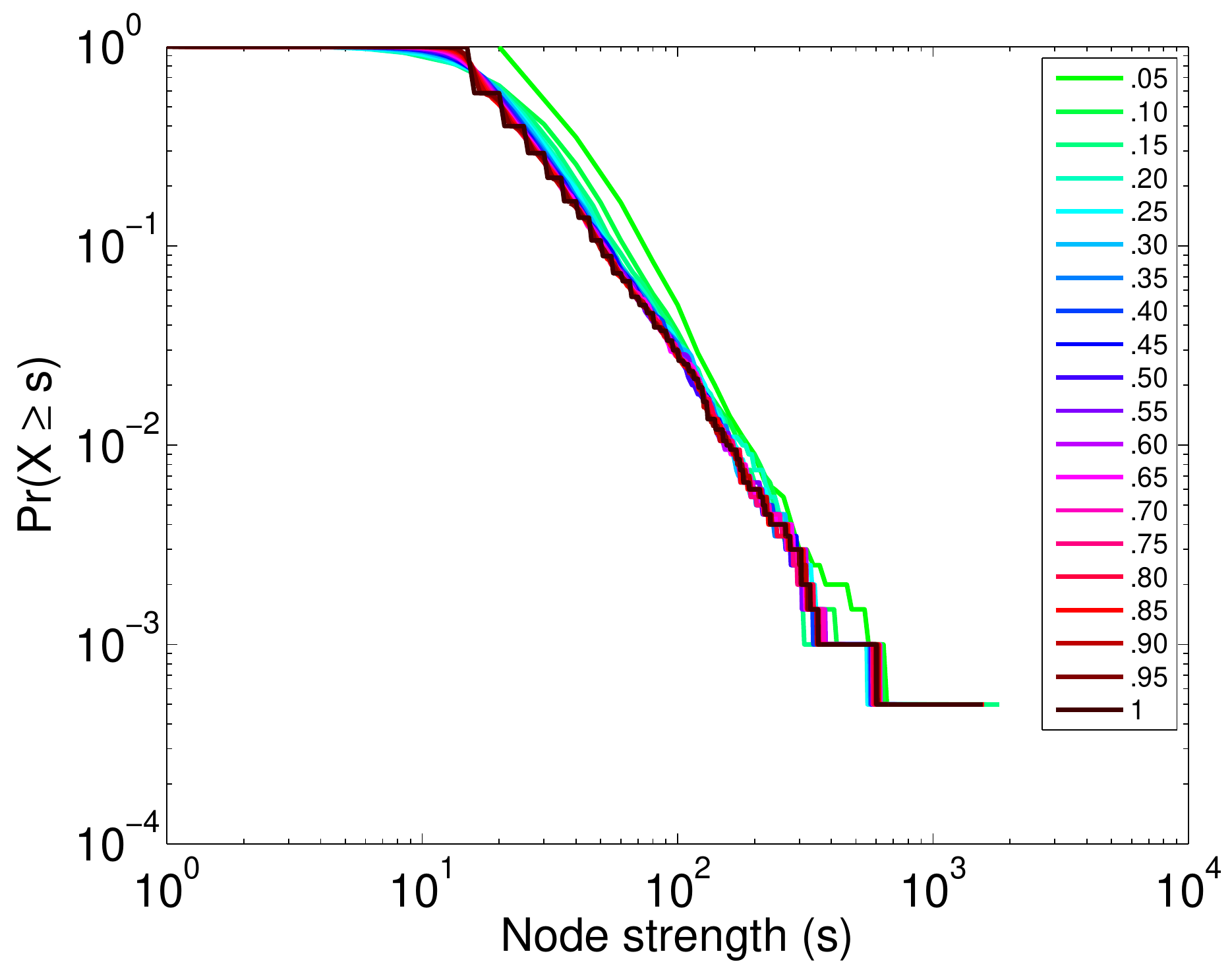}}\\
\subfigure[Pref, Case 6]{\includegraphics[width=.19\textwidth]{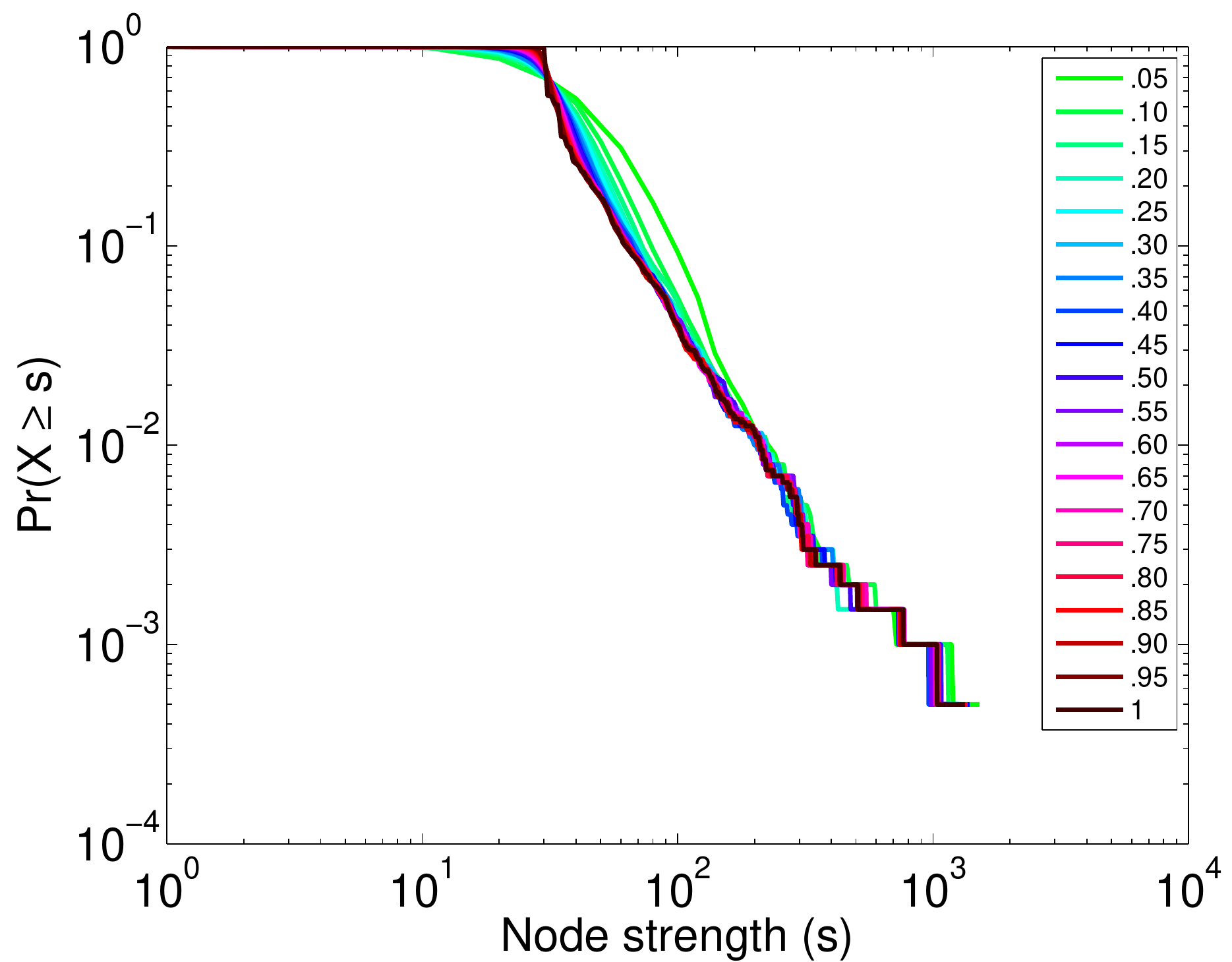}}
\subfigure[Pref, Case 7]{\includegraphics[width=.19\textwidth]{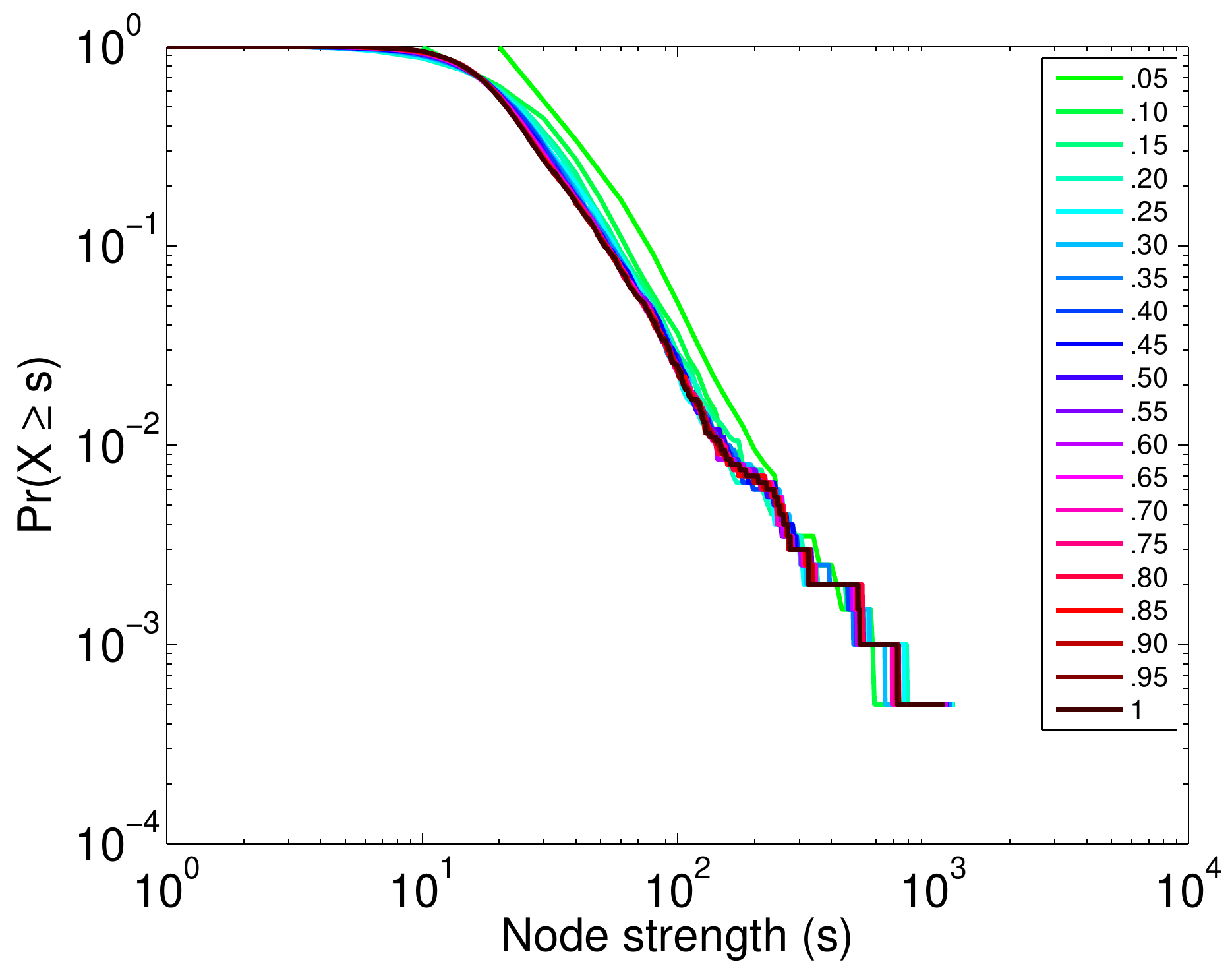}}\\
\caption[Predicted node strength distribution for weighted, simulated networks]{Predicted node strength distribution for weighted, simulated networks.}
\label{fig:Predicted_node_strength_one_over_q}
\end{figure*}
\newpage
\setcounter{equation}{15}
\begin{figure*}[!ht]
\centering
\subfigure[Erdrey, Case 1]{\includegraphics[width=.19\textwidth]{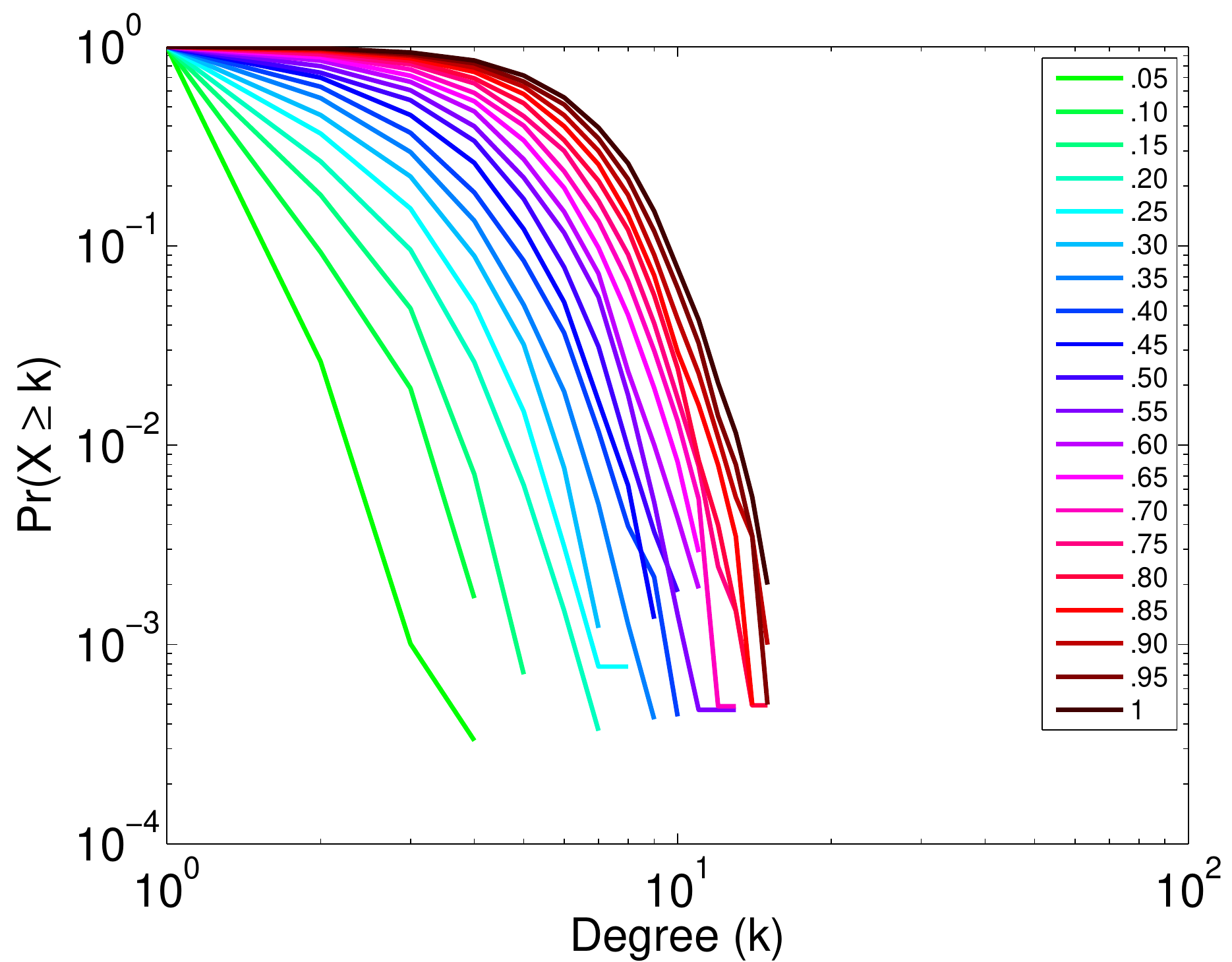}}
\subfigure[Erdrey, Case 2]{\includegraphics[width=.19\textwidth]{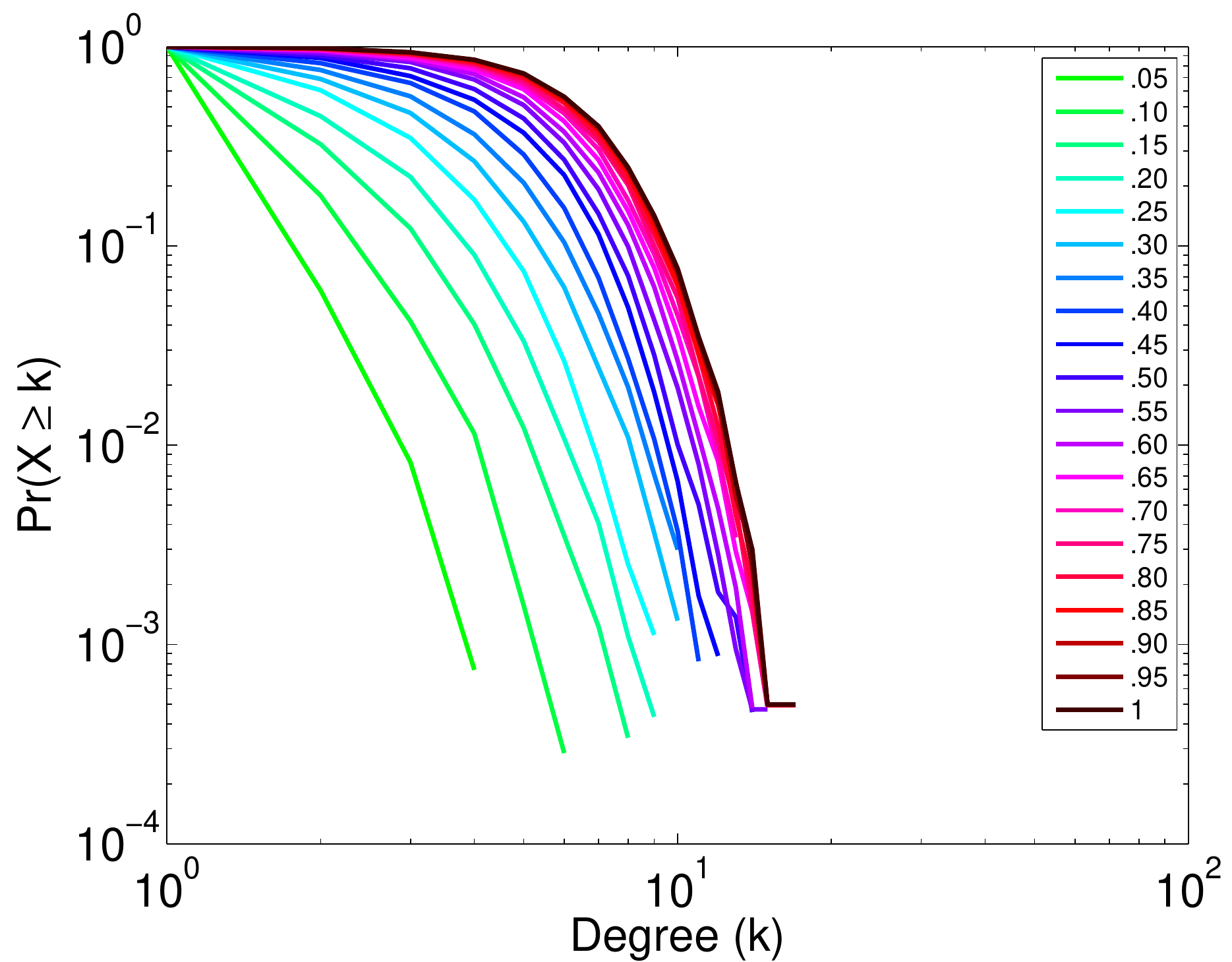}}
\subfigure[Erdrey, Case 3]{\includegraphics[width=.19\textwidth]{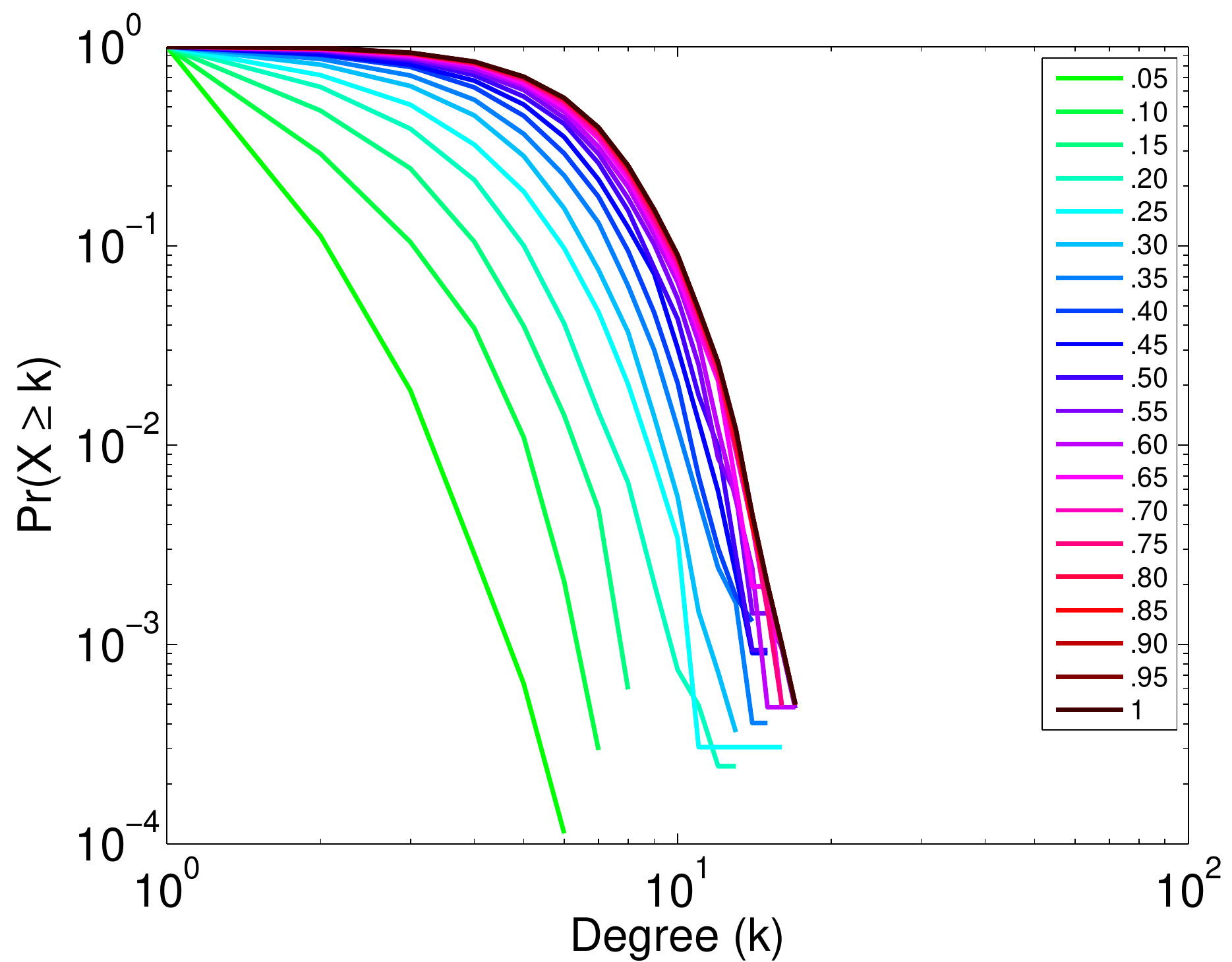}}
\subfigure[Erdrey, Case 4]{\includegraphics[width=.19\textwidth]{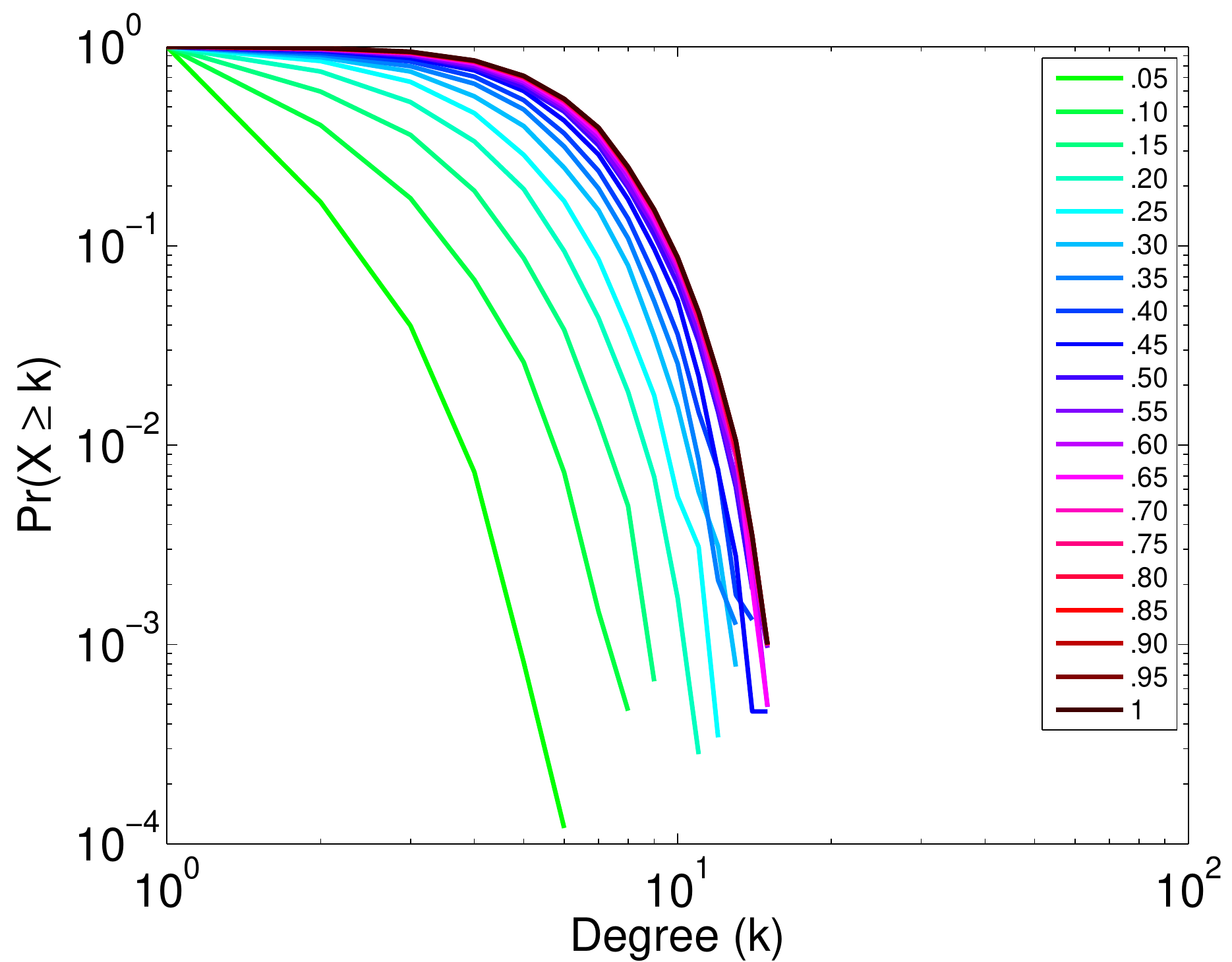}}
\subfigure[Erdrey, Case 5]{\includegraphics[width=.19\textwidth]{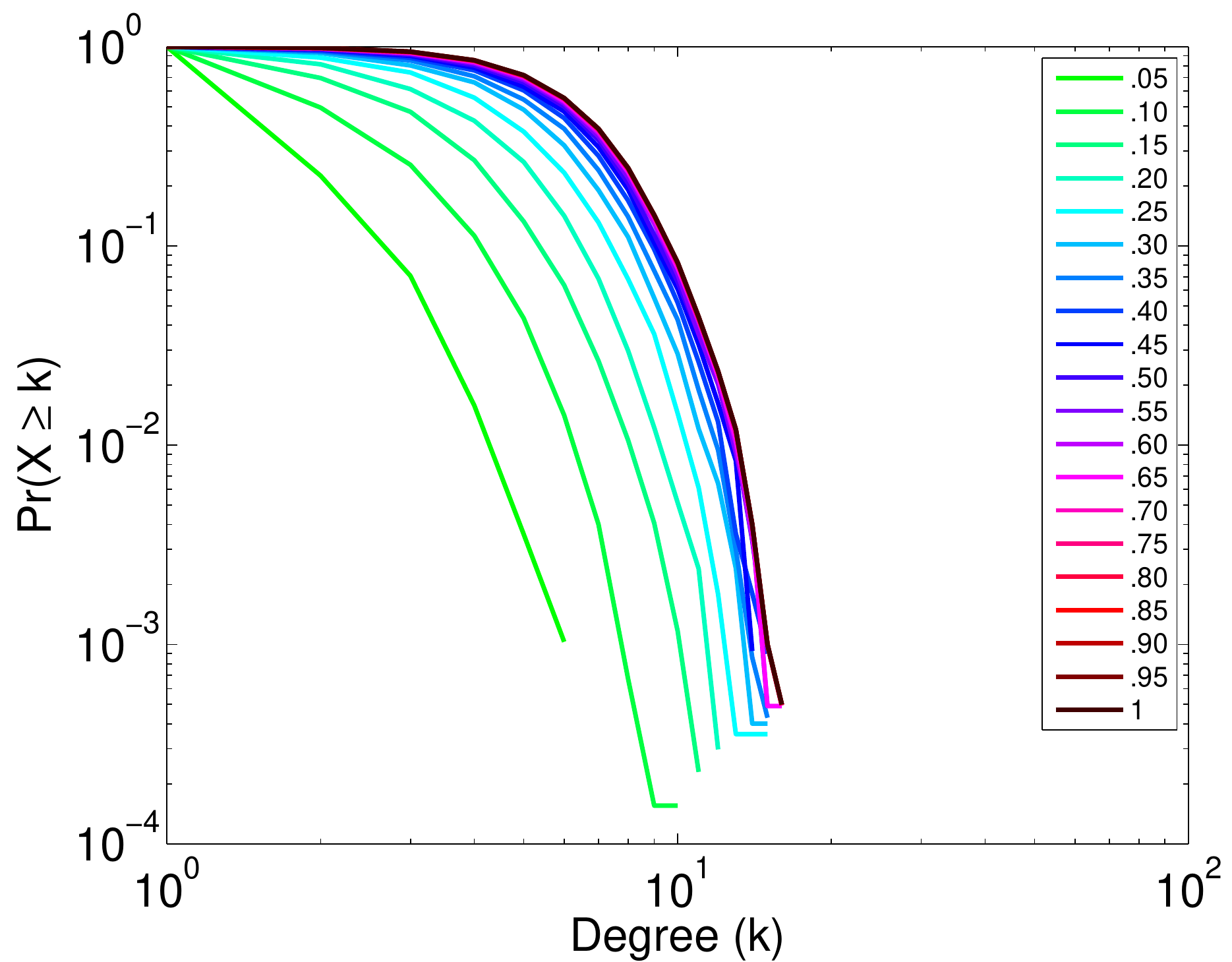}}\\
\subfigure[Erdrey, Case 6]{\includegraphics[width=.19\textwidth]{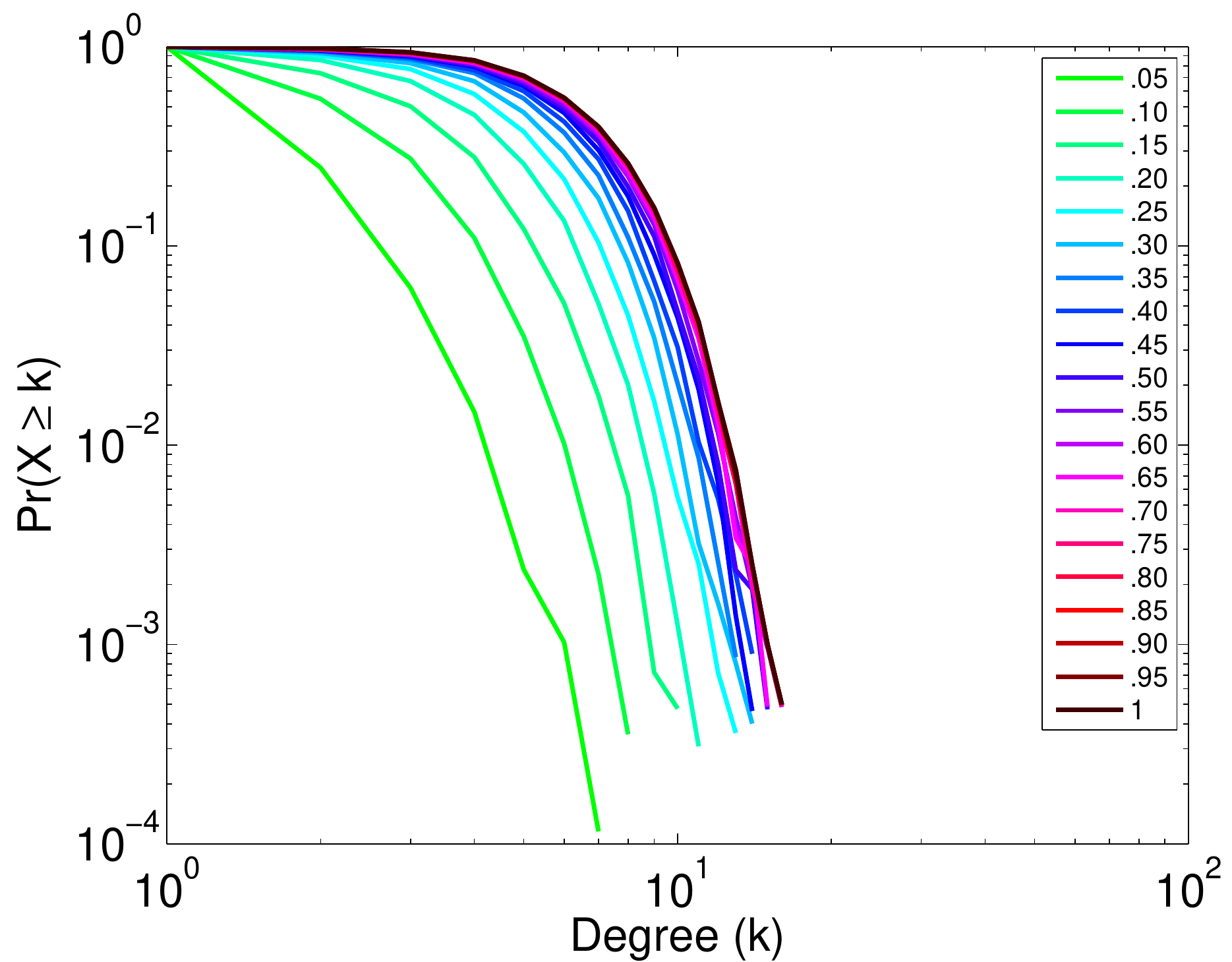}}
\subfigure[Erdrey, Case 7]{\includegraphics[width=.19\textwidth]{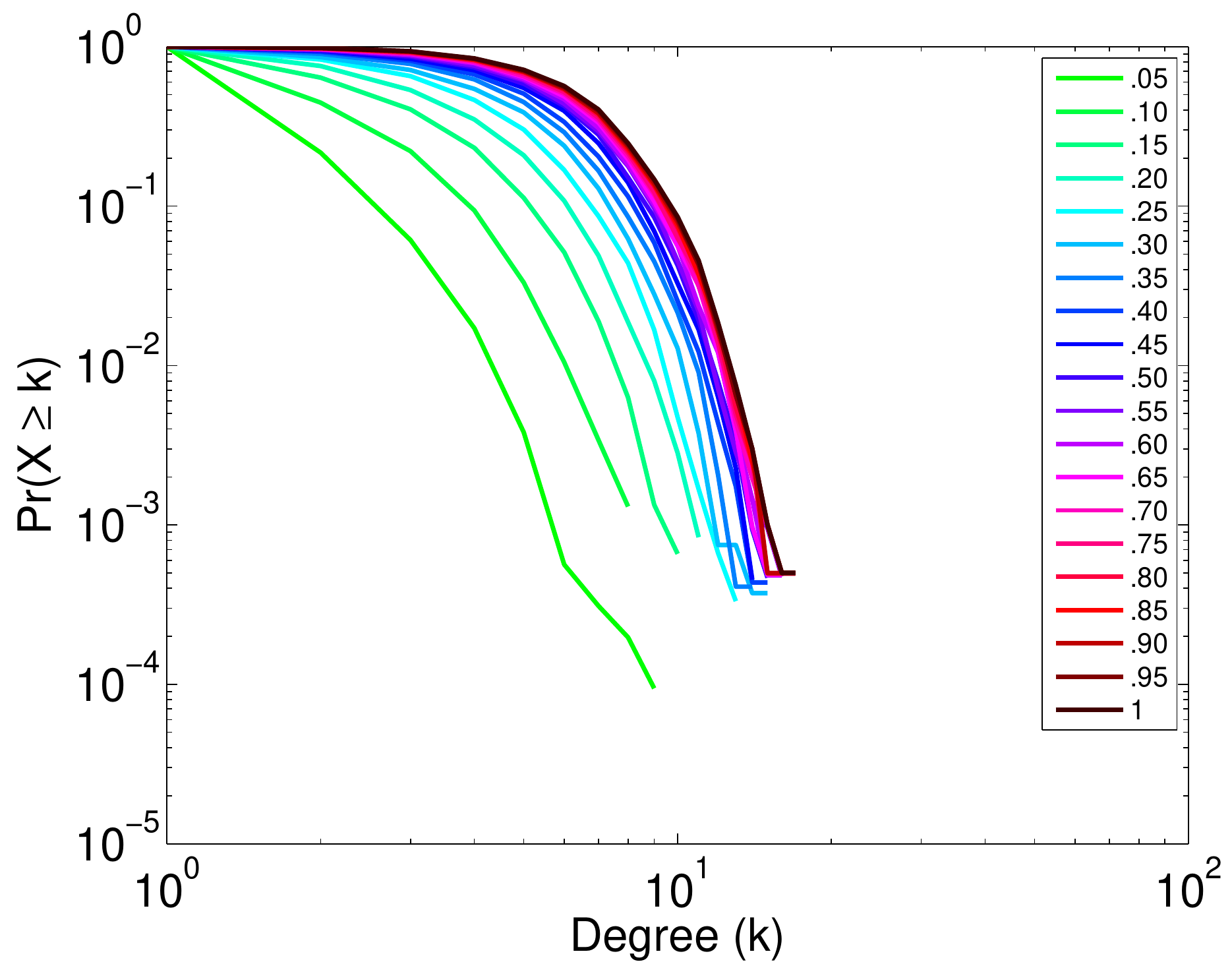}}\\
\subfigure[Pref, Case 1]{\includegraphics[width=.19\textwidth]{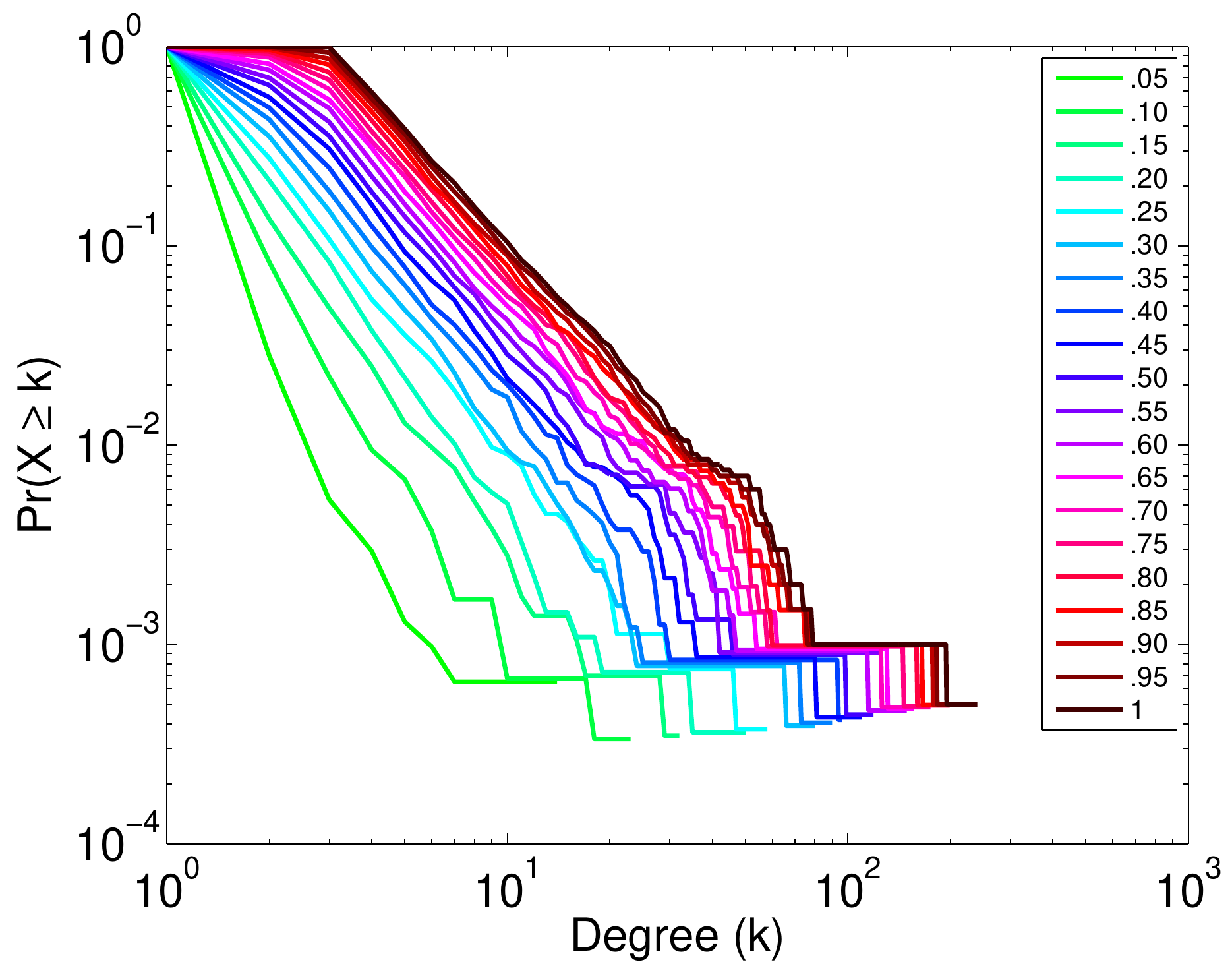}}
\subfigure[Pref, Case 2]{\includegraphics[width=.19\textwidth]{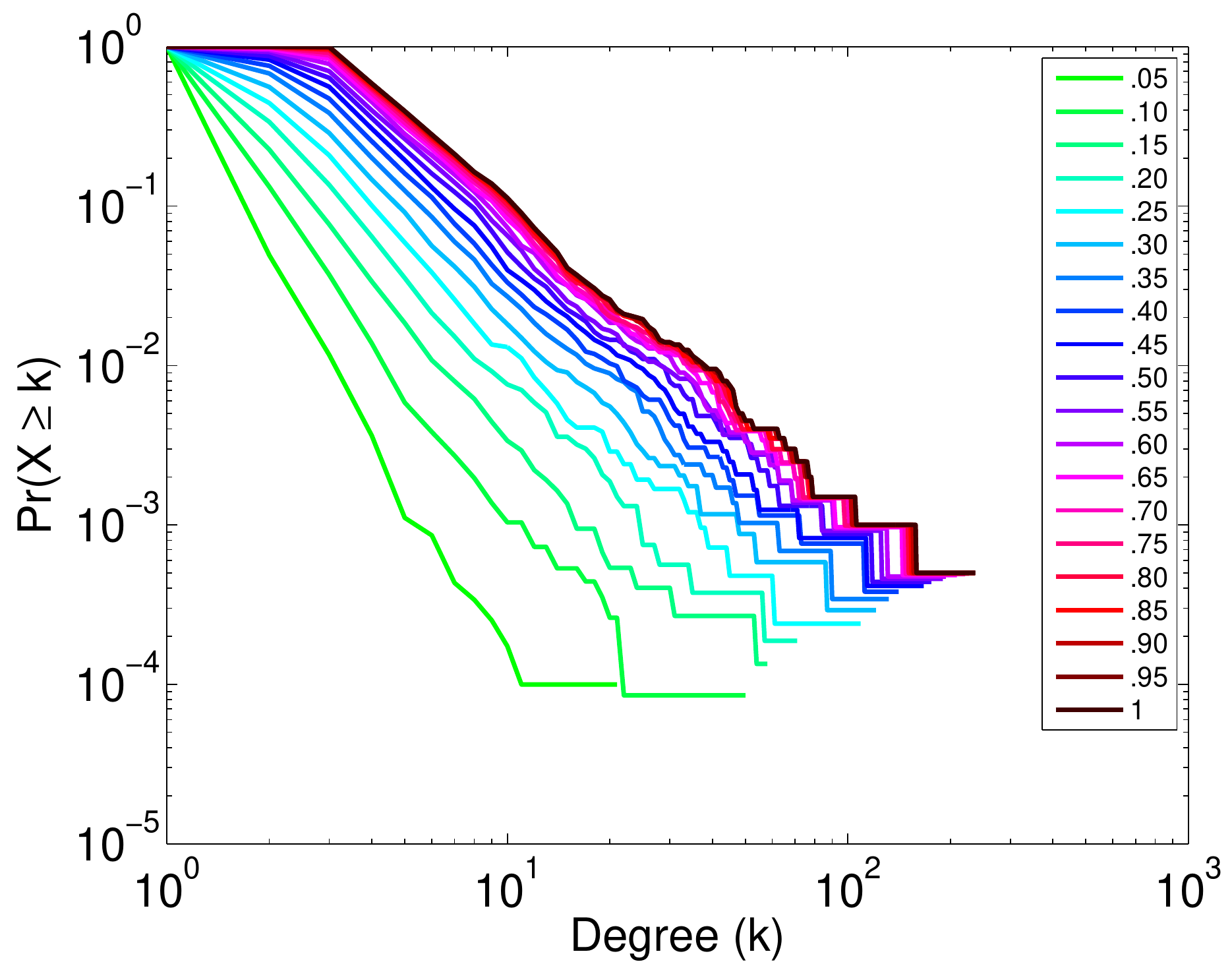}}
\subfigure[Pref, Case 3]{\includegraphics[width=.19\textwidth]{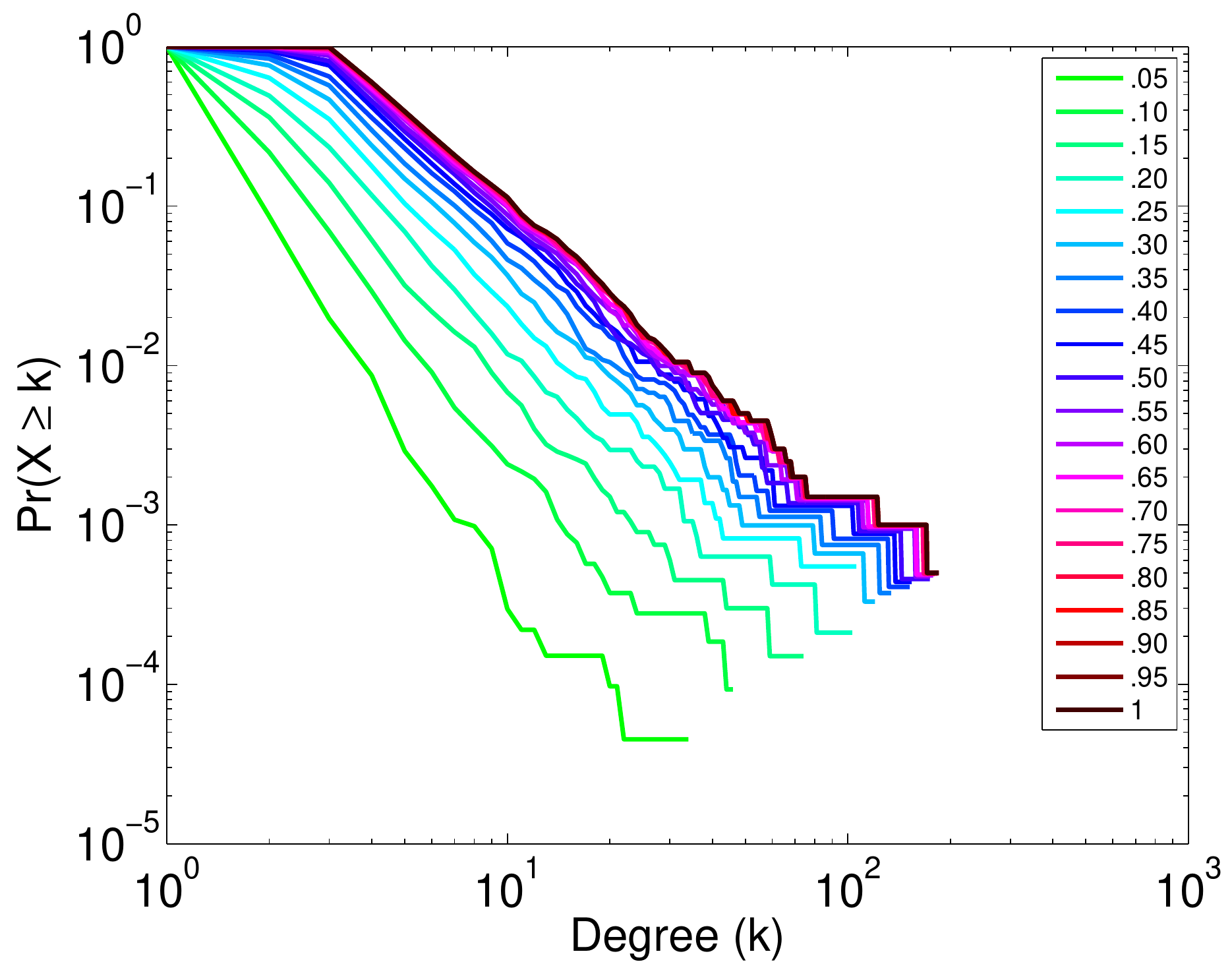}}
\subfigure[Pref, Case 4]{\includegraphics[width=.19\textwidth]{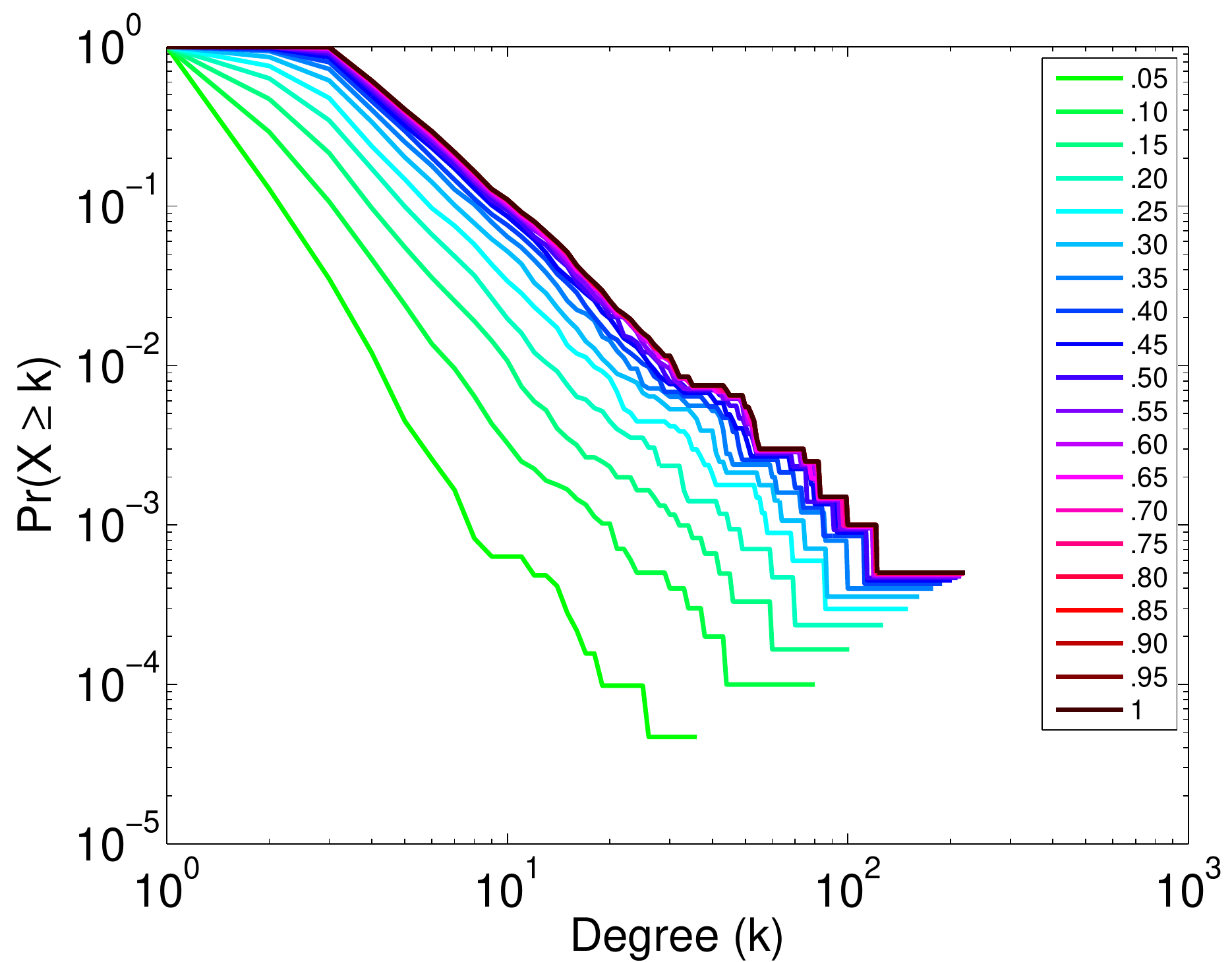}}
\subfigure[Pref, Case 5]{\includegraphics[width=.19\textwidth]{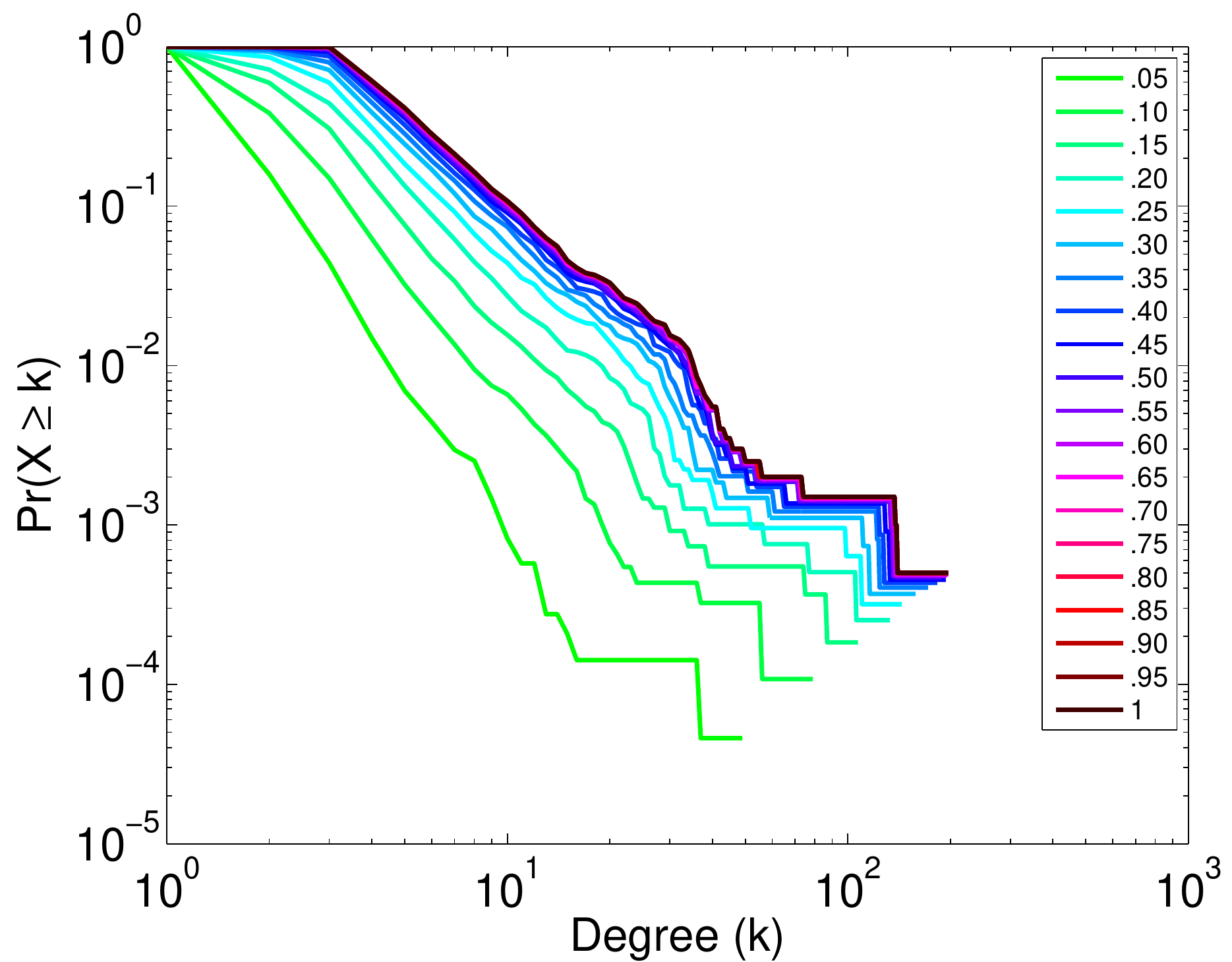}}\\
\subfigure[Pref, Case 6]{\includegraphics[width=.19\textwidth]{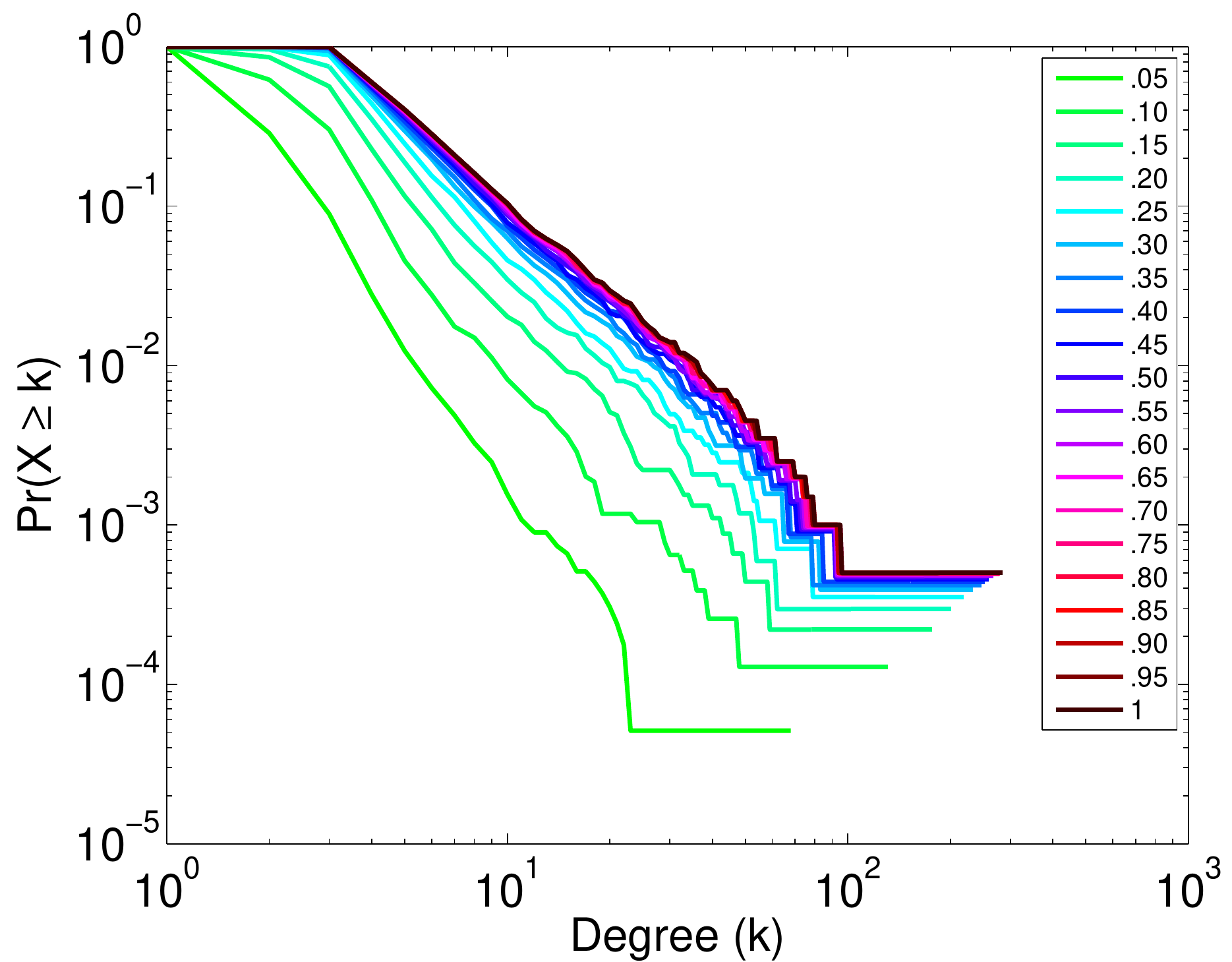}}
\subfigure[Pref, Case 7]{\includegraphics[width=.19\textwidth]{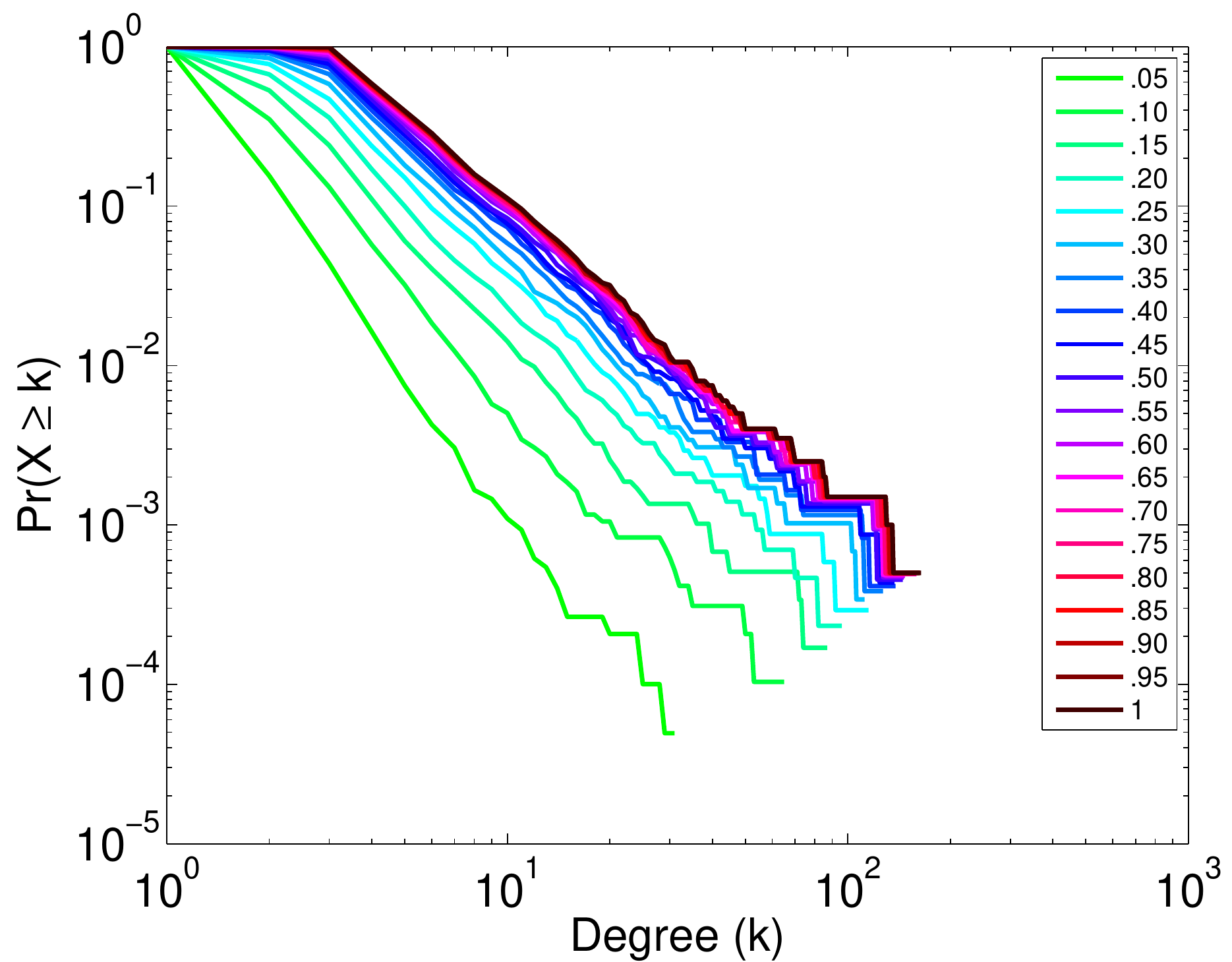}}\\
\caption[Predicted degree distribution for weighted, simulated networks]{Predicted degree distribution for weighted, simulated networks.}
\label{fig:Predicted_degree_one_over_q_exp}
\end{figure*}
\FloatBarrier
\newpage
  \setcounter{equation}{1}
\begin{table*}[htp!] \small
\caption[Error in $\hat{N}$ when sampling by nodes]{Error in $\hat{N}$ when sampling by nodes. }
\centering
\begin{tabular}{|l | l l l l l l l l l l|}
\hline
$q$ & Erdrey & Pref & Smallw & Renga & C.elegans & Airlines & Karate & Dolphins & Condmat & Power \\ \hline
  0.05 &   0.00 &   0.00 &   0.00 &   0.00 &   0.00 &   0.00 &   0.00 &   0.00 &   0.00 &   0.00 \\ 
  0.10 &   0.00 &   0.00 &   0.00 &   0.00 &   0.00 &   0.00 &   0.00 &   0.00 &   0.00 &   0.00 \\ 
  0.15 &   0.00 &   0.00 &   0.00 &   0.00 &   0.00 &   0.00 &   0.00 &   0.00 &   0.00 &   0.00 \\ 
  0.20 &   0.00 &   0.00 &   0.00 &   0.00 &   0.00 &   0.00 &   0.00 &   0.00 &   0.00 &   0.00 \\ 
  0.25 &   0.00 &   0.00 &   0.00 &   0.00 &   0.00 &   0.00 &   0.00 &   0.00 &   0.00 &   0.00 \\ 
  0.30 &   0.00 &   0.00 &   0.00 &   0.00 &   0.00 &   0.00 &   0.00 &   0.00 &   0.00 &   0.00 \\ 
  0.35 &   0.00 &   0.00 &   0.00 &   0.00 &   0.00 &   0.00 &   0.00 &   0.00 &   0.00 &   0.00 \\ 
  0.40 &   0.00 &   0.00 &   0.00 &   0.00 &   0.00 &   0.00 &   0.00 &   0.00 &   0.00 &   0.00 \\ 
  0.45 &   0.00 &   0.00 &   0.00 &   0.00 &   0.00 &   0.00 &   0.00 &   0.00 &   0.00 &   0.00 \\ 
  0.50 &   0.00 &   0.00 &   0.00 &   0.00 &   0.00 &   0.00 &   0.00 &   0.00 &   0.00 &   0.00 \\ 
  0.55 &   0.00 &   0.00 &   0.00 &   0.00 &   0.00 &   0.00 &   0.00 &   0.00 &   0.00 &   0.00 \\ 
  0.60 &   0.00 &   0.00 &   0.00 &   0.00 &   0.00 &   0.00 &   0.00 &   0.00 &   0.00 &   0.00 \\ 
  0.65 &   0.00 &   0.00 &   0.00 &   0.00 &   0.00 &   0.00 &   0.00 &   0.00 &   0.00 &   0.00 \\ 
  0.70 &   0.00 &   0.00 &   0.00 &   0.00 &   0.00 &   0.00 &   0.00 &   0.00 &   0.00 &   0.00 \\ 
  0.75 &   0.00 &   0.00 &   0.00 &   0.00 &   0.00 &   0.00 &   0.00 &   0.00 &   0.00 &   0.00 \\ 
  0.80 &   0.00 &   0.00 &   0.00 &   0.00 &   0.00 &   0.00 &   0.00 &   0.00 &   0.00 &   0.00 \\ 
  0.85 &   0.00 &   0.00 &   0.00 &   0.00 &   0.00 &   0.00 &   0.00 &   0.00 &   0.00 &   0.00 \\ 
  0.90 &   0.00 &   0.00 &   0.00 &   0.00 &   0.00 &   0.00 &   0.00 &   0.00 &   0.00 &   0.00 \\ 
  0.95 &   0.00 &   0.00 &   0.00 &   0.00 &   0.00 &   0.00 &   0.00 &   0.00 &   0.00 &   0.00 \\ 
  1.00 &   0.00 &   0.00 &   0.00 &   0.00 &   0.00 &   0.00 &   0.00 &   0.00 &   0.00 &   0.00 \\ 
  \hline
\end{tabular}
\label{table:error_N_sample_by_nodes}
\end{table*}
  
\newpage
  \setcounter{equation}{2}
\begin{table*}[htp!] \small
\caption[Error in $\hat{M}$ when sampling by nodes]{Error in $\hat{M}$ when sampling by nodes. The percent error in the number of predicted nodes is nearly zero when, except in the small empirical networks where for small $q$, we violate the assumption that $n \gg 1$ and incur large errors.}
\centering
\begin{tabular}{|l | l l l l l l l l l l|}
\hline
$q$ & Erdrey & Pref & Smallw & Renga & C. elegans & Airlines & Karate & Dolphins & Condmat & Power \\ \hline
  0.05 &   0.00 &   0.00 &   0.00 &   0.00 &   0.08 &   0.02 &   2.71 &   2.04 &   0.00 &   0.01 \\ 
  0.10 &   0.00 &   0.00 &   0.00 &   0.00 &   0.02 &   0.03 &   1.04 &   0.28 &   0.00 &   0.00 \\ 
  0.15 &   0.00 &   0.00 &   0.00 &   0.00 &   0.01 &   0.00 &   0.24 &   0.04 &   0.00 &   0.01 \\ 
  0.20 &   0.00 &   0.00 &   0.00 &   0.00 &   0.01 &   0.01 &   0.03 &   0.02 &   0.00 &   0.00 \\ 
  0.25 &   0.00 &   0.00 &   0.00 &   0.00 &   0.06 &   0.01 &   0.08 &   0.06 &   0.01 &   0.00 \\ 
  0.30 &   0.00 &   0.00 &   0.00 &   0.00 &   0.02 &   0.03 &   0.05 &   0.02 &   0.00 &   0.00 \\ 
  0.35 &   0.00 &   0.00 &   0.00 &   0.00 &   0.02 &   0.01 &   0.02 &   0.04 &   0.00 &   0.00 \\ 
  0.40 &   0.00 &   0.00 &   0.00 &   0.00 &   0.00 &   0.01 &   0.01 &   0.01 &   0.00 &   0.00 \\ 
  0.45 &   0.00 &   0.00 &   0.00 &   0.00 &   0.01 &   0.01 &   0.04 &   0.03 &   0.00 &   0.00 \\ 
  0.50 &   0.00 &   0.00 &   0.00 &   0.00 &   0.00 &   0.00 &   0.07 &   0.01 &   0.00 &   0.00 \\ 
  0.55 &   0.00 &   0.00 &   0.00 &   0.00 &   0.00 &   0.00 &   0.05 &   0.03 &   0.00 &   0.00 \\ 
  0.60 &   0.00 &   0.00 &   0.00 &   0.00 &   0.00 &   0.02 &   0.01 &   0.01 &   0.00 &   0.00 \\ 
  0.65 &   0.00 &   0.00 &   0.00 &   0.00 &   0.01 &   0.00 &   0.01 &   0.00 &   0.00 &   0.00 \\ 
  0.70 &   0.00 &   0.00 &   0.00 &   0.00 &   0.01 &   0.01 &   0.00 &   0.02 &   0.00 &   0.00 \\ 
  0.75 &   0.00 &   0.00 &   0.00 &   0.00 &   0.00 &   0.01 &   0.00 &   0.01 &   0.00 &   0.00 \\ 
  0.80 &   0.00 &   0.00 &   0.00 &   0.00 &   0.01 &   0.00 &   0.02 &   0.01 &   0.00 &   0.00 \\ 
  0.85 &   0.00 &   0.00 &   0.00 &   0.00 &   0.00 &   0.00 &   0.00 &   0.00 &   0.00 &   0.00 \\ 
  0.90 &   0.00 &   0.00 &   0.00 &   0.00 &   0.00 &   0.00 &   0.00 &   0.01 &   0.00 &   0.00 \\ 
  0.95 &   0.00 &   0.00 &   0.00 &   0.00 &   0.00 &   0.00 &   0.00 &   0.00 &   0.00 &   0.00 \\ 
  1.00 &   0.00 &   0.00 &   0.00 &   0.00 &   0.00 &   0.00 &   0.00 &   0.00 &   0.00 &   0.00 \\  \hline
\end{tabular}
\label{table:error_M_sample_by_nodes}
\end{table*}
\newpage
    
  \newpage
  \setcounter{equation}{3}
\begin{table*}[htp!] \small
\caption[Error in $\hat{k}_{\rm avg}$ when sampling by nodes]{Error in $\hat{k}_{\rm avg}$ when sampling by nodes.  Errors in $\hat{M}$ are largely responsible for errors in $\hat{k}_{\rm avg}$.}
\centering
\begin{tabular}{|l | l l l l l l l l l l|}
\hline
$q$ & Erdrey & Pref & Smallw & Renga & C.elegans & Airlines & Karate & Dolphins & Condmat & Power \\ \hline
  0.05 &   0.00 &   0.00 &   0.00 &   0.00 &   0.08 &   0.02 &   2.71 &   2.04 &   0.00 &   0.01 \\ 
  0.10 &   0.00 &   0.00 &   0.00 &   0.00 &   0.02 &   0.03 &   1.04 &   0.28 &   0.00 &   0.00 \\ 
  0.15 &   0.00 &   0.00 &   0.00 &   0.00 &   0.01 &   0.00 &   0.24 &   0.04 &   0.00 &   0.01 \\ 
  0.20 &   0.00 &   0.00 &   0.00 &   0.00 &   0.01 &   0.01 &   0.03 &   0.02 &   0.00 &   0.00 \\ 
  0.25 &   0.00 &   0.00 &   0.00 &   0.00 &   0.06 &   0.01 &   0.08 &   0.06 &   0.01 &   0.00 \\ 
  0.30 &   0.00 &   0.00 &   0.00 &   0.00 &   0.02 &   0.03 &   0.05 &   0.02 &   0.00 &   0.00 \\ 
  0.35 &   0.00 &   0.00 &   0.00 &   0.00 &   0.02 &   0.01 &   0.02 &   0.04 &   0.00 &   0.00 \\ 
  0.40 &   0.00 &   0.00 &   0.00 &   0.00 &   0.00 &   0.01 &   0.01 &   0.01 &   0.00 &   0.00 \\ 
  0.45 &   0.00 &   0.00 &   0.00 &   0.00 &   0.01 &   0.01 &   0.04 &   0.03 &   0.00 &   0.00 \\ 
  0.50 &   0.00 &   0.00 &   0.00 &   0.00 &   0.00 &   0.00 &   0.07 &   0.01 &   0.00 &   0.00 \\ 
  0.55 &   0.00 &   0.00 &   0.00 &   0.00 &   0.00 &   0.00 &   0.05 &   0.03 &   0.00 &   0.00 \\ 
  0.60 &   0.00 &   0.00 &   0.00 &   0.00 &   0.00 &   0.02 &   0.01 &   0.01 &   0.00 &   0.00 \\ 
  0.65 &   0.00 &   0.00 &   0.00 &   0.00 &   0.01 &   0.00 &   0.01 &   0.00 &   0.00 &   0.00 \\ 
  0.70 &   0.00 &   0.00 &   0.00 &   0.00 &   0.01 &   0.01 &   0.00 &   0.02 &   0.00 &   0.00 \\ 
  0.75 &   0.00 &   0.00 &   0.00 &   0.00 &   0.00 &   0.01 &   0.00 &   0.01 &   0.00 &   0.00 \\ 
  0.80 &   0.00 &   0.00 &   0.00 &   0.00 &   0.01 &   0.00 &   0.02 &   0.01 &   0.00 &   0.00 \\ 
  0.85 &   0.00 &   0.00 &   0.00 &   0.00 &   0.00 &   0.00 &   0.00 &   0.00 &   0.00 &   0.00 \\ 
  0.90 &   0.00 &   0.00 &   0.00 &   0.00 &   0.00 &   0.00 &   0.00 &   0.01 &   0.00 &   0.00 \\ 
  0.95 &   0.00 &   0.00 &   0.00 &   0.00 &   0.00 &   0.00 &   0.00 &   0.00 &   0.00 &   0.00 \\ 
  1.00 &   0.00 &   0.00 &   0.00 &   0.00 &   0.00 &   0.00 &   0.00 &   0.00 &   0.00 &   0.00 \\   \hline
\end{tabular}
\label{table:error_avk_sample_by_nodes}
\end{table*}
\newpage
  \setcounter{equation}{4}
\begin{table*}[!htp] \small
\caption[Error in $\hat{k}_{\max}$ when sampling by nodes]{Error in $\hat{k}_{\max}$ when sampling by nodes. The percent error in the predicted max degree is nearly zero for large $q$. In general, predicting the max. degree is difficult due to the dependence on network structure. }
\centering
\begin{tabular}{|l | l l l l l l l l l l|}
\hline
$q$ & Erdrey & Pref & Smallw & Renga & C. elegans & Airlines & Karate & Dolphins & Condmat & Power \\ \hline
 0.05 &   2.70 &   0.67 &   7.70 &   3.73 &   0.59 &   0.39 &   0.00 &   0.08 &   0.13 &   0.14 \\ 
  0.10 &   1.60 &   0.54 &   4.94 &   2.26 &   0.52 &   0.28 &   0.18 &   0.02 &   0.09 &   0.08 \\ 
  0.15 &   1.13 &   0.49 &   3.73 &   1.67 &   0.46 &   0.21 &   0.31 &   0.01 &   0.05 &   0.05 \\ 
  0.20 &   0.89 &   0.42 &   2.96 &   1.29 &   0.46 &   0.18 &   0.30 &   0.01 &   0.06 &   0.04 \\ 
  0.25 &   0.72 &   0.38 &   2.46 &   1.06 &   0.44 &   0.15 &   0.28 &   0.01 &   0.05 &   0.03 \\ 
  0.30 &   0.57 &   0.33 &   2.09 &   0.87 &   0.33 &   0.12 &   0.21 &   0.01 &   0.05 &   0.02 \\ 
  0.35 &   0.48 &   0.27 &   1.77 &   0.73 &   0.33 &   0.12 &   0.18 &   0.01 &   0.06 &   0.01 \\ 
  0.40 &   0.40 &   0.24 &   1.50 &   0.62 &   0.30 &   0.09 &   0.10 &   0.01 &   0.05 &   0.01 \\ 
  0.45 &   0.34 &   0.21 &   1.22 &   0.52 &   0.25 &   0.07 &   0.17 &   0.01 &   0.02 &   0.01 \\ 
  0.50 &   0.29 &   0.19 &   1.00 &   0.44 &   0.20 &   0.07 &   0.13 &   0.01 &   0.01 &   0.01 \\ 
  0.55 &   0.24 &   0.16 &   0.82 &   0.38 &   0.20 &   0.07 &   0.11 &   0.01 &   0.01 &   0.01 \\ 
  0.60 &   0.21 &   0.15 &   0.67 &   0.33 &   0.19 &   0.04 &   0.04 &   0.01 &   0.00 &   0.01 \\ 
  0.65 &   0.17 &   0.13 &   0.54 &   0.27 &   0.16 &   0.04 &   0.03 &   0.01 &   0.00 &   0.00 \\ 
  0.70 &   0.14 &   0.10 &   0.43 &   0.23 &   0.10 &   0.04 &   0.02 &   0.00 &   0.01 &   0.00 \\ 
  0.75 &   0.11 &   0.07 &   0.33 &   0.18 &   0.12 &   0.04 &   0.01 &   0.00 &   0.01 &   0.00 \\ 
  0.80 &   0.09 &   0.05 &   0.25 &   0.13 &   0.10 &   0.02 &   0.01 &   0.00 &   0.01 &   0.00 \\ 
  0.85 &   0.07 &   0.04 &   0.18 &   0.10 &   0.07 &   0.01 &   0.01 &   0.00 &   0.00 &   0.00 \\ 
  0.90 &   0.04 &   0.03 &   0.11 &   0.07 &   0.04 &   0.01 &   0.00 &   0.00 &   0.00 &   0.00 \\ 
  0.95 &   0.02 &   0.02 &   0.05 &   0.04 &   0.03 &   0.01 &   0.01 &   0.00 &   0.00 &   0.00 \\ 
  1.00 &   0.00 &   0.00 &   0.00 &   0.00 &   0.00 &   0.00 &   0.00 &   0.00 &   0.00 &   0.00 \\ \hline
  \end{tabular}
\label{table:error_kmax_sample_by_nodes}
\end{table*}

\newpage
  \setcounter{equation}{5}
\begin{table*}[!htp] \small
\caption[Error in $\hat{C}$ when sampling by nodes]{Error in $\hat{C}$ when sampling by nodes. For some small networks with a small portion of nodes sampled $q$, no paths of length three occurred and the clustering coefficient was not computed in these cases. }
\centering
\begin{tabular}{|l | l l l l l l l l l l|}
\hline
$q$ & Erdrey & Pref & Smallw & Renga & C. elegans & Airlines & Karate & Dolphins & Condmat & Power \\ \hline
  0.05 &   0.01 &   0.33 &   0.00 &   0.00 &    -- &   0.04 &    -- &    -- &   0.00 &    -- \\ 
  0.10 &   0.07 &   0.15 &   0.00 &   0.00 &   0.09 &   0.02 &    -- &    -- &   0.01 &   0.21 \\ 
  0.15 &   0.04 &   0.10 &   0.00 &   0.00 &   0.08 &   0.02 &    -- &    -- &   0.01 &   0.03 \\ 
  0.20 &   0.03 &   0.07 &   0.00 &   0.00 &   0.01 &   0.02 &    -- &    -- &   0.01 &   0.02 \\ 
  0.25 &   0.05 &   0.06 &   0.00 &   0.00 &   0.01 &   0.01 &    -- &    -- &   0.00 &   0.07 \\ 
  0.30 &   0.04 &   0.05 &   0.00 &   0.00 &   0.05 &   0.01 &    -- &   0.15 &   0.00 &   0.03 \\ 
  0.35 &   0.05 &   0.03 &   0.00 &   0.00 &   0.03 &   0.01 &    -- &   0.14 &   0.00 &   0.06 \\ 
  0.40 &   0.05 &   0.02 &   0.00 &   0.00 &   0.00 &   0.01 &   0.06 &   0.12 &   0.00 &   0.02 \\ 
  0.45 &   0.03 &   0.02 &   0.00 &   0.00 &   0.01 &   0.01 &   0.13 &   0.02 &   0.00 &   0.05 \\ 
  0.50 &   0.01 &   0.02 &   0.00 &   0.00 &   0.01 &   0.01 &   0.20 &   0.04 &   0.00 &   0.02 \\ 
  0.55 &   0.00 &   0.01 &   0.00 &   0.00 &   0.01 &   0.00 &   0.12 &   0.05 &   0.00 &   0.00 \\ 
  0.60 &   0.02 &   0.00 &   0.00 &   0.00 &   0.01 &   0.00 &   0.08 &   0.02 &   0.00 &   0.02 \\ 
  0.65 &   0.01 &   0.00 &   0.00 &   0.00 &   0.01 &   0.00 &   0.04 &   0.01 &   0.00 &   0.01 \\ 
  0.70 &   0.01 &   0.00 &   0.00 &   0.00 &   0.00 &   0.00 &   0.04 &   0.01 &   0.00 &   0.00 \\ 
  0.75 &   0.00 &   0.01 &   0.00 &   0.00 &   0.01 &   0.00 &   0.03 &   0.00 &   0.00 &   0.00 \\ 
  0.80 &   0.00 &   0.01 &   0.00 &   0.00 &   0.00 &   0.00 &   0.01 &   0.00 &   0.00 &   0.00 \\ 
  0.85 &   0.00 &   0.01 &   0.00 &   0.00 &   0.00 &   0.00 &   0.02 &   0.01 &   0.00 &   0.00 \\ 
  0.90 &   0.00 &   0.01 &   0.00 &   0.00 &   0.00 &   0.00 &   0.01 &   0.01 &   0.00 &   0.01 \\ 
  0.95 &   0.00 &   0.00 &   0.00 &   0.00 &   0.00 &   0.00 &   0.01 &   0.00 &   0.00 &   0.00 \\ 
  1.00 &   0.00 &   0.00 &   0.00 &   0.00 &   0.00 &   0.00 &   0.00 &   0.00 &   0.00 &   0.00 \\   \hline
\end{tabular}
\label{table:error_cluster_sample_by_nodes}
\end{table*}

\newpage
    \setcounter{equation}{6}
\begin{table*}[!htp] \small
\caption[Error in $\hat{N}$ when sampling by failing links]{Error in $\hat{N}$ when sampling by failing links. No error is encountered because all nodes remain in the subnetwork.}
\centering
\begin{tabular}{|l | l l l l l l l l l l|}
\hline
$q$ & Erdrey & Pref & Smallw & Renga & C. elegans & Airlines & Karate & Dolphins & Condmat & Power \\ \hline
  0.05 &   0.00 &   0.00 &   0.00 &   0.00 &   0.00 &   0.00 &   0.00 &   0.00 &   0.00 &   0.00 \\ 
  0.10 &   0.00 &   0.00 &   0.00 &   0.00 &   0.00 &   0.00 &   0.00 &   0.00 &   0.00 &   0.00 \\ 
  0.15 &   0.00 &   0.00 &   0.00 &   0.00 &   0.00 &   0.00 &   0.00 &   0.00 &   0.00 &   0.00 \\ 
  0.20 &   0.00 &   0.00 &   0.00 &   0.00 &   0.00 &   0.00 &   0.00 &   0.00 &   0.00 &   0.00 \\ 
  0.25 &   0.00 &   0.00 &   0.00 &   0.00 &   0.00 &   0.00 &   0.00 &   0.00 &   0.00 &   0.00 \\ 
  0.30 &   0.00 &   0.00 &   0.00 &   0.00 &   0.00 &   0.00 &   0.00 &   0.00 &   0.00 &   0.00 \\ 
  0.35 &   0.00 &   0.00 &   0.00 &   0.00 &   0.00 &   0.00 &   0.00 &   0.00 &   0.00 &   0.00 \\ 
  0.40 &   0.00 &   0.00 &   0.00 &   0.00 &   0.00 &   0.00 &   0.00 &   0.00 &   0.00 &   0.00 \\ 
  0.45 &   0.00 &   0.00 &   0.00 &   0.00 &   0.00 &   0.00 &   0.00 &   0.00 &   0.00 &   0.00 \\ 
  0.50 &   0.00 &   0.00 &   0.00 &   0.00 &   0.00 &   0.00 &   0.00 &   0.00 &   0.00 &   0.00 \\ 
  0.55 &   0.00 &   0.00 &   0.00 &   0.00 &   0.00 &   0.00 &   0.00 &   0.00 &   0.00 &   0.00 \\ 
  0.60 &   0.00 &   0.00 &   0.00 &   0.00 &   0.00 &   0.00 &   0.00 &   0.00 &   0.00 &   0.00 \\ 
  0.65 &   0.00 &   0.00 &   0.00 &   0.00 &   0.00 &   0.00 &   0.00 &   0.00 &   0.00 &   0.00 \\ 
  0.70 &   0.00 &   0.00 &   0.00 &   0.00 &   0.00 &   0.00 &   0.00 &   0.00 &   0.00 &   0.00 \\ 
  0.75 &   0.00 &   0.00 &   0.00 &   0.00 &   0.00 &   0.00 &   0.00 &   0.00 &   0.00 &   0.00 \\ 
  0.80 &   0.00 &   0.00 &   0.00 &   0.00 &   0.00 &   0.00 &   0.00 &   0.00 &   0.00 &   0.00 \\ 
  0.85 &   0.00 &   0.00 &   0.00 &   0.00 &   0.00 &   0.00 &   0.00 &   0.00 &   0.00 &   0.00 \\ 
  0.90 &   0.00 &   0.00 &   0.00 &   0.00 &   0.00 &   0.00 &   0.00 &   0.00 &   0.00 &   0.00 \\ 
  0.95 &   0.00 &   0.00 &   0.00 &   0.00 &   0.00 &   0.00 &   0.00 &   0.00 &   0.00 &   0.00 \\ 
  1.00 &   0.00 &   0.00 &   0.00 &   0.00 &   0.00 &   0.00 &   0.00 &   0.00 &   0.00 &   0.00 \\ \hline
  \end{tabular}
\label{table:error_N_failed_links}
\end{table*}

\newpage
  \setcounter{equation}{16}
\begin{figure*}[!ht]
\centering
\subfigure[Erdrey]{\includegraphics[width=.24\textwidth]{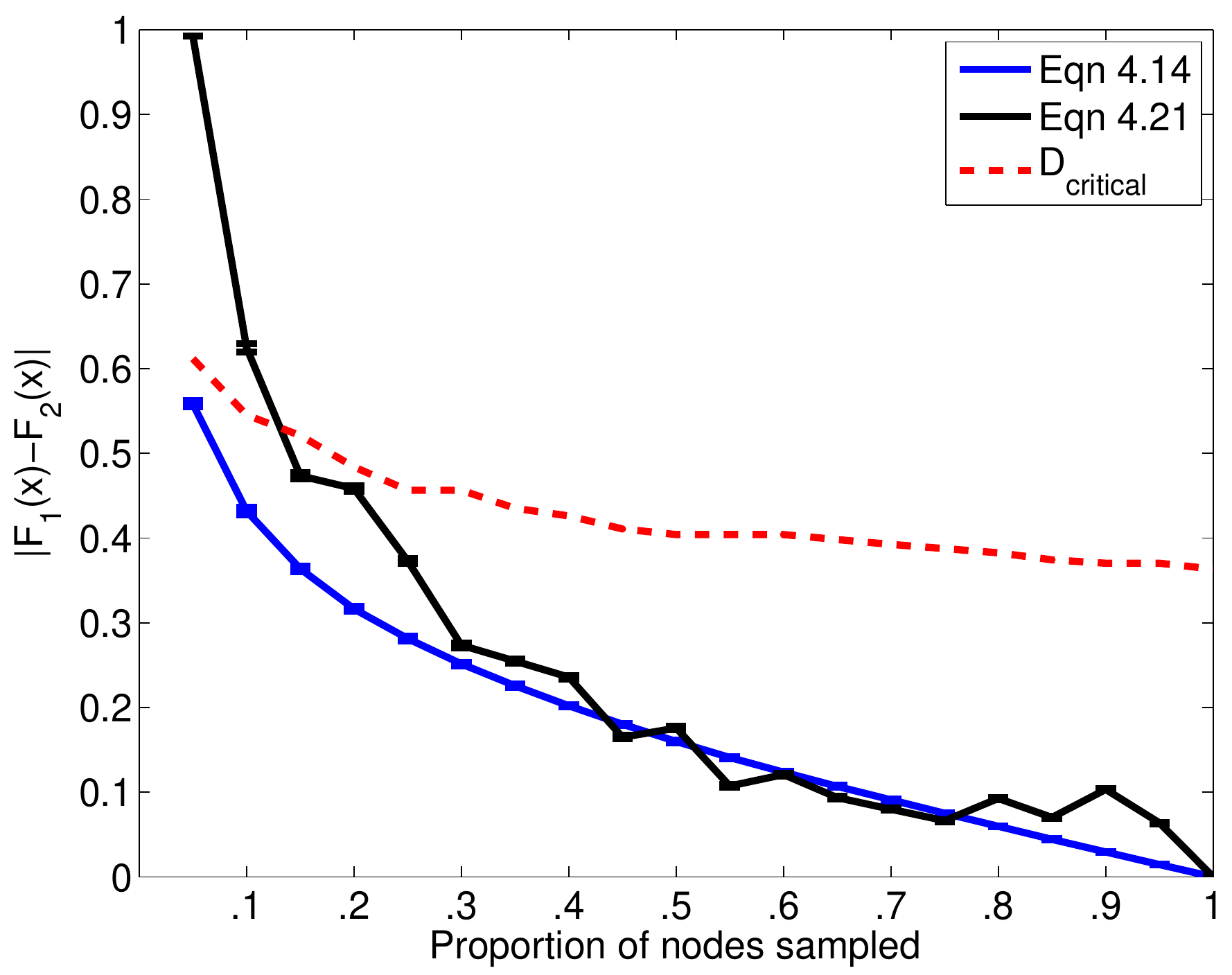}}
\subfigure[Pref]{\includegraphics[width=.24\textwidth]{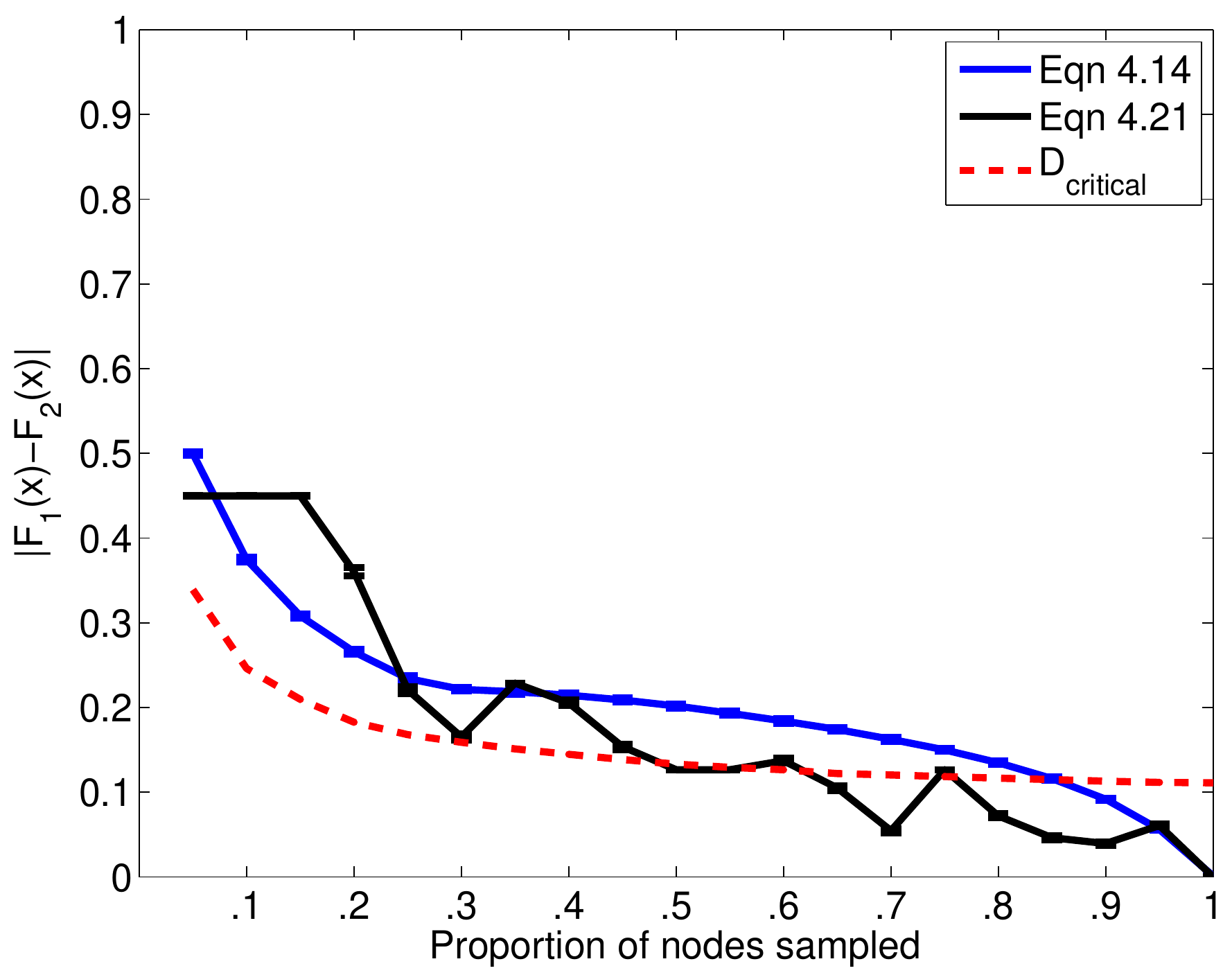}}
\subfigure[Smallworld]{\includegraphics[width=.24\textwidth]{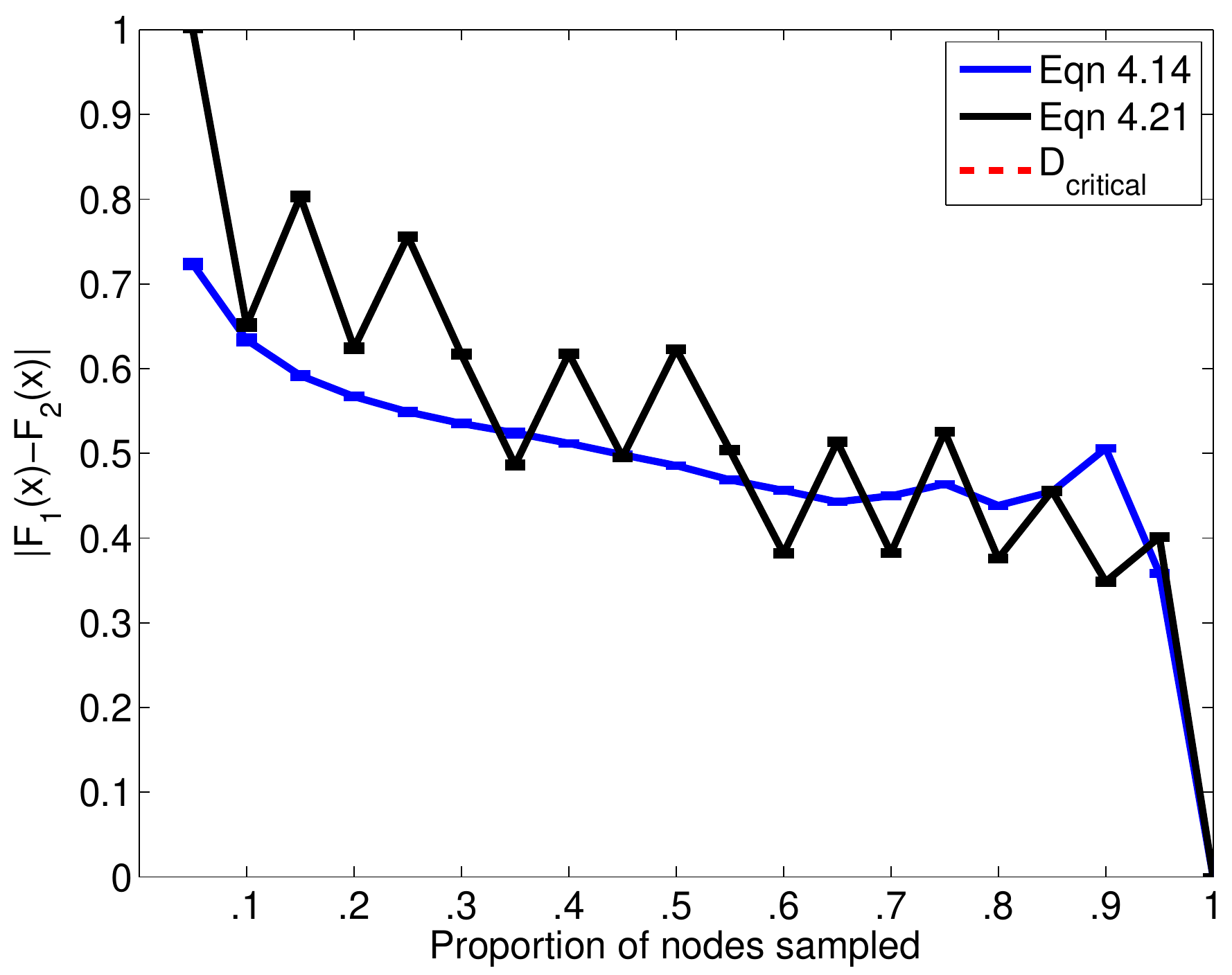}}
\subfigure[Renga]{\includegraphics[width=.24\textwidth]{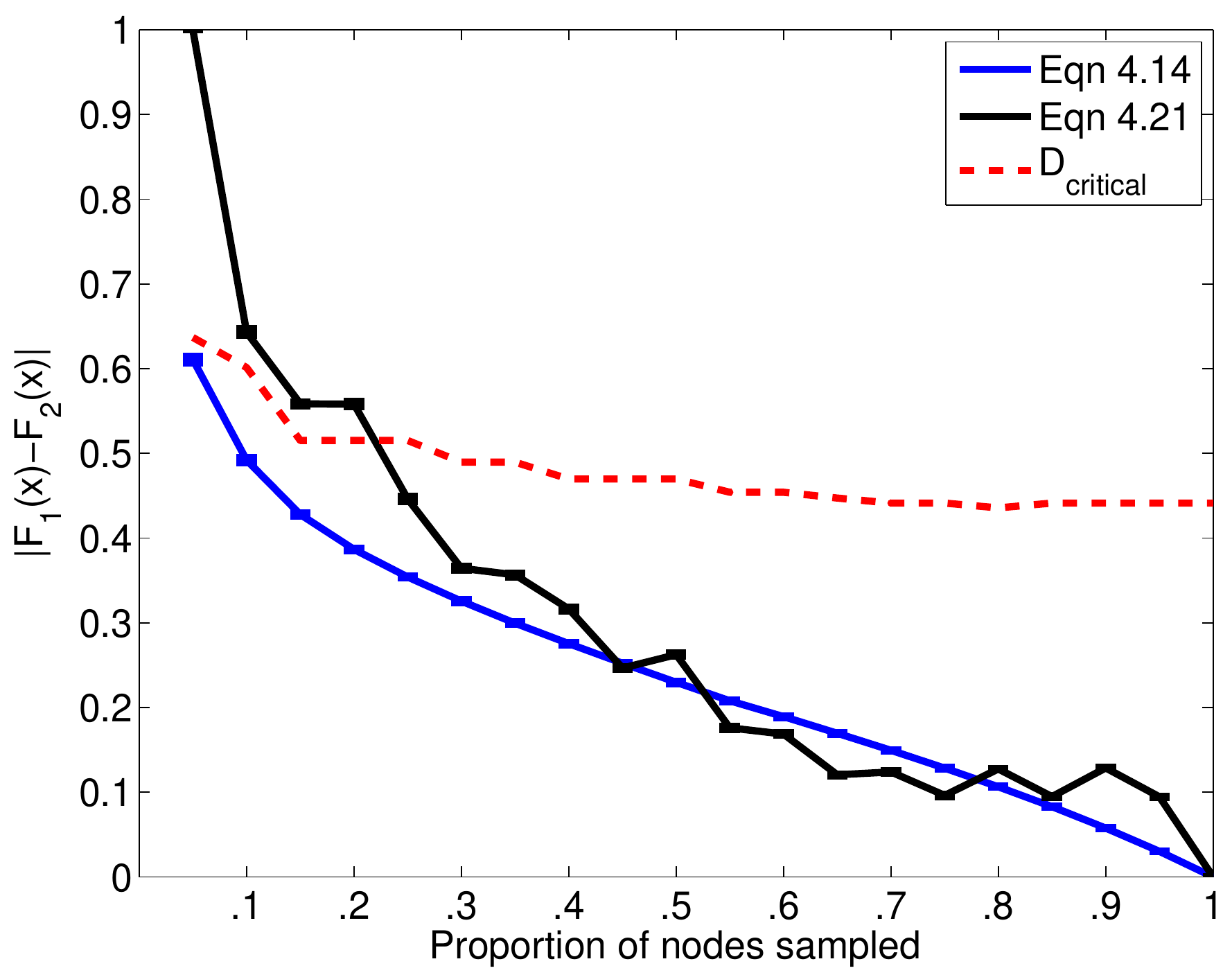}}\\
\subfigure[C. elgegans]{\includegraphics[width=.24\textwidth]{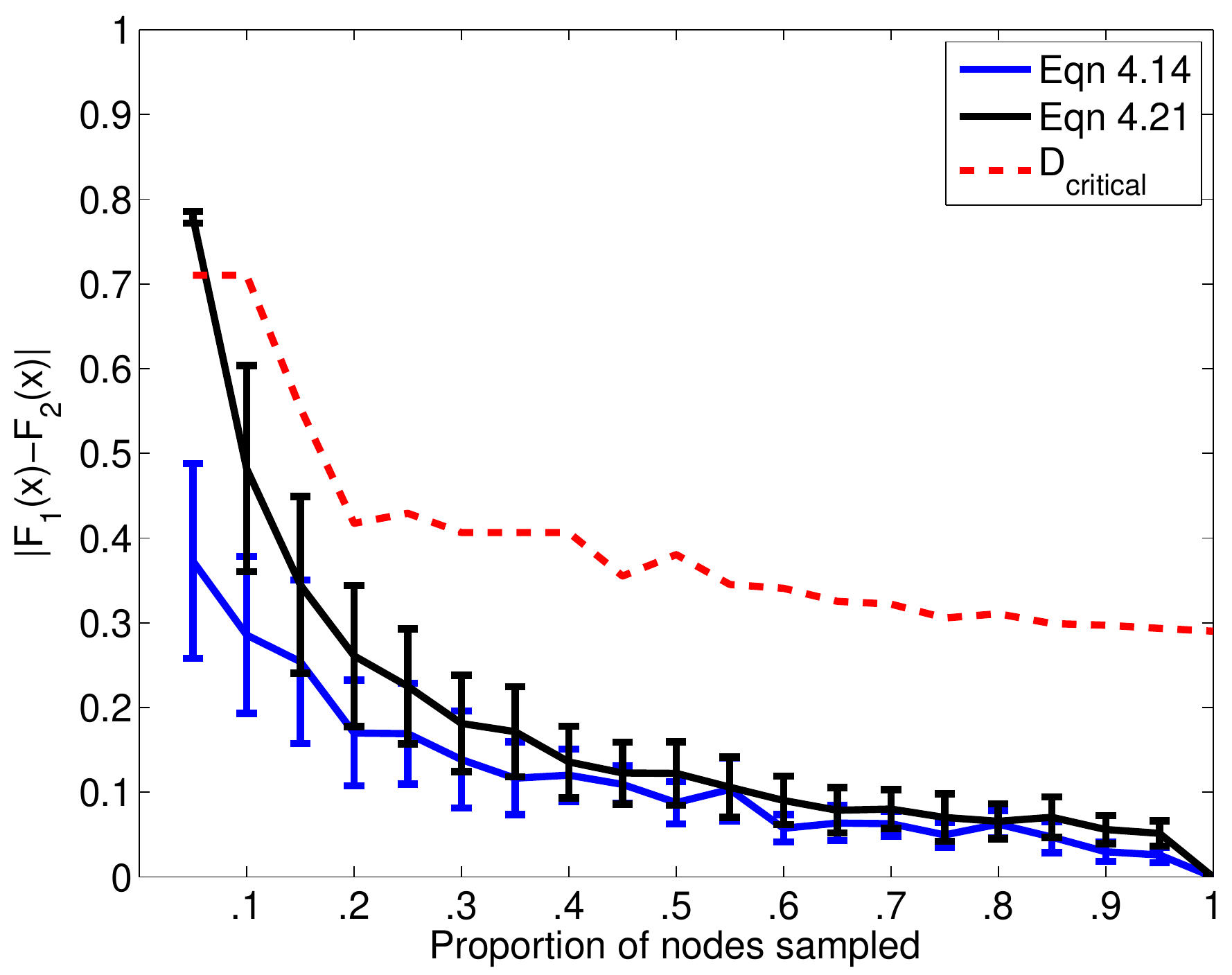}}
\subfigure[Airlines]{\includegraphics[width=.24\textwidth]{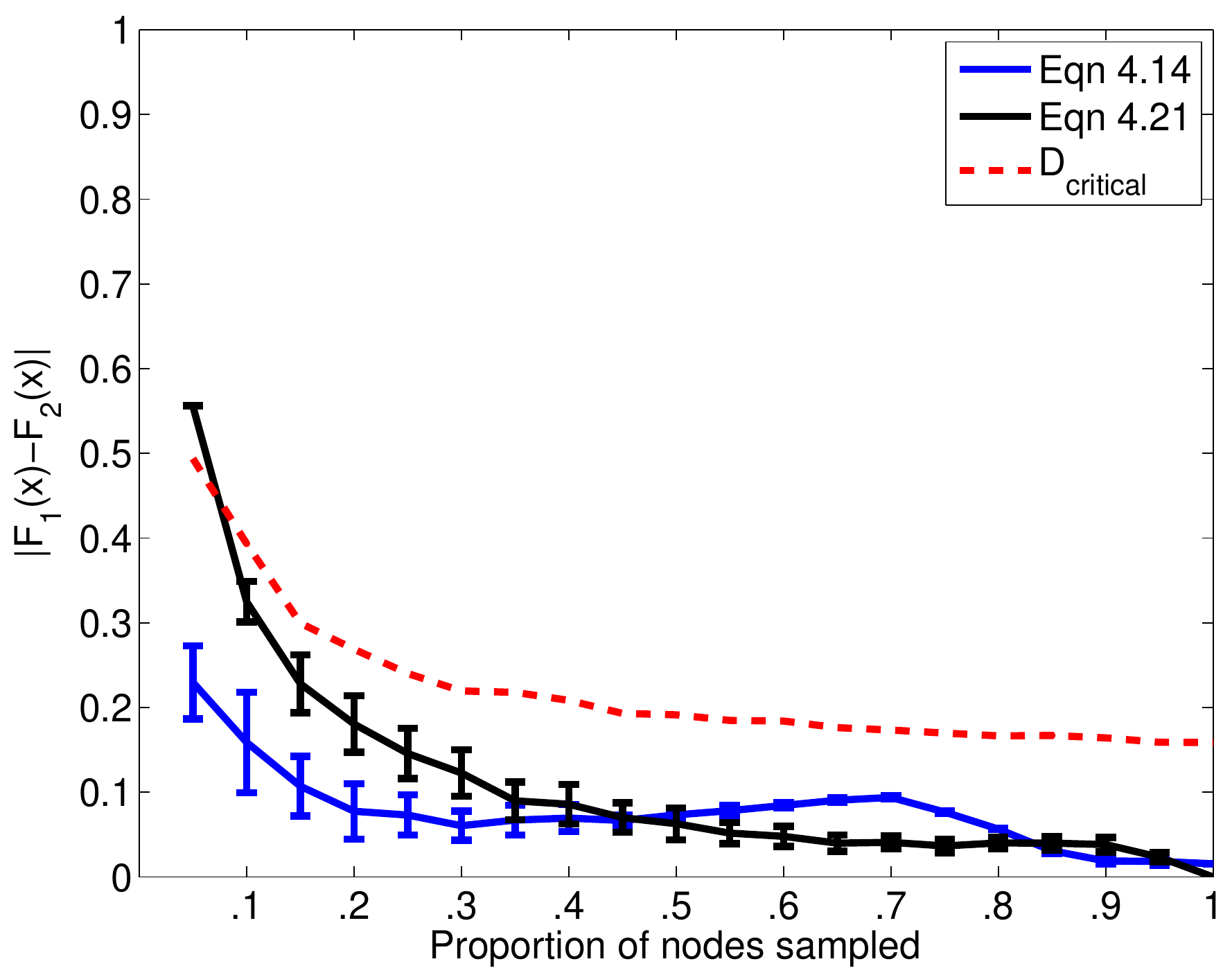}}
\subfigure[Karate]{\includegraphics[width=.24\textwidth]{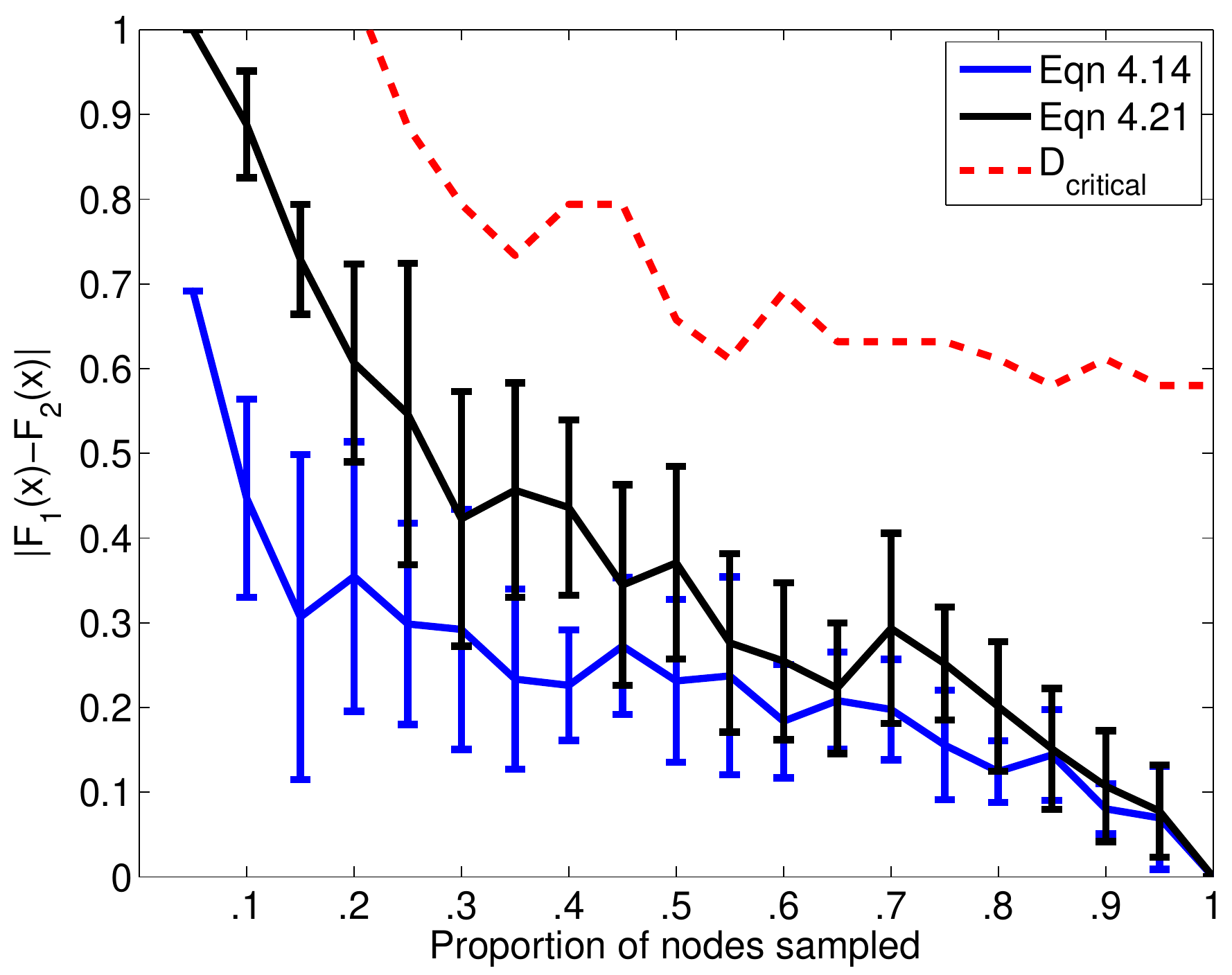}}
\subfigure[Dolphins]{\includegraphics[width=.24\textwidth]{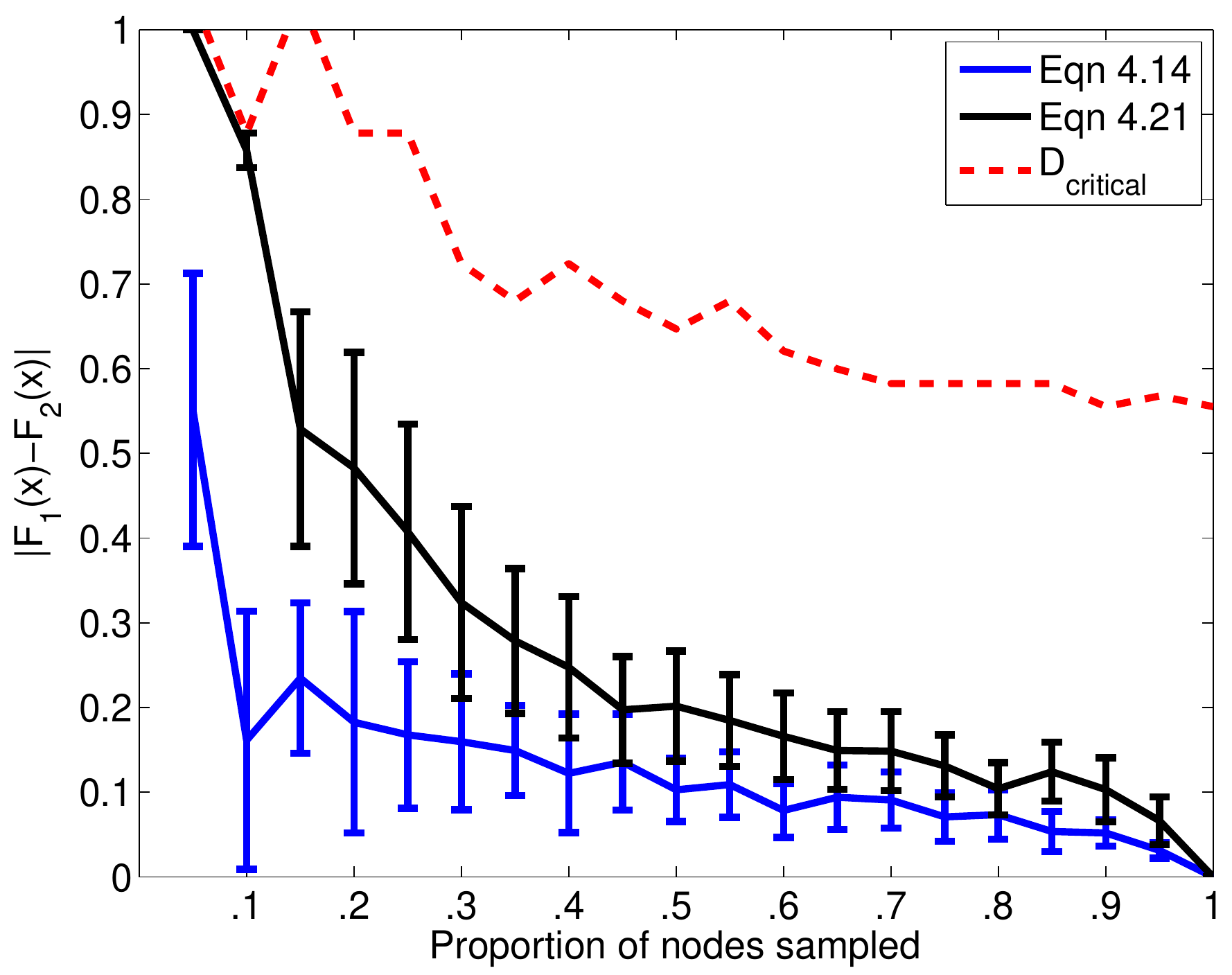}}\\
\subfigure[Condmat]{\includegraphics[width=.24\textwidth]{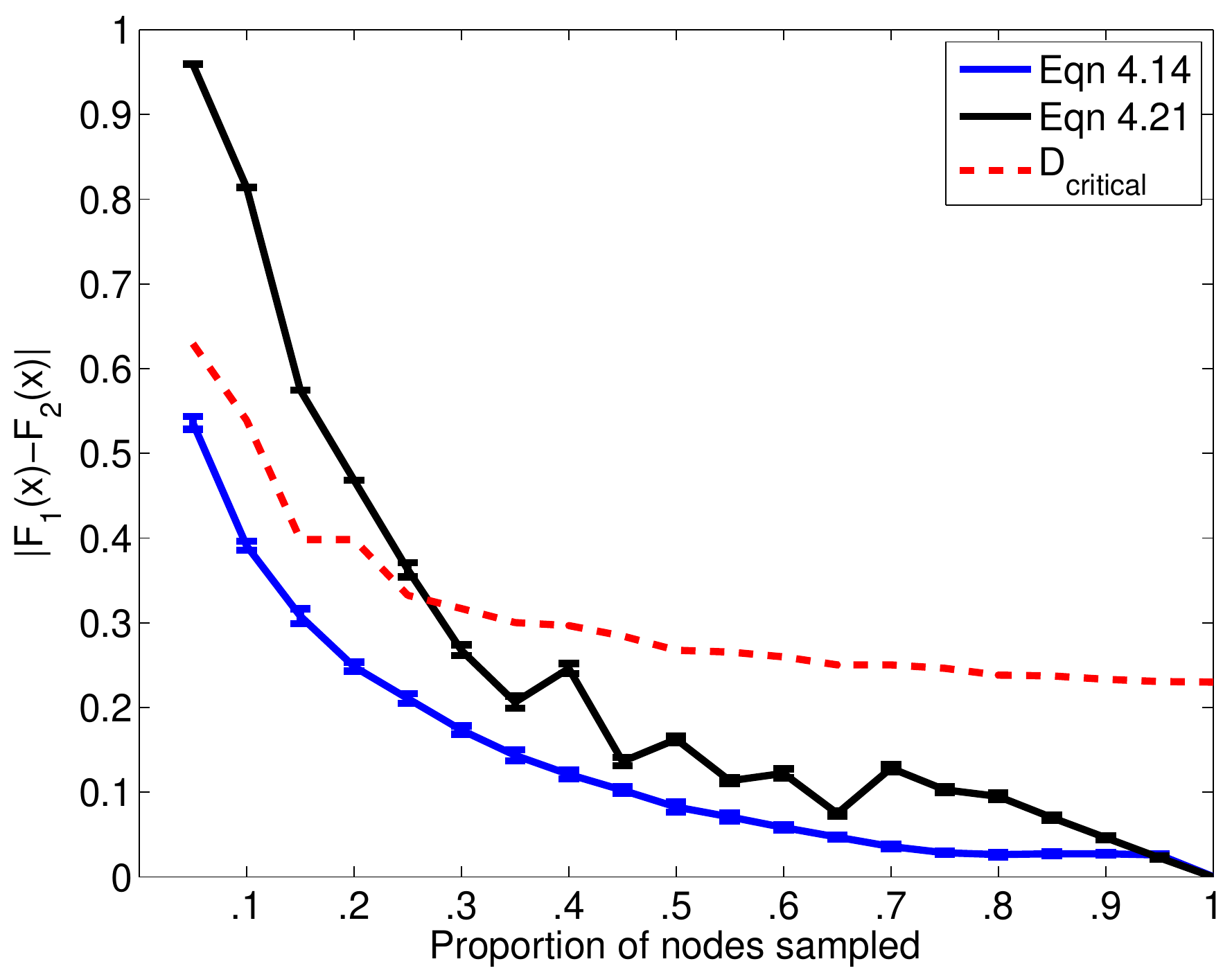}}
\subfigure[Powergrid]{\includegraphics[width=.24\textwidth]{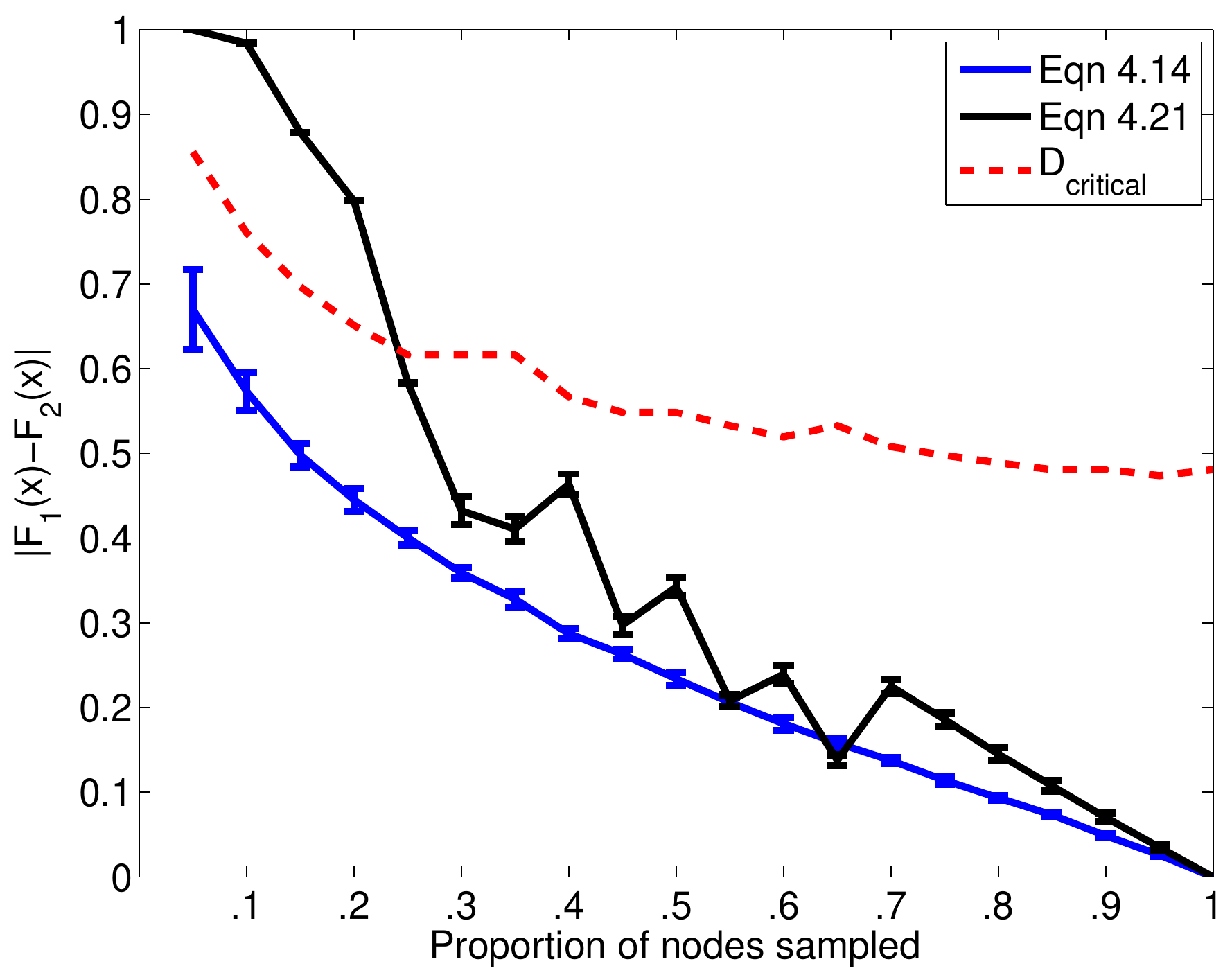}}
\caption[Kolmogorov-Smirnov two sample test for true CDF and predicted CDF from subnetworks induced on sampled nodes]{Kolmogorov-Smirnov two sample test for true CDF and predicted CDF from subnetworks induced on sampled nodes. The red line represents $D_{\text{crit}}$ for $\alpha=0.05$ and sample sizes $n_1=\kmax$ of the true CDF and $n_2=\kmax$ of the observed CDF. The predicted CDFs for for most networks are statistically indistinguishable from the true CDF for these networks for $q>0.3$. Due to the presence of large hubs in Pref, $n_1$ and $n_2$ are quite large leading to  high statistical power in the KS test. Thus, even very small differences between the true and predicted CDFs result in a statistically significant difference and rejection of the null hypothesis, even though the curves show relatively good agreement.}
\label{fig:one_over_q_nodes_D_stat_plot}
\end{figure*}

\FloatBarrier
\newpage
    \setcounter{equation}{7}
\begin{table*}[!htp] \small
\caption[Error in $\hat{M}$ when sampling by failing links]{Error in $\hat{M}$ when sampling by failing links. Since we are sampling $qM$ links, errors in predicting the true number of links are quite small and nonzero only due to roundoff error (e.g., $m=$round($qM$)).}
\centering
\begin{tabular}{|l | l l l l l l l l l l|}
\hline

$q$ & Erdrey & Pref & Smallw & Renga & C. elegans & Airlines & Karate & Dolphins & Condmat & Power \\ \hline
  0.05 &   0.00 &   0.00 &   0.00 &   0.00 &   0.00 &   0.00 &   0.03 &   0.01 &   0.00 &   0.00 \\ 
  0.10 &   0.00 &   0.00 &   0.00 &   0.00 &   0.00 &   0.00 &   0.03 &   0.01 &   0.00 &   0.00 \\ 
  0.15 &   0.00 &   0.00 &   0.00 &   0.00 &   0.00 &   0.00 &   0.03 &   0.01 &   0.00 &   0.00 \\ 
  0.20 &   0.00 &   0.00 &   0.00 &   0.00 &   0.00 &   0.00 &   0.03 &   0.01 &   0.00 &   0.00 \\ 
  0.25 &   0.00 &   0.00 &   0.00 &   0.00 &   0.00 &   0.00 &   0.03 &   0.01 &   0.00 &   0.00 \\ 
  0.30 &   0.00 &   0.00 &   0.00 &   0.00 &   0.00 &   0.00 &   0.02 &   0.01 &   0.00 &   0.00 \\ 
  0.35 &   0.00 &   0.00 &   0.00 &   0.00 &   0.00 &   0.00 &   0.01 &   0.01 &   0.00 &   0.00 \\ 
  0.40 &   0.00 &   0.00 &   0.00 &   0.00 &   0.00 &   0.00 &   0.01 &   0.01 &   0.00 &   0.00 \\ 
  0.45 &   0.00 &   0.00 &   0.00 &   0.00 &   0.00 &   0.00 &   0.00 &   0.01 &   0.00 &   0.00 \\ 
  0.50 &   0.00 &   0.00 &   0.00 &   0.00 &   0.00 &   0.00 &   0.00 &   0.01 &   0.00 &   0.00 \\ 
  0.55 &   0.00 &   0.00 &   0.00 &   0.00 &   0.00 &   0.00 &   0.00 &   0.01 &   0.00 &   0.00 \\ 
  0.60 &   0.00 &   0.00 &   0.00 &   0.00 &   0.00 &   0.00 &   0.00 &   0.00 &   0.00 &   0.00 \\ 
  0.65 &   0.00 &   0.00 &   0.00 &   0.00 &   0.00 &   0.00 &   0.01 &   0.00 &   0.00 &   0.00 \\ 
  0.70 &   0.00 &   0.00 &   0.00 &   0.00 &   0.00 &   0.00 &   0.01 &   0.00 &   0.00 &   0.00 \\ 
  0.75 &   0.00 &   0.00 &   0.00 &   0.00 &   0.00 &   0.00 &   0.01 &   0.00 &   0.00 &   0.00 \\ 
  0.80 &   0.00 &   0.00 &   0.00 &   0.00 &   0.00 &   0.00 &   0.01 &   0.00 &   0.00 &   0.00 \\ 
  0.85 &   0.00 &   0.00 &   0.00 &   0.00 &   0.00 &   0.00 &   0.00 &   0.00 &   0.00 &   0.00 \\ 
  0.90 &   0.00 &   0.00 &   0.00 &   0.00 &   0.00 &   0.00 &   0.00 &   0.00 &   0.00 &   0.00 \\ 
  0.95 &   0.00 &   0.00 &   0.00 &   0.00 &   0.00 &   0.00 &   0.00 &   0.00 &   0.00 &   0.00 \\ 
  1.00 &   0.00 &   0.00 &   0.00 &   0.00 &   0.00 &   0.00 &   0.00 &   0.00 &   0.00 &   0.00 \\ \hline
    \end{tabular}
\label{table:error_M_failed_links}
\end{table*}

\newpage
    \setcounter{equation}{8}
\begin{table*}[!htp] \small
\caption[Error in $\hat{k}_{\rm avg}$ when sampling by failing links]{Error in $\hat{k}_{\rm avg}$ when sampling by failing links. The predicted average degree is computed from $\hat{N}$ and $\hat{M}$. Error in the predicted average agree are small and only occur due to rounding errors in the selecting an integer number of $qM$ edges in the random sample.}
\centering
\begin{tabular}{|l | l l l l l l l l l l|}
\hline
$q$ & Erdrey & Pref & Smallw & Renga & C. elegans & Airlines & Karate & Dolphins & Condmat & Power \\ \hline
  0.05 &   0.00 &   0.00 &   0.00 &   0.00 &   0.03 &   0.03 &   0.06 &   0.09 &   0.00 &   0.00 \\ 
  0.10 &   0.00 &   0.00 &   0.00 &   0.00 &   0.02 &   0.02 &   0.03 &   0.05 &   0.00 &   0.00 \\ 
  0.15 &   0.00 &   0.00 &   0.00 &   0.00 &   0.02 &   0.01 &   0.03 &   0.03 &   0.00 &   0.00 \\ 
  0.20 &   0.00 &   0.00 &   0.00 &   0.00 &   0.01 &   0.01 &   0.03 &   0.02 &   0.00 &   0.00 \\ 
  0.25 &   0.00 &   0.00 &   0.00 &   0.00 &   0.01 &   0.01 &   0.03 &   0.02 &   0.00 &   0.00 \\ 
  0.30 &   0.00 &   0.00 &   0.00 &   0.00 &   0.01 &   0.01 &   0.02 &   0.02 &   0.00 &   0.00 \\ 
  0.35 &   0.00 &   0.00 &   0.00 &   0.00 &   0.01 &   0.00 &   0.01 &   0.01 &   0.00 &   0.00 \\ 
  0.40 &   0.00 &   0.00 &   0.00 &   0.00 &   0.00 &   0.00 &   0.01 &   0.01 &   0.00 &   0.00 \\ 
  0.45 &   0.00 &   0.00 &   0.00 &   0.00 &   0.00 &   0.00 &   0.00 &   0.01 &   0.00 &   0.00 \\ 
  0.50 &   0.00 &   0.00 &   0.00 &   0.00 &   0.00 &   0.00 &   0.00 &   0.01 &   0.00 &   0.00 \\ 
  0.55 &   0.00 &   0.00 &   0.00 &   0.00 &   0.00 &   0.00 &   0.00 &   0.00 &   0.00 &   0.00 \\ 
  0.60 &   0.00 &   0.00 &   0.00 &   0.00 &   0.00 &   0.00 &   0.00 &   0.00 &   0.00 &   0.00 \\ 
  0.65 &   0.00 &   0.00 &   0.00 &   0.00 &   0.00 &   0.00 &   0.01 &   0.00 &   0.00 &   0.00 \\ 
  0.70 &   0.00 &   0.00 &   0.00 &   0.00 &   0.00 &   0.00 &   0.01 &   0.00 &   0.00 &   0.00 \\ 
  0.75 &   0.00 &   0.00 &   0.00 &   0.00 &   0.00 &   0.00 &   0.01 &   0.00 &   0.00 &   0.00 \\ 
  0.80 &   0.00 &   0.00 &   0.00 &   0.00 &   0.00 &   0.00 &   0.01 &   0.00 &   0.00 &   0.00 \\ 
  0.85 &   0.00 &   0.00 &   0.00 &   0.00 &   0.00 &   0.00 &   0.00 &   0.00 &   0.00 &   0.00 \\ 
  0.90 &   0.00 &   0.00 &   0.00 &   0.00 &   0.00 &   0.00 &   0.00 &   0.00 &   0.00 &   0.00 \\ 
  0.95 &   0.00 &   0.00 &   0.00 &   0.00 &   0.00 &   0.00 &   0.00 &   0.00 &   0.00 &   0.00 \\ 
  1.00 &   0.00 &   0.00 &   0.00 &   0.00 &   0.00 &   0.00 &   0.00 &   0.00 &   0.00 &   0.00 \\ \hline
    \end{tabular}
\label{table:error_avk_failed_links}
\end{table*}

\newpage
    \setcounter{equation}{9}
\begin{table*}[!htp]\small
\caption[Error in $\hat{C}$ when sampling by failing links]{Error in $\hat{C}$ when sampling by failing links.}
\centering
\begin{tabular}{|l | l l l l l l l l l l|}
\hline
$q$ & Erdrey & Pref & Smallw & Renga & C. elegans & Airlines & Karate & Dolphins & Condmat & Power \\ \hline
  0.05 &   0.50 &   0.11 &   0.00 &   0.01 &   0.13 &   0.18 &    -- &    -- &   0.06 &   0.28 \\ 
  0.10 &   0.66 &   0.05 &   0.00 &   0.00 &   0.05 &   0.03 &   0.36 &   0.04 &   0.02 &   0.18 \\ 
  0.15 &   0.18 &   0.02 &   0.00 &   0.00 &   0.00 &   0.01 &   0.18 &   0.24 &   0.06 &   0.05 \\ 
  0.20 &   0.06 &   0.02 &   0.00 &   0.00 &   0.00 &   0.04 &   0.45 &   0.18 &   0.03 &   0.01 \\ 
  0.25 &   0.05 &   0.00 &   0.00 &   0.00 &   0.02 &   0.00 &   0.01 &   0.09 &   0.10 &   0.04 \\ 
  0.30 &   0.03 &   0.01 &   0.00 &   0.00 &   0.01 &   0.00 &   0.21 &   0.02 &   0.03 &   0.00 \\ 
  0.35 &   0.01 &   0.01 &   0.00 &   0.00 &   0.01 &   0.01 &   0.02 &   0.02 &   0.01 &   0.02 \\ 
  0.40 &   0.02 &   0.01 &   0.00 &   0.00 &   0.01 &   0.01 &   0.15 &   0.07 &   0.00 &   0.01 \\ 
  0.45 &   0.01 &   0.01 &   0.00 &   0.00 &   0.01 &   0.01 &   0.05 &   0.02 &   0.04 &   0.00 \\ 
  0.50 &   0.01 &   0.00 &   0.00 &   0.00 &   0.00 &   0.01 &   0.07 &   0.01 &   0.02 &   0.01 \\ 
  0.55 &   0.01 &   0.00 &   0.00 &   0.00 &   0.00 &   0.00 &   0.03 &   0.01 &   0.00 &   0.01 \\ 
  0.60 &   0.00 &   0.00 &   0.00 &   0.00 &   0.01 &   0.00 &   0.05 &   0.03 &   0.00 &   0.00 \\ 
  0.65 &   0.00 &   0.01 &   0.00 &   0.00 &   0.00 &   0.01 &   0.06 &   0.02 &   0.00 &   0.00 \\ 
  0.70 &   0.00 &   0.00 &   0.00 &   0.00 &   0.00 &   0.00 &   0.01 &   0.02 &   0.01 &   0.00 \\ 
  0.75 &   0.01 &   0.00 &   0.00 &   0.00 &   0.01 &   0.01 &   0.01 &   0.00 &   0.01 &   0.00 \\ 
  0.80 &   0.01 &   0.00 &   0.00 &   0.00 &   0.00 &   0.01 &   0.00 &   0.02 &   0.00 &   0.00 \\ 
  0.85 &   0.00 &   0.00 &   0.00 &   0.00 &   0.00 &   0.00 &   0.02 &   0.01 &   0.00 &   0.00 \\ 
  0.90 &   0.00 &   0.00 &   0.00 &   0.00 &   0.00 &   0.00 &   0.02 &   0.00 &   0.01 &   0.00 \\ 
  0.95 &   0.01 &   0.00 &   0.00 &   0.00 &   0.00 &   0.00 &   0.00 &   0.01 &   0.00 &   0.00 \\ 
  1.00 &   0.00 &   0.00 &   0.00 &   0.00 &   0.00 &   0.00 &   0.00 &   0.00 &   0.00 &   0.00 \\ \hline
    \end{tabular}
\label{table:error_cluster_failed_links}
\end{table*}
\newpage
    \setcounter{equation}{10}
\begin{table*}[!htp]\small
\caption[Error in $\hat{k}_{\max}$ when sampling by failing links]{Error in $\hat{k}_{\max}$ when sampling by failing links.}
\centering
\begin{tabular}{|l | l l l l l l l l l l|}
\hline
$q$ & Erdrey & Pref & Smallw & Renga & C. elegans & Airlines & Karate & Dolphins & Condmat & Power \\ \hline
 0.05 &   0.47 &   0.01 &   0.33 &   0.23 &   0.07 &   0.17 &   0.21 &   0.18 &   0.06 &   0.24 \\ 
  0.10 &   0.29 &   0.00 &   0.02 &   0.39 &   0.02 &   0.17 &   0.05 &   0.13 &   0.15 &   0.37 \\ 
  0.15 &   0.26 &   0.00 &   0.11 &   0.32 &   0.01 &   0.13 &   0.17 &   0.10 &   0.12 &   0.19 \\ 
  0.20 &   0.26 &   0.00 &   0.09 &   0.32 &   0.01 &   0.09 &   0.08 &   0.00 &   0.12 &   0.07 \\ 
  0.25 &   0.09 &   0.01 &   0.09 &   0.33 &   0.01 &   0.05 &   0.11 &   0.11 &   0.08 &   0.13 \\ 
  0.30 &   0.24 &   0.00 &   0.02 &   0.26 &   0.01 &   0.06 &   0.03 &   0.07 &   0.09 &   0.01 \\ 
  0.35 &   0.21 &   0.00 &   0.00 &   0.21 &   0.01 &   0.03 &   0.10 &   0.10 &   0.08 &   0.02 \\ 
  0.40 &   0.09 &   0.00 &   0.00 &   0.18 &   0.01 &   0.05 &   0.08 &   0.02 &   0.07 &   0.01 \\ 
  0.45 &   0.09 &   0.00 &   0.00 &   0.19 &   0.00 &   0.04 &   0.07 &   0.06 &   0.07 &   0.01 \\ 
  0.50 &   0.00 &   0.00 &   0.00 &   0.13 &   0.01 &   0.04 &   0.13 &   0.05 &   0.05 &   0.04 \\ 
  0.55 &   0.06 &   0.00 &   0.00 &   0.09 &   0.00 &   0.01 &   0.10 &   0.02 &   0.05 &   0.03 \\ 
  0.60 &   0.06 &   0.00 &   0.00 &   0.12 &   0.01 &   0.02 &   0.08 &   0.04 &   0.03 &   0.05 \\ 
  0.65 &   0.02 &   0.00 &   0.00 &   0.14 &   0.00 &   0.03 &   0.08 &   0.03 &   0.03 &   0.01 \\ 
  0.70 &   0.02 &   0.00 &   0.00 &   0.11 &   0.00 &   0.02 &   0.06 &   0.02 &   0.02 &   0.02 \\ 
  0.75 &   0.05 &   0.00 &   0.00 &   0.09 &   0.00 &   0.02 &   0.06 &   0.03 &   0.01 &   0.02 \\ 
  0.80 &   0.00 &   0.00 &   0.00 &   0.06 &   0.01 &   0.01 &   0.04 &   0.02 &   0.02 &   0.04 \\ 
  0.85 &   0.02 &   0.00 &   0.00 &   0.04 &   0.01 &   0.00 &   0.03 &   0.02 &   0.02 &   0.04 \\ 
  0.90 &   0.00 &   0.00 &   0.06 &   0.03 &   0.00 &   0.00 &   0.02 &   0.01 &   0.01 &   0.02 \\ 
  0.95 &   0.00 &   0.00 &   0.05 &   0.01 &   0.00 &   0.00 &   0.01 &   0.01 &   0.00 &   0.01 \\ 
  1.00 &   0.00 &   0.00 &   0.00 &   0.00 &   0.00 &   0.00 &   0.00 &   0.09 &   0.00 &   0.00 \\  \hline
    \end{tabular}
\label{table:error_kmax_failed_links}
\end{table*}

\newpage

 \setcounter{equation}{17}
\begin{figure*}[!hT]
\centering
\subfigure[Erdrey]{\includegraphics[width=.24\textwidth]{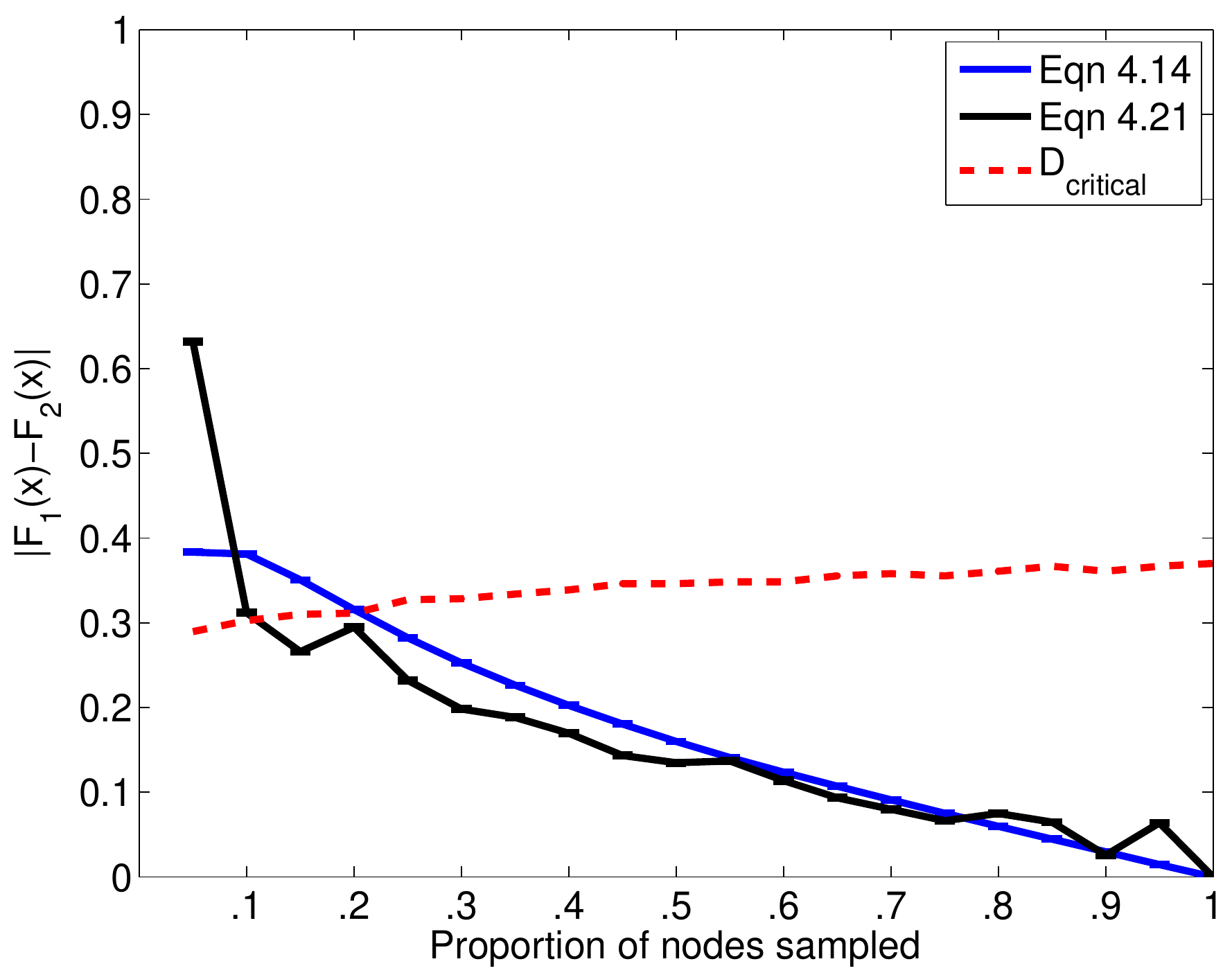}}
\subfigure[Pref]{\includegraphics[width=.24\textwidth]{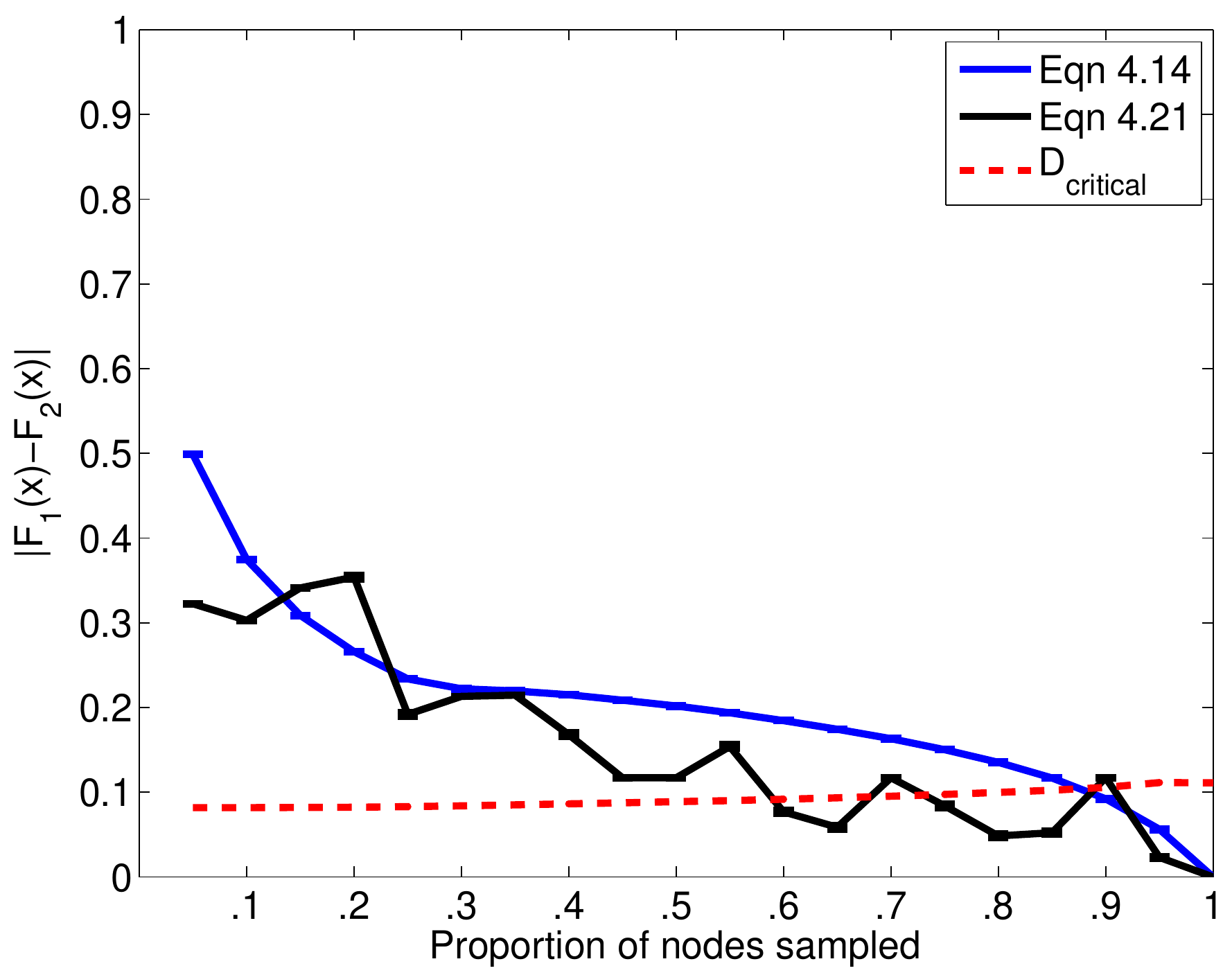}}
\subfigure[Smallworld]{\includegraphics[width=.24\textwidth]{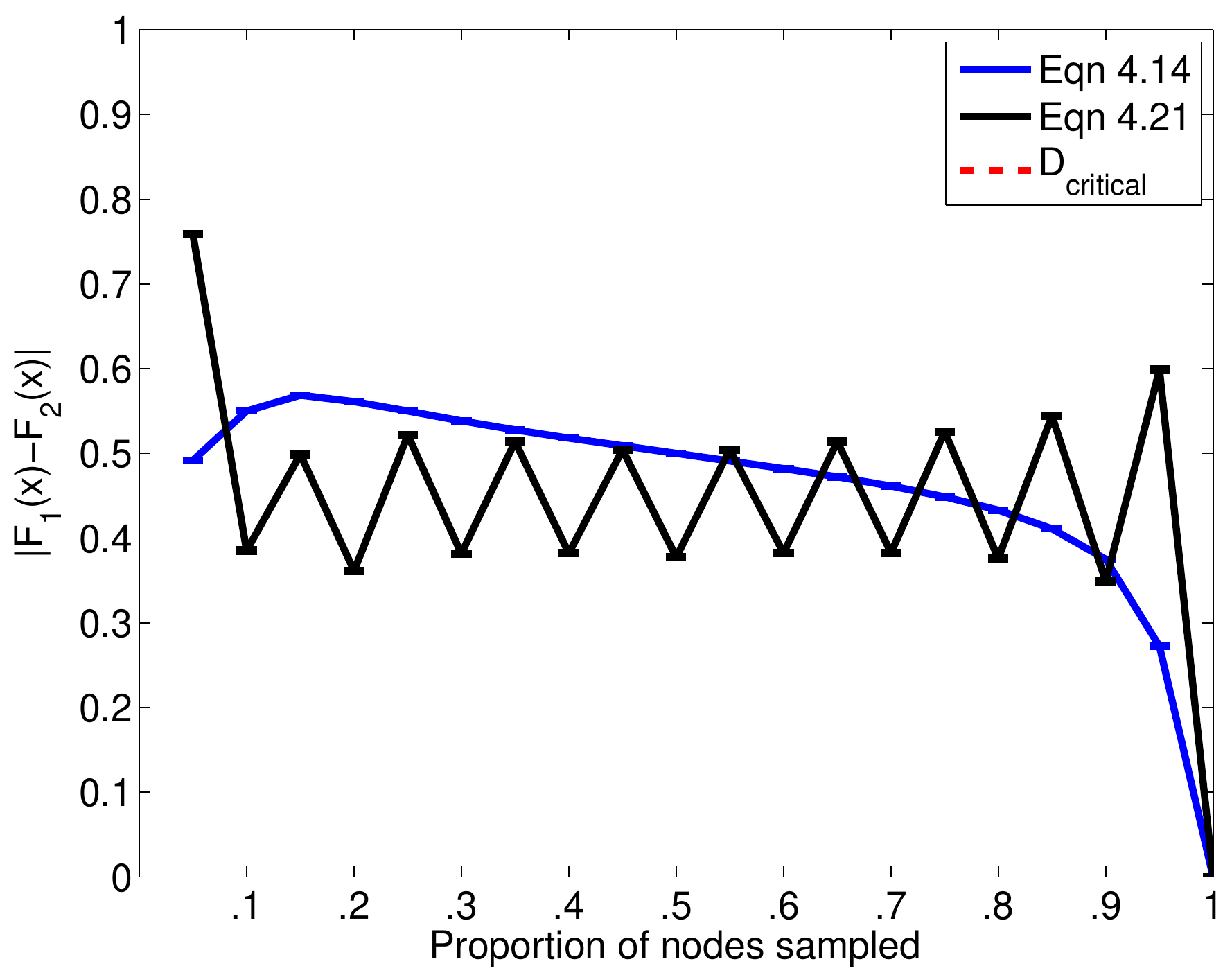}}
\subfigure[Renga]{\includegraphics[width=.24\textwidth]{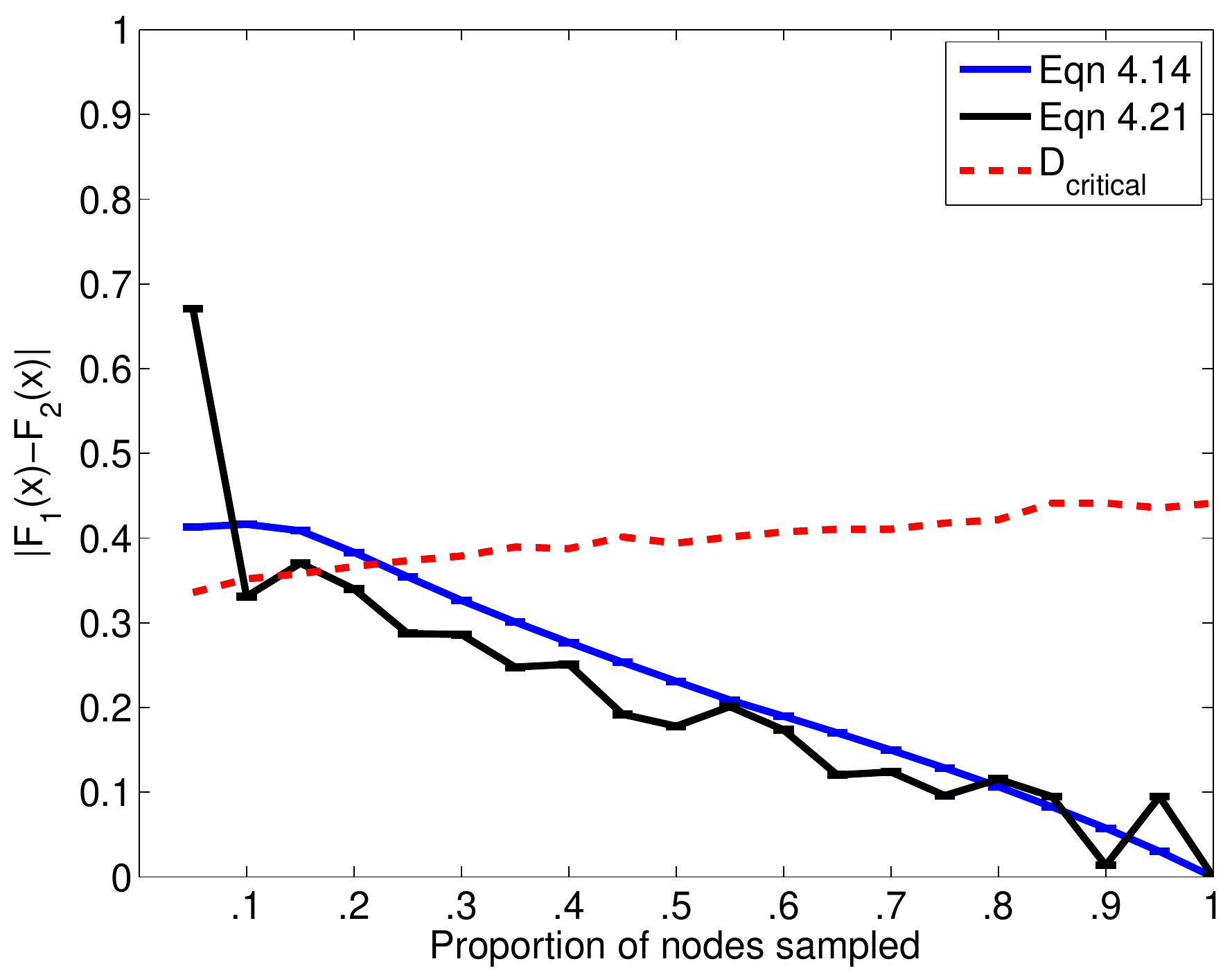}}\\
\subfigure[C. elgegans]{\includegraphics[width=.24\textwidth]{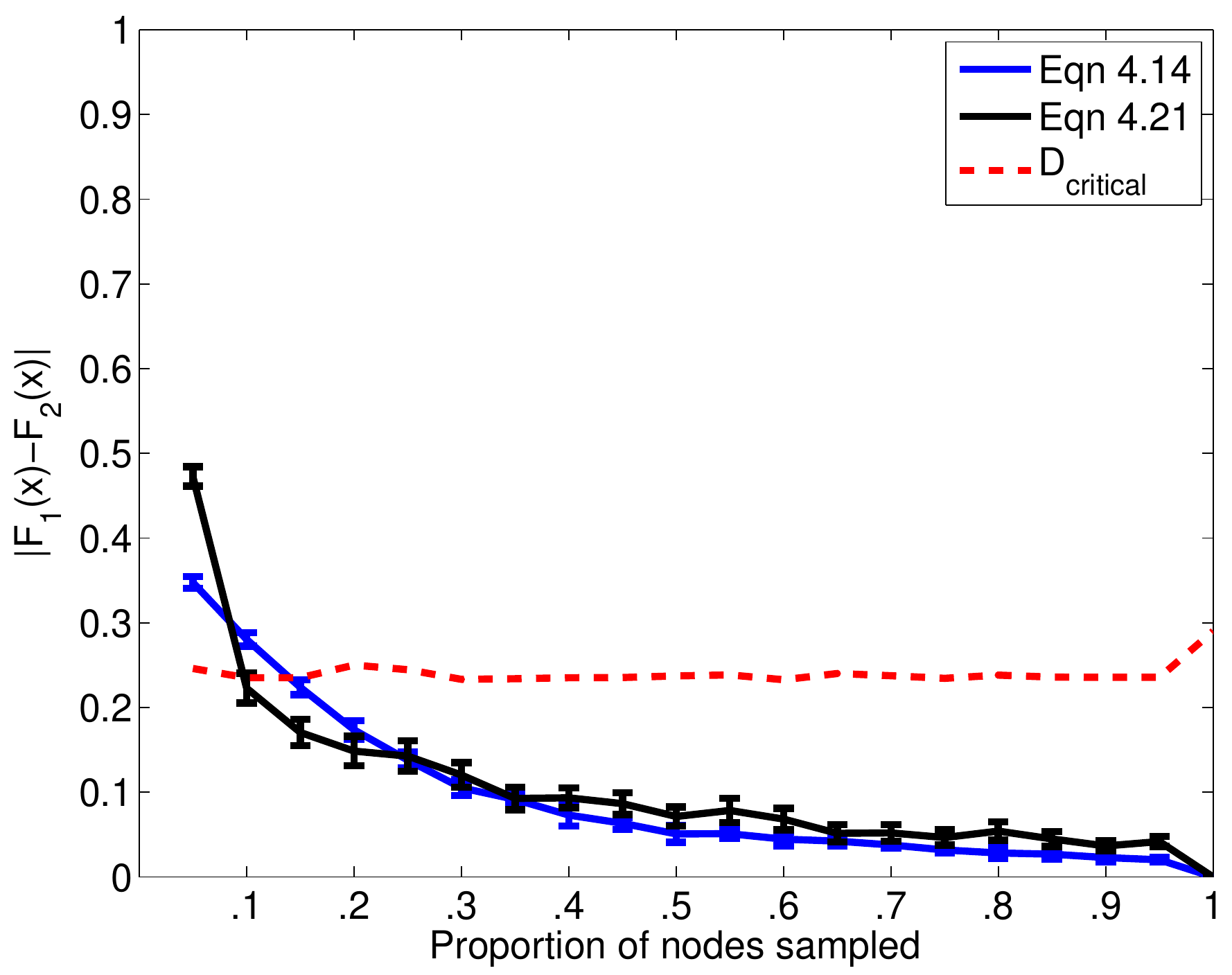}}
\subfigure[Airlines]{\includegraphics[width=.24\textwidth]{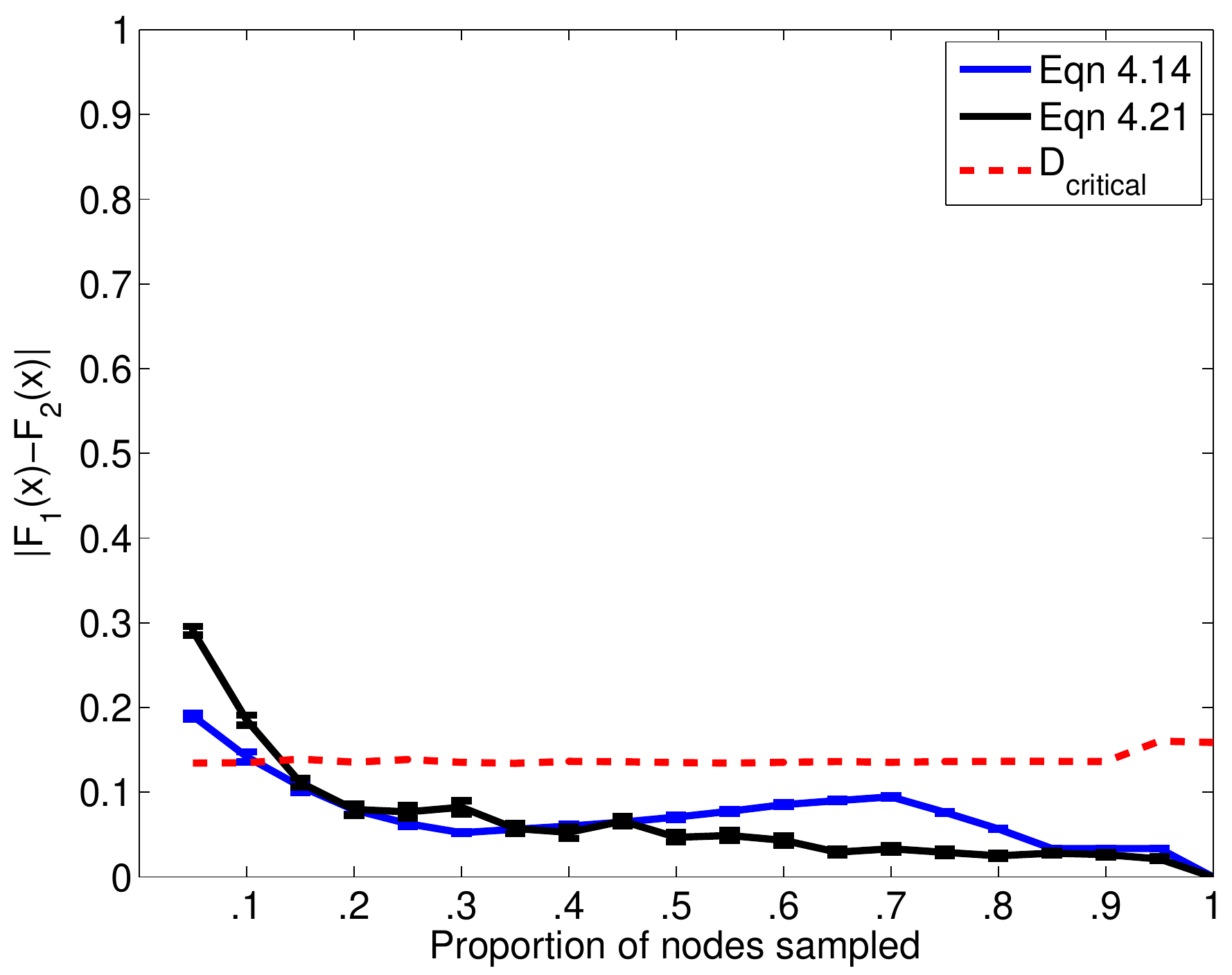}}
\subfigure[Karate]{\includegraphics[width=.24\textwidth]{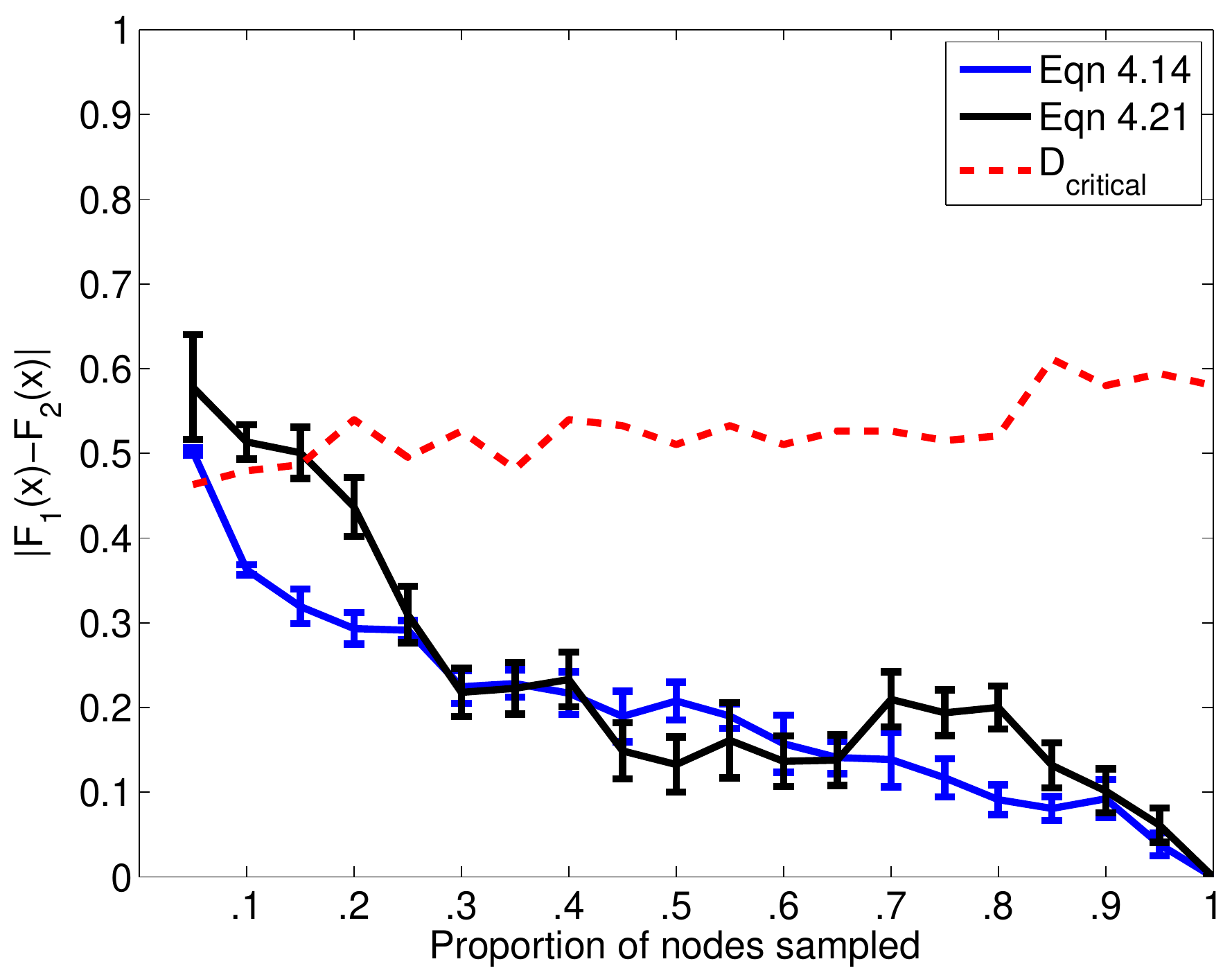}}
\subfigure[Dolphins]{\includegraphics[width=.24\textwidth]{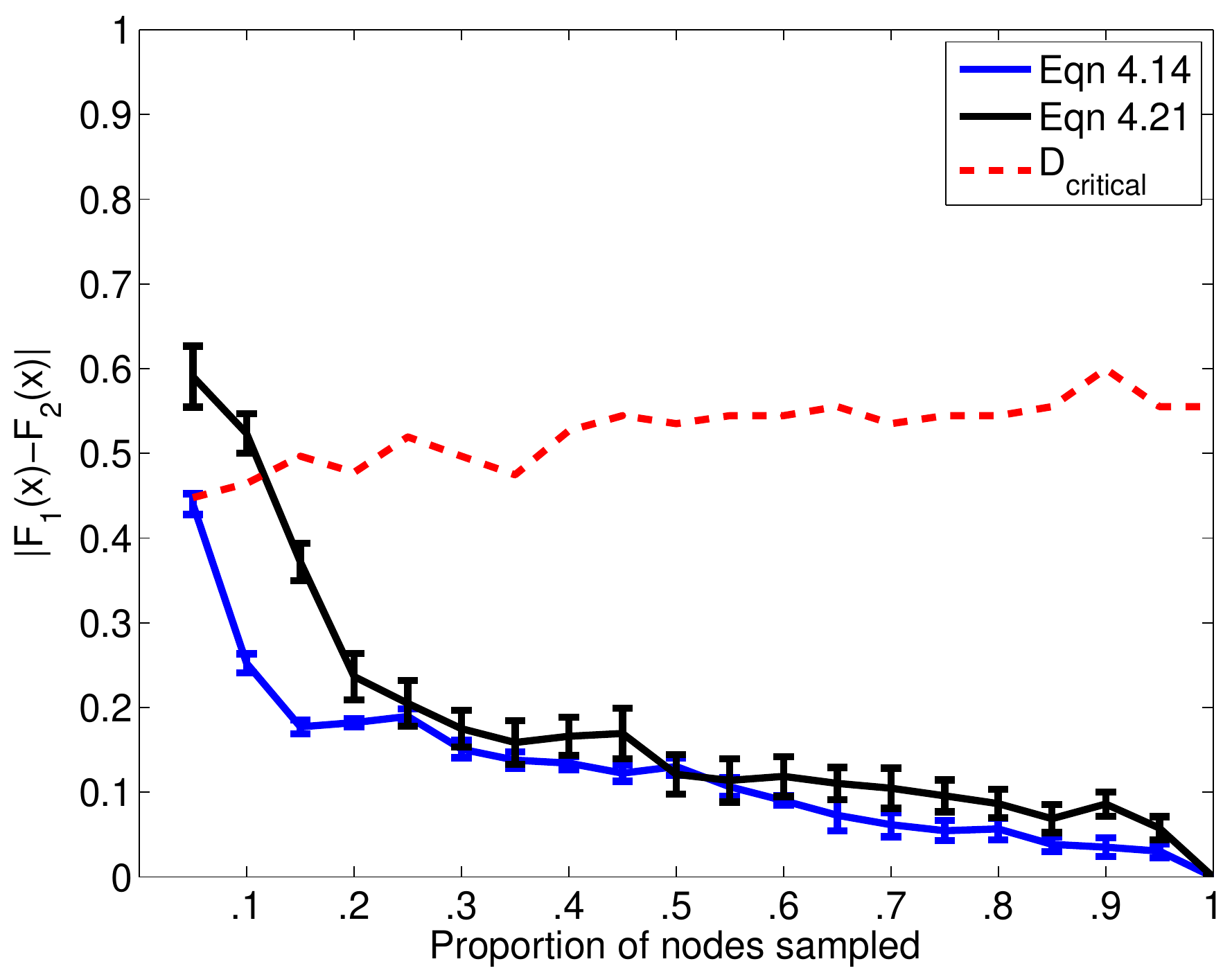}}\\
\subfigure[Condmat]{\includegraphics[width=.24\textwidth]{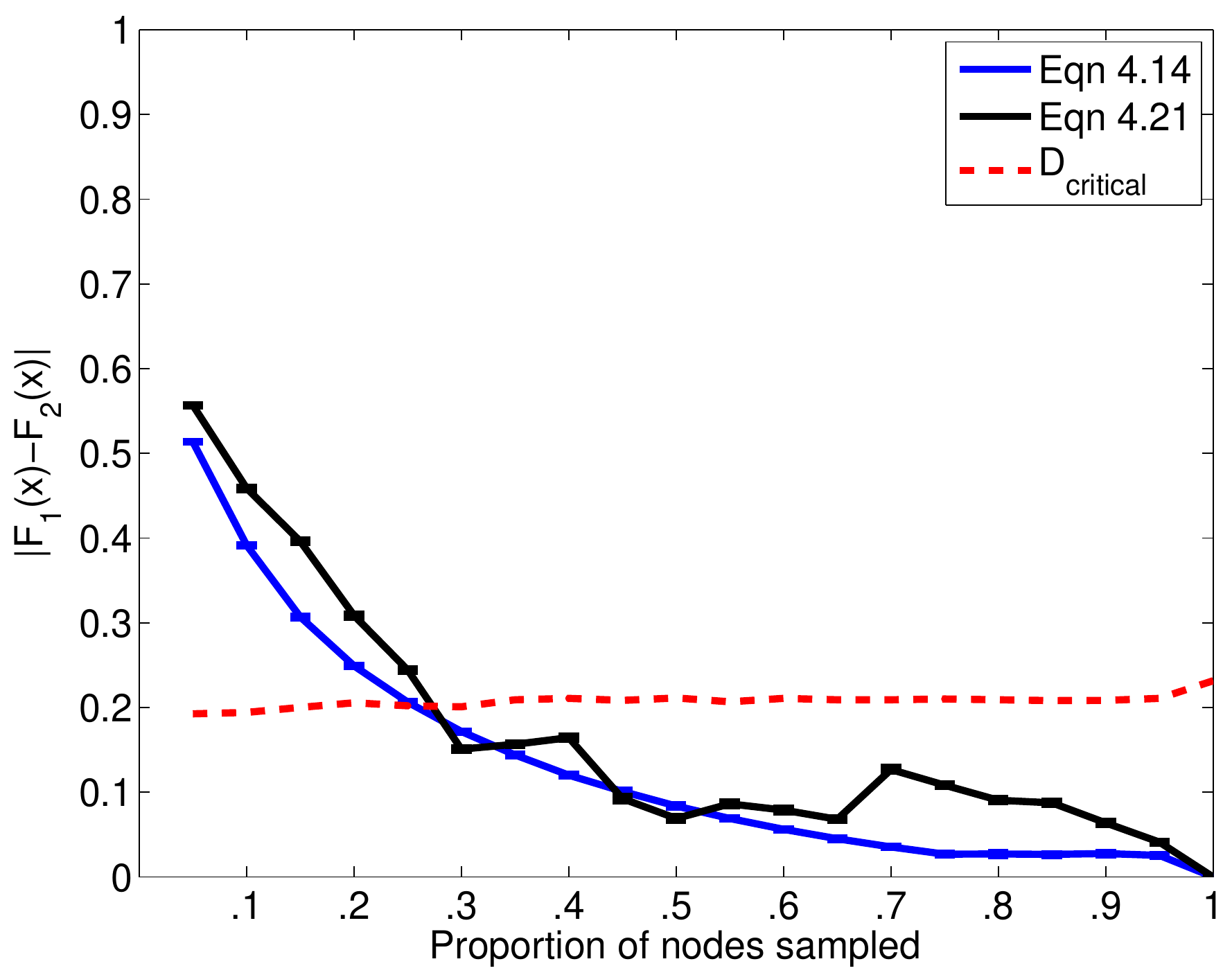}}
\subfigure[Powergrid]{\includegraphics[width=.24\textwidth]{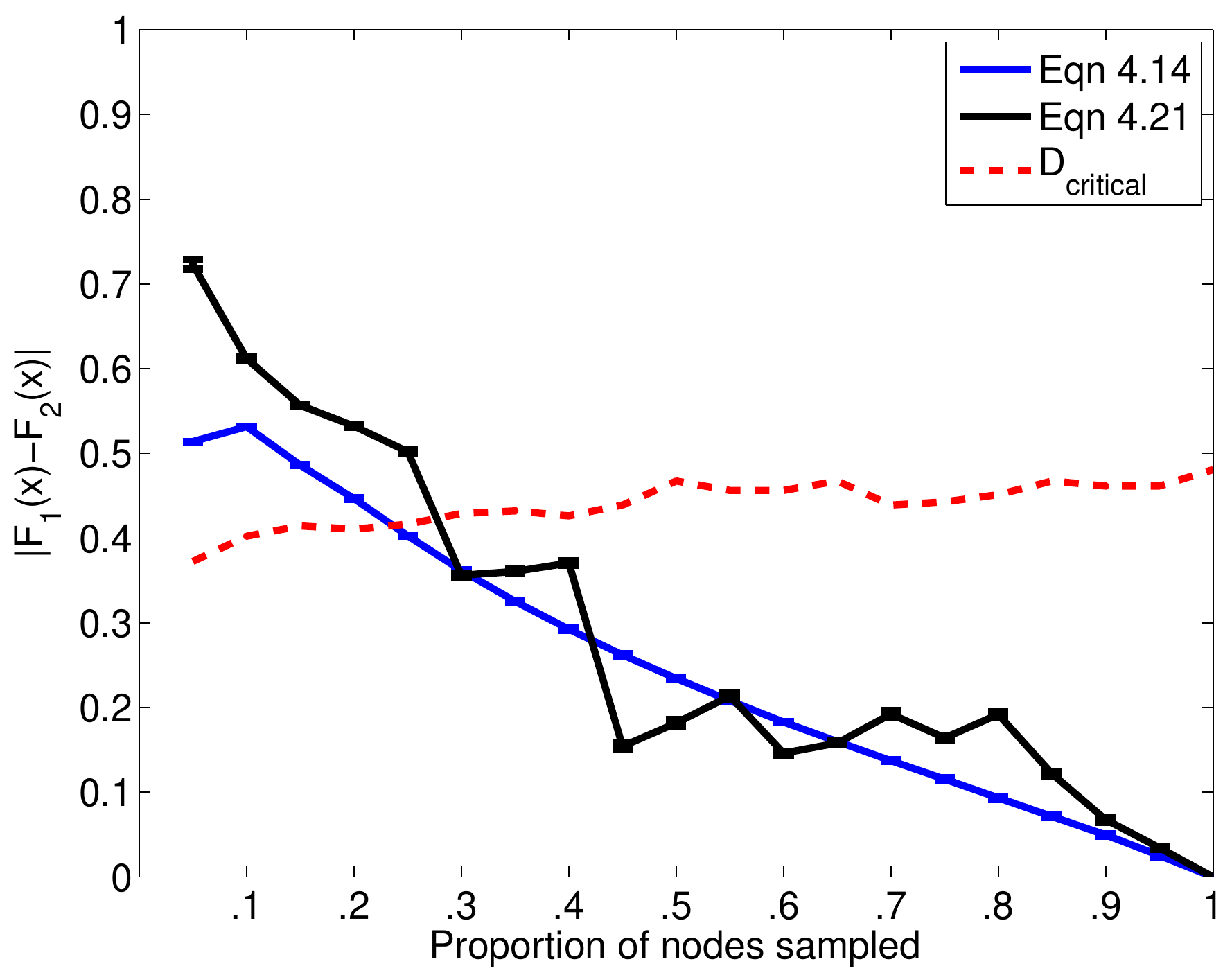}}
\caption[Kolmogorov-Smirnov two sample test for true CDF and predicted CDF from subnetworks obtained by failing links]{Kolmogorov-Smirnov two sample test for true CDF and predicted CDF from subnetworks obtained by failing links. The red line represents $D_{\text{crit}}$ for $\alpha=0.05$ and sample sizes $n_1=\kmax$ of the true CDF and $n_2=\kmax$ of the observed CDF. The predicted CDFs for for most networks are statistically indistinguishable from the true CDF for these networks for $q>0.3$. Due to the presence of large hubs in Pref, $n_1$ and $n_2$ are quite large leading to  high statistical power in the KS test. Thus, even very small differences between the true and predicted CDFs result in a statistically significant difference and rejection of the null hypothesis, even though the curves show relatively good agreement.}
\label{fig:one_over_q_failed_links_D_stat_plot}
\end{figure*}
   
\newpage
\setcounter{equation}{11}
\begin{table*}[!htp]\small
\caption[Error in $\hat{N}$ when sampling by links]{Error in $\hat{N}$ when sampling by links. Predictors show good agreements with true values, except for low values of $q$. In these cases, errors in the predicted degree distribution contribute to errors in the predicted number of nodes. Future improvements in the predicted degree distribution would improve $\hat{N}$.}
\centering
\begin{tabular}{|l | l l l l l l l l l l|}
\hline
$q$ & Erdrey & Pref & Smallw & Renga & C. elegans & Airlines & Karate & Dolphins & Condmat & Power \\ \hline
   0.05 &    0.40 &   0.47 &   0.38 &   0.39 &   0.34 &   0.53 &   0.68 &   0.64 &   0.65 &   0.80 \\ 
  0.10 &    0.11 &   0.21 &   0.08 &   0.09 &   0.11 &   0.34 &   0.46 &   0.41 &   0.44 &   0.64 \\ 
  0.15 &    0.02 &   0.06 &   0.06 &   0.04 &   0.02 &   0.23 &   0.33 &   0.26 &   0.31 &   0.51 \\ 
  0.20 &    0.07 &   0.02 &   0.10 &   0.09 &   0.01 &   0.17 &   0.23 &   0.16 &   0.22 &   0.40 \\ 
  0.25 &    0.08 &   0.05 &   0.10 &   0.09 &   0.01 &   0.12 &   0.15 &   0.10 &   0.15 &   0.31 \\ 
  0.30 &    0.07 &   0.07 &   0.08 &   0.08 &   0.01 &   0.10 &   0.10 &   0.06 &   0.11 &   0.24 \\ 
  0.35 &    0.05 &   0.07 &   0.06 &   0.06 &   0.01 &   0.07 &   0.06 &   0.04 &   0.07 &   0.18 \\ 
  0.40 &    0.04 &   0.06 &   0.04 &   0.04 &   0.00 &   0.06 &   0.04 &   0.03 &   0.05 &   0.14 \\ 
  0.45 &    0.03 &   0.05 &   0.02 &   0.03 &   0.00 &   0.04 &   0.00 &   0.02 &   0.03 &   0.10 \\ 
  0.50 &    0.02 &   0.04 &   0.01 &   0.02 &   0.00 &   0.03 &   0.01 &   0.02 &   0.02 &   0.07 \\ 
  0.55 &    0.01 &   0.03 &   0.01 &   0.01 &   0.00 &   0.03 &   0.01 &   0.01 &   0.01 &   0.05 \\ 
  0.60 &    0.01 &   0.02 &   0.00 &   0.00 &   0.00 &   0.01 &   0.02 &   0.01 &   0.01 &   0.03 \\ 
  0.65 &    0.00 &   0.01 &   0.00 &   0.00 &   0.00 &   0.01 &   0.02 &   0.01 &   0.00 &   0.02 \\ 
  0.70 &    0.00 &   0.01 &   0.00 &   0.00 &   0.00 &   0.00 &   0.02 &   0.00 &   0.00 &   0.01 \\ 
  0.75 &    0.00 &   0.00 &   0.00 &   0.00 &   0.00 &   0.01 &   0.02 &   0.00 &   0.00 &   0.00 \\ 
  0.80 &    0.00 &   0.00 &   0.00 &   0.00 &   0.00 &   0.00 &   0.01 &   0.00 &   0.00 &   0.00 \\ 
  0.85 &    0.00 &   0.00 &   0.00 &   0.00 &   0.00 &   0.00 &   0.01 &   0.00 &   0.00 &   0.00 \\ 
  0.90 &    0.00 &   0.00 &   0.00 &   0.00 &   0.00 &   0.00 &   0.00 &   0.00 &   0.00 &   0.00 \\ 
  0.95 &    0.00 &   0.00 &   0.00 &   0.00 &   0.00 &   0.00 &   0.00 &   0.00 &   0.00 &   0.00 \\ 
  1.00 &    0.00 &   0.00 &   0.00 &   0.00 &   0.00 &   0.00 &   0.00 &   0.00 &   0.00 &   0.00 \\ \hline
    \end{tabular}
\label{table:error_N_incident_links}
\end{table*}

\newpage
    \setcounter{equation}{12}
\begin{table*}[!htp]\small
\caption[Error in $\hat{M}$ when sampling by links]{Error in $\hat{M}$ when sampling by links. Error is nonzero only because of roundoff errors when selecting an integer number of edges to sample.}
\centering
\begin{tabular}{|l | l l l l l l l l l l|}
\hline
$q$ & Erdrey & Pref & Smallw & Renga & C. elegans & Airlines & Karate & Dolphins & Condmat & Power \\ \hline
  0.05 &   0.00 &   0.00 &   0.00 &   0.00 &   0.00 &   0.00 &   0.03 &   0.01 &   0.00 &   0.00 \\ 
  0.10 &   0.00 &   0.00 &   0.00 &   0.00 &   0.00 &   0.00 &   0.03 &   0.01 &   0.00 &   0.00 \\ 
  0.15 &   0.00 &   0.00 &   0.00 &   0.00 &   0.00 &   0.00 &   0.03 &   0.01 &   0.00 &   0.00 \\ 
  0.20 &   0.00 &   0.00 &   0.00 &   0.00 &   0.00 &   0.00 &   0.03 &   0.01 &   0.00 &   0.00 \\ 
  0.25 &   0.00 &   0.00 &   0.00 &   0.00 &   0.00 &   0.00 &   0.03 &   0.01 &   0.00 &   0.00 \\ 
  0.30 &   0.00 &   0.00 &   0.00 &   0.00 &   0.00 &   0.00 &   0.02 &   0.01 &   0.00 &   0.00 \\ 
  0.35 &   0.00 &   0.00 &   0.00 &   0.00 &   0.00 &   0.00 &   0.01 &   0.01 &   0.00 &   0.00 \\ 
  0.40 &   0.00 &   0.00 &   0.00 &   0.00 &   0.00 &   0.00 &   0.01 &   0.01 &   0.00 &   0.00 \\ 
  0.45 &   0.00 &   0.00 &   0.00 &   0.00 &   0.00 &   0.00 &   0.00 &   0.01 &   0.00 &   0.00 \\ 
  0.50 &   0.00 &   0.00 &   0.00 &   0.00 &   0.00 &   0.00 &   0.00 &   0.01 &   0.00 &   0.00 \\ 
  0.55 &   0.00 &   0.00 &   0.00 &   0.00 &   0.00 &   0.00 &   0.00 &   0.01 &   0.00 &   0.00 \\ 
  0.60 &   0.00 &   0.00 &   0.00 &   0.00 &   0.00 &   0.00 &   0.00 &   0.00 &   0.00 &   0.00 \\ 
  0.65 &   0.00 &   0.00 &   0.00 &   0.00 &   0.00 &   0.00 &   0.01 &   0.00 &   0.00 &   0.00 \\ 
  0.70 &   0.00 &   0.00 &   0.00 &   0.00 &   0.00 &   0.00 &   0.01 &   0.00 &   0.00 &   0.00 \\ 
  0.75 &   0.00 &   0.00 &   0.00 &   0.00 &   0.00 &   0.00 &   0.01 &   0.00 &   0.00 &   0.00 \\ 
  0.80 &   0.00 &   0.00 &   0.00 &   0.00 &   0.00 &   0.00 &   0.01 &   0.00 &   0.00 &   0.00 \\ 
  0.85 &   0.00 &   0.00 &   0.00 &   0.00 &   0.00 &   0.00 &   0.00 &   0.00 &   0.00 &   0.00 \\ 
  0.90 &   0.00 &   0.00 &   0.00 &   0.00 &   0.00 &   0.00 &   0.00 &   0.00 &   0.00 &   0.00 \\ 
  0.95 &   0.00 &   0.00 &   0.00 &   0.00 &   0.00 &   0.00 &   0.00 &   0.00 &   0.00 &   0.00 \\ 
  1.00 &   0.00 &   0.00 &   0.00 &   0.00 &   0.00 &   0.00 &   0.00 &   0.00 &   0.00 &   0.00 \\ \hline
    \end{tabular}
\label{table:error_M_incident_links}
\end{table*} 

\newpage
    \setcounter{equation}{13}
\begin{table*}[!htp]\small
\caption[Error in $\hat{k}_{\rm avg}$ when sampling by links]{Error in $\hat{k}_{\rm avg}$ when sampling by links.}
\centering
\begin{tabular}{|l | l l l l l l l l l l|}
\hline
$q$ & Erdrey & Pref & Smallw & Renga & C. elegans & Airlines & Karate & Dolphins & Condmat & Power \\ \hline
 0.05 &   0.66 &   0.89 &   0.61 &   0.63 &   0.50 &   1.13 &   2.18 &   1.79 &   1.83 &   4.03 \\ 
  0.10 &   0.12 &   0.26 &   0.08 &   0.10 &   0.12 &   0.51 &   0.90 &   0.71 &   0.79 &   1.79 \\ 
  0.15 &   0.02 &   0.07 &   0.05 &   0.04 &   0.02 &   0.30 &   0.52 &   0.36 &   0.45 &   1.04 \\ 
  0.20 &   0.06 &   0.02 &   0.09 &   0.08 &   0.01 &   0.20 &   0.32 &   0.20 &   0.28 &   0.67 \\ 
  0.25 &   0.07 &   0.05 &   0.09 &   0.08 &   0.01 &   0.14 &   0.21 &   0.12 &   0.18 &   0.46 \\ 
  0.30 &   0.06 &   0.06 &   0.07 &   0.07 &   0.01 &   0.11 &   0.09 &   0.07 &   0.12 &   0.32 \\ 
  0.35 &   0.05 &   0.06 &   0.05 &   0.05 &   0.01 &   0.08 &   0.06 &   0.05 &   0.08 &   0.22 \\ 
  0.40 &   0.04 &   0.05 &   0.04 &   0.04 &   0.00 &   0.06 &   0.04 &   0.03 &   0.05 &   0.16 \\ 
  0.45 &   0.03 &   0.04 &   0.02 &   0.03 &   0.00 &   0.05 &   0.00 &   0.03 &   0.03 &   0.11 \\ 
  0.50 &   0.02 &   0.03 &   0.01 &   0.02 &   0.00 &   0.03 &   0.01 &   0.02 &   0.02 &   0.07 \\ 
  0.55 &   0.01 &   0.02 &   0.01 &   0.01 &   0.00 &   0.03 &   0.01 &   0.00 &   0.01 &   0.05 \\ 
  0.60 &   0.01 &   0.02 &   0.00 &   0.00 &   0.00 &   0.01 &   0.01 &   0.01 &   0.01 &   0.03 \\ 
  0.65 &   0.00 &   0.01 &   0.00 &   0.00 &   0.00 &   0.01 &   0.02 &   0.00 &   0.00 &   0.02 \\ 
  0.70 &   0.00 &   0.01 &   0.00 &   0.00 &   0.00 &   0.00 &   0.02 &   0.00 &   0.00 &   0.01 \\ 
  0.75 &   0.00 &   0.00 &   0.00 &   0.00 &   0.00 &   0.01 &   0.01 &   0.00 &   0.00 &   0.00 \\ 
  0.80 &   0.00 &   0.00 &   0.00 &   0.00 &   0.00 &   0.00 &   0.02 &   0.00 &   0.00 &   0.00 \\ 
  0.85 &   0.00 &   0.00 &   0.00 &   0.00 &   0.00 &   0.00 &   0.01 &   0.00 &   0.00 &   0.00 \\ 
  0.90 &   0.00 &   0.00 &   0.00 &   0.00 &   0.00 &   0.00 &   0.01 &   0.00 &   0.00 &   0.00 \\ 
  0.95 &   0.00 &   0.00 &   0.00 &   0.00 &   0.00 &   0.00 &   0.00 &   0.00 &   0.00 &   0.00 \\ 
  1.00 &   0.00 &   0.00 &   0.00 &   0.00 &   0.00 &   0.00 &   0.00 &   0.00 &   0.00 &   0.00 \\ \hline
    \end{tabular}
\label{table:error_avk_incident_links}
\end{table*}  

\newpage
  \setcounter{equation}{14}
\begin{table*}[!htp]\small
\caption[Error in $\hat{C}$ when sampling by links]{Error in $\hat{C}$ when sampling by links.}
\centering
\begin{tabular}{|l | l l l l l l l l l l|}
\hline
$q$ & Erdrey & Pref & Smallw & Renga & C. elegans & Airlines & Karate & Dolphins & Condmat & Power \\ \hline
 0.05 &   0.51 &   0.20 &   0.00 &   0.01 &   0.05 &   0.15 &    -- &    --&   0.02 &   0.05 \\ 
  0.10 &   0.36 &   0.05 &   0.00 &   0.01 &   0.04 &   0.05 &   -- &   -- &   0.00 &   0.27 \\ 
  0.15 &   0.21 &   0.00 &   0.00 &   0.00 &   0.00 &   0.06 &   -- &   -- &   0.01 &   0.03 \\ 
  0.20 &   0.20 &   0.02 &   0.00 &   0.00 &   0.02 &   0.01 &   -- &   -- &   0.00 &   0.02 \\ 
  0.25 &   0.01 &   0.00 &   0.00 &   0.00 &   0.01 &   0.00 &   -- &   0.19 &   0.00 &   0.04 \\ 
  0.30 &   0.00 &   0.00 &   0.00 &   0.00 &   0.02 &   0.02 &   0.16 &   0.06 &   0.00 &   0.01 \\ 
  0.35 &   0.05 &   0.00 &   0.00 &   0.00 &   0.00 &   0.01 &   0.05 &   0.07 &   0.00 &   0.00 \\ 
  0.40 &   0.03 &   0.01 &   0.00 &   0.00 &   0.00 &   0.01 &   0.08 &   0.07 &   0.00 &   0.01 \\ 
  0.45 &   0.02 &   0.01 &   0.00 &   0.00 &   0.00 &   0.01 &   0.05 &   0.03 &   0.00 &   0.02 \\ 
  0.50 &   0.01 &   0.00 &   0.00 &   0.00 &   0.00 &   0.00 &   0.08 &   0.02 &   0.00 &   0.03 \\ 
  0.55 &   0.01 &   0.00 &   0.00 &   0.00 &   0.01 &   0.00 &   0.01 &   0.02 &   0.00 &   0.01 \\ 
  0.60 &   0.01 &   0.00 &   0.00 &   0.00 &   0.01 &   0.01 &   0.06 &   0.02 &   0.00 &   0.01 \\ 
  0.65 &   0.01 &   0.00 &   0.00 &   0.00 &   0.00 &   0.00 &   0.00 &   0.02 &   0.00 &   0.00 \\ 
  0.70 &   0.00 &   0.01 &   0.00 &   0.00 &   0.01 &   0.00 &   0.02 &   0.01 &   0.00 &   0.00 \\ 
  0.75 &   0.00 &   0.00 &   0.00 &   0.00 &   0.00 &   0.01 &   0.02 &   0.01 &   0.00 &   0.00 \\ 
  0.80 &   0.00 &   0.00 &   0.00 &   0.00 &   0.00 &   0.00 &   0.05 &   0.02 &   0.00 &   0.00 \\ 
  0.85 &   0.01 &   0.00 &   0.00 &   0.00 &   0.00 &   0.00 &   0.01 &   0.01 &   0.00 &   0.00 \\ 
  0.90 &   0.00 &   0.00 &   0.00 &   0.00 &   0.00 &   0.00 &   0.01 &   0.01 &   0.00 &   0.00 \\ 
  0.95 &   0.00 &   0.00 &   0.00 &   0.00 &   0.00 &   0.00 &   0.00 &   0.00 &   0.00 &   0.00 \\ 
  1.00 &   0.00 &   0.00 &   0.00 &   0.00 &   0.00 &   0.00 &   0.00 &   0.00 &   0.00 &   0.00 \\  \hline
    \end{tabular}
\label{table:error_cluster_incident_links}
\end{table*}  
\newpage
  \setcounter{equation}{15}

\begin{table*}[!htp]\small
\caption[Error in $\hat{k}_{\max}$ when sampling by links]{Error in $\hat{k}_{\max}$ when sampling by links.}
\centering
\begin{tabular}{|l | l l l l l l l l l l|}
\hline
$q$ & Erdrey & Pref & Smallw & Renga & C. elegans & Airlines & Karate & Dolphins & Condmat & Power \\ \hline
 0.05 &   0.67 &   0.00 &   0.20 &   0.16 &   0.11 &   0.18 &   1.14 &   2.38 &   0.06 &   0.24 \\ 
  0.10 &   0.33 &   0.00 &   0.05 &   0.37 &   0.01 &   0.09 &   0.62 &   1.39 &   0.15 &   0.37 \\ 
  0.15 &   0.30 &   0.00 &   0.10 &   0.18 &   0.00 &   0.14 &   0.42 &   0.02 &   0.12 &   0.19 \\ 
  0.20 &   0.28 &   0.01 &   0.10 &   0.40 &   0.02 &   0.10 &   0.36 &   0.02 &   0.12 &   0.07 \\ 
  0.25 &   0.17 &   0.00 &   0.05 &   0.23 &   0.03 &   0.06 &   0.32 &   0.10 &   0.08 &   0.13 \\ 
  0.30 &   0.17 &   0.00 &   0.03 &   0.24 &   0.01 &   0.07 &   0.16 &   0.11 &   0.09 &   0.01 \\ 
  0.35 &   0.15 &   0.00 &   0.00 &   0.27 &   0.00 &   0.04 &   0.15 &   0.03 &   0.08 &   0.02 \\ 
  0.40 &   0.19 &   0.00 &   0.00 &   0.20 &   0.01 &   0.04 &   0.11 &   0.05 &   0.07 &   0.01 \\ 
  0.45 &   0.11 &   0.00 &   0.00 &   0.11 &   0.00 &   0.05 &   0.13 &   0.04 &   0.07 &   0.01 \\ 
  0.50 &   0.07 &   0.00 &   0.00 &   0.16 &   0.01 &   0.03 &   0.13 &   0.04 &   0.05 &   0.04 \\ 
  0.55 &   0.01 &   0.00 &   0.00 &   0.15 &   0.00 &   0.03 &   0.09 &   0.06 &   0.05 &   0.03 \\ 
  0.60 &   0.09 &   0.00 &   0.00 &   0.17 &   0.01 &   0.03 &   0.06 &   0.02 &   0.03 &   0.05 \\ 
  0.65 &   0.08 &   0.00 &   0.00 &   0.13 &   0.01 &   0.01 &   0.09 &   0.04 &   0.03 &   0.01 \\ 
  0.70 &   0.07 &   0.00 &   0.00 &   0.10 &   0.00 &   0.02 &   0.06 &   0.02 &   0.02 &   0.02 \\ 
  0.75 &   0.02 &   0.00 &   0.00 &   0.07 &   0.01 &   0.02 &   0.06 &   0.02 &   0.01 &   0.02 \\ 
  0.80 &   0.01 &   0.00 &   0.00 &   0.03 &   0.00 &   0.00 &   0.04 &   0.03 &   0.02 &   0.04 \\ 
  0.85 &   0.00 &   0.00 &   0.00 &   0.04 &   0.00 &   0.00 &   0.03 &   0.03 &   0.02 &   0.04 \\ 
  0.90 &   0.01 &   0.00 &   0.05 &   0.02 &   0.01 &   0.00 &   0.01 &   0.01 &   0.01 &   0.02 \\ 
  0.95 &   0.01 &   0.00 &   0.05 &   0.00 &   0.00 &   0.00 &   0.00 &   0.02 &   0.00 &   0.01 \\ 
  1.00 &   0.00 &   0.00 &   0.00 &   0.00 &   0.00 &   0.00 &   0.00 &   0.09 &   0.00 &   0.00 \\ \hline
    \end{tabular}
\label{table:error_kmax_incident_links}  
\end{table*}  
\FloatBarrier
\newpage
  \setcounter{equation}{18}
\begin{figure*}[!ht]
\centering
\subfigure[Erdrey]{\includegraphics[width=.24\textwidth]{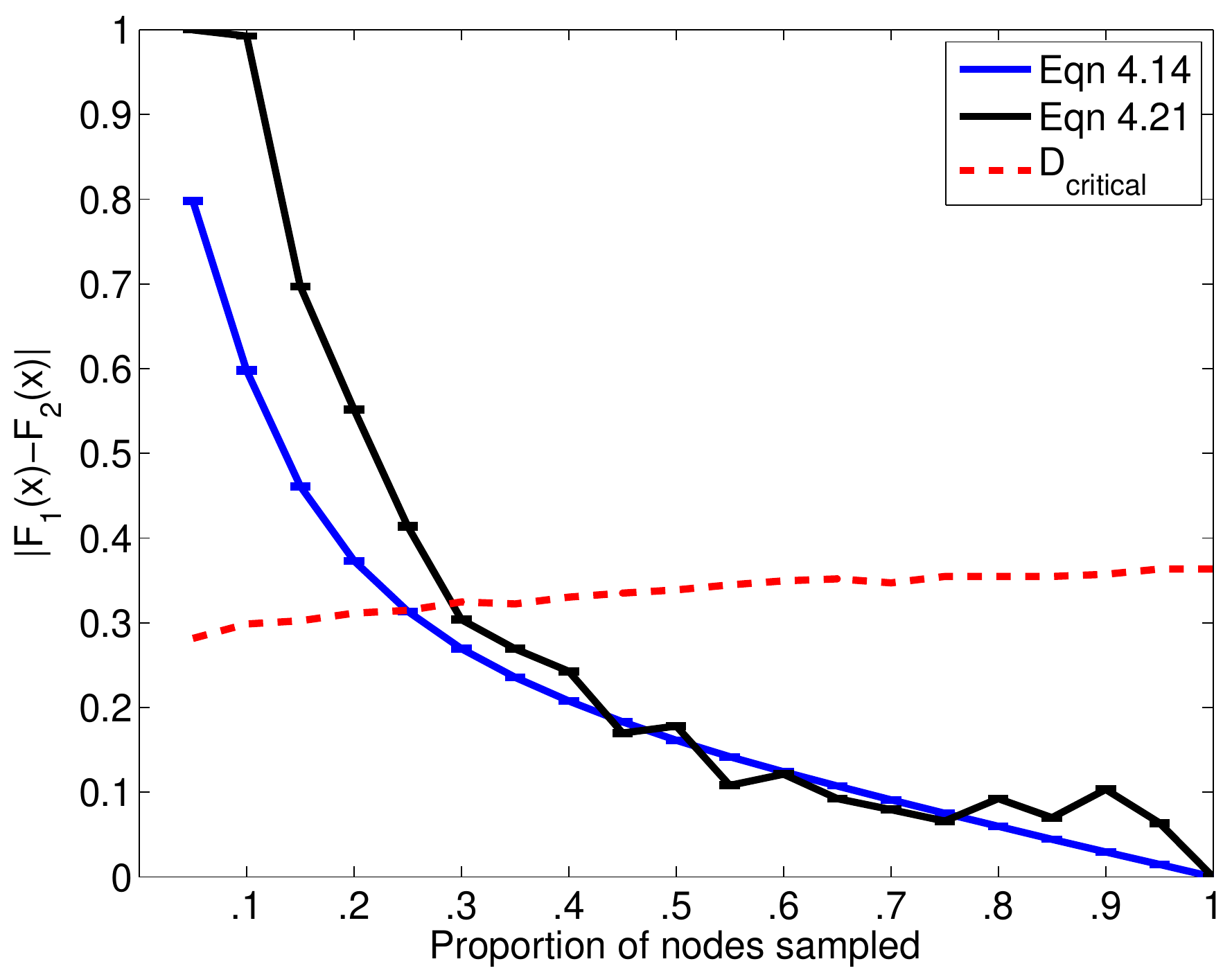}}
\subfigure[Pref]{\includegraphics[width=.24\textwidth]{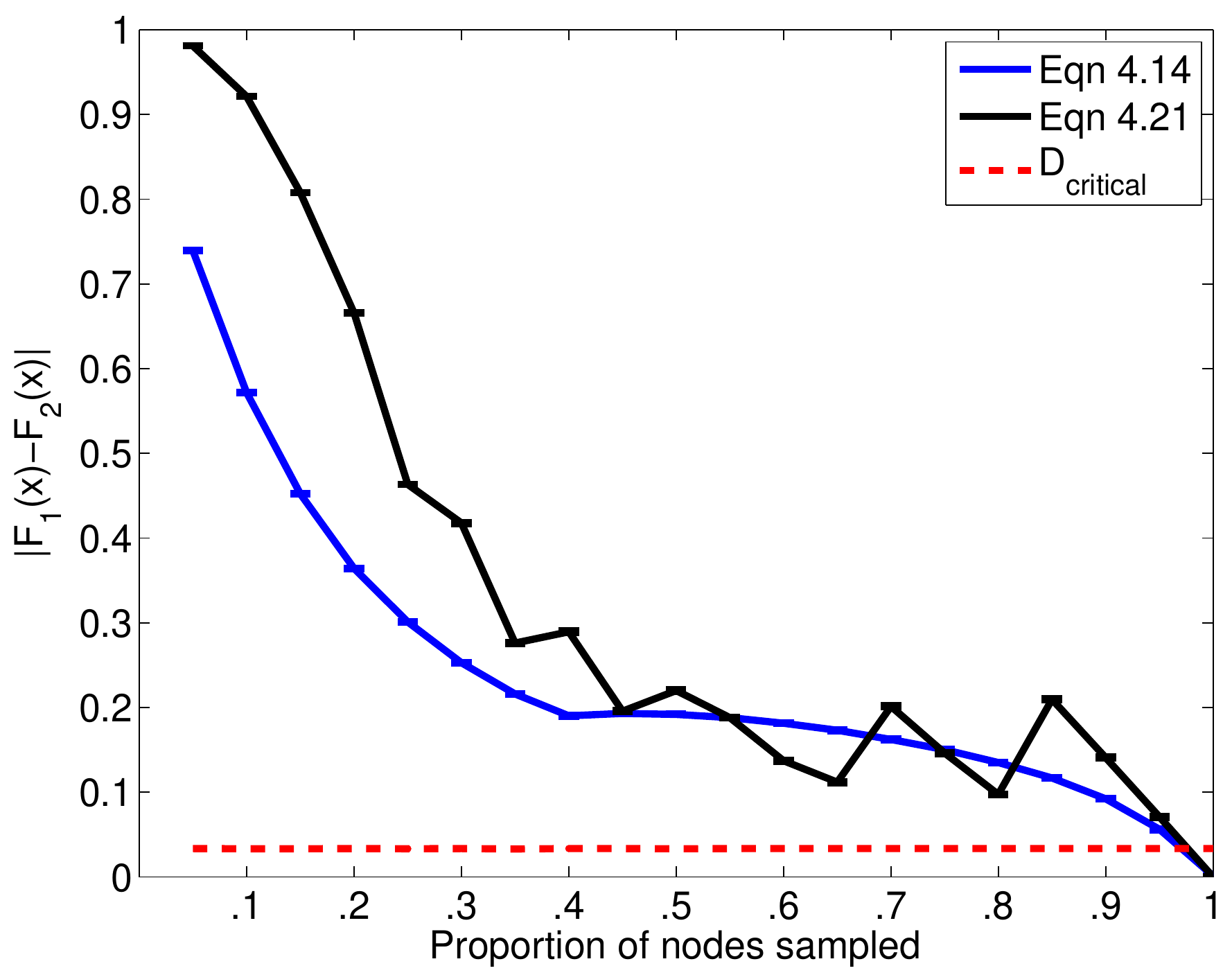}}
\subfigure[Smallworld]{\includegraphics[width=.24\textwidth]{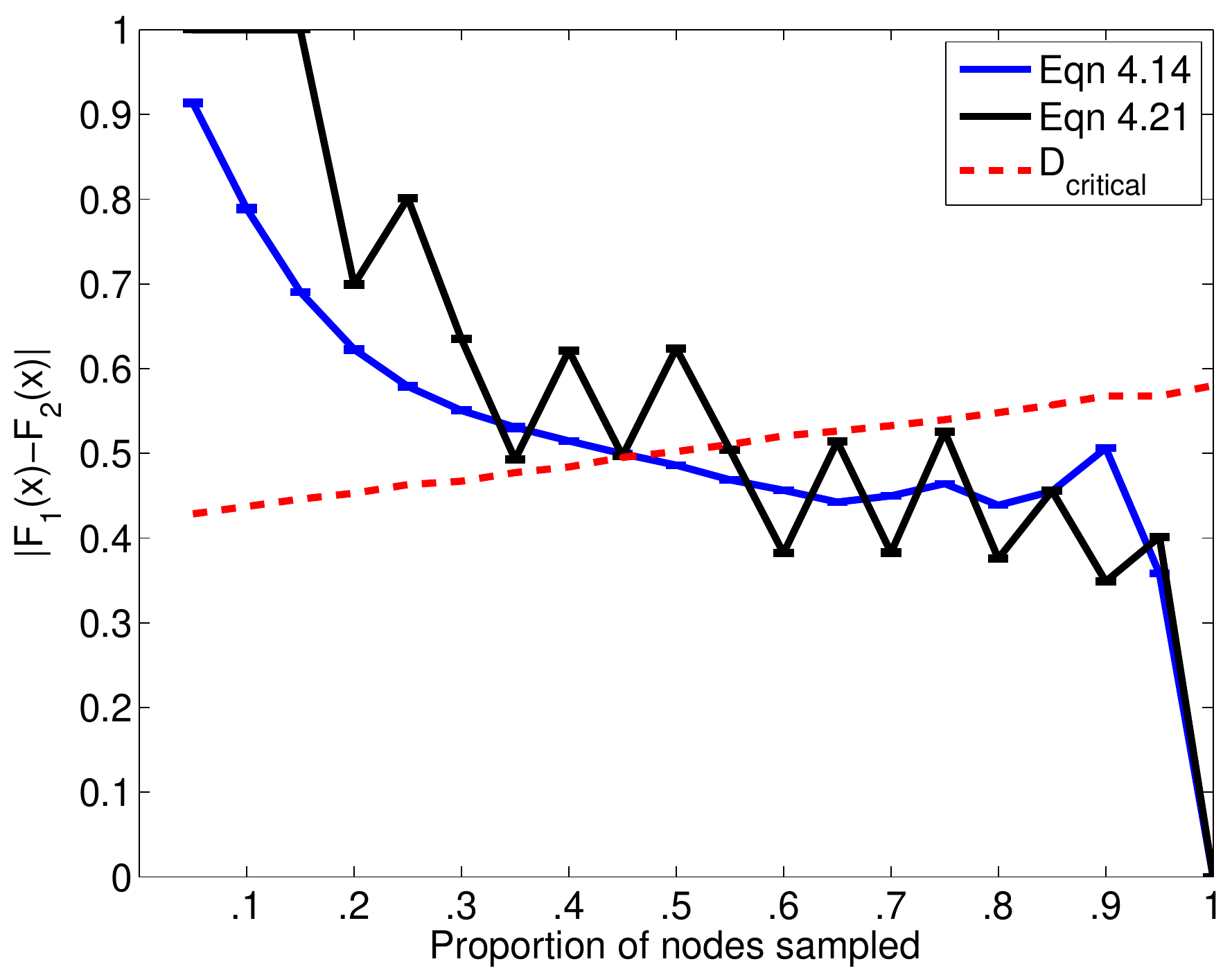}}
\subfigure[Renga]{\includegraphics[width=.24\textwidth]{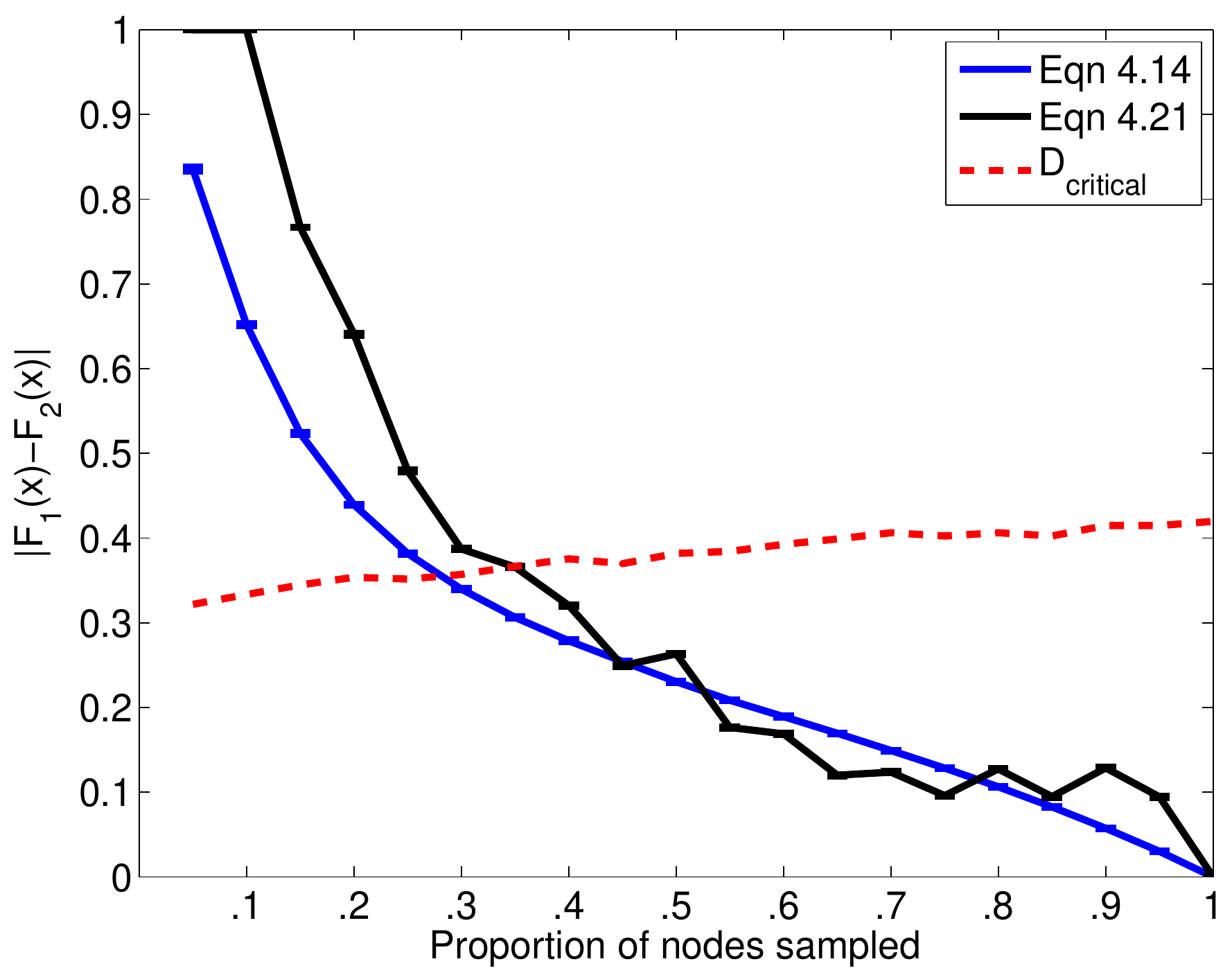}}\\
\subfigure[C. elgegans]{\includegraphics[width=.24\textwidth]{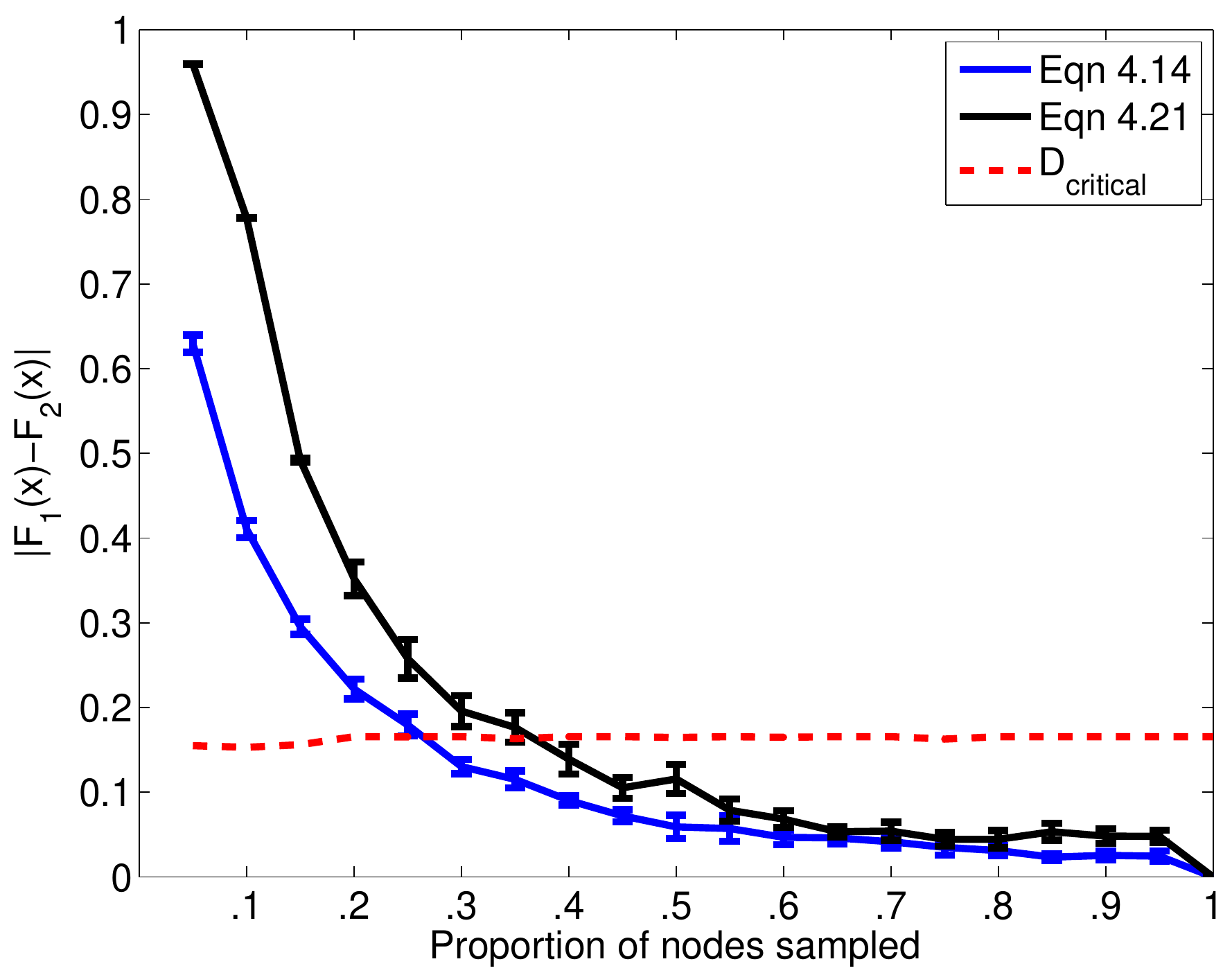}}
\subfigure[Airlines]{\includegraphics[width=.24\textwidth]{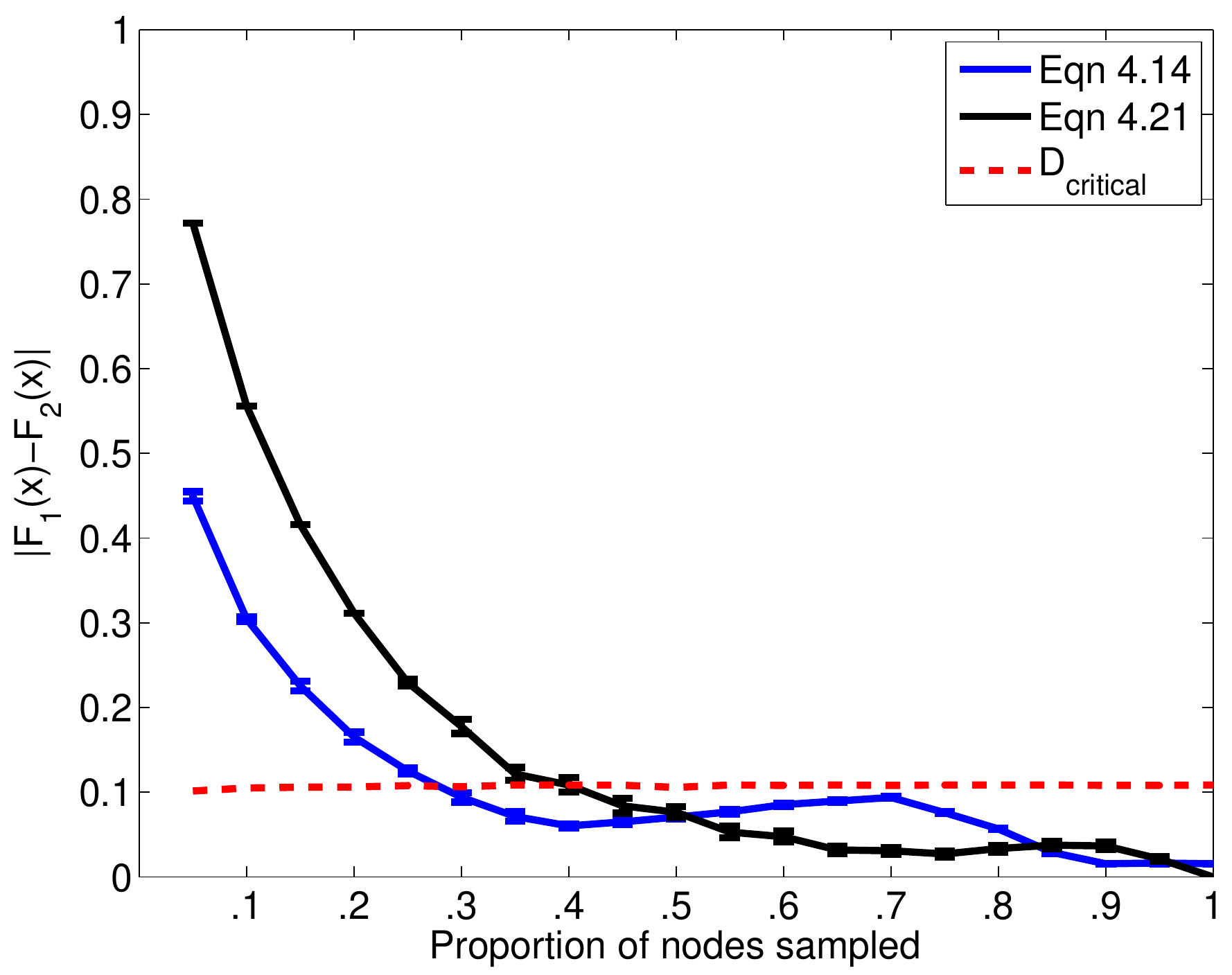}}
\subfigure[Karate]{\includegraphics[width=.24\textwidth]{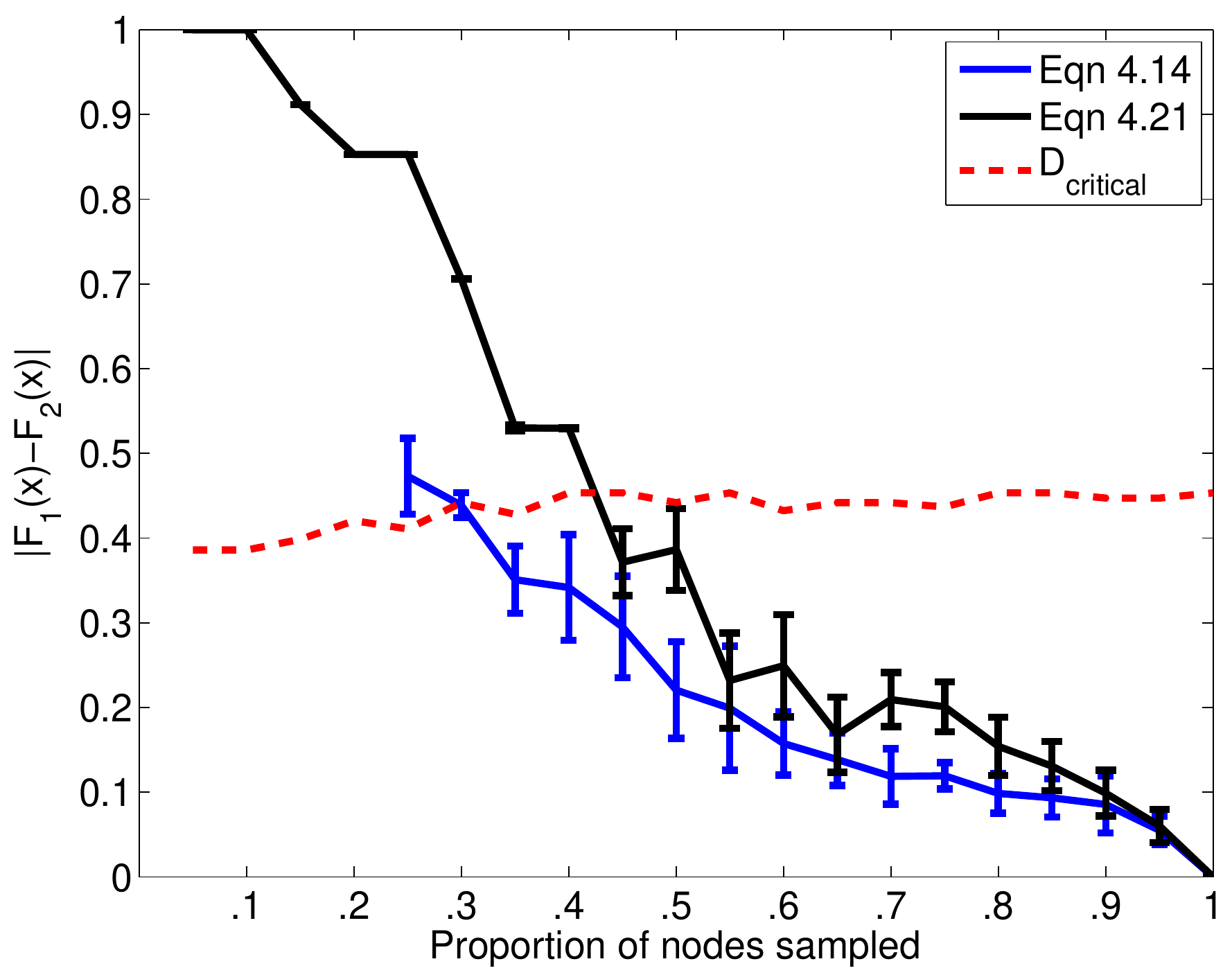}}
\subfigure[Dolphins]{\includegraphics[width=.24\textwidth]{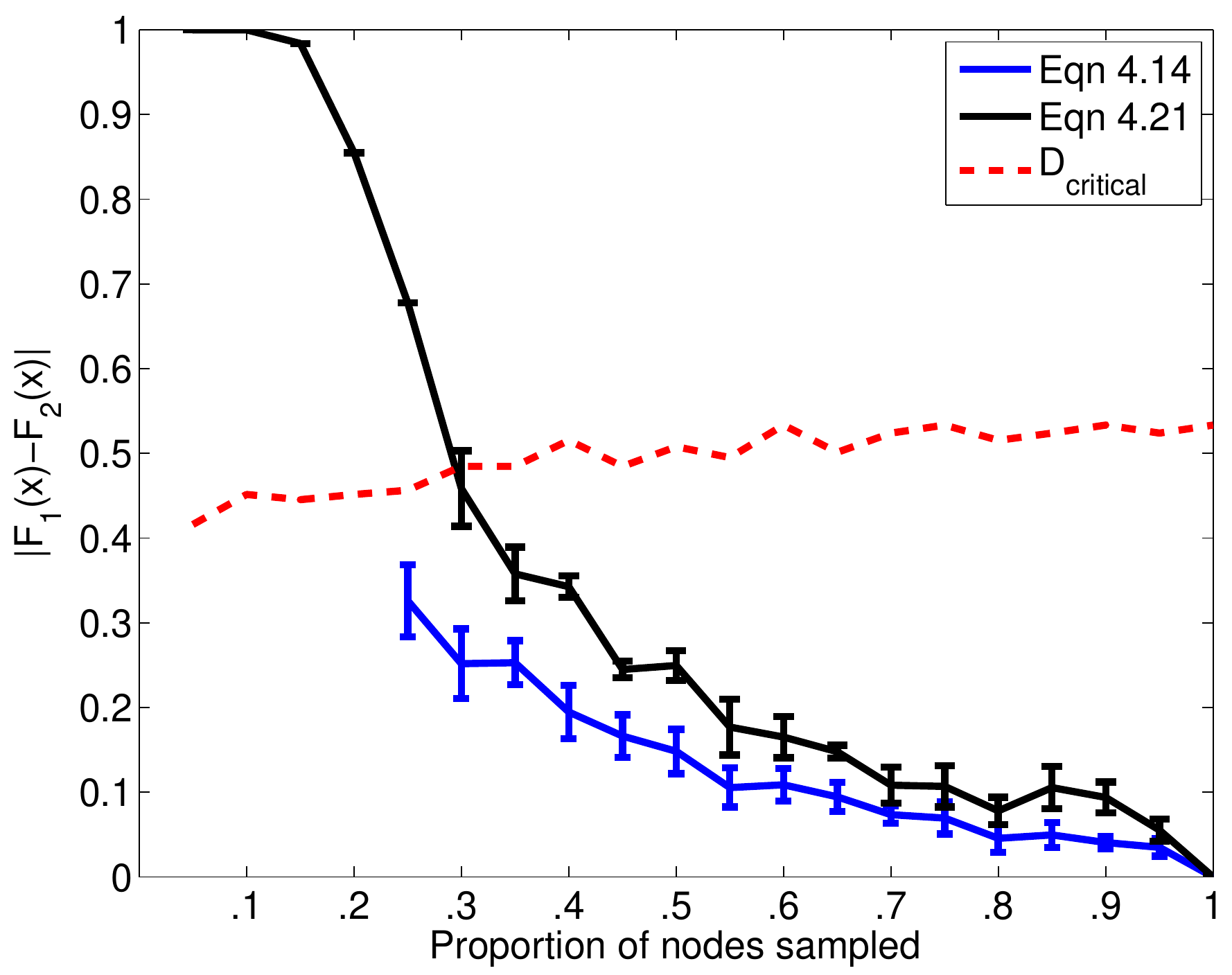}}\\
\subfigure[Condmat]{\includegraphics[width=.24\textwidth]{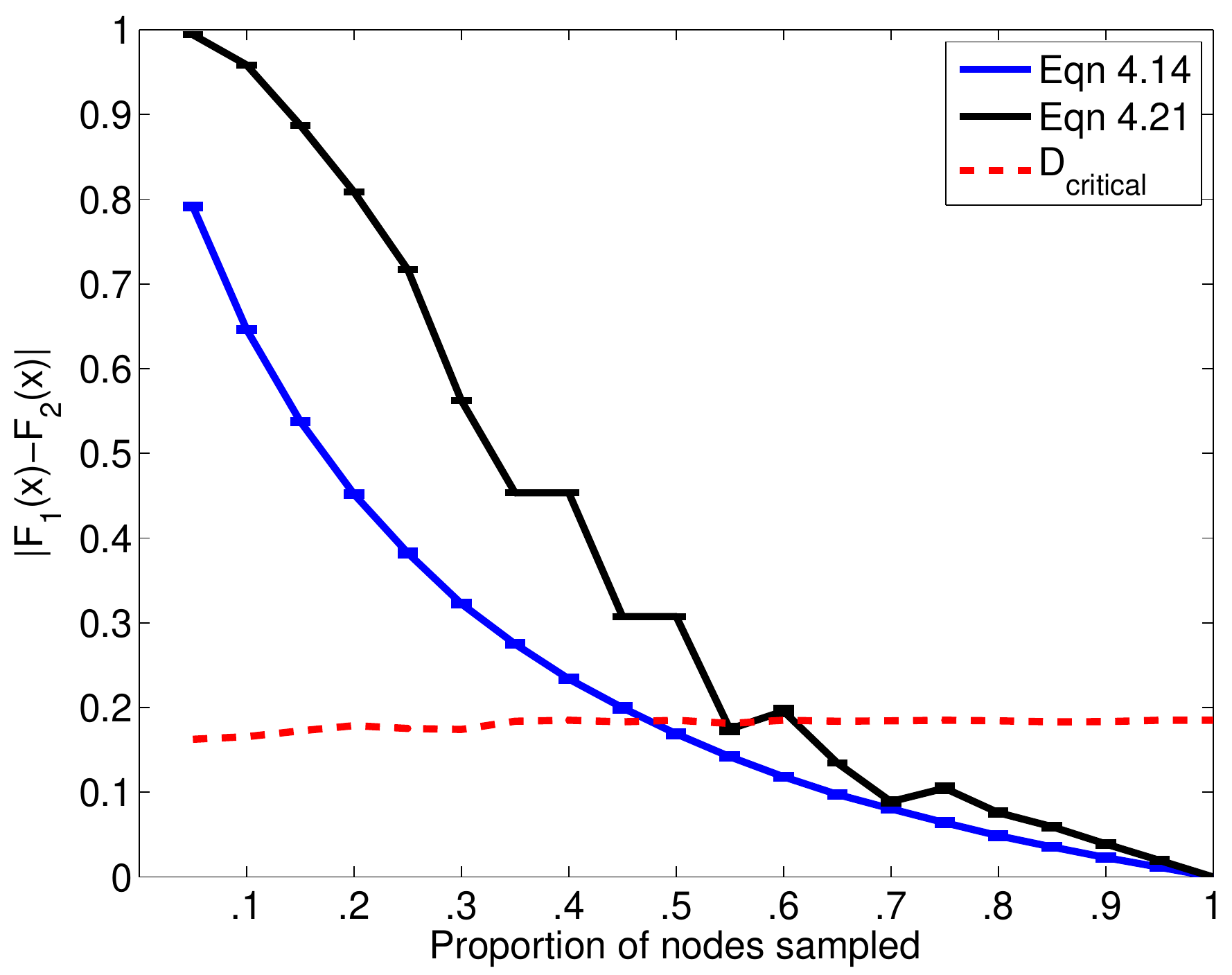}}
\subfigure[Powergrid]{\includegraphics[width=.24\textwidth]{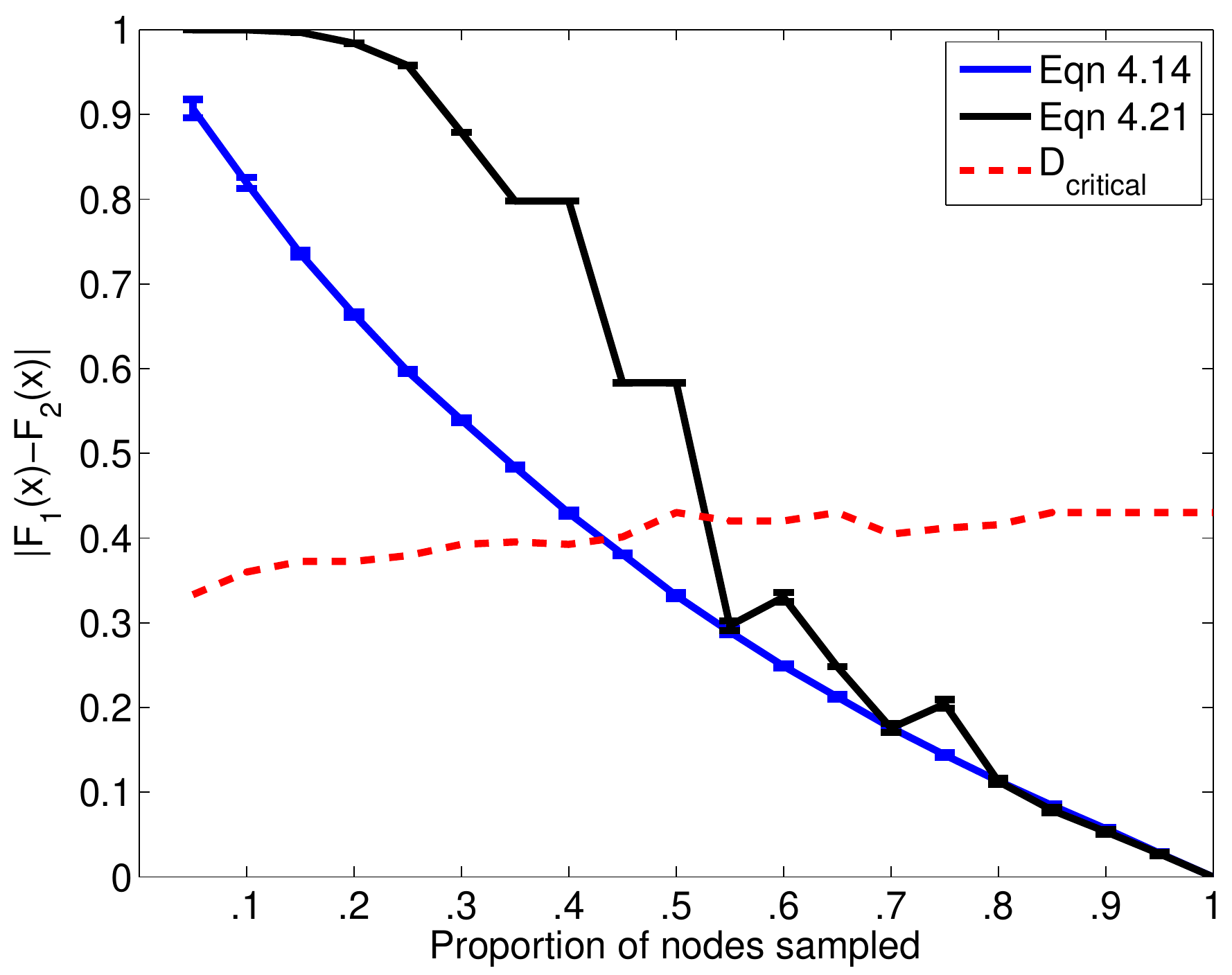}}
\caption[Kolmogorov-Smirnov two sample test for true CDF and predicted CDF from subnetworks generated by sampled links]{Kolmogorov-Smirnov two sample test for true CDF and predicted CDF from subnetworks generated by sampled links. The red line represents $D_{\text{crit}}$ for $\alpha=0.05$ and sample sizes $n_1=\kmax$ of the true CDF and $n_2=\kmax$ of the observed CDF. The predicted CDFs for for most networks are statistically indistinguishable from the true CDF for these networks for $q>0.3$. Due to the presence of large hubs in Pref, $n_1$ and $n_2$ are quite large leading to  high statistical power in the KS test. Thus, even very small differences between the true and predicted CDFs result in a statistically significant difference and rejection of the null hypothesis, even though the curves show relatively good agreement.}
\label{fig:one_over_q_incident_links_D_stat_plot}
\end{figure*}

\newpage
    \setcounter{equation}{16}
\begin{table*}[!htp]\small
\caption[Error in $\hat{N}$ when sampling by interactions on an Erd{\"o}s-R\'{e}nyi random graph.]{Error in $\hat{N}$ when sampling by interactions on an Erd{\"o}s-R\'{e}nyi random graph.}
\centering
\begin{tabular}{|l | l l l l l l l |}
\hline
q & I & II & III & IV & V & VI & VII \\ \hline
  0.05 &   0.54 &   0.50 &   0.46 &   0.41 &   0.36 &   0.34 &   0.36\\ 
  0.10 &   0.48 &   0.39 &   0.30 &   0.23 &   0.18 &   0.14 &   0.18\\ 
  0.15 &   0.42 &   0.28 &   0.19 &   0.12 &   0.09 &   0.06 &   0.10\\ 
  0.20 &   0.35 &   0.20 &   0.12 &   0.07 &   0.05 &   0.03 &   0.05\\ 
  0.25 &   0.29 &   0.14 &   0.07 &   0.04 &   0.02 &   0.01 &   0.03\\ 
  0.30 &   0.24 &   0.10 &   0.05 &   0.02 &   0.01 &   0.01 &   0.02\\ 
  0.35 &   0.19 &   0.07 &   0.03 &   0.02 &   0.01 &   0.00 &   0.01\\ 
  0.40 &   0.14 &   0.05 &   0.02 &   0.01 &   0.01 &   0.00 &   0.01\\ 
  0.45 &   0.11 &   0.03 &   0.01 &   0.01 &   0.00 &   0.00 &   0.01\\ 
  0.50 &   0.08 &   0.02 &   0.01 &   0.00 &   0.00 &   0.00 &   0.00\\ 
  0.55 &   0.06 &   0.01 &   0.01 &   0.00 &   0.00 &   0.00 &   0.00\\ 
  0.60 &   0.05 &   0.01 &   0.00 &   0.00 &   0.00 &   0.00 &   0.00\\ 
  0.65 &   0.03 &   0.01 &   0.00 &   0.00 &   0.00 &   0.00 &   0.00\\ 
  0.70 &   0.02 &   0.00 &   0.00 &   0.00 &   0.00 &   0.00 &   0.00\\ 
  0.75 &   0.01 &   0.00 &   0.00 &   0.00 &   0.00 &   0.00 &   0.00\\ 
  0.80 &   0.01 &   0.00 &   0.00 &   0.00 &   0.00 &   0.00 &   0.00\\ 
  0.85 &   0.01 &   0.00 &   0.00 &   0.00 &   0.00 &   0.00 &   0.00\\ 
  0.90 &   0.00 &   0.00 &   0.00 &   0.00 &   0.00 &   0.00 &   0.00\\ 
  0.95 &   0.00 &   0.00 &   0.00 &   0.00 &   0.00 &   0.00 &   0.00\\ 
  1.00 &   0.00 &   0.00 &   0.00 &   0.00 &   0.00 &   0.00 &   0.00\\ \hline
    \end{tabular}
\label{table:error_N_weighted_ER}  
\end{table*}  

\newpage
\setcounter{equation}{17}
\begin{table*}[!htp]\small
\caption[Error in $\hat{N}$ when sampling by interactions from a Scale-free weighted network]{Error in $\hat{N}$ when sampling by interactions from a Scale-free weighted network.}
\centering
\begin{tabular}{|l | l l l l l l l |}
\hline
q & I & II & III & IV & V & VI & VII\\ \hline
  0.05 &   0.36 &   0.36 &   0.36 &   0.36 &   0.36 &   0.36 &   0.36\\ 
  0.10 &   0.18 &   0.18 &   0.18 &   0.18 &   0.18 &   0.18 &   0.18\\ 
  0.15 &   0.10 &   0.10 &   0.10 &   0.10 &   0.10 &   0.10 &   0.10\\ 
  0.20 &   0.05 &   0.05 &   0.05 &   0.05 &   0.05 &   0.05 &   0.05\\ 
  0.25 &   0.03 &   0.03 &   0.03 &   0.03 &   0.03 &   0.03 &   0.03\\ 
  0.30 &   0.02 &   0.02 &   0.02 &   0.02 &   0.02 &   0.02 &   0.02\\ 
  0.35 &   0.01 &   0.01 &   0.01 &   0.01 &   0.01 &   0.01 &   0.01\\ 
  0.40 &   0.01 &   0.01 &   0.01 &   0.01 &   0.01 &   0.01 &   0.01\\ 
  0.45 &   0.01 &   0.01 &   0.01 &   0.01 &   0.01 &   0.01 &   0.01\\ 
  0.50 &   0.00 &   0.00 &   0.00 &   0.00 &   0.00 &   0.00 &   0.00\\ 
  0.55 &   0.00 &   0.00 &   0.00 &   0.00 &   0.00 &   0.00 &   0.00\\ 
  0.60 &   0.00 &   0.00 &   0.00 &   0.00 &   0.00 &   0.00 &   0.00\\ 
  0.65 &   0.00 &   0.00 &   0.00 &   0.00 &   0.00 &   0.00 &   0.00\\ 
  0.70 &   0.00 &   0.00 &   0.00 &   0.00 &   0.00 &   0.00 &   0.00\\ 
  0.75 &   0.00 &   0.00 &   0.00 &   0.00 &   0.00 &   0.00 &   0.00\\ 
  0.80 &   0.00 &   0.00 &   0.00 &   0.00 &   0.00 &   0.00 &   0.00\\ 
  0.85 &   0.00 &   0.00 &   0.00 &   0.00 &   0.00 &   0.00 &   0.00\\ 
  0.90 &   0.00 &   0.00 &   0.00 &   0.00 &   0.00 &   0.00 &   0.00\\ 
  0.95 &   0.00 &   0.00 &   0.00 &   0.00 &   0.00 &   0.00 &   0.00\\ 
  1.00 &   0.00 &   0.00 &   0.00 &   0.00 &   0.00 &   0.00 &   0.00\\  \hline
    \end{tabular}
\label{table:error_N_weighted_pref}  
\end{table*}  

\newpage
\setcounter{equation}{18}
\begin{table*}[!htp]\small
\caption[Error in $\hat{M}$ when sampling by interactions from an Erd{\"o}s-R\'{e}nyi weighted network]{Error in $\hat{M}$ when sampling by interactions from an Erd{\"o}s-R\'{e}nyi weighted network.}
\centering
\begin{tabular}{|l | l l l l l l l  |}
\hline
q & I & II & III & IV & V & VI & VII\\ \hline
  0.05 &   0.00 &   0.85 &   0.78 &   0.72 &   0.66 &   0.65 &   0.67\\ 
  0.10 &   0.00 &   0.71 &   0.60 &   0.49 &   0.40 &   0.41 &   0.44\\ 
  0.15 &   0.00 &   0.59 &   0.44 &   0.32 &   0.22 &   0.24 &   0.29\\ 
  0.20 &   0.00 &   0.48 &   0.31 &   0.19 &   0.10 &   0.12 &   0.19\\ 
  0.25 &   0.00 &   0.38 &   0.21 &   0.10 &   0.02 &   0.05 &   0.13\\ 
  0.30 &   0.00 &   0.30 &   0.13 &   0.03 &   0.02 &   0.00 &   0.08\\ 
  0.35 &   0.00 &   0.22 &   0.07 &   0.01 &   0.05 &   0.02 &   0.06\\ 
  0.40 &   0.00 &   0.16 &   0.02 &   0.04 &   0.06 &   0.04 &   0.04\\ 
  0.45 &   0.00 &   0.11 &   0.01 &   0.05 &   0.06 &   0.04 &   0.03\\ 
  0.50 &   0.00 &   0.07 &   0.03 &   0.05 &   0.05 &   0.04 &   0.02\\ 
  0.55 &   0.00 &   0.03 &   0.04 &   0.04 &   0.04 &   0.03 &   0.01\\ 
  0.60 &   0.00 &   0.01 &   0.04 &   0.04 &   0.03 &   0.02 &   0.01\\ 
  0.65 &   0.00 &   0.01 &   0.04 &   0.03 &   0.02 &   0.02 &   0.01\\ 
  0.70 &   0.00 &   0.02 &   0.03 &   0.02 &   0.01 &   0.01 &   0.01\\ 
  0.75 &   0.00 &   0.02 &   0.02 &   0.01 &   0.01 &   0.01 &   0.01\\ 
  0.80 &   0.00 &   0.02 &   0.02 &   0.01 &   0.00 &   0.00 &   0.00\\ 
  0.85 &   0.00 &   0.02 &   0.01 &   0.00 &   0.00 &   0.00 &   0.00\\ 
  0.90 &   0.00 &   0.01 &   0.00 &   0.00 &   0.00 &   0.00 &   0.00\\ 
  0.95 &   0.00 &   0.00 &   0.00 &   0.00 &   0.00 &   0.00 &   0.00\\ 
  1.00 &   0.00 &   0.00 &   0.00 &   0.00 &   0.00 &   0.00 &   0.00\\    \hline
    \end{tabular}
\label{table:error_M_weighted_ER}  
\end{table*}  

\newpage
\setcounter{equation}{19}
\begin{table*}[!htp]\small
\caption[Error in $\hat{M}$ when sampling by interactions from a Scale-free weighted network]{Error in $\hat{M}$ when sampling by interactions Scale-free weighted network.}
\centering
\begin{tabular}{|l | l l l l l l l |}
\hline
q & I & II & III & IV & V & VI & VII \\ \hline
  0.05 &   0.67 &   0.67 &   0.67 &   0.67 &   0.67 &   0.67 &   0.67\\ 
  0.10 &   0.44 &   0.44 &   0.44 &   0.44 &   0.44 &   0.44 &   0.44\\ 
  0.15 &   0.29 &   0.29 &   0.29 &   0.29 &   0.29 &   0.29 &   0.29\\ 
  0.20 &   0.19 &   0.19 &   0.19 &   0.19 &   0.19 &   0.19 &   0.19\\ 
  0.25 &   0.13 &   0.13 &   0.13 &   0.13 &   0.13 &   0.13 &   0.13\\ 
  0.30 &   0.08 &   0.08 &   0.08 &   0.08 &   0.08 &   0.08 &   0.08\\ 
  0.35 &   0.06 &   0.06 &   0.06 &   0.06 &   0.06 &   0.06 &   0.06\\ 
  0.40 &   0.04 &   0.04 &   0.04 &   0.04 &   0.04 &   0.04 &   0.04\\ 
  0.45 &   0.03 &   0.03 &   0.03 &   0.03 &   0.03 &   0.03 &   0.03\\ 
  0.50 &   0.02 &   0.02 &   0.02 &   0.02 &   0.02 &   0.02 &   0.02\\ 
  0.55 &   0.01 &   0.01 &   0.01 &   0.01 &   0.01 &   0.01 &   0.01\\ 
  0.60 &   0.01 &   0.01 &   0.01 &   0.01 &   0.01 &   0.01 &   0.01\\ 
  0.65 &   0.01 &   0.01 &   0.01 &   0.01 &   0.01 &   0.01 &   0.01\\ 
  0.70 &   0.01 &   0.01 &   0.01 &   0.01 &   0.01 &   0.01 &   0.01\\ 
  0.75 &   0.01 &   0.01 &   0.01 &   0.01 &   0.01 &   0.01 &   0.01\\ 
  0.80 &   0.00 &   0.00 &   0.00 &   0.00 &   0.00 &   0.00 &   0.00\\ 
  0.85 &   0.00 &   0.00 &   0.00 &   0.00 &   0.00 &   0.00 &   0.00\\ 
  0.90 &   0.00 &   0.00 &   0.00 &   0.00 &   0.00 &   0.00 &   0.00\\ 
  0.95 &   0.00 &   0.00 &   0.00 &   0.00 &   0.00 &   0.00 &   0.00\\ 
  1.00 &   0.00 &   0.00 &   0.00 &   0.00 &   0.00 &   0.00 &   0.00\\   \hline
    \end{tabular}
\label{table:error_M_weighted_pref}  
\end{table*}  

\newpage
    \setcounter{equation}{20}
\begin{table*}[!htp]\small
\caption[Error in $\hat{k}_{\rm avg}$ when sampling by interactions from an Erd{\"o}s-R\'{e}nyi weighted network]{Error in $\hat{k}_{\rm avg}$ when sampling by interactions from an Erd{\"o}s-R\'{e}nyi weighted network.}
\centering
\begin{tabular}{|l | l l l l l l l  |}
\hline
q & I & II & III & IV & V & VI & VII \\ \hline 
  0.05 &   0.35 &   0.90 &   0.85 &   0.80 &   0.75 &   0.74 &   0.76\\ 
  0.10 &   0.33 &   0.79 &   0.69 &   0.59 &   0.49 &   0.48 &   0.53\\ 
  0.15 &   0.29 &   0.68 &   0.53 &   0.40 &   0.28 &   0.28 &   0.35\\ 
  0.20 &   0.26 &   0.57 &   0.38 &   0.24 &   0.14 &   0.14 &   0.23\\ 
  0.25 &   0.23 &   0.46 &   0.26 &   0.13 &   0.04 &   0.06 &   0.15\\ 
  0.30 &   0.19 &   0.36 &   0.17 &   0.05 &   0.01 &   0.01 &   0.10\\ 
  0.35 &   0.16 &   0.27 &   0.10 &   0.00 &   0.04 &   0.02 &   0.07\\ 
  0.40 &   0.13 &   0.20 &   0.04 &   0.03 &   0.05 &   0.03 &   0.05\\ 
  0.45 &   0.10 &   0.14 &   0.01 &   0.04 &   0.05 &   0.04 &   0.03\\ 
  0.50 &   0.08 &   0.09 &   0.02 &   0.05 &   0.04 &   0.04 &   0.02\\ 
  0.55 &   0.06 &   0.05 &   0.03 &   0.04 &   0.03 &   0.03 &   0.02\\ 
  0.60 &   0.04 &   0.02 &   0.04 &   0.04 &   0.03 &   0.02 &   0.01\\ 
  0.65 &   0.03 &   0.00 &   0.03 &   0.03 &   0.02 &   0.02 &   0.01\\ 
  0.70 &   0.02 &   0.01 &   0.03 &   0.02 &   0.01 &   0.01 &   0.01\\ 
  0.75 &   0.01 &   0.02 &   0.02 &   0.01 &   0.00 &   0.01 &   0.01\\ 
  0.80 &   0.01 &   0.02 &   0.01 &   0.01 &   0.00 &   0.00 &   0.00\\ 
  0.85 &   0.01 &   0.01 &   0.01 &   0.00 &   0.00 &   0.00 &   0.00\\ 
  0.90 &   0.00 &   0.01 &   0.00 &   0.00 &   0.00 &   0.00 &   0.00\\ 
  0.95 &   0.00 &   0.00 &   0.00 &   0.00 &   0.00 &   0.00 &   0.00\\ 
  1.00 &   0.00 &   0.00 &   0.00 &   0.00 &   0.00 &   0.00 &   0.00\\  \hline
    \end{tabular}
\label{table:error_avk_weighted_ER}  
\end{table*}  
\newpage
\setcounter{equation}{21}
\begin{table*}[!htp]\small
\caption[Error in $\hat{k}_{\rm avg}$ when sampling by interactions from a Scale-free weighted network]{Error in $\hat{k}_{\rm avg}$ when sampling by interactions from a Scale-free weighted network.}
\centering
\begin{tabular}{|l | l l l l l l l |}
\hline
q & I & II & III & IV & V & VI & VII \\ \hline
 0.05 &   0.76 &   0.76 &   0.76 &   0.76 &   0.76 &   0.76 &   0.76\\ 
  0.10 &   0.53 &   0.53 &   0.53 &   0.53 &   0.53 &   0.53 &   0.53\\ 
  0.15 &   0.35 &   0.35 &   0.35 &   0.35 &   0.35 &   0.35 &   0.35\\ 
  0.20 &   0.23 &   0.23 &   0.23 &   0.23 &   0.23 &   0.23 &   0.23\\ 
  0.25 &   0.15 &   0.15 &   0.15 &   0.15 &   0.15 &   0.15 &   0.15\\ 
  0.30 &   0.10 &   0.10 &   0.10 &   0.10 &   0.10 &   0.10 &   0.10\\ 
  0.35 &   0.07 &   0.07 &   0.07 &   0.07 &   0.07 &   0.07 &   0.07\\ 
  0.40 &   0.05 &   0.05 &   0.05 &   0.05 &   0.05 &   0.05 &   0.05\\ 
  0.45 &   0.03 &   0.03 &   0.03 &   0.03 &   0.03 &   0.03 &   0.03\\ 
  0.50 &   0.02 &   0.02 &   0.02 &   0.02 &   0.02 &   0.02 &   0.02\\ 
  0.55 &   0.02 &   0.02 &   0.02 &   0.02 &   0.02 &   0.02 &   0.02\\ 
  0.60 &   0.01 &   0.01 &   0.01 &   0.01 &   0.01 &   0.01 &   0.01\\ 
  0.65 &   0.01 &   0.01 &   0.01 &   0.01 &   0.01 &   0.01 &   0.01\\ 
  0.70 &   0.01 &   0.01 &   0.01 &   0.01 &   0.01 &   0.01 &   0.01\\ 
  0.75 &   0.01 &   0.01 &   0.01 &   0.01 &   0.01 &   0.01 &   0.01\\ 
  0.80 &   0.00 &   0.00 &   0.00 &   0.00 &   0.00 &   0.00 &   0.00\\ 
  0.85 &   0.00 &   0.00 &   0.00 &   0.00 &   0.00 &   0.00 &   0.00\\ 
  0.90 &   0.00 &   0.00 &   0.00 &   0.00 &   0.00 &   0.00 &   0.00\\ 
  0.95 &   0.00 &   0.00 &   0.00 &   0.00 &   0.00 &   0.00 &   0.00\\ 
  1.00 &   0.00 &   0.00 &   0.00 &   0.00 &   0.00 &   0.00 &   0.00\\   \hline
\end{tabular}
\label{table:error_avk_weighted_pref}  
\end{table*}  

\newpage
\setcounter{equation}{22}
\begin{table*}[!htp]\small
\caption[Error in $\kmax$ when sampling by interactions from an Erd{\"o}s-R\'{e}nyi weighted network]{Error in $\kmax$ when sampling by interactions from an Erd{\"o}s-R\'{e}nyi weighted network.}
\centering
\begin{tabular}{|l | l l l l l l l  |}
\hline
q & I & II & III & IV & V & VI & VII\\ \hline
  0.05 &   3.00 &   0.76 &   0.81 &   0.84 &   0.83 &   0.84 &   0.85\\ 
  0.10 &   1.82 &   0.66 &   0.73 &   0.76 &   0.76 &   0.77 &   0.80\\ 
  0.15 &   1.27 &   0.60 &   0.67 &   0.71 &   0.69 &   0.70 &   0.73\\ 
  0.20 &   0.82 &   0.53 &   0.60 &   0.66 &   0.62 &   0.66 &   0.69\\ 
  0.25 &   0.72 &   0.48 &   0.56 &   0.59 &   0.59 &   0.60 &   0.63\\ 
  0.30 &   0.49 &   0.44 &   0.51 &   0.52 &   0.54 &   0.55 &   0.58\\ 
  0.35 &   0.51 &   0.35 &   0.50 &   0.52 &   0.49 &   0.53 &   0.55\\ 
  0.40 &   0.35 &   0.36 &   0.42 &   0.49 &   0.47 &   0.47 &   0.49\\ 
  0.45 &   0.29 &   0.28 &   0.39 &   0.44 &   0.41 &   0.44 &   0.45\\ 
  0.50 &   0.20 &   0.31 &   0.37 &   0.37 &   0.37 &   0.39 &   0.42\\ 
  0.55 &   0.17 &   0.26 &   0.32 &   0.36 &   0.34 &   0.34 &   0.37\\ 
  0.60 &   0.16 &   0.22 &   0.33 &   0.33 &   0.31 &   0.32 &   0.33\\ 
  0.65 &   0.13 &   0.20 &   0.31 &   0.28 &   0.23 &   0.27 &   0.30\\ 
  0.70 &   0.10 &   0.19 &   0.26 &   0.26 &   0.20 &   0.23 &   0.26\\ 
  0.75 &   0.08 &   0.15 &   0.21 &   0.23 &   0.17 &   0.17 &   0.23\\ 
  0.80 &   0.01 &   0.13 &   0.21 &   0.18 &   0.14 &   0.14 &   0.18\\ 
  0.85 &   0.02 &   0.12 &   0.17 &   0.14 &   0.08 &   0.08 &   0.14\\ 
  0.90 &   0.01 &   0.09 &   0.12 &   0.11 &   0.04 &   0.04 &   0.11\\ 
  0.95 &   0.04 &   0.08 &   0.10 &   0.06 &   0.01 &   0.02 &   0.06\\ 
  1.00 &   0.05 &   0.04 &   0.06 &   0.02 &   0.07 &   0.05 &   0.03\\   \hline
  \end{tabular}
\label{table:error_kmax_weighted_ER}  
\end{table*} 
\newpage
\setcounter{equation}{23}
\begin{table*}[!htp]\small
\caption[Error in $\kmax$ when sampling by interactions from a Scale-free weighted network]{Error in $\kmax$ when sampling by interactions from a Scale-free weighted network.}
\centering
\begin{tabular}{|l | l l l l l l l  |}
\hline
q & I & II & III & IV & V & VI & VII \\ \hline  
 0.05 &   0.85 &   0.85 &   0.85 &   0.85 &   0.85 &   0.85 &   0.85\\ 
  0.10 &   0.80 &   0.80 &   0.80 &   0.80 &   0.80 &   0.80 &   0.80\\ 
  0.15 &   0.73 &   0.73 &   0.73 &   0.73 &   0.73 &   0.73 &   0.73\\ 
  0.20 &   0.69 &   0.69 &   0.69 &   0.69 &   0.69 &   0.69 &   0.69\\ 
  0.25 &   0.63 &   0.63 &   0.63 &   0.63 &   0.63 &   0.63 &   0.63\\ 
  0.30 &   0.58 &   0.58 &   0.58 &   0.58 &   0.58 &   0.58 &   0.58\\ 
  0.35 &   0.55 &   0.55 &   0.55 &   0.55 &   0.55 &   0.55 &   0.55\\ 
  0.40 &   0.49 &   0.49 &   0.49 &   0.49 &   0.49 &   0.49 &   0.49\\ 
  0.45 &   0.45 &   0.45 &   0.45 &   0.45 &   0.45 &   0.45 &   0.45\\ 
  0.50 &   0.42 &   0.42 &   0.42 &   0.42 &   0.42 &   0.42 &   0.42\\ 
  0.55 &   0.37 &   0.37 &   0.37 &   0.37 &   0.37 &   0.37 &   0.37\\ 
  0.60 &   0.33 &   0.33 &   0.33 &   0.33 &   0.33 &   0.33 &   0.33\\ 
  0.65 &   0.30 &   0.30 &   0.30 &   0.30 &   0.30 &   0.30 &   0.30\\ 
  0.70 &   0.26 &   0.26 &   0.26 &   0.26 &   0.26 &   0.26 &   0.26\\ 
  0.75 &   0.23 &   0.23 &   0.23 &   0.23 &   0.23 &   0.23 &   0.23\\ 
  0.80 &   0.18 &   0.18 &   0.18 &   0.18 &   0.18 &   0.18 &   0.18\\ 
  0.85 &   0.14 &   0.14 &   0.14 &   0.14 &   0.14 &   0.14 &   0.14\\ 
  0.90 &   0.11 &   0.11 &   0.11 &   0.11 &   0.11 &   0.11 &   0.11\\ 
  0.95 &   0.06 &   0.06 &   0.06 &   0.06 &   0.06 &   0.06 &   0.06\\ 
  1.00 &   0.03 &   0.03 &   0.03 &   0.03 &   0.03 &   0.03 &   0.03\\   \hline
\end{tabular}
\label{table:error_kmax_weighted_pref}  
\end{table*}  

\newpage
\setcounter{equation}{24}
\begin{table*} [!ht]\small
\caption[Number of messages from September 2008-November 2009]{Number of messages from September 2008-November 2009. The number of ``observed'' messages in our database comprise a fraction of the total number of Twitter messages made during period of this study (September 2008 through November 2009). While our feed from the Twitter API remains fairly constant, the total \# of tweets grows, thus reducing the \% of all tweets observed in our database. We calculate the total $\#$ of messages as the difference between the last message id and the first message id that we observe for a given month. This provides a reasonable estimation of the number of tweets made per month as message ids were assigned (by Twitter) sequentially during the time period of this study. The \% observed represent the percent of messages observed out of the estimated total. We also report the number observed messages that are replies to specific messages and the percentage of our observed messages which constitute replies. }
\centering
\begin{tabular}{lllllll}
\hline
Week & Start date	& \# Obsvd.	Msgs. & \# Total Msgs.	& \% Obsvd. & \# Replies	& \%  Replies  \\
& & $\times 10^6$ & $\times 10^6$ &    &   $\times 10^6$ &    \\[1ex]\hline\hline                            
 1 & 09.09.08 & 3.14 & 7.26 & 43.2 &  0.88 & 28.1 \\ 
 2 & 09.16.08 & 3.36 & 8.31 & 40.4 &  0.90 & 26.9 \\ 
 3 & 09.23.08 & 3.43 & 8.89 & 38.6 &  0.90 & 26.2 \\ 
 4 & 09.30.08 & 3.33 & 9.06 & 36.8 &  0.89 & 26.6 \\ 
 5 & 10.07.08 & 2.33 & 9.38 & 24.8 &  0.64 & 27.5 \\ 
 6 & 10.14.08 & 4.39 & 9.87 & 44.4 &  1.24 & 28.3 \\ 
 7 & 10.21.08 & 4.70 & 10.01 & 47.0 &  1.35 & 28.8 \\
 8 & 10.28.08 & 5.74 & 10.34 & 55.5 &  1.64 & 28.5 \\ 
 9 & 11.04.08 & 5.58 & 11.14 & 50.1 &  1.63 & 29.3 \\ 
10 & 11.11.08 & 4.70 & 9.88 & 47.6 &  1.42 & 30.2 \\ 
\hline
\end{tabular}
\label{table:datawehave}
\end{table*}

\end{document}